\documentclass[review]{elsarticle} %%for review

\usepackage{amsmath,newtxmath}
\usepackage{mathtools}

\usepackage{lineno,hyperref}
\modulolinenumbers[5]

\journal{Engineering Analysis with Boundary Elements}

\begin{document}

\begin{frontmatter}

\title{A log-linear time algorithm for the elastodynamic boundary integral equation method}

%\author{
%Dye SK Sato\fnref{myfootnote}}
%\author{Ryosuke Ando}

\author[mymainaddress]{Dye SK Sato\corref{mycorrespondingauthor}}
\ead{sato.daisuke.44p@st.kyoto-u.ac.jp}
%\ead[url]{sato.daisuke.6r@kyoto-u.ac.jp}

\author[mysecondaryaddress]{Ryosuke Ando}
\cortext[mycorrespondingauthor]{Corresponding author}
\ead{ando@eps.s.u-tokyo.ac.jp}

\address[mymainaddress]{Disaster Prevention Research Institute, Kyoto University, Gokasho, Uji, Kyoto 611-0011, Japan}
\address[mysecondaryaddress]{The University of Tokyo, 7-3-1 Hongo, Bunkyo-ku, Tokyo 113-8654, Japan}

\begin{abstract}
We present a fast and memory-efficient algorithm for transient, space-time-domain, and elastodynamic boundary-integral analysis. Associated data-sparse approximations and operations are named {\it fast domain 
partitioning hierarchical matrices} (FDP=H-matrices). The {\it fast domain partitioning method} (the FDPM) solves a known problem of {\it hierarchical matrices} (H-matrices) in compressing discretized elastodynamic kernel functions. A novel set of plane-wave approximations then unites the FDPM and H-matrices in an accurate analytic manner. 
Memory usage is $\mathcal O(N \log N)$ and computation time $\mathcal O(NM\log N)$ 
in our algorithm throughout one run with $N$ boundary elements and $M$ time steps. 
The amount of associated cost reduction is remarkable, as the memory usage and computational time have been originally $\mathcal O(N^2M)$ and  $\mathcal O(N^2M^2)$, respectively, to run the orthodox time-marching implementation. 
Numerical experiments indicate that FDP=H-matrices achieve $\mathcal O(NM/\log N)$ times smaller memory and computation time while ensuring the accuracy of the analyses. 
\end{abstract}


\end{frontmatter}

%\linenumbers

\section{Introduction}
\label{FDPH1}

Wave-radiation and wave-scattering phenomena extend over various scientific fields, such as electromagnetics~\cite{wannamaker1984electromagnetic}, acoustics~\cite{jones1974integral}, solid mechanics~\cite{schanz1999boundary}, and geophysics~\cite{rice1993spatio,ando2018dynamic}. The latter two of them are computed often coupled with the dynamical crack (the dynamic rupture) problems of the fracture mechanics~\cite{nishimura1989regularized}. 

One of common solvers for these problems will be the boundary integral equation method (the BIEM)~\cite{beskos1987boundary,bonnet1999boundary,aliabadi2002boundary}. 
Its formulation starts with rewriting the governing equation of an original problem to an integral equation of boundary variables, namely a boundary integral equation (a BIE). 
The BIE convolves the boundary variables and integral kernel (hereafter, the ``kernel'') over all the boundaries (``sources'') and time histories. Evaluating the BIE on each boundary element (on each ``receiver''), the BIEM determines the boundary values that fit the given boundary conditions. The evaluation of the BIE repeats at respective time steps in the transient problems~\cite{bonnet1999boundary,zhang1991novel}. 
We refer to this space-time-domain BIEM, especially for the transient problems, as the spatiotemporal BIEM (the ST-BIEM). 
A series of these problem reductions has established the reputation of the ST-BIEM for reducing the number of elements~\cite{nishimura2002fast}, numerical dispersions~\cite{day2005comparison}, and spatial discretization errors in handling complex objects and open spaces~\cite{ando2018dynamic}. Analytical expressions, known as semi-analytic BIEs, of the discretized kernels further contribute to the accuracy of the BIEM~\cite{tada1997non}. 

Despite these merits, the usability of the ST-BIEM is often degraded due to the computational expense to multiply the third- (the kernel) and second-rank (the history of the boundary variables) tensors at every time step~\cite{bonnet1999boundary,zhang1991novel}.
A discrete kernel is a dense tensor of the $N^2M$ components for $N$ boundary elements and $M$ time steps~\cite{nishimura2002fast}, due to the $N^2M$ combinations of the $N$ receivers, $N$ sources, and $M$ time steps. 
Then if we convolve it naively in the BIE, it yields the $\mathcal O(N^2M)$ computation time per time step, which amounts to $\mathcal O(N^2M^2)$ for a single run of the ST-BIEM~\cite{takahashi2003fast}. 
It also requires a considerable memory capacity to store the $\mathcal O(N^2M)$ components of the discrete kernel, as well as to store the $\mathcal O(NM)$ time histories of the boundary variables on the elements. These numerical costs of the ST-BIEM can be easily enormous for large $N$ and $M$. In contrast, the volume-based methods, such as the finite-difference and the finite-element methods, require only the $\mathcal O(N_v)$ computation time per time step [$\mathcal O(N_vM)$ in total]
and $\mathcal O(N_v)$ total memory usage for $N_v$ volume elements~\cite{nishimura2002fast}. 
As seen above, even though drastically reducing the number of elements ($N\ll N_v$), the ST-BIEM is originally $\mathcal O(N^2M/N_v)$ times inferior to the volume-based methods in terms of the numerical cost. 

Developing the fast algorithms is hence a major need in the use of the ST-BIEM. 
One widely-known versatile algorithm is the plane-wave time-domain (PWTD) method~\cite{ergin1999plane}. It can reduce the total computation time to $\mathcal O(NM\log^2N)$ [$\mathcal O(N\log^2N)$ per time step]. The foundation of the PWTD method is similar to that of the fast multipole method (the FMM)~\cite{rokhlin1985rapid} that accelerates the convergence of basis function expansions of the kernel involved with the discretization of the boundary variables. Undesirable requirements of the PWTD method are then also inherited from the FMM, such as intractable analytic calculation and numerical integration to obtain the expanded discretized kernel, which complicate the application of the PWTD method; the difficulty of its formulation has interfered its widespread use~\cite{mavaleix2020fast}. 
Besides, the associated memory reduction is less remarkable, as the PWTD method requires $\mathcal O(NM)$ memory to store the time history of the boundary variables~\cite{ergin1999plane} while the compressed kernel of the PWTD method only has $\mathcal O(N\log^2N)$ components. 

The current state-of-the-art algorithm for solving the BIE of the wave-equation will be the convolution quadrature methods (CQMs)~\cite{lubich1988convolutionI,lubich1988convolutionII}. This fast solves a transient problem in the complex frequency domain (the Laplace-domain)~\cite{mavaleix2020fast,banjai2009rapid}. The CQM then evades the complicated formulation in the analytic time-domain expansion of the PWTD method and utilizes a tractable Laplace-domain BIE. The CQM has also been applied to the elastodynamic problems~\cite{chaillat2017fast}. Ref.~\cite{mavaleix2020fast} reported that the CQM achieves the $\mathcal O(N\log N)$ time complexity per time step with the use of the high-frequency approximations. Meanwhile, the use of the high-frequency approximations raises another issue in involving the low-frequency motions for its versatility. 

A versatile yet analytically simple algorithm is still required for computing the ST-BIEM, and the total memory usage should reduce to $\mathcal O(N\log N)$. However, even other existing versatile methods, including the above-mentioned CQM~\cite{maruyama2016transient}, 
such as the fast domain partitioning method (the FDPM)~\cite{ando2007efficient,ando2016fast} and hierarchical matrices (H-matrices)~\cite{hackbusch1999sparse} later-mentioned, do not simultaneously achieve the $\mathcal O(N\log N)$ computation time per time step and $\mathcal O(N\log N)$ total memory usage. 
On the planar boundary with structured elements, the spectral method reduces both the total memory usage 
and computation time per time step to $\mathcal O(N\log N)$ by truncating the temporal convolution after the characteristic time-scale of respective wavenumbers~\cite{lapusta2000elastodynamic}. Nonetheless, the spectral method does not apply to various nonplanar boundary shapes at the same efficiency as to a planar boundary.
Although the frequency-domain FMM reduces both the total memory and computation time per iteration to $\mathcal O(N\log^2 N)$ in time-harmonic problems~\cite{chaillat2008multi}, it does not work for the transient problems at the same efficiency. 

In this study, we develop a versatile fast algorithm for the ST-BIEM of the elastodynamic problems to accomplish the $\mathcal O(N\log N)$ total memory usage and $\mathcal O(N\log N)$ computation time per time step [$\mathcal O(NM\log N)$ computation time in total]. 
This proposal for the transient elastodynamic problems functions on arbitrary boundary geometry and is also applicable to a simple wave equation as it can be solved as a special case of an elastodynamic equation. 
The algorithm incorporates an ordinary time-marching scheme with our new methods of data-sparse (low-rank) approximations and operations. 
Their large part comprises the FDPM and H-matrices, and we name them fast domain partitioning hierarchical matrices (FDP=H-matrices). 

H-matrices (detailed in \S\ref{FDPH22}) is an efficient computational technique for a dense yet data-sparse tensor, a tensor that can be expressed in a low-ranked manner, such as the discretized kernel of the elliptic BIEs~\citep{hackbusch1999sparse}. 
H-matrices are similar to the FMM in the formulation but are known by their practicality: the module algorithm of H-matrices for the low-rank approximation, typified by the adaptive cross approximation (the ACA)~\cite{bebendorf2003adaptive}, enables simple numerical low-rank approximations of the kernels without analytical efforts. 
The low-ranked kernel generated by the H-matrix technique has reported to have $\mathcal O(N^2)$ components in the elastodynamic (the hyperbolic) ST-BIEM, requiring $\mathcal O(N^2)$ memory and $\mathcal O(N^2)$ computation time per time step [$\mathcal O(N^2M)$ total computation time]~\cite{yoshikawa20152}. 
It is relatively higher than the $\mathcal O(N\log N)$ scaling desired in this paper.
As suggested in Ref.~\cite{borm2003hierarchical}, 
the rank (i.e. the number of the effectively independent components) of the low-ranked kernel in H-matrices is bounded by the number of the discretized kernel components that involve the singular points of the original continuous kernel. This lower bound is scaled by $\mathcal O(N^2)$ for wave equations, where the kernel is singular at any location at the wave arrival time although their static limits, Poisson's equation, localizes the singular point exactly at the source location. 
These suggest that the difficulty in applying H-matrices to the ST-BIEM is incurred by the low-rank approximation near the singular points distributed along the wave arrival time in the space-time domain. Indeed, the ACA and H-matrices work well to some extent for the above-mentioned frequency- and Laplace-domain elastodynamic BIEs~\cite{maruyama2016transient,chaillat2017theory}, where the singularity due to the impulsive waves cancels. 

The FDPM (detailed in \S\ref{FDPH23}) is a fast algorithm for the elastodynamic ST-BIEM leveraging the analytic character of the fundamental solution (also called ``Green's function''). 
The elastodynamic Green's function comprises the longitudinal wave (the ``P-wave''), the transverse wave (the ``S-wave''), and the near-field term in-between the P- and S-waves. 
The temporally-integrated spatial-derivative of the Green's function is associated with the kernel (of the non-hypersingular formulation)~\cite{bonnet1999boundary}, 
and
the FDPM suitably divides the time domain of the BIE into three domains: 1) Domain F that fully involves wave arrival times of the P- and S-waves, 2) Domain I in-between P- and S-waves, and 3) Domain S after S-waves. 
The discretized kernel in Domain I or S separates into a matrix representing its source- and receiver-dependence and a vector representing its time-dependence as explicitly shown in the semi-analytic BIE schemes~\cite{ando2007efficient,ando2016fast}, like the temporally integrated Green's function spatiotemporally separating in these domains ~\cite{ando2007efficient}. 
This factorization of the kernel makes the required memory and computation time per time step of $\mathcal O(N^2)$ and the total computation time of $\mathcal O(N^2M)$. 
Furthermore, geometrical spreading~\cite{aki2002quantitative}, attenuation expressed by a power function of distance, holds in the kernel within Domain F~\cite{ando2016fast}. 
This suggests the expandability of the kernel in Domain F, so is remarkable as the Domain F fully involves the $\mathcal O(N^2)$ components that cannot be expanded by the previous techniques of H-matrices. 
The expansion in Domain F theoretically corresponds to the expansion along the wavefront, an isochronous surface drawn by a wave radiated by a source location in a snapshot~\cite{aki2002quantitative}, which has been successful in the context of the PWTD method~\cite{takahashi2003fast,ergin1999plane}. 
This attenuating nature of the kernel along the wavefront motivates us to integrate the FDPM with H-matrices in the present study and necessarily resolves the above-mentioned problem of H-matrices in the ST-BIEM. 
 
The main challenge for this study will be to deal with the singular points distributed along the wavefronts. 
This purpose led us to further develop two modules for this purpose, called the averaged reduced time (the ART) and the quantization method (Quantization) (both introduced in \S\ref{FDPH3}). The ART, applied to the respective above-mentioned domains, is a kind of plane-wave approximations that utilizes the averaged value of so-called ``reduced time''~\cite{aki2002quantitative}, elapsed time from the wave arrival. The ART is based on the spatial sorting of boundary elements and does not impose hierarchical division in the time domain of the BIE, unlike the PWTD method that divides the domain spatiotemporally~\cite{ergin1999plane}. Consequently, as detailed in \S\ref{FDPH5}, the ART provides an arithmetic of FDP=H-matrices that does not necessitate the memory to store the history of the boundary variables. It then accomplishes the desired memory order of $\mathcal O(N\log N)$, and gives an advantage to FDP=H-matrices over the PWTD method that requires the $\mathcal O(NM)$ memory concerning the time history of the boundary variables. 
Quantization reduces the memory to store the kernel and time to compute the BIE with the help of the quantization technique, a sparse resampling technique common in the signal-processing literature~\cite{gonzalez2002digital}.
Quantization samples the kernel temporally sparsely 
and deals with the indirect source- and receiver-dependence of the time definition range of Domain I that can inhibit $\mathcal O(N\log N)$ memory (mentioned in \S\ref{FDPH3}). 

This paper is organized as follows.
First, we describe the ST-BIEM with the FDPM and H-matrices in a formulation provided by the previous studies (Section \ref{FDPH2}). Second, we introduce the basic concepts and structure of our new method by outlining the key features and the relationships between the incorporated module algorithms (the FDPM, H-matrices, Quantization, and the ART) of FDP=H-matrices (Section \ref{FDPH3}); this section is intended to provide sufficient information to understand the basics of FDP=H-matrices.
Third, we detail a technique for incorporating H-matrices and the FDPM (Section \ref{FDPH4}).
Fourth, we construct the arithmetic of FDP=H-matrices (Section \ref{FDPH5}). 
Finally, we demonstrate the cost reduction and computational accuracy of FDP=H-matrices (Section 6). 

To guide the reader, we list frequently used variables and parameters in Tables~\ref{SymbolTable1}, \ref{SymbolTable2}, and \ref{SymbolTable3}.  
Tables~\ref{SymbolTable1} and \ref{SymbolTable2} show the variables and parameters given by the previous studies in the standard nomenclature. Table~\ref{SymbolTable3} shows newly defined variables and parameters to implement FDP=H-matrices.  
Key formulas will be summarized in \ref{sec:keyformulas}. 

%\input{variablelistFDPH.tex}

%%%%%%variable list
\begin{table}[tb]
%\begin{table}[H] %% usepackage{here}
   \caption{List of frequently used variables and parameters. The list contains the spaces to which the variables and parameters belong. 
$\mathbb N$, $\mathbb Z$, and $\mathbb R$ 
represent the sets of natural, integer, and real numbers, respectively. $D_v$(spatial dimension of the given problem)-dependences of $T$, $D$, and $K$ are omitted in the list. 
%In the actual application, the fast algorithms (the FDPM, H-matrices, Quantization, and FDP=H-matrices) are applied to each component pair of the stress $\sigma$ and the slip $\Delta u$ in the given undiscretized problem.
}
  \begin{tabular}{r|l}
%\hline
%\multicolumn{2}{l}{Given problem}
%\\
%\hline
%$({\bf x},\boldsymbol \xi)\in \mathbb{ R}^{D_v}$ &position 
%in $D_v(=2,3)$-dimension 
%\\
%$r \in \mathbb R$& distance
%\\
%$(t,\tau) \in \mathbb R$ & time 
%\\
%$u({\bf x},t) \in \mathbb R^{D_v}$ & displacement at ${\bf x}$ at $t$
%\\
%$\rho \in \mathbb R$ & mass density 
%\\
%$(\lambda,\mu)\in\mathbb R$ & elastic constants
%\\
%$\sigma({\bf x},t)\in\mathbb R^{D_v\times D_v}$  & stress at ${\bf x}$ at $t$
%\\
%$\Gamma$ & boundary area 
%\\
%$\nu (\boldsymbol \xi) \in \mathbb R^{D_v}$ & normal vector at $\xi$ (on $\Gamma$)
%\\
%$T(\boldsymbol \xi,\tau) \in \mathbb R^{D_v}$ & traction at $\xi$ at $\tau$
%\\
%$\Delta  u(\boldsymbol\xi,\tau) \in \mathbb R^{D_v}$ & slip distance at $\boldsymbol\xi$ at $\tau$. 
%\\
%$\Delta \dot u  (\boldsymbol\xi,\tau) \in \mathbb R^{D_v}$ & slip rate  at $\boldsymbol\xi$ at $\tau$. 
%\\
%$K({\bf x},\xi,t-\tau)\in \mathbb R^{D_v^3}$ 
%& kernel connecting $\Delta u(\boldsymbol\xi,\tau)$ and $\sigma({\bf x},t)$
%\\
%$L\in\mathbb R$& characteristic length of the discretized object
\hline
\multicolumn{2}{l}{Original ST-BIEM}\\
\hline
$N\in \mathbb N$
& numbers of elements
\\
$M\in \mathbb N$
& numbers of time steps
\\
$i=1,..., N$&  receiver number
\\
$j=1,..., N$&  source number
\\
$n\in[0,M)$ 
&the latest time step 
\\
$m\in\mathbb Z $ 
& relative time step 
\\
$\Delta x_j\in \mathbb R$& spatial discretization length of $j$ 
\\
$\Delta t \in \mathbb R$& temporal discretization length
\\
%${\bf x}_i\in\mathbb R^{D_v}$ 
%&spatial collocation point of element $i$ 
%\\
%$\nu_i\in\mathbb R^{D_v}$ 
%&nomal vector of element $i$ 
%\\
$T_i(t)\in \mathbb R$& stress of receiver $i$ at time $t$  
\\
$T_{i,n}\in \mathbb R$& discretized $T_i$ at time step $n$  
\\
$D_j(\tau)\in \mathbb R$& slip-/opening-rate of $j$ at time $\tau$ 
\\
$D_{j,n-m}\in \mathbb R$& discretized $D_j$ at step $n-m$ 
\\
$K_{i,j}(t-\tau)\in\mathbb R$& kernel of $T_i(t)$ incurred by $D_{j}(\tau)$ 
\\
$K_{i,j,m}\in\mathbb R$& kernel of $T_{i,n}$ incurred by $D_{j,n-m}$
\\
%${\bf T}(t)\in \mathbb R^N$& 
%vector placing $T_i(t)$ at the $i$ component 
%\\
%${\bf T}_n\in \mathbb R^N$& 
%vector placing $T_{i,n}$ at the $i$ component 
%\\
%${\bf D}(\tau)\in \mathbb R^N$& vector placing $D_j(\tau)$ at the $j$ component 
%\\
%${\bf D}_{n-m}\in \mathbb R^N$& vector placing $D_{j,n-m}$ at the  $j$ component
%\\
%$K(\tau)\in \mathbb R^{N\times N}$& matrix placing $K_{i,j}(\tau)$ at the $i,j$ component
%\\ 
%$K_m \in \mathbb R^{N\times N}$& matrix placing $K_{i,j,m}$ at the $i,j$ component
%\\ 
%  %ãããã
%    \end{tabular} 
%   %\captionsetup{labelformat=empty,labelsep=none}
%   %\caption{}
%   \caption{List of frequently used variables. The list contains the spaces to which the variables belong. $\mathbb R$ and $\mathbb N$ respectively represent the sets of real numbers and natural numbers. 
%$D_v$-dependencies of $T$, $D$, and $K$ are omitted, where $D_v$ denotes the dimension of the given problem. 
% In the actual application, 
%the fast algorithms (the FDPM, H-matrices, Quantization, and FDP=H-matrices) are applied to each component pair of the stress $\sigma$ and the slip $\Delta u$ in the given undiscretized problem.} 
%   %\caption{Symbols 1.}
%   \label{SymbolTable1}
%  \end{table}
%  
%  %\setcounter{table*}{0}
%  \begin{table*}[htb]
%\begin{center}
%    \begin{tabular}{c|l}
%%  \hline
%  %ãããŸã§FDP=Håºççã§ã¯æ¶å»
 %ãããã
\hline
    \end{tabular} 
%\end{center}
   %\captionsetup{labelformat=empty,labelsep=none}
   %\caption{}
   %\caption{Symbols 1.}
   \label{SymbolTable1}
  \end{table}

  \begin{table}[tbp]
     \caption{List of frequently used variables and parameters (continued). 
%The list contains the spaces to which the variables belong. $\mathbb R$ and $\mathbb N$ respectively represent the sets of real numbers and natural numbers. $D_v$(spatial dimension of the given problem)-dependences of $T$, $D$, and $K$ are omitted.
%In the actual application, the fast algorithms (the FDPM, H-matrices, Quantization, and FDP=H-matrices) are applied to each component pair of the stress $\sigma$ and the slip $\Delta u$ in the given undiscretized problem.
}
    \begin{tabular}{r|l}
%  \hline
  %ãããŸã§FDP=Håºççã§ã¯æ¶å»
\hline
\multicolumn{2}{l}{FDPM}\\
\hline
$c(=\alpha,\beta)\in\mathbb R$ 
& phase speed (of the P-/S-wave)
\\
$t_{ij}\in\mathbb R$ & collocated travel time of $i$ and $j$
\\
$t_{ij}^-,t_{ij}^+\in\mathbb R$& wave arrival/passage time of ($i$, $j$)
\\
$\Delta t_j^{\pm}\in\mathbb R$ & absolute difference of $t_{ij}^\pm$ and $t_{ij}$
\\
$\Delta t_j\in\mathbb R$ & duration of Domain F 
%\\
%$m_{ij}^-,m_{ij}^+\in\mathbb R$& time steps respectively experiencing wave arrival and wave passage completion
\\
${\bf K}^W(t) \in\mathbb R^{N\times N}$
&kernel of Domain W = F, I, S 
\\
${\bf T}^W \in\mathbb R^N$
&stress associated with Domain W
\\
${\bf \hat K}^{I}\in \mathbb R^{N\times N}$& 
space-dependent part of ${\bf K}^{I}(t)$
\\
${\bf \hat K}^{S}\in \mathbb R^{N\times N}$& 
space-dependent part of ${\bf K}^{S}(t)$ 
\\
$h^I(t)\in \mathbb R$& time-$t$-dependent part of ${\bf K}^I$. 
\\
\hline
\multicolumn{2}{l}{H-matrices}\\
\hline
$diam\in\mathbb R$
&diameter of a given cluster
\\
$dist \in\mathbb R$
&distance between given two clusters
\\
$l_{min}\in\mathbb R$&admissible minimum of $diam$
\\
$\eta\in\mathbb R$&admissible maximum of $diam/dist$
\\
$a\in\mathbb N$&block cluster number
\\
$\epsilon_H,\epsilon_{ACA}\in\mathbb R$
&tolerance in the LRA and ACA
\\
$N_{r,a}\in\mathbb N$&number of receivers in $a$
\\
$N_{s,a}\in\mathbb N$&number of sources in $a$
\\
$l_a^*\in\mathbb N$&rank of the low-ranked kernel in $a$
\\
${\bf f}_{al}\in\mathbb R^{N_{r,a}}$ 
&$l$-th $i$-dependence of subkernel in $a$
\\
${\bf g}_{al}\in\mathbb R^{N_{s,a}}$
&$l$-th $j$-dependence of subkernel in $a$
\\
\hline
%\hline
 %ãããã
    \end{tabular} 
%\end{center}
   %\captionsetup{labelformat=empty,labelsep=none}
   %\caption{}
   %\caption{Symbols 1.}
   \label{SymbolTable2}
  \end{table}

  \begin{table}[tbp]
  \caption{List of frequently used variables and parameters (continued). 
The leaf-number $a$ dependencies of the variables and parameters in FDP=H-matrices, all depending on $a$, are omitted in the list for brevity.
Maximum $\max[\delta m_i+\bar m_j]$ appearing in the dimension of ${\bf \bar T}$ is taken over each leaf $a$.
%, S^{source},S^{receriver},{\bf F},{\bf G}$, and $\mathcal M$ such as $\max[\delta m_i+\bar m_j]$ are 
}
    \begin{tabular}{r|l}
%  \hline
  %ãããŸã§FDP=Håºççã§ã¯æ¶å»
\hline
\multicolumn{2}{l}{Quantization}\\
\hline
$\epsilon_Q,\epsilon_{st}\in \mathbb R$ &relative and absolute error bounds 
\\
$q \in \mathbb N$ &quantization number 
\\
$b_q\in\mathbb Z$& sampled time step for $q$
\\
\hline
\multicolumn{2}{l}{FDP=H-matrices}\\
\hline
$\hat K^F_{ij}\in \mathbb R^{N\times N}$ & amplitude term
\\
$h_{ij}^F(t)\in \mathbb R$& normalized waveform
\\
$i_*, j_*$
&representative receiver and source 
\\
$\delta t_i\in \mathbb R$
&travel-time difference 
\\
$\bar t_j\in\mathbb R$
&receiver-averaged travel time
\\
$h_{j}^F(t)\in \mathbb R$& degenerating normalized waveform
%\\
%$c_{ij} \in\mathbb R$&effective wave speed of $i$ and $j$
%\\
%$\eta_0\in\mathbb R$& $\eta$ at $\delta r=l_{min}$
\\
$\bar m_j^-\in\mathbb Z$ & receiver-averaged travel time step 
\\
$\Delta m_j\in \mathbb Z$ & discretized duration of Domain F
\\
$h^F_{j,m}\in \mathbb R$& temporally discretized $h_j^F(t)$
\\
$\delta m_i\in \mathbb Z$
&travel-time-step difference 
\\
$\hat D_{j,n}^F\in\mathbb R$ & convolution of $D_{j,n-m}$ and $h^F_{j,m}$
\\
$\bar T_m \in\mathbb R$& representative stress at time step $m$ 
\\
%${\bf \bar T}_n\in \mathbb R^{\max[\delta m_i+\bar m_j]} $& vector storing $\bar T_{n-m}$ in the $m$-th component 
%\\
\hline
   %\caption{Symbols 1.}
    \end{tabular} 
   \label{SymbolTable3}
  \end{table}
  
%  %\setcounter{table*}{0}
%  \begin{table*}[htb]
%\begin{center}
%    \begin{tabular}{c|l}
%%  \hline
%  %ãããŸã§FDP=Håºççã§ã¯æ¶å»
%\multicolumn{2}{l}{FDP=H-matrices (continued)}\\
%\hline
%$S^{source}\in \mathbb R^{\max[\bar m_j]\times N_{s,a}}$ &time shift matrix for sources
%\\
%$S^{receiver}\in \mathbb R^{N_{r,a}\times\max[\delta m_i]}$ &time shift matrix for receivers 
%\\
%${\bf F}\in\mathbb R^{N_{r,a}\times \max[\delta m_i]}$
%&sparse matrix embedded with $f$ along nonzero components of $S^{receiver}$ 
%\\
%${\bf G}\in\mathbb R^{\max[\bar m_j]\times N_{s,a}}$
%&sparse matrix embedded with $g$ along nonzero components of $S^{source}$ 
%\\
%$\mathcal M\in\mathbb R^{\max[\delta m_i+\bar m_j]\times\max[\delta m_i+\bar m_j]}$ & matrix the $m,m^\prime$ component of which is $\delta_{m,m^\prime+1}$.
%\\
%\multicolumn{2}{l}{}\\
%    \end{tabular} 
%\end{center}
%  %\captionsetup{labelformat=empty,labelsep=none}
%   %\caption{}
%   \caption{List of variables (continued). The maximums to determine the dimensions of ${\bf \bar T}, S^{source},S^{receriver},{\bf F},{\bf G}$, and $\mathcal M$ such as $\max[\delta m_i]$ are taken in each leaf $a$. The rank dependence of ${\bf F},{\bf G},{\bf \bar T}_n$ is omitted for brevity.}
%   %\caption{Symbols 1.}
%   \label{SymbolTable4}
%  \end{table*}
%%%%%%end of variable list

\section{Problem Setting and Previously Proposed Techniques Used in FDP=H-Matrices}
\label{FDPH2}
We solve a transient elastodynamic problem as an initial boundary value problem in a $D_v$-dimensional linear elastic volume $V\subseteq\mathbb R^{D_v}$. Three-dimensional (3D) cases ($D_v=3$) are our main concern in the formulation phase, as they give two-dimensional (2D) cases ($D_v=2$) in certain limits. For simplicity, we assume an isotropic homogeneous medium of infinite volume ($V=\mathbb R^{D_v}$) with buried smooth crack interfaces (``faults'') $\Gamma\subset R^{D_v}$ without any sources of single force. 
In the following formulation, $\Gamma$ can be multiple unconnected faces and includes a kinked fault as long as a set of jointed smooth boundaries represent it. 
More general applications of FDP=H-matrices will be mentioned in \S\ref{FDPH73}. 

We first obtain the formulation of the ST-BIEM for the above setting in \S\ref{FDPH21}. 
We then outline the FDPM in \S\ref{FDPH22} and H-matrices in \S\ref{FDPH23} for later development of FDP=H-matrices.

\subsection{Spatiotemporal Boundary Integral Equation Method}
\label{FDPH21}
Based on Refs.~\cite{tada1997non,tada2000non}, 
we introduce a boundary integral equations (a BIE), which describes the dynamic stress field raised by dislocations (associated with displacement discontinuities) on boundary surfaces in an elastic volume. 

\subsubsection{Definition of the Boundary Integral Equation}
Assume the equation of motion,
\begin{equation}
\rho\partial_t^2 {\bf u} ({\bf x},t)= 
(\lambda+\mu){\bf \nabla}({\bf \nabla}\cdot {\bf u} ({\bf x},t))
+\mu({\bf \nabla}\cdot {\bf \nabla}) {\bf u} ({\bf x},t),
\nonumber
\end{equation}
for displacements ${\bf u}({\bf x},t)\in \mathbb R^{D_v}$ at location ${\bf x}=(x_1,x_2,x_3) \in V$ in a 3D volume ($D_v=3$) and time $t\in (0,t_{end}]$ with certain initial and boundary conditions, where constant $\rho\in\mathbb R$ is the density of mass, constants $\lambda\in\mathbb R$ and $\mu\in\mathbb R$ are Lame's parameters, and $t_{end}\in\mathbb R$ denotes the physical ending time of the simulation. 
Further, $\partial_t=\partial/(\partial t)$ and ${\bf \nabla}=(\partial/(\partial x_1)$, $\partial/(\partial x_2)$, $\partial/(\partial x_3))$ denote the temporal and spatial partial derivatives, respectively. A special constraint $\partial {\bf u}/\partial x_3=0$ gives the 2D problems from the 3D settings. 

We suppose the initial conditions, 
\begin{equation}
{\bf u}({\bf x}, 0)=\dot {\bf u}({\bf x}, 0)=0 \mbox{ in } V,
\nonumber
\end{equation}
where $\dot {\bf u}:=\partial_t {\bf u}$ is introduced for brevity. 
Besides, we consider mixed boundary conditions that involve the displacement discontinuity ${\bf \Delta u}\in \mathbb R^{D_v}$ (called ``slip'' for shear dislocations and ``opening'' for dilatational dislocations) and traction ${\bf T}\in \mathbb R^{D_v}$ on the fault $\Gamma$: 
\begin{flalign}
{\bf \Delta u}({\bf x},t) &= 
\lim_{\delta\to0}
[{\bf u}({\bf x}+\boldsymbol\nu({\bf x})\delta ,t) 
-
{\bf u}({\bf x}-\boldsymbol\nu({\bf x})\delta ,t)] 
\\
{\bf T}({\bf x},t) &= \boldsymbol\sigma({\bf x},\tau) \boldsymbol\nu({\bf x}),
\end{flalign}
where $\boldsymbol \nu({\bf x})\in \mathbb R^{D_v}$ represents the normal vector of the fault (pointing from its lower face to its upper face) at location ${\bf x}$ on $\Gamma$, and $\boldsymbol\sigma({\bf x},\tau)\in \mathbb R^{D_v\times D_v}$ denotes the stress tensor. 
Hereafter, the time invariance of $\boldsymbol \nu$ is assumed for simplicity. The $a,b$ component of $\boldsymbol\sigma$ is computed as $\sigma_{ab}=C_{abcd}(\partial u_c)/(\partial x_d)$ via $C_{abcd}:=\lambda \delta_{a,b}\delta_{c,d}+\mu(\delta_{a,c}\delta_{b,d}+\delta_{a,d}\delta_{b,c})$, where $\delta_{a,b}$ ($=1$ if $a=b$ and $=0$ otherwise) denotes the Kronecker delta. Summation over the repeated indices is implied wherever necessary. 
The above-mentioned mixed boundary conditions are imposed as
\begin{flalign}
\Delta {\bf u}({\bf x}, t)&= {\bf f}_{\Delta u}({\bf x}, t) \mbox{ at } {\bf x}\in\Gamma_{\Delta u},
\nonumber
\\
{\bf T}({\bf x}, t)&= {\bf f}_{T}({\bf x}, t) \mbox{ at }{\bf x}\in \Gamma_T,
\nonumber
\end{flalign}
by given functions ${\bf f}_{\Delta u},{\bf f}_{T}\in \mathbb R^{D_v}$ on two parts, $\Gamma_{\Delta u}$ and $\Gamma_T$, of $\Gamma$ ($\Gamma=\Gamma_{\Delta u}+\Gamma_T$).
Typically, ${\bf f}_{\Delta u}$ and ${\bf f}_T$ at location ${\bf x}$ at time $t$ are functions of ${\bf \Delta u}$ and ${\bf T}$ at the same ${\bf x}$ and $t$. We show later an example of such boundary conditions in the numerical experiments of the dynamic rupture problems (\S\ref{FDPH62}).

The solution over the entire volume is in general a function of the slip and opening in the above-mentioned initial boundary value problem. Its functional form is given by the representation theorem for the adjacent multiple faces, that is the fault(s) $\Gamma$ \cite{aki2002quantitative,tada2000non}:
\begin{flalign}
u_d({\bf x},t) =& \int_\Gamma d\Sigma (\boldsymbol\xi) \int^{t_{end}}_0 d\tau \Delta u_e(\boldsymbol\xi,\tau)\nu_f(\boldsymbol\xi)C_{efgh}
\nonumber\\&
\times
\frac{\partial G_{dg}}{\partial\xi_h} ({\bf x}-\boldsymbol\xi,t-\tau), 
\label{eq:Betti}
\end{flalign}
where $G_{dg}({\bf x},t)$ $\in \mathbb R$ denotes the $dg$ component of the associated Green's function; in a 3D space, it is given as 
\begin{flalign}
&G_{dg}({\bf x},t) 
\nonumber\\
=&
\frac 1{4\pi\rho\alpha^2} \frac{\gamma_d\gamma_g}{r}\delta(t-r/\alpha)
-\frac 1{4\pi\rho\beta^2} \frac{\gamma_d\gamma_g-\delta_{d,g}}{r}\delta(t-r/\beta)
\nonumber\\&
+\frac 1{4\pi\rho\alpha^2} \frac{3\gamma_d\gamma_g-\delta_{d,g}}{r^3}t[
H(t-r/\alpha)-H(t-r/\beta)
],
\label{eq:3DGreen}
\end{flalign}
where Euclidean norm $r:=|{\bf x}|\in\mathbb R$ is the distance, constants $\alpha:=\sqrt{(\lambda+2\mu)/\rho}\in\mathbb R$ and $\beta:=\sqrt{\mu/\rho}\in\mathbb R$ denote the P- and S-wave speeds,  respectively, and $\delta(\cdot)$ and $H(\cdot)$ respectively the Dirac delta and Heaviside functions. Integration of Eq.~(\ref{eq:3DGreen}) along the $x_3$ direction gives the 2D Green's function~\cite{eringen1975elastodynamics}.
The elastodynamic Green's function comprises the interactions of the impulsive P- and S-waves [the first and second terms in Eq.~(\ref{eq:3DGreen}), respectively] and the near-field term (the third term)~\cite{aki2002quantitative}.
%The contribution from the boundary $\Gamma$ vanishes due to the stress continuity~\cite{aki2002quantitative}. The quiescent past $\Delta u=0$ at $t<0$ is used in Eq.~(\ref{eq:Betti}) given the initial condition; the temporal integral is truncated at $t>t_{end}$ given the causality, $G_{dg}({\bf x},t)=0$ at $t<0$. 

Since the displacement field and thus the traction field are the explicit functions of slip and opening ${\bf \Delta u}$, we can reduce the original problem to the time evolution problem of ${\bf \Delta u}$ under the given mixed boundary condition. 
The traction incurred by the boundary motion is evaluable by using the space derivative of Eq.~(\ref{eq:Betti}), which gives a BIE for evaluating the stress field: 
\begin{equation}
\sigma_{ab}({\bf x},t)
=\int_\Gamma d
\Sigma (\boldsymbol\xi) \int^{t_{end}}_0 d\tau \Delta \dot u_e(\boldsymbol\xi,\tau)K_{abe}({\bf x},\boldsymbol\xi,t-\tau), 
\label{FDPHeq:1}
\end{equation}
with a kernel function $K_{abe}: \mathbb R^{D_v}\times \Gamma\times(0,t_{end}] \to \mathbb R^{D_v\times D_v}$ of a convolution operator, s.t.,
\begin{equation}
K_{abe} ({\bf x},\boldsymbol\xi,t-\tau) 
:=
C_{abcd}\nu_f(\boldsymbol\xi) 
C_{efgh}
\frac{
\partial }{\partial x_c}
\int^{t-\tau}_{-\infty} d\tau^\prime
\frac{
\partial G_{dg}}{\partial\xi_h}
({\bf x}-\boldsymbol\xi,\tau^\prime), 
\nonumber
\end{equation} 
where we introduced the temporal partial derivative $\Delta \dot u:=\partial_t\Delta u$ of the slip and opening (called the slip- and opening-rates) along the line of the conventional regularized BIEs~\cite{bonnet1999boundary,tada1997non,tada2000non}. Eq.~(\ref{FDPHeq:1}) is known to be hypersingular (for the slip and opening) yet regularizable (becoming evaluable in the sense of Cauchy integrals for the slip- and opening-rates)~\cite{bonnet1999boundary,tada2000non}, 
and hereafter, we suppose to use the regularized expression of $K$, the explicit form of which is found in the previous studies, e.g., Refs.~\cite{tada2001dynamic,tada2006stress}.

For simplifying the notation, hereafter, we omit the subscripts of spatiotemporally continuous variables, such as $\Delta u_a({\bf x},t)$ and $T_a({\bf x},t)$. 
The fast algorithms in the present study are supposed to apply to each pair of components of the stress $\boldsymbol\sigma$ and the slip- and opening-rate $\Delta \dot {\bf u}$.
Please refer to \S\ref{FDPHC1} for the handling of numerical errors associated with the projection of stress tensor $\boldsymbol\sigma$ to traction vector ${\bf T}$.

\subsubsection{Discretization of BIE}
Numerical evaluation of Eq.~(\ref{FDPHeq:1}) is the main computational object of the ST-BIEM. Eq.~(\ref{FDPHeq:1}) is spatiotemporally discretized for numerical analysis~\cite{tada2001dynamic,tada2006stress,cochard1994dynamic}. In this paper, we impose the spatial discretization and the temporal discretization separately. The temporally continuous BIE is found to be useful in reducing the error in the temporal interpolation of FDP=H-matrices in \S\ref{FDPH4} and \ref{FDPHB}.

Boundary area $\Gamma$ is subdivided into small patches $\Gamma_i$ of
the elements $i(=1,...,N)$ that satisfy $\sum_i \Gamma_i=\Gamma$ and $\Gamma_i\cap\Gamma_j=\emptyset$ for $i\neq j$. 
It gives the expanded (the discrete) forms of the boundary variables, such as $\Delta \dot  u$ and $T$. 
For the basis function of the slip and opening, we consider a piecewise-constant interpolation \cite{cochard1994dynamic}, 
\begin{equation}
\Delta \dot u ({\bf x},t) \approx D_i (t) \mbox{ at } {\bf x}\in\Gamma_i, 
\end{equation}
where $D_i(t)$ represents an expansion coefficients of the spatial basis for $\Delta \dot u$ of element $i$, depending on time $t$. 
We also consider that the associated traction is collocated at collocation point ${\bf x}_i\in \Gamma_i$ on each element $i$:
\begin{equation}
T_i (t)=T({\bf x}_i,t).
\end{equation}
Eq.~(\ref{FDPHeq:1}) is then spatially discretized as 
\begin{equation}
T_i(t)=  \sum_{j=1}^{N} \int_0^{t_{end}} d\tau K_{i,j} (t-\tau)D_j(t),
\label{FDPHeq:2}
\end{equation}
where $K_{i,j}(t-\tau):=\int_{\Gamma_j}d\Sigma(\boldsymbol\xi) K({\bf x}_i-\boldsymbol\xi,t-\tau)\in \mathbb R$ is the spatially discretized kernel for receiver $i$ and source $j$. 
Eq.~(\ref{FDPHeq:2}) is shortened to a matrix-vector form:
\begin{equation} 
{\bf T}(t)= \int_0^{t_{end}}d\tau  {\bf K}(t-\tau){\bf D}(\tau), 
\label{FDPHeq:3}
\end{equation}
where ${\bf T}(t)$ $= (T_1 (t),$ $T_2(t), ...,$ $T_N (t))^{\rm T}$$\in\mathbb R^N$ and ${\bf D}(\tau)$ $= (D_1(\tau),$ $ D_2(\tau), ...,$ $D_N (\tau))^{\rm T}$$\in\mathbb R^N$ denote vectors such that their $i$-th components store $T_i(t)$ and $D_i(\tau)$ of element $i$ at corresponding time ($t$, $\tau$), respectively; ${\bf K}(t)$$\in\mathbb R^{N\times N}$ denotes the matrix the $i,j$ entry of which is $[{\bf K}(t)] _{i,j}$ $:=K_{i,j}(t)$, and superscript $^{\rm T}$ represents the transpose.
%(The temporally continuous convolution, Eq.~(\ref{FDPHeq:3}), is recalled in \S\ref{FDPH4} and \ref{FDPHB}.) 

We then subdivide given time range $(0,t_{end}]$ into small ranges
$t\in(m\Delta t, (m+1)\Delta t)$ of time steps $m=0,...,M-1$ assuming constant time interval $\Delta t$. 
We interpolate the slip- and opening-rates in a piecewise-constant manner:
\begin{equation}
{\bf D} (t) \approx \sum_m {\bf D}_{m} [H(t-m\Delta t)-H(t-(m+1)\Delta t)] .
\label{FDPHeq:Dinterpolation}
\end{equation}
The traction is evaluated at the corresponding collocation time $t_n=(n+\epsilon_t)\Delta t$ at time step $n$ with parameter $\epsilon_t\in\mathbb R$ as
\begin{equation}
{\bf T}_n={\bf T}((n+\epsilon_t)\Delta t).
\label{FDPHeq:Tcollocation}
\end{equation}
Collocation time $t_n$ is within the $n$-th interval $t\in(n\Delta t, (n+1)\Delta t)$ as far as $\epsilon_t\in(0,1)$ is met. 
Throughout the paper, $\epsilon_t$ is assumed to be a constant. 

The spatiotemporally discretized form of Eq.~(\ref{FDPHeq:1}) is then expressed as
\begin{equation}
{\bf T}_n=\sum_{m=0}^{M-1}{\bf K}_m{\bf D}_{n-m},
\label{FDPHeq:4}
\end{equation}
where ${\bf K}_{m}:=\int^{t_m}_{t_{m-1}} d\tau {\bf K}(\tau)\in\mathbb R^{N\times N}$ $(m=0,....,M-1)$ represents the spatiotemporally discretized kernel. Summation $\sum_{m=0}^{M-1}$ in Eq.~(\ref{FDPHeq:4}) represents the discretized temporal convolution while $\sum_{j=1}^N$ in Eq.~(\ref{FDPHeq:2}) does the spatial one. Hereafter, $n=0,...,N-1$ denotes the current time step [associated with $t$ in Eq.~(\ref{FDPHeq:1})]. 
In the summation, $m\in\mathbb Z$ is also used in a limited way to represent the elapsed time step [associated with $t-\tau$ in Eq.~(\ref{FDPHeq:1})] from the initial time step of the discretized temporal convolution. 

\begin{figure*}[bt]
%\begin{figure*}[H] %% usepackage{here}
   %\begin{center}
  \includegraphics[width=120mm]{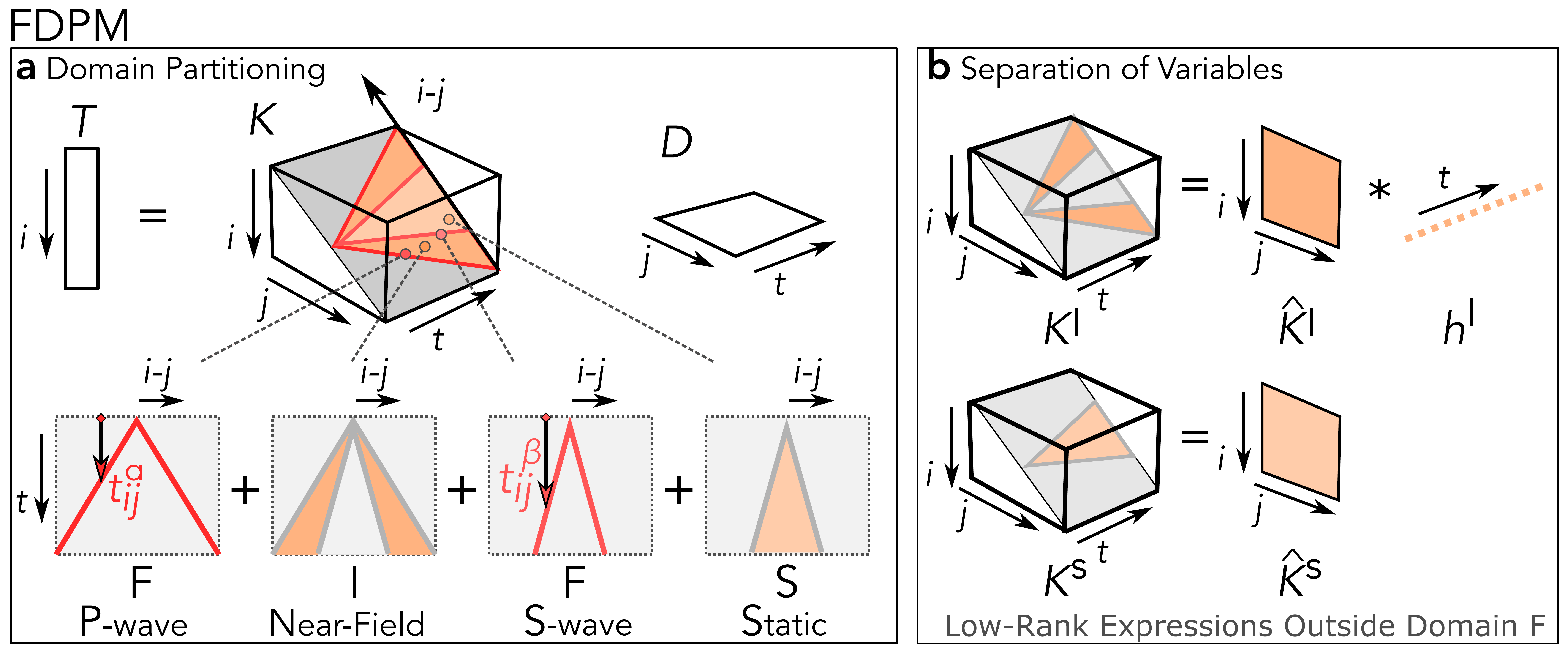}
   %\end{center}
\caption{
Schematic of the FDPM. A 3D elastodynamic example problem of a linear boundary is considered in the figure. 
{\bf a}, Schematic of the domain partitioning. 
The panel depicts a spatiotemporal BIE that convolves $K$ and $D$ over sources $j=1,...,N$ and time $t\in (0,M\Delta t)$ for evaluating $T$ of respective receivers $i=1,...,N$. 
The domain of kernel $K$ is partitioned into subdomains. 
Domain F (the red parts) fully encloses the wavefronts of the P- and S-waves. (Fp and Fs, respectively). 
The separators of the subdomains are the propagation times (the travel times) of the P- and S-waves ($t_{ij}^\alpha$ and $t_{ij}^\beta$, respectively) assigned to the collocation points of receiver $i$ and source $j$. 
Domain I (the orange part) is in-between Fp and Fs (the P- and S-wave parts of Domain F, respectively). 
Domain S (the ivory part) is after Fs. 
{\bf b}, Schematic of the separation of variables. 
The kernel tensor $K^I$ in Domain I separates into the spatially-varying part and time-dependent part, expressed by ($i,j$)-dependent matrices $\hat K^I$ and ($t$-dependent) vectors $h^I$. 
The kernel tensor $K^S$ in Domain S is time-invariant, expressed by an ($i,j$)-dependent matrix $\hat K^S$.
}
\label{FDPHfig:1_a}
\end{figure*}

Fully discretized kernel $K_{i,j,m}$ (denoted by ${\bf K}\in \mathbb R^{N\times N\times M}$ symbolically) is illustrated by a cuboid spanned by the axes of source number $i$, receiver number $j$, and time step number $m$ (Fig.~\ref{FDPHfig:1_a}). 
The volume of the discretized kernel describes the number of elements in the discretized kernel scaled by $N^2M$, which corresponds to the memory usage to store them and the computation time per time step to evaluate Eq.~(\ref{FDPHeq:4}). The computation time, intrinsically the complexity, of the original ST-BIEM is $\mathcal O(N^2M^2)$, due to the computationally dominant operation to evaluate Eq.~(\ref{FDPHeq:4}) repeated $M$-times. 
%Our aim is to reduce such computation costs per time step and total memory cost to be of $\mathcal O(N\log N)$ and total computation time to $\mathcal O(NM\log N)$. 
%The cost reduction of FDP=H-matrices also allows us to eliminate 
Memory usage to store all the entries of the slip- and opening-rate $D_{j,n-m}$ of $O(NM)$, required in the original ST-BIEM, is expressed by the area of $D_{j,n-m}$ spanned by the source- and receiver-number axes in Fig.~\ref{FDPHfig:1_a}. 
Our algorithm begins with reducing these huge costs of the ST-BIEM with the FDPM. 

\subsection{Outline of the FDPM}
\label{FDPH22}

We saw in the previous subsection that the ST-BIEM entails the costly dense kernel tensor. On the other hand, the 3D elastodynamic fundamental solution (Green's function) [Eq.~(\ref{eq:3DGreen})] separates into the impulsive P- and S-waves and the near-field term; further favorably, only the near-field term occupies most of the time domain, and it is factorized into the spatial part and the temporal part. 
As the kernel of the BIE Eq.~(\ref{FDPHeq:1}) is given by the Green's function with the spatiotemporal integrodifferential operator, 
we can expect a similar decomposition for the kernel, that is a natural low-rank expression of the kernel tensor. 
The FDPM expresses this by partitioning the time domain (Fig.~\ref{FDPHfig:1_a}a) and accelerates the computation by the factorization of the kernel (Fig.~\ref{FDPHfig:1_a}b). 
Here, we outline the FDPM by focusing on its domain-partitioning technique, which becomes crucial for developing FDP=H-matrices. 
Please refer 
to Table~\ref{SymbolTable2} for the relevant parameters of the FDPM, 
and 
to Refs.~\cite{ando2007efficient} and \cite{ando2016fast} for the analytic expressions (the semi-analytic BIEs) of the associated discretized kernel implementing the separation of variables in the FDPM. 
The illustration of Fig.~\ref{FDPHfig:1_a} is supposing the case of linearly aligned same-shaped boundary elements in a 3D space, solely for explanatory simplicity; the following formulation of the FDPM applies to nonplanar boundary geometries in both the 2D and 3D problems without any modifications. 

The idea of the domain partitioning can be grasped by using a simple convolution of the Green's function and single force $f$ like $f G$, which corresponds to the case of the single-layer potential convolved with the boundary traction~\cite{bonnet1999boundary}. 
For this case, the explicit form of the Green's function Eq.~(\ref{eq:3DGreen}) crudely yields
\begin{equation}
    G=
    \begin{dcases}
        \frac 1{4\pi\rho\alpha^2}\frac {\gamma_d\gamma_g} r \delta(t-r/\alpha)
        &
        (t= r/\alpha)
        \\
        \frac 1{4\pi\rho\alpha^2}\frac{3\gamma_d\gamma_g-\delta_{d,g}}{r^3}t 
        &
        (r/\alpha<t<r/\beta)
        \\
        -\frac 1{4\pi\rho\beta^2} \frac{\gamma_d\gamma_g-\delta_{d,g}}{r}\delta(t-r/\beta)
         &
         (t=r/\beta)
    \end{dcases}
\end{equation}
The above treatment of the delta function is not mathematically precise, but this sketches out the concept of the domain partitioning. 
We have the time domain involving the impulsive waves, which is called Domain F in the FDPM~\cite{ando2016fast}. The P-wave part ($t=r/\alpha$ in the above) is Domain Fp, the S-wave part ($t=r/\beta$) Domain Fs, and the sum of them constitutes Domain F.
The domain in-between Domains Fp and Fs is called Domain I. 
The most of the time range that gives the non-zero kernel values belongs to Domain I, and there the kernel separates into the spatial part $...r^{-3}$ and the temporal part $t$ without any approximations. 

The domain partitioning also holds for the discretized cases. For the boundary integral $\int_{\Gamma_j} d\Sigma G$ of $G$ on $\Gamma_j$ that has the characteristic length $\Delta x_j\in\mathbb R$ such that $\Delta x_j:=2\max_{{\bf x}\in \Gamma_j}|{\bf x}-{\bf x}_j|$, 
we have
\begin{equation}
\int_{\Gamma_j} d\Sigma G=
\begin{dcases}\!% alignment adjustment
  \begin{aligned}[b]
    +&\int_{\Gamma_j} \frac {d\Sigma}{4\pi\rho\alpha^2}\frac {\gamma_d\gamma_g} r \delta(t-r/\alpha) 
    \\ 
    +&\int_{\Gamma_j} \frac {d\Sigma}{4\pi\rho\alpha^2} \frac{3\gamma_d\gamma_g-\delta_{d,g}}{r^3}t
    \\
    &\times H(t-r/\alpha)
  \end{aligned} 
  &
  \left(\left|t- \frac r \alpha\right|<\frac{\Delta x_j}{2 \alpha}\right)
  \\
  \left(\int_{\Gamma_j} \frac {d\Sigma}{4\pi\rho\alpha^2} \frac{3\gamma_d\gamma_g-\delta_{d,g}}{r^3}\right)t
  &
  \left(\frac{r+\Delta x_j/2}{\alpha}<t<\frac{r-\Delta x_j/2}{\beta}\right)
  \\
  \begin{aligned}[b]
    -&\int_{\Gamma_j}\frac {d\Sigma}{4\pi\rho\beta^2} \frac{\gamma_d\gamma_g-\delta_{d,g}}{r}\delta(t-r/\beta)
    \\
    +&\int_{\Gamma_j} \frac {d\Sigma}{4\pi\rho\alpha^2} \frac{3\gamma_d\gamma_g-\delta_{d,g}}{r^3}t
    \\
    &\times[1-H(t-r/\beta)]
  \end{aligned}
  &
  \left(\left|t- \frac r \beta\right|<\frac{\Delta x_j}{2 \beta}\right)
\end{dcases}
\end{equation}
where we assumed $\alpha^{-1}(r+\Delta x_j/2)<\beta^{-1}(r-\Delta x_j/2)$ for brevity; it corresponds to assuming a certain distance between receiver $i$ and source $j$, and the most part of the kernel tensor except the part for neighboring elements follows it. 
The above conditional branching gives Domains Fp, I, and Fs in order, and the sum of Doamins Fp and Fs gives Domain F, as in the continuous case.
Domain F for the discrete case occupies the finite time range because of the finite size of the source element $j$. 
The value of $\Delta x_j$ is twice the maximum distance between collocation point ${\bf x}_j$ and the position within element $j$, which provides the upper bound ($\Delta x_j/c$ for $c=\alpha,\beta$) of the duration of the wave for spatially integrated $\int_{\Gamma_j} d\Sigma G$. 
The complicatedness of the above expression is largely due to the fraction of the near-field term in Domain F, not separating into the spatial part and the temporal part due to the spatiotemporal dependence of the step functions. Meanwhile, the near-field term in Domain I simply separates as in the undiscretized case, and hence Domain I is still the time domain that gives the low-rank expression to the kernel tensor even in the discrete space. 

We then go into the formalism of the FDPM. 
As in the above example, the same factorization applies to the regularized double-layer potential $K$~\cite{ando2016fast}, defined around Eq.~(\ref{FDPHeq:1}). 
Although the functional form of $K$ is much more complicated~\cite{tada2006stress} than the single-layer potential $G$ (the Green's function), 
the formalism of the domain partitioning is the same, intrinsically because the kernel is given as an integrodifferential form of the Green's function. Please refer to Refs.~\cite{ando2007efficient,ando2016fast} for analytical details. 
In light of that factorization, the FDPM introduces three subdomains, which are shown by different colors on the cross-section in Fig.~\ref{FDPHfig:1_a}a. 
The red, orange, and ivory represent the domain of the waves (Domain F), that of the near-field term (and the static term due to the P-waves) (Domain I), and that of the static equilibrium (Domain S), respectively. 
Here, as the kernel $K$ involves the time integration of the Green's function, we further introduced Domain S in addition to the aforementioned Domains F and I; the terms in Domain I is also subtly modified due to that time integration as it involves the static term incurred by the temporally integrated P-wave~\cite{ando2016fast}. 
The gray area is the outside of the causal cone and is excluded from the computation as the kernel is zero there. 

Domain F (red lines on the cross-section) is a time domain defined such that it fully involves the P- and S-waves. 
For defining it precisely, we introduce the propagation time of the wave from a source to a receiver, called ``travel time''~\cite{aki2002quantitative}. 
The travel time $t_{ij}^c\in\mathbb R$ between the source and receiver collocation points is given as 
\begin{equation}
t_{ij}^c := r_{ij} / c 
\label{FDPHeq:5}
\end{equation} 
for receiver $i$ and source $j$, where $r_{ij}\in\mathbb R$ is the distance between the collocation points of $i$ and $j$; $c\in\mathbb R$ represents the phase speed of the P-wave (denoted by $\alpha$) or the S-wave (denoted by $\beta$). 
Hereafter, the travel time between source-receiver collocation points is called ``travel time'' for brevity. The travel times of the P- and S-waves are respectively denoted by $t_{ij}^\alpha: = r_{ij} / \alpha$ and $t_{ij}^\beta: = r_{ij} / \beta$.

Domain F occupies the finite time range due to the spatiotemporal discretization of the boundary variables. 
We parametrize the duration of Domain F by using the characteristic length $\Delta x_j\in\mathbb R$ of element $j$, defined as $\Delta x_j:=2\max_{{\bf x}\in \Gamma_j}|{\bf x}-{\bf x}_j|$. The value of $\Delta x_j$ is twice the maximum distance between collocation point ${\bf x}_j$ and the position within element $j$. As for $\int d\Sigma G$ treated earlier, $\Delta x_j$ provides the upper bound ($\Delta x_j/c$) of the nominal duration of the waveform for the spatially discretized (yet temporally continuous) BIE, Eq~(\ref{FDPHeq:2}). 
By using this bound, we define the temporal distances from the travel time to the leading- and trailing-edges of the wave (denoted by $\Delta t_j^{c-}\in\mathbb R$ and $\Delta t_j^{c+}\in\mathbb R$, respectively): 
%In 3D problems, %, when we adopt $T_{i,n}=$$T_i((n+1)\Delta t)$ and $D_j(t)=$ $\sum_m$$D_{j,m}$$[H(m\Delta t-t)$$-H((m+1)\Delta t-t)]$, 
\begin{flalign}
\Delta t_j^{c+} &:= \Delta x_j / (2 c)  
+\delta C_{j}^{c+}\Delta t 
\label{FDPHeq:6}
\\
\Delta t_j^{c-} &:= \Delta x_j / (2c)
+\delta C_{j}^{c-}\Delta t,
\label{FDPHeq:7}
\end{flalign}
where we introduced non-negative safe coefficients $\delta C_j^{c\pm}\geq 0(\in\mathbb R)$ 
for the later imposed temporal discretization of Domain F, like in Ref.~\cite{ando2016fast}; we note that the above sketch of the domain partitioning using $\int d\Sigma G$ skipped this bothersome temporal discretization. 
The duration of the waveform, denoted by $\Delta t_j^c\in\mathbb R$ for each source $j$, is expressed as the sum of $\Delta t_j^{c\pm}$: 
\begin{equation}
\Delta t_j^c:=\Delta t_j^{c-}+\Delta t_j^{c+}.
\label{FDPHeq:durationofDomF}
\end{equation}

The time range involving P-waves (called Domain Fp) and S-waves (called Domain Fs) are defined as $t-\tau$ [in Eq.~(\ref{FDPHeq:1})] such that $t-\tau\in (t_{ij}^\alpha-\Delta t_j ^{\alpha-},t_{ij}^\alpha+ \Delta t_j^{\alpha+})$ and $t-\tau\in(t_{ij}^\beta-\Delta t_j ^{\beta-},t_{ij}^\beta+\Delta t_j^{\beta+}$), respectively. 
Further, we define the time-step definition ranges of the temporally discretized Domains Fp and Fs such that $m_{ij}^{\alpha-}\leq m < m_{ij}^{\alpha+}$ and $m_{ij}^{\beta-}\leq m <m_{ij}^{\beta+}$, respectively, where time steps $m_{ij}^{c-}\in\mathbb Z$ and $m_{ij}^{c+}-1\in\mathbb Z$ are respectively defined as the time steps that enclose the collocation time minus $t_{ij}^{c-}:=t_{ij}^c-\Delta t_{j}^{c-}\in\mathbb R$ and $t_{ij}^{c+}:=t_{ij}^c+\Delta t_{j}^{c+}\in\mathbb R$. 
For both the continuous time ranges and discrete time step ranges,
Domain F (red in Fig.~\ref{FDPHfig:1_a}a) is the union of Domains Fp and Fs,
Domain I (orange) 
between Domains Fp and Fs, 
and Domain S (ivory)
after Domain Fs. 

In the later algorithm development, we refer to the kernel corresponding to Domain W = F (Fp, Fs), I, S as 
${\bf K}^W(t)\in\mathbb R^{N\times N}$ (also as the kernel of Domain W). 
The explicit forms of their $ij$ entries are as follows for the case of $t^{\alpha+}_{ij}<t^{\beta-}_{ij}$: 
\begin{flalign}
K^F_{i,j}(t)&:=K_{i,j}(t) 
[H(t-t_{ij}^{c-})-
H(t-t_{ij}^{c+})]
\nonumber
\\
K^I_{i,j}(t)&:=K_{i,j}(t) 
[H(t-t_{ij}^{\alpha+})-
H(t-t_{ij}^{\beta-})]
\nonumber
\\
K^S_{i,j}(t)&:=K_{i,j}(t) 
H(t-t_{ij}^{\beta +}),
\nonumber
\end{flalign}
where F=Fp, Fs for $c=\alpha,\beta$, respectively. When $t^{\alpha+}_{ij}\leq t^{\beta-}_{ij}$, Domain I vanishes for such a  receiver-source $i$-$j$ pair, and we set ${K}^F_{i,j}(t):=K_{i,j} 
[H(t-t_{ij}^{\alpha-})-
H(t-t_{ij}^{\beta+})]$ without distinction between Fp and Fs while the definition of ${\bf K}^S$ is kept. 
We also refer to the convolution of those kernel and the slip- and opening-rate as ${\bf T}^W(t)\in\mathbb R^N$ (also as the stress associated with Domain W). 
They constitute the kernel and the stress computed in the original ST-BIEM as
\begin{flalign}
{\bf K}(t)&={\bf K}^F(t)+{\bf K}^I(t)+{\bf K}^S(t)
\nonumber
\\
{\bf T}(t)&={\bf T}^F(t)+{\bf T}^I(t)+{\bf T}^S(t).
\nonumber
\end{flalign}
Their temporally discretized expressions ${\bf K}^W_m$ and ${\bf T}_n^W$ are also defined as for the original ones of the ST-BIEM. 

We have finished defining the domain partitioning and below mentions the left separation-of-variable part of the FDPM (Fig.~\ref{FDPHfig:1_a}b). We limit the explanation to its minimum necessary part for developing FDP=H-matrices. Please refer to Refs.~\cite{ando2007efficient,ando2016fast} for detail. 
Like the near-field term of $G$, the kernel of Domain I separates into its space-dependent part (denoted by ${\bf \hat K}^I\in\mathbb{R}^{N^2}$) and its time-dependent part (denoted by ${\bf h}^I(t)$, discretized to ${\bf h}^I\in\mathbb{R}^{\min[M,L/(\beta\Delta t)]}$)~\cite{ando2007efficient,ando2016fast}: 
\begin{equation}
{\bf K}^I_{i,j,m}= {\bf \hat K}^I_{i,j} h^I_m 
[H(m-m_{ij}^{\alpha+}+0)-
H(m-m_{ij}^{\beta-}+0)],
\label{FDPHeq:factorizedformofkernelofDomainI}
\end{equation}
where scalar $L:=\max_{{\bf x},\boldsymbol\xi\in \Gamma}|{\bf x}-\boldsymbol\xi|\in\mathbb R$ represents the characteristic size of the fault areas $\Gamma$ [$L/(\beta\Delta t)\lesssim M$].
The Heaviside functions represent the time range of Domain I where ${\bf K}^I$ becomes nonzero. 
Since the kernel $K$ (the double-layer potential) is proportional to $\partial\int dt G$, 
$K$ contains the contribution from the P-wave as well as from the near-field term. 
${\bf \hat K}^I$ then involves two kinds of ${\bf \hat K}^I$ and $h^I$ in the stress nucleus $K$ to express the time-invariant contribution (giving $h^I\propto t^0$) of the passed P-wave (and a time-invariant contribution from the near-field term) and temporally parabolic contribution (giving $h^I\propto t^2$) from the near-field-term; we omitted the summation about them for brevity in the above expression. 
On the other hand, the elastodynamic kernel $K$ converges to the elastostatic one, and consequently the kernel of Domain S also reduces to a time-invariant form (denoted by ${\bf \hat K}^S\in\mathbb{R}^{N^2}$) after S-wave-passage completion as 
\begin{equation}
{\bf K}^S_{i,j,m}= {\bf \hat K}^S_{i,j} 
H(m-m_{ij}^{\beta+}+0).
\label{FDPHeq:factorizedformofkernelofDomainS}
\end{equation}
On the other hand, the kernel in Domain F is directly evaluated in the FDPM by using its definitional identity. 
%As all the terms contribute to the wavefronts as seen in the case of $\int_{\Gamma_j} d\Sigma G$, and further because these step- and delta-functional motions are smoothed due to the discretization of the boundary and time, the kernel of Domain F does not have simple spatiotemporal separation.
%As implied from the above functional form of $\partial\int dt G$, the kernel of Domains F and S is contributed to by all the P- and S-waves and near-field term, and the kernel of Domain I is contributed to by the P-wave and near-field term of the Green's function. Although it means that the kernel in different domains is contributed to by mostly the same terms of the Green's function, the above-mentioned clear transition of the time-dependence in these terms gives different natures to the kernel of Domains F, I, and S [Fig.~\ref{FDPHfig:1_a} (labeled as separation of variables)]. 

As duration $\Delta t_j^c$ of the time definition range of Domain F is $\mathcal O(\Delta t)$ for any source $j$, the number of components in discretized kernel ${\bf K}^F$ of Domain F is independent from the number of total time steps $M$ and is thus $\mathcal O(N^2)$. The number of components to express the discretized kernel of Domain I, separating into ${\bf \hat K}^I$ (a matrix) and ${\bf h}^I$ (a vector), is $\mathcal O[N^2+L/(\beta \Delta t)]$. Likewise, the number of components to express the discretized kernel of Domain S, reduced to ${\bf \hat K}^S$, is $\mathcal O(N^2)$. These show the kernel is expressed in a low-ranked form in the FDPM. 

We note that the separation of variables in Domains I and S does not induce any accuracy deterioration in the 3D kernel due to the finiteness of the wavefront phases while it is an asymptotic expansion in the 2D cases~\cite{ando2016fast}. 
Such a dimension dependence appears due to that a point source of the 2D problems is an infinitely long 3D line source resulting in long temporal tails of the wavefront phases~\cite{ando2007efficient}. 
The temporal distance $\Delta t_j^{c+}$ (or equivalently $\delta C_j^{c+}$) between the trailing edge and the travel time is then an error-control parameter in the 2D problems, taking a moderately large value compared with $\Delta x_j/(2c)$ to deal with this point~\cite{ando2007efficient}. 

The semi-analytic BIE performs the LRA in the FDPM described above by analytically deriving the spatiotemporally separated forms of the kernel~\cite{ando2007efficient,ando2016fast}. In this sense, the LRA in the FDPM is similar to the analytical method in the FMM. On the other hand, it is useful to explicitly isolate the impulsive domain as Domain F for the later algorithm development of FDP=H-matrices. 
After all, the motivation of the present study is to find a way to algebraically handle the impulsive parts that cannot be handled algebraically in the ordinary way (i.e. in the ordinary H-matrices), and separating the tensor components representing the waves as Domain F, or equivalently storing them as matrices, is the first step in formulating the present method. 

Hereinafter, superscript $c$ in $t_{ij}$ and $\Delta t_j^\pm$ will be omitted unless necessary. 

\subsection{Outline of H-Matrices}
\label{FDPH23}
\begin{figure}[bt]
%\begin{figure*}[H] %% usepackage{here}
   %\begin{center}
  \includegraphics[width=78mm]{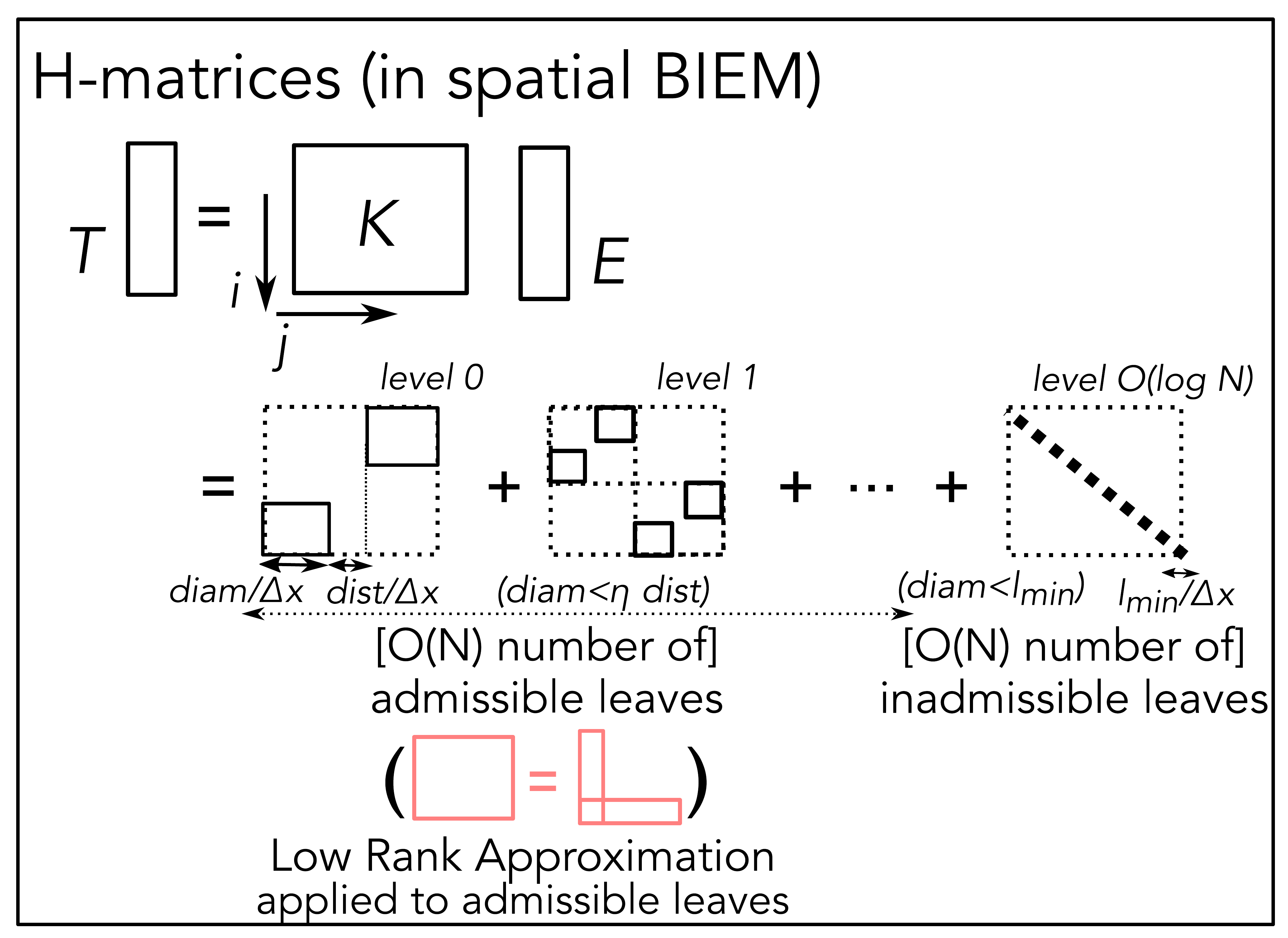}
   %\end{center}
\caption{
Schematic of H-matrices, illustrating an example case of linearly aligned structured boundary elements in a static problem, convolving $K$ and $E$ to evaluate $T$. 
Kernel matrix $K$ is subdivided into submatrices as the associated pairs of source clusters and receiver clusters are divided. 
The levels of the source-receiver clusters represent their number of divisions. 
The figure also shows the two division-stopping conditions: $diam<\eta dist$ for admissibly distant source- and receiver-cluster pairs and $diam<l_{min}$ for inadmissibly small ones under given parameters $\eta$ and $l_{min}$. 
The size $diam$ and distance $dist$ of clusters are indicated in the matrix, particularly for the above-mentioned boundary geometry, after divided by element length $\Delta x$. 
The low rank approximation of the kernel for an admissible leaf is also described. 
}
\label{FDPHfig:1_b}
\end{figure}

We next overview the two main procedures of H-matrices: a clustering of the source-receiver pairs (Fig.~\ref{FDPHfig:1_b}) and the low-rank approximation (the LRA) applied to certain subsets of the discretized kernel. We will subsume both of them into FDP=H-matrices. 
Please refer to Ref.~\cite{borm2003hierarchical} for details of H-matrices. 
For explanatory simplicity, we suppose a static problem, $T_i^{stat} = \sum_j K^{stat}_{i,j} E_j$ (illustrated in Fig.~\ref{FDPHfig:1_b}), 
where $E_j\in\mathbb R$ denotes the slip and opening of source $j$, and $K^{stat}_{i,j}\in\mathbb R$ denotes the matrix component of static kernel ${\bf K}^{stat}\in \mathbb R^{N\times N}$, connecting $T_i^{stat}\in\mathbb R$ and $E_j$. Traction $T_i^{stat}$ of receiver $i$ is here treated as a time-invariant.

%Figure 1_separated_b. 

As in the FMM, H-matrices first cluster the source elements and the receiver elements, to set the cluster pairs to which the LRA applies. 
The clustering procedure follows a hierarchical decomposition of the pairs (called ``block clusters''~\cite{borm2003hierarchical}) of neighboring elements. 
There are various clustering methods, and we adopted a spatial sorting using coordinates of the centers of elements, as in the pioneering work of Ref.~\cite{hackbusch1999sparse}. Our implementation is shown below, and a similar implementation can be found in Ref.~\cite{ida2014parallel}. 
Initially, a bounding box is configured to enclose all the locations of the centers of masses of boundary elements. Recursively bisecting the side of the maximal length of a bounding box, we then create bounding boxes of different sizes hierarchically. Each bounding box gives a subset, called a cluster, of boundary elements, the centers of masses of which are enclosed in the bounding box. 
The number of bisecting operations that a bounding box is subjected to is called the level of the corresponding cluster~\cite{borm2003hierarchical}. 

Such a hierarchical sorting of elements produces a tree structure of the clusters pairs, called the block clusters, and the tree of the block cluster is called the block cluster tree~\cite{borm2003hierarchical}; 
as a cluster may be expressed as a vector comprising an element subset, the pair of them depicts a ``block'' that is a submatrix in the discretized kernel matrix (Fig.~\ref{FDPHfig:1_b}). 
The recursive division of the block clusters continues until one of the following stop conditions is satisfied: 
\begin{flalign}
diam &< \eta \cdot dist 
\label{FDPHeq:8}
\\
diam &< l_{min}, 
\label{FDPHeq:9}
\end{flalign}
where $diam\in\mathbb R$ is the maximum distance between the center of mass of the boundary elements contained in each cluster, and $dist\in\mathbb R$ is the shortest distance between 1) the center of mass of the boundary elements contained in the receiver cluster and 2) that for the source cluster; $\eta\in\mathbb R$ and $l_{min}\in\mathbb R$ are the accuracy controlling parameters of the clustering. 
An intuitive example can be seen in the case of linearly aligned same-shaped elements, sketched in Fig.~\ref{FDPHfig:1_b}, where values of $diam$ and $dist$ can be associated with sizes and distances of submatrices in the original matrix. 
In general, the values or bounds of $dist$ and $diam$ in our implementation using the coordinate values of the centers of elements can be parametrized solely by the arrangement of the bounding boxes rather than by those of elements, as detailed later in \S\ref{FDPH42}.  

The condition in Eq.~(\ref{FDPHeq:8}) is for detecting a sufficiently distant cluster pair, and is called an admissibility condition. Eq.~(\ref{FDPHeq:9}) is for unacceptably small clusters, and is called an inadmissibility condition. A pair of the source and receiver clusters that satisfies one of the stop conditions, Eqs.~(\ref{FDPHeq:8}) or (\ref{FDPHeq:9}), is a leaf of the graph formed by this clustering process, called an admissible leaf or an inadmissible leaf, respectively. As arbitrariness exists in these definitions of stop conditions, the stop conditions used throughout the paper are detailed and investigated in \S\ref{FDPH42}, with the introduction of the ART. 

Tracing the block cluster tree, we obtain an appropriate set of the disjoint block clusters. Accordingly, 
discretized static kernel ${\bf K}^{stat}$ separates into submatrices ${\bf K}^{stat}_a\in \mathbb R^{N_{a}\times N_{a}}$ of leaves $a$ (of $N_{a}$ receivers and $N_{a}$ sources) in the block cluster tree as ${\bf K}^{stat} = \sum_a {\bf K}^{stat}_a$; 
although the number of sources and that of receivers can be different in a block cluster in general (See Table~\ref{SymbolTable2}), we can identify them as far as we use this simple example problem. 
Submatrix ${\bf K}^{stat}_a$ for an admissible leaf $a$ is approximated to a low-ranked expression, ${\bf K}^{stat}_{a,LRA}$ (illustrated by a red square and two bars in Fig.~\ref{FDPHfig:1_b}). 
${\bf K}^{stat}_{a,LRA}$ for $a$ can be given as ${\bf K}^{stat}_{a,LRA}:= \sum_{l=0}^{l_a^*-1} {\bf f}_{a,l} {\bf g}_{a,l}^T$ using its rank $l_a^*\in \mathbb N$ and vectors ${\bf f}_{a,l}\in \mathbb R^{N_{a}}$ (column) and ${\bf g}^T_{a,l}\in \mathbb R^{N_{a}}$ (row) associated with the $l$-th largest singular value of ${\bf K}^{stat}_a$. The error of the LRA is regulated so as to satisfy $|{\bf K}^{stat}_a - {\bf K}^{stat}_{a,LRA}| < \epsilon_H|{\bf K}^{stat}_a|$ in each leaf $a$, where $\epsilon_H < 1$ is a given constant, and $|{\bf K}^{stat}|$ denotes the Frobenius norm of matrix ${\bf K}^{stat}$. The LRA is commonly implemented with fast algorithms of approximately executing the singular value decomposition, such as the adaptive cross approximation (the ACA)~\cite{bebendorf2003adaptive} of the partially-pivoting implementation. 

After the LRA, the convolution of the above-mentioned spatial BIE is evaluated as 
\begin{equation}
T^{stat}_i= \sum_{a\in A_{adm},l} f_{a,l,i}\sum_j g_{a,l,j}E_j+\sum_{a\in A_{inadm}} K^{stat}_{a,i,j} E_j, 
\label{FDPHeq:10}
\end{equation}
where $A_{adm}$ and $A_{inadm}$ denote the sets of admissible leaves and inadmissible leaves, respectively. 
Note that this style of treating the integral kernel (including the clustering, the LRA, and the multiplication of the hierarchically low-ranked matrix and a vector) are conventionally referred to as ``H-matrices'', while the approximated matrix is referred to as an ``H-matrix''~\cite{borm2003hierarchical}. 

The above series of the data-compression techniques works well in the spatial BIE of the elastostaics. 
Intrinsically, the elastostatic Green's function, and thus the continuous static kernel consisting of its spatial differentiation, are expressed by the products of 1) power functions of the source-receiver distance and 2) functions depending only on the source-receiver azimuth (the orientation) (e.g., shown in  Ref.~\cite{segall2010earthquake}). Therefore, the discretized kernel takes similar values in an admissible leaf pairing distant source and receiver clusters. 
The distance between the clusters ($dist$) is relatively larger than cluster sizes ($diam$) in the admissible leaves, and this scale separation gives an expansion of the kernel in $2^{-1}diam/(diam +dist)<2^{-1}/(1+1/\eta)$]; $2^{-1}diam$ here corresponds to the maximum of the variations in the source or receiver locations, and $diam +dist$ corresponds to the minimum of the distance between the centers of the associated bounding boxes. 
The same expansion applies to the orientational variations in an admissible leaf also being of $\mathcal O[2^{-1}/(1+1/\eta)]$. 
We see from these that $\eta$ in the admissibility condition gives a perturbation parameter distancing sources and receivers.
The perturbation series in $2^{-1}diam/(diam +dist)$ bounded by $2^{-1}/(1+1/\eta)$ is uniformly a convergent series (as long as $\eta<\infty$). Furthermore, such a Taylor series in the locations ${\bf x}_i$ and ${\bf x}_j$ of receiver $i$ and source $j$, around ${\bf x}_{i0}$ and ${\bf x}_{j0}$, respectively, can be expressed as $\sum_l c_l({\bf x}_i-{\bf x}_{i0})^{p_{1,l}} ({\bf x}_j-{\bf x}_{j0})^{p_{2,l}}$, with some constants $c_l\in\mathbb R$, $p_{1,l}\in\mathbb R$, and $p_{2,l}\in\mathbb R$ at respective effective ranks $l\in\mathbb Z$ (See Ref.~\cite{borm2003hierarchical}). This parallels the above-mentioned low-ranked expression of the kernel.
The existence of such a separate (and fast convergent) expansion, called a degenerate form~\cite{borm2003hierarchical}, gives the basis for rank $l_a^*$ to reach $\mathcal O(1)$ after the LRA of H-matrices~\cite{bebendorf2003adaptive}. 

The cost reduction of H-matrices is evaluable as follows. 
The costs, namely the computational complexity and memory usage, 
are 
originally $\mathcal O($$\sum_a $$N_a^2)$ for the admissible leaves in the spatial BIEM. 
These become $\mathcal O$$(\sum_a $$2N_a l_a ^*)$ by the LRA. 
Besides, given the existence of the above-mentioned degenerate form of the kernel, $l_a^*$ is $\mathcal O(1)$, and hence 
the costs of the admissible leaves are estimated to be $\mathcal O(\sum_a 2N_a l_ a ^*)$ $=\mathcal O(\sum_a N_a)$ $=\mathcal O(N\log N)$; for counting this cost, it is helpful that the number of block clusters and the number $N_a$ of the source or receiver elements in a block cluster are $\mathcal O (2^c)$ and $\mathcal O(N/2^c)$, respectively, at each level $c=1,2,...,\mathcal O(\log N)$ (Please refer to Fig.~\ref{FDPHfig:1_b}). 
%In addition, the maximum level is of $\mathcal O(\log N)$. The number of same-size block clusters is of $\mathcal O(N/N^\prime)$ when the number of contained sources (or receivers) is  $N^\prime(=N/2,N/4,...)$ (See Fig.~\ref{FDPHfig:1_b}). 
On the other hand, the costs of diagonally distributed inadmissible leaves are strictly $\mathcal O(N)$. 
These $\mathcal O(N\log N,N)$ costs are much smaller than the $\mathcal O(N^2)$ costs required to evaluate the original spatial BIE. 

\section{Architecture of FDP=H-Matrices}
\label{FDPH3}
This section is organized to introduce the basic structure and concepts, the architecture of our new method. 
FDP=H-matrices are first outlined in \S\ref{FDPH31} to relate four modules of FDP=H-matrices, named the FDPM, H-matrices, Quantization and the ART. 
The roles of the individual modules to reduce the numerical cost are shown in \S\ref{FDPH32}. 
This section is intended to be self-contained for highlighting the basics, and point-by-point guides to technical details in Sections \ref{FDPH4} and \ref{FDPH5} (the key formulas of which are listed in \ref{sec:keyformulas}) are also provided for readability. 

%Figure 2.
\begin{figure*}[tb]
   \begin{center}
  \includegraphics[width=120mm]{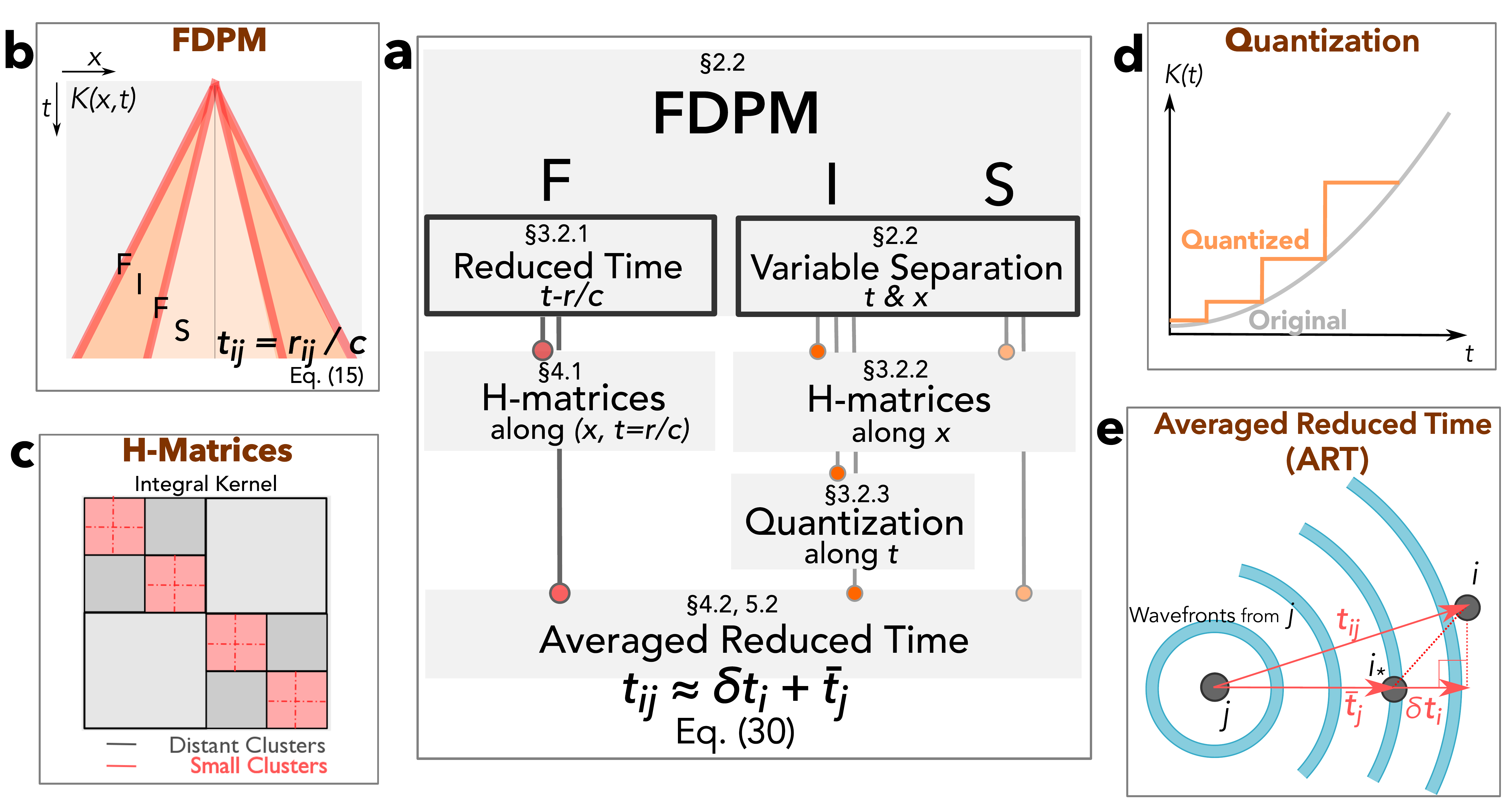}
   \end{center}
\caption{ 
Structure of FDP=H-matrices. 
{\bf a}, Flowchart of the subtasks in FDP=H-matrices, introduced in \S\ref{FDPH31}. 
The subtasks are executed in order from the top and are assembled in respective temporal subdomains of the FDPM: Domains F, I, and S. 
The numbers assigned to the subtasks express their associated sections. 
{\bf b}-{\bf e}, Diagrams showing the four module algorithms of FDP=H-matrices. 
{\bf b}, The FDPM (simplifying Fig.~\ref{FDPHfig:1_a}). The partitions of Domains F, I, and S are given by travel time $t_{ij}=r_{ij}/c$, the ratio of source($j$)-receiver($i$) distance $r_{ij}$ to phase velocity $c$. 
{\bf c}, H-matrices (corresponding to Fig.~\ref{FDPHfig:1_b}). 
{\bf d}, Quantization (simplifying Fig.~\ref{FDPHfig:5}), illustrating the employed staircase approximation of the kernel $K_m$ depending on time $m$. 
{\bf e}, The ART (simplifying Fig.~\ref{FDPHfig:7_a}), illustrating the plane wave approximation, which approximates travel time $t_{ij}$ as the sum of receiver-$i$-dependent part $\delta t_i$ and source-$j$-dependent part $\bar t_j$ as Eq.~(\ref{FDPHeq:tijsep}) via relay point $i_*$. 
} 
\label{FDPHfig:2}
\end{figure*}

\subsection{Outline and Relationship of Modules in FDP=H-Matrices}
\label{FDPH31}

The algorithm of FDP=H-matrices is developed as a hybrid of four module algorithms: the FDPM, H-matrices, Quantization and the ART. 
Fig.~\ref{FDPHfig:2} shows a schematic diagram to relate these four modules in FDP=H-matrices. 

Fig.~\ref{FDPHfig:2}a lists the subtasks executed in the algorithm. They are executed in order from the top, and the operations performed in each domain are independent of each other; as mentioned earlier, three subdomains are introduced by the FDPM as Domain F, Domain I, and Domain S (red, orange, and ivory parts in Fig.~\ref{FDPHfig:2}b, respectively). The details of the operations are explained in the body texts corresponding to the numbers assigned to each subtask in the figure. 

The left parts Fig.~\ref{FDPHfig:2}b-e roughly sketches the four modules, intended to guide the readers to the corresponding figures and texts; please refer to them for details. 
The most challenging portion of the method development is to run the H-matrix technique (\S\ref{FDPH23}) on the impulsive wave part of the elastodynamic integral kernel. 
FDP=H-matrices first extract such an intractable time domain as Domain F of the FDPM (Fig.~\ref{FDPHfig:2}b, related to \S\ref{FDPH22}). 
H-matrices (Fig.~\ref{FDPHfig:2}c, \S\ref{FDPH23}) work on respective subdomains partitioned by the FDPM. 
Furthermore, a plane wave approximation is required as in the PWTD method, and the ART (Fig.~\ref{FDPHfig:2}e, detailed in \S\ref{FDPH42}) plays the role of it. 
Additionally, Quantization (Fig.~\ref{FDPHfig:2}d, detailed in \S\ref{FDPH323}) sparsely resamples the non-impulsive part of the kernel in Domain I in a quantizing manner and accelerates the computation. 

%Fig.~\ref{FDPHfig:2} is a schematic diagram to relate these four modules in FDP=H-matrices. Fig.~\ref{FDPHfig:2}a is the core of Fig.~\ref{FDPHfig:2}, and Fig.~\ref{FDPHfig:2}b-e %the other panels at four corners, 
%illustrates the four modules. %connect Fig.~\ref{FDPHfig:2} to related other figures similar to these panels. 
%The FDPM divides the time domain of the BIE, as mentioned earlier, into three subdomains: Domain F, Domain I, and Domain S (red, orange, and ivory parts in Fig.~\ref{FDPHfig:2}b, respectively). We keep this way of domain partitioning, and apply H-matrices (Fig.~\ref{FDPHfig:2}c) to respective domains, considering their differences in the space and time dependences. Such differences lead to the different application of the modules as described in Fig.~\ref{FDPHfig:2}a, showing the related flows of subtasks in the respective domains. %, detailed in the following sub-subsections. 
%Subtasks include Quantization (Fig.~\ref{FDPHfig:2}d) and the ART (Fig.~\ref{FDPHfig:2}e).

In the following, we outline the algorithm by supplementing Fig.~\ref{FDPHfig:2}a. 
We focus on the admissible leaves being computationally demanding, considering the application of H-matrices. 
Please refer to \ref{FDPHE} for the handling of the inadmissible leaves, which is relatively computationally trivial. 

\subsubsection{Domain F}
The data-sparse approximation in Domain F comprises the following three procedures as illustrated in the chart of Fig.~\ref{FDPHfig:2}a. 
1) The FDPM first gathers a set of singular points of the impulsive P- and S-waves in the kernel as dense matrices. 2) H-matrices are applied to them and express the kernel values in a low-rank manner. 
3) The ART approximates the onset of Domain F (the travel time) in a memory-efficient manner by using a sort of plane wave approximations. 
%further invented to specify the wavefront locations in the original space-time domain with lowered memory. 
We overview respective subtasks below. 

The FDPM expresses 
the time position inside Domain F by time $t-r_{ij}/c$ onset at wave arrival time $r_{ij}/c$ (called reduced time \cite{aki2002quantitative}) for each source-receiver pair distanced by $r_{ij}$ with wave speed $c$. 
At the same reduced time ($t-r_{ij}/c=const.$), the time variation of the wave is similar, and we can expect the geometrical-spreading nature to the corresponding kernel values~\cite{ando2016fast}. 
This structure is robust for the corresponding terms in the elastodynamic (or widely, hyperbolic) integral kernel [e.g. $r^{-1}\delta(t-r/c)$ in Eq.~(\ref{eq:3DGreen})]. 

Consequently, we can gather the tensor components of the kernel representing singular waves as smoothly-varying matrices expected in H-matrices, by using the domain partitioning of the FDPM. 
We apply H-matrices to such matrices. 
The gathered kernel values spread geometrically, and thus
the ranks of the associated matrices are $\mathcal O(1)$ as in the case of the elastostatic kernel. 
%This allows us to avoid the known difficulty~\cite{yoshikawa20152,borm2003hierarchical} in applying H-matrices around the singular points of the P- and S-waves. 

The wave arrival time $t_{ij}=r_{ij}/c$ (called the travel time \cite{aki2002quantitative}) that determines the onset of the reduced time takes different values for $\mathcal O(N^2)$ combinations of receivers $i$ and sources $j$. 
Under the plane wave approximation~\cite{aki2002quantitative}, the ART approximates these travel time values in each admissible leaf of H-matrices and separate their $i$ and $j$ dependencies. 
As illustrated in Fig.~\ref{FDPHfig:2}e, 
the travel time $t_{ij}$ is reduced as $t_{ij}\approx \delta t_i+\bar t_j$ in each leaf to the sum of the travel time $\bar t_j$ between the relay point $i_*$ and source $j$ and effective travel time difference $\delta t_{i}$ between receiver $i$ and $i_*$ (given their distance by the projected line, along the path from $j$ to $i_*$ under the plane wave approximation, as in Fig.~\ref{FDPHfig:2}e). 
Technical details will appear in \S\ref{FDPH42}. %Values of $\delta t_{i}$ and $\bar t_j$ are assigned to respective receivers $i$ and sources $j$ the numbers of which are of order the number of elements in each admissible leaf, that is, totally of $\mathcal O(N\log N)$. 
%Then, %like the kernel value of Domain F, 
%the time definition range [($t_{ij}-\Delta t_j ^{-},t_{ij}+ \Delta t_j^{+}$), defined around Eq.~(\ref{FDPHeq:durationofDomF}) in \S\ref{FDPH22}] of Domain F can also be stored with the $\mathcal O(N\log N)$ memory. 
The computations of Domain F can finally be performed in $\mathcal O(N\log N)$ time in each time step with $\mathcal O(N\log N)$ memory, under the sparse-matrix arithmetic developed in \S\ref{FDPH5}.

\subsubsection{Domain I}
The data-sparse approximation reduces to the following four procedures in Domain I. 
1) The FDPM reduces the kernel in Domain I into a matrix-vector form without analytical errors, and 2) H-matrices reduce the matrix parts into low-ranked forms. 3) Quantization is used supplementarily for the related arithmetic of Domain I. 
4) The ART is also applied as in Domain F. 
These subtasks are overviewed below.

The FDPM separates the kernel into time-dependent functions represented by vectors and space-dependent functions represented by matrices (\S\ref{FDPH22}). 

The space-dependent parts follow the geometrical spreading as the elastostatic kernel does~\cite{ando2016fast}, 
and hence H-matrices apply to the space-dependence of the kernel; so to speak, we apply H-matrices along the spatial ${\bf x}$ axes. The receiver($i$)-source($j$)-dependent matrix becomes low-ranked one of the $\mathcal O(1)$ rank. 

Quantization, used solely for Domain I, executes the staircase approximation of the kernel along the time $t$ axis in an adaptive time stepping manner, which is exactly the quantization in the signal processing~\cite{gonzalez2002digital}, as illustrated in Fig.~\ref{FDPHfig:2}d. This reduces the memory usage required in the computation in Domain I. 
Please refer to \S\ref{FDPH323} for additional descriptions and \S\ref{FDPHB2} for details. 

The ART separates the receiver- and source-dependent travel time determining the time definition range of Domain I, as in Domain F. 

The sparse-matrix arithmetic of Domain I is described in  \S\ref{FDPHB2}.

\subsubsection{Domain S}
In Domain S giving the time-independent kernel values, the data-sparse approximation
is similar to that in Domain I, excluding the use of Quantization. 
1) The FDPM reduces the kernel to a time-invariant spatially-varying function represented by a matrix (\S\ref{FDPH22}). 
2) An H-matrix is introduced along the spatial axes similarly to the widely used elastostatic ones. %(\S\ref{FDPH322}), and works as well as for the elastostatic cases. 
3) The ART is introduced as in Domain I. 
The sparse-matrix arithmetic of Domain S is described in  \S\ref{FDPHB1}. 

\subsection{Cost Reduction Procedure: Roles of the FDPM, H-Matrices and Quantization}
\label{FDPH32}
This subsection outlines the implementation process of the data-sparse approximations by combining the subtasks introduced in the previous subsection. 
We start from the cost order of the FDPM and focus specifically on the cost reduction by H-matrices. The role of Quantization is also mentioned. 
We do not mention the role of the ART here to avoid intricacies. We will go back to it in \S\ref{FDPH42}. 

The following considers only the admissible leaves. Please refer to \ref{FDPHE} for the cost of the inadmissible leaves, which is shown to be $\mathcal O(N)$ in that appendix.

\subsubsection{Role of H-Matrices Applied to the Spatiotemporally-Varying Wavefronts of the Kernel in Domain F}
\label{FDPH321}
%Figure 3. âš
\begin{figure*}[tb]
   %\begin{center}
  \includegraphics[width=120mm]{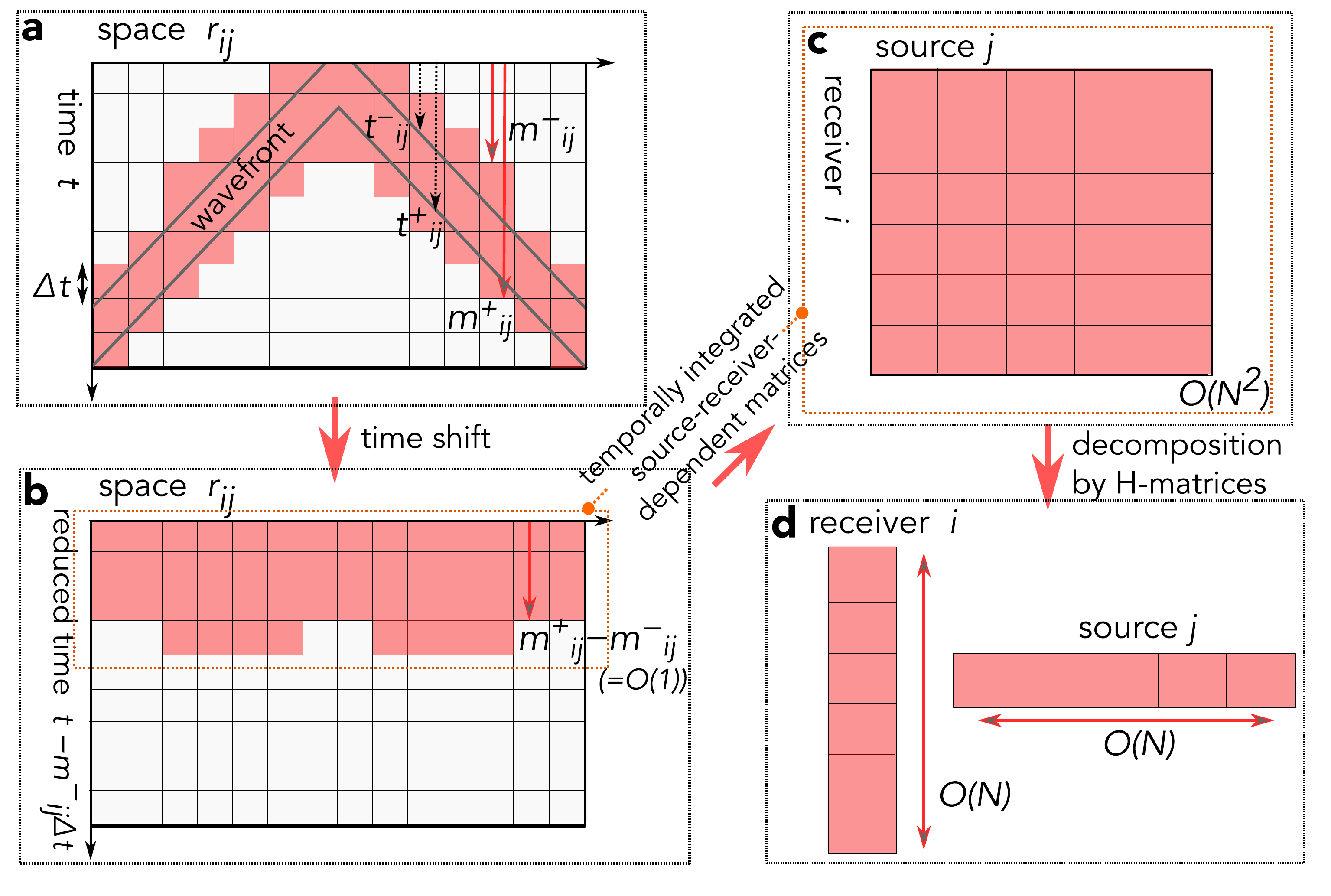}
  %\includegraphics[width=78mm]{DomainFred5.pdf}
   %\end{center}
\caption{
The approximation procedure in Domain F. 
{\bf a}, Spatiotemporal area belonging to Domain F. The start and the end of Domain F are respectively wave-arrival time $t_{ij}^-$ and wave-passage-completion time $t_{ij}^+$ in the continuous time scale. The corresponding time steps are denoted by $m_{ij}^-$ and $m_{ij} ^+$, respectively. %Handling of $t_{ij}^\pm$ is explored in \S\ref{FDPH43} related to the ART. 
{\bf b}, Time shift of Domain F. The kernel is densely aligned to the spatial direction. The number of time steps in Domain F is $m_{ij}^+ - m_{ij}^-=\mathcal O (1)$. 
{\bf c}, A matrix structure representing the source-receiver dependence of the kernel temporally integrated over Domain F. The number of entries in this matrix is $\mathcal O (N ^ 2)$. 
{\bf d}, An approximate (sub)matrix generated by H-matrices. 
The source-receiver-dependent matrix is separated into two vectors depending on the sources or receivers. The number of components to express the kernel becomes almost $\mathcal O (N)$.
}
\label{FDPHfig:3}
\end{figure*}

H-matrices in elastostatics owe its theoretical basis to the perturbation expansion in the source-receiver distance like the FMM. In this case (giving e.g. $1/r$ for Poisson's equation), the number of basis functions are at least as many as the number of perturbation parameters, essentially the number of the singular points ($r=0$) contained in the kernel of the BIE. 
This means that the number $N$ of the source elements is the lower cost bound in the elastostatic problem. 
On the other hand, the elastodynamic kernel (giving e.g. $r^{-1}\delta(t-r/c)$ for the wave equation) is singular also at the wave arrival time ($t=r/c$) even at a distance. Therefore, if we estimate the cost using the same logic as for the elastostatics, the lower bound of the elastodynamic case would be
the number of singular points ($t=r/c$) in the kernel, which are the $\mathcal O(N^2)$ combinations of $N$ sources and $N$ receivers. 
This naive cost estimates is indeed consistent with the previous reports of the elastodynamic application of H-matrices, e.g., Ref.~\cite{yoshikawa20152}. 
However, as shown below, we can reduce this cost further by gathering the set of the singular points distributed along the wavefronts ($r=ct$), an isochrone drawn by a wave radiated by a source location \cite{aki2002quantitative}. 
Because they obey the geometrical spreading ($\propto 1/r$) as the elastostotic kernel does, 
we can expect H-matrices work efficiently to approximate these $\mathcal O(N^2)$ components along Domain F fully involving the wavefronts (within the range s.t. $|t-r/c|<const.$)~\cite{ando2016fast}. 
Consequently, we can store even such singular wavefront components as low-rank matrices with $\mathcal O(N\log N)$ costs by incorporating H-matrices with the FDPM. 

Fig.~\ref{FDPHfig:3} illustrates the way of applying H-matrices along Domain F. %(detailed in \S\ref{FDPH41}-\ref{FDPH43}). 
%The figure is shown with the discretized time for brevity.
First, the FDPM specifies the temporal location $t$ (Fig.~\ref{FDPHfig:3}a) by using reduced time $t-t_{ij}$, namely the time elapsed from the wave arrival (Fig.~\ref{FDPHfig:3}b). 
It gathers the kernel in the same reduced-time region and makes a matrix (along the horizontal axis in Fig.~\ref{FDPHfig:3}b, detailed in \S\ref{FDPH41}); note that the time axis in Fig.~\ref{FDPHfig:3} is illustrated with discrete time steps using $m^\pm_{ij}$, which are the discretized counterparts of $t^{\pm}_{ij}$ of receiver $i$ and source $j$ introduced in \S\ref{FDPH22}. 
%This temporal discretization will be developed in \S\ref{FDPH43} with its intricacies. 
H-matrices are then applied to such a time-shifted matrix (from Fig.~\ref{FDPHfig:3}c to Fig.~\ref{FDPHfig:3}d, detailed in \S\ref{FDPH41}). 
Except that the time position is specified by the reduced time instead of the original time, the above procedure is almost parallel to the conventional H-matrices in the ST-BIEM, e.g. Ref.~\cite{yoshikawa20152}, where the LRA is applied to the components of the kernel tensor of the same time step. 

%Such approximations are implemented in the following way.
%1) The Domain F kernel, ${\bf K}^F$, is separated into the time integral of the Domain F kernel ${\bf \hat K}^F:=\int dt {\bf K}^F$ and the normalized ${\bf K}^F$ by ${\bf \hat K}^F$  $h^F_{ij}:=K^F_{ij}/\hat K^F_{ij}$ called normalized waveform (detailed in \S\ref{FDPH41}). 
%${\bf \hat K}^F$ corresponds to the impulse of the kernel in Domain F and is approximated by H-matrices (detailed in \S\ref{FDPH43}). 
%The impulsive time-dependence of $K^F$ is fully expressed by $h_{ij}(t)$ and not contained in ${\bf \hat K}^F$.
%2) The ART reduces $t_{ij}$ and $h_{ij}$ respectively to $t_{ij} \to \delta t_i + \bar t_j$, $h^F_{ij}(t)\to h^F_j(t)$ for each pair of the receiver $i$ and source $j$ (detailed in \S\ref{FDPH42}) in each admissible leaf. 3) The discretized time definition range of Domain F is given by the discretization of $\delta m_i$ and $\bar m_j^\pm$ and $h_j(t)$ is discretized as $h_{j,m}$ (detailed in \S\ref{FDPH43}).

%$\hat K^F$ is described by the geometrical spreading and orientation dependence (radiation pattern) as in the static kernel (detailed in \S\ref{FDPH41}). Hence the rank of $\hat K^F$ can be reduced to be of $\mathcal O(1)$ in each admissible leaf by H-matrices. The $m$ (representing the discretized reduced time) in $h_{j,m}$ is of $\mathcal O(1)$ by considering the duration of the time definition range of Domain F. 
The source- and receiver-dependence of the kernel in Domain F is expressed by an $\mathcal O(1)$-rank matrix in each admissible leaf, owing to the geometrical-spreading nature of the elastodynamic kernel along the wavefront. 
Such a matrix structure is fully stored in the $\mathcal O (N\log N)$ memory space in contrast to its original memory requirement of $\mathcal O(N^2)$. Note that in Fig.~\ref{FDPHfig:3}, the matrix and submatrix were undistinguished for brevity, and the log factor is omitted [i.e. $\mathcal O(N\log N)\approx\mathcal O(N)$ in the figure].

%The following omitted subtasks are detailed in the associated places in the paper. 
As an intricacy, we would add that the kernel in Domain F, analogous with the fundamental solution $(4\pi r)^{-1}\delta(t-r/c)$ of the wave equation, comprises a geometrically spreading part [like $1/(4\pi r)$] and a impulsive part [$\delta(t-r/c)$]. The former is efficiently approximated by H-matrices as seen above, and as detailed in \S\ref{FDPH42}, the latter is treated by a sort of plane wave approximations, the ART ($t_{ij}:=r_{ij}/c\approx\delta t_i+\bar t_j$ for receiver $i$ and source $j$, mentioned earlier in \S\ref{FDPH31}). 
The kernel is then fully stored in the $\mathcal O (N\log N)$ memory space by the use of H-matrices and the ART on the framework of the FDPM; accordingly, the arithmetic for the discretized convolution in Domain F reduces both the time complexity per time step 
and the total memory required for simulating the ST-BIEM to $\mathcal O (N\log N)$, with obviating the $\mathcal O (NM)$ memory to store the history of the boundary variables (e.g., the slip- and opening-rates). Please refer to \S\ref{FDPH5} for details. 

\subsubsection{Role of H-Matrices Applied to the Spatial Part of the Kernel in Domains I and S}
\label{FDPH322}
%Figure 4. âš
\begin{figure*}[tb]
   %\begin{center}
  \includegraphics[width=120mm]{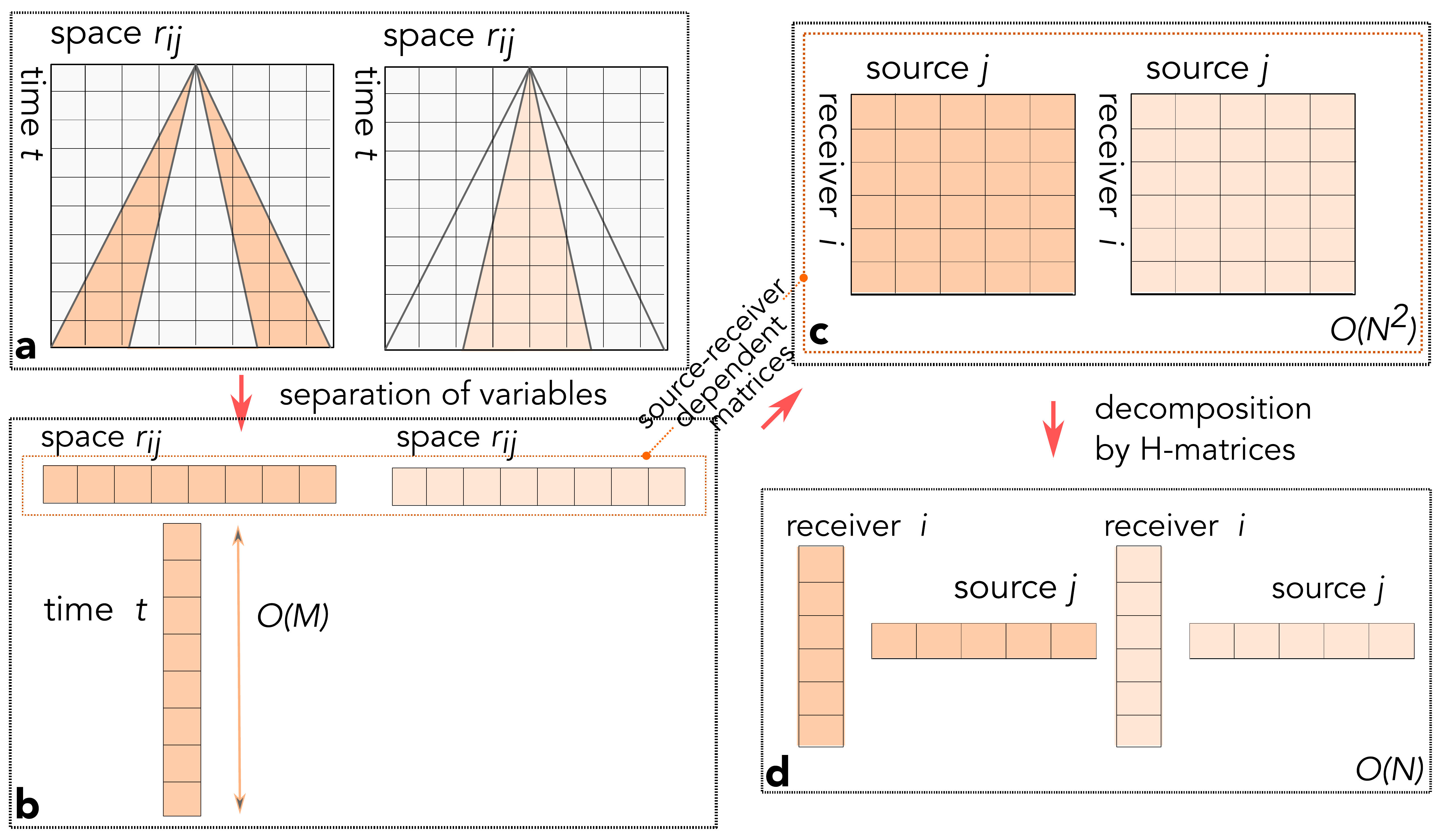}
  %\includegraphics[width=78mm]{DomainIandSred4.pdf}
   %\end{center}
\caption{
The approximation procedure in Domains I and S. 
{\bf a}, Spatiotemporal area belonging to Domains I and S. 
{\bf b}, Spatiotemporal separation of the kernel by the FDPM. 
The kernel in Domain I is separated into a matrix representing the space dependence of the kernel and a vector representing the time dependence. The kernel in Domain S is reduced to a matrix depending on the space. 
{\bf c}, Matrix structures representing the space dependencies of the kernel in Domains I and S. Their numbers of entries are $\mathcal O (N ^ 2)$. 
{\bf d}, The kernel after approximated by H-matrices. Each source-receiver-dependent matrix in Fig.~\ref{FDPHfig:4}c is separated into two vectors depending on the sources or receivers. The numbers of components to express the kernel in Domains I and S become almost $\mathcal O (N)$. 
}
\label{FDPHfig:4}
\end{figure*}

The FDPM separates the kernel ${\bf K}^I$ of Domain I into space-dependent terms ${\bf \hat K}^I$ and time-dependent terms $h^I$ in-between the P- and S-waves [Fig.~\ref{FDPHfig:4}a to Fig.~\ref{FDPHfig:4}b, and also as Eq.~(\ref{FDPHeq:factorizedformofkernelofDomainI})]. The kernel ${\bf K}^S$ of Domain S takes a time-invariant form ${\bf\hat K}^S$ after the passage of the S-wave [Eq.~(\ref{FDPHeq:factorizedformofkernelofDomainS})]. 
${\bf \hat K}^I$ and ${\bf \hat K}^S$ are both the matrices that depend on receivers $i$ and sources $j$, 
and H-matrices separate them into receiver-$i$-dependent vectors and source-$j$-dependent vectors (Fig.~\ref{FDPHfig:4}c). 

The rank of ${\bf\hat K}^S$ lowers to $\mathcal O(1)$ with H-matrices given its elastostatic nature.  
The rank of ${\bf \hat K}^I$ also lowers to $\mathcal O(1)$ given a numerical observation in Ref.~\cite{ando2016fast} that ${\bf\hat K}^I$ is a geometrically spreading function as ${\bf \hat K}^S$ is; it follows directly from the geometrical-spreading natures of the P-wave and near-field term \cite{aki2002quantitative} that constitute ${\bf \hat K}^I$. 
After the LRA, the memory to store ${\bf K}^I$ and ${\bf K}^S$, which is $\mathcal O [N ^ 2+L/(\beta\Delta t)]$ in the FDPM, reduces to $\mathcal O (N\log N)$ (Fig.~\ref{FDPHfig:4}d). 
The time complexity for the associated tensor-matrix products per time step is also reduced to $\mathcal O(N\log N)$ by the use of certain arithmetics, detailed in \S\ref{FDPHB1} and \S\ref{FDPHB2}. 

%The Domain I kernel is originally convolved over the sources and the time steps for respective receivers and takes the $\mathcal O (N ^ 2 )$ computation time. When H-matrices and the ART separates the source-receiver dependence of the kernel, the time taken to compute the stress of Domain I is reduced to be of $\mathcal O(N)$ (detailed in \S\ref{FDPHB2}). %The $\mathcal O(NM)$ cost corresponds to the convolution of the slip rate and the kernel in which the spatial convolution is made efficient by H-matrices (Fig.~\ref{FDPHfig:4}c) while the temporal convolution remains (Fig.~\ref{FDPHfig:4}b). 

%The time complexity of Domain S becomes of almost $\mathcal O (N)$ by using H-matrices and the ART. The independence from the factor $M$ is resulted from the time independence of the kernel in Domain S (detailed in \S\ref{FDPHB1}). 

\subsubsection{Role of Temporal Quantization in Domain I}
\label{FDPH323}
%Figure 5. âš
\begin{figure*}[tbp]
   \begin{center}
  \includegraphics[width=100mm]{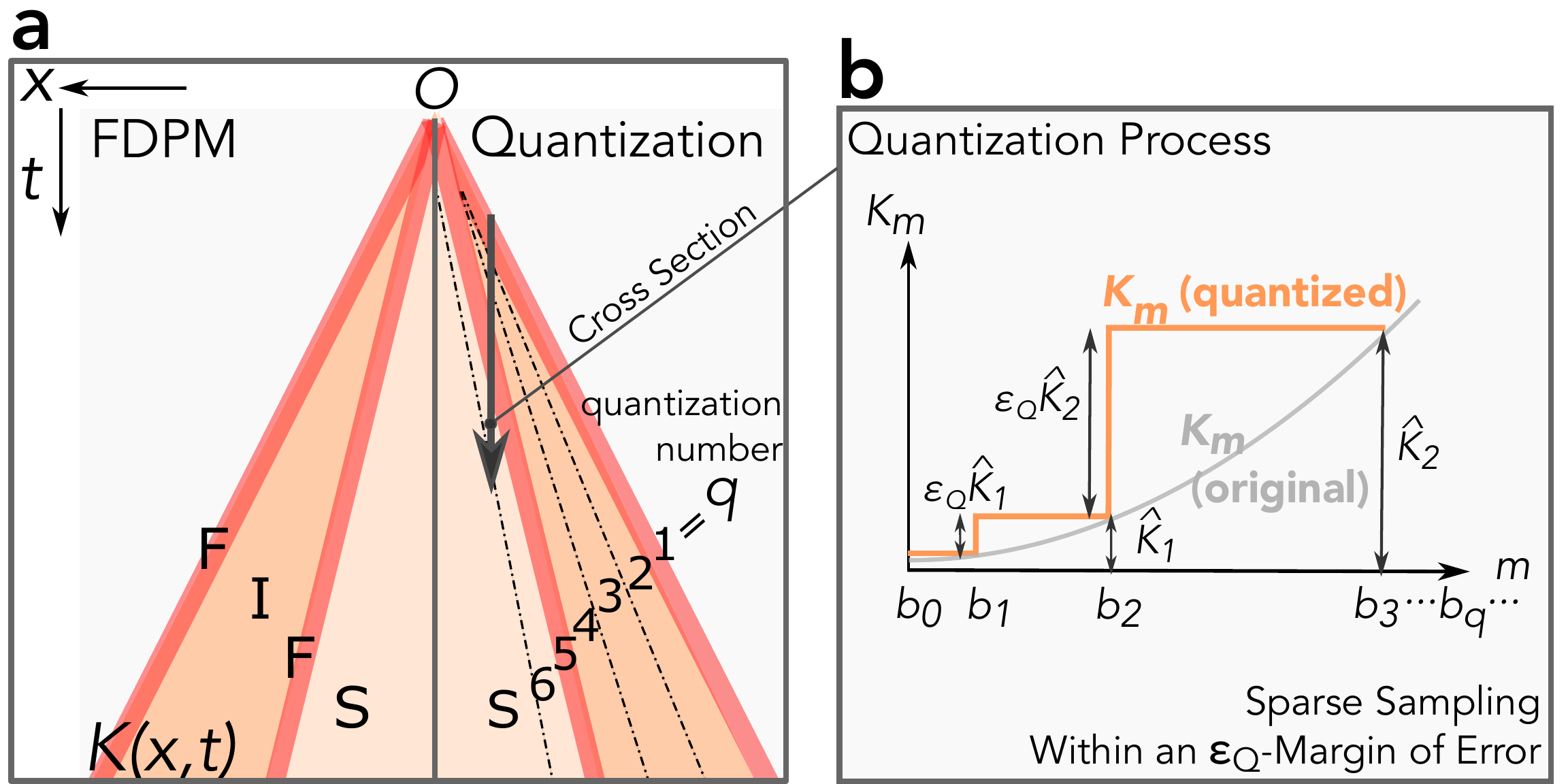}
   \end{center}
\caption{
Schematic of Quantization. {\bf a}, Quantization compared to the FDPM. The domain partitioning of the FDPM is shown in the left side, setting Domains F, I, F, and S. The associated partitioning of Quantization is in the right side, setting multiple time segments until the time step allowing the kernel to be replaced with its static limit. {\bf b}, Approximation of Quantization. The illustrated kernel corresponds to that in Domain I. 
The panel depicts the replacement of the original kernel value $K$ by the representative one $\hat K$ in each time step range of quantization number $q$ within relative error bound $\epsilon_Q$. 
$\hat K$ is given as $K$ at the end of each range in the panel. 
The time step $b_q$ partitioning the subdomains of Quantization is also indicated for each $q$ value. 
}
\label{FDPHfig:5}
\end{figure*}

%The memory consumption required to evaluate the stress in Domain I is reduced to be of almost $\mathcal O (N)$ by Quantization in FDP=H-matrices (detailed in \S\ref{FDPHB23}). 
The kernel outside Domain F becomes a sum of power functions of time like the near-field term proportional to time. 
In such a case, the LRA works as efficiently as in the case of a geometrically spreading kernel being a power function of distance. 
Then, like the PWTD method introducing the hierarchical decomposition of time~\cite{ergin1998fast}, 
we can consider some efficiently-working temporal LRA supposing subdomains adapting their intervals to the number of the elapsed time step ($m$) [Fig.~\ref{FDPHfig:5}a]. 
Quantization determines such subdomains by using an error criterion and executes the LRA in a piecewise-constant manner [Fig.~\ref{FDPHfig:5}b]. 
Quantization can be used additionally for reducing the memory consumption in Domain I in the algorithm of FDP=H-matrices. 
%, detailed in \ref{FDPHA}, 

The sampling interval of Quantization is maximized provided that the relative error is within $\epsilon_Q$. 
The original kernel is replaced with a sampled value $\hat K_q\in\mathbb R$ in each interval for quantization number $q$. 
For the case where the kernel is a power function of time (e.g., $t^\gamma$ with a constant $\gamma\in\mathbb R$),  
this sampling becomes sparse as the elapsed time step increases  
because the rate of the relative change of the kernel is a decreasing function of time 
[$(dt^\gamma/dt)/t^\gamma=\gamma/t$]; consequently, the assigned time domain decomposition becomes similar to the hierarchical decomposition supposed in the PWTD method widening the interval of the subdomain at a large time step. 
The kernel of Domain I, being a sum of functions proportional to $t^2$ or $t^0$ (for the regularized double-layer potential) ~\cite{ando2016fast}, gives such an example. 
We can also impose bound $\epsilon_{st}$ on the absolute error without changing the asymptotic cost order (\ref{FDPHA}). 

The above staircase approximation of the kernel reduces temporal convolution $\sum_{m=b_q}^{b_{q+1}-1} K_m D_m$ to the product of $\hat K_q$ and the slip and opening $\hat D_q:=\sum_{m=b_q}^{b_{q+1}-1} D_m(\in\mathbb R)$ in time-step range 
$b_q \leq m <b_{q+1}$:
\begin{equation}
\sum_{m=b_q}^{b_{q+1}-1} K_m D_m\simeq \hat K_q \sum_{m=b_q}^{b_{q+1}-1} D_m=\hat K_q\hat D_q, 
\label{FDPHeq:11}
\end{equation}
where the trivial suffixes about $n$ are omitted. By storing $\hat D_q$ over $q$ and evolving $\hat D_q$ at each time step under its incremental updating rule, Quantization makes the direct temporal convolution over $m$ unnecessary (\ref{FDPHA1}). 

%After the above-mentioned, the time-step-dependence of the numerical costs are reduced to the dependence of the number of sampling. 
%This situation is the same as Domain S in 3D problems whose kernel is time-independent. 
 
When the time-dependent parts of the kernel separate into the power functions of time as in Domain I, the number of the piecewise-constant basis made by Quantization is scaled by the logarithm of the time range to be quantized. Please refer to \ref{FDPHA} for details. 
This reduces the memory area required by the computation of Domain I to a quqsilinear order for various boundary geometries, as detailed in \ref{FDPHB23}.

\section{Data-Sparse Approximations in Domain F Using H-Matrices, ART, and Discretization}
\label{FDPH4}
We get into the detail of FDP=H-matrices overviewed in the previous section. 
As various fast BIE algorithms do, FDP=H-matrices also comprise a data-sparse approximation of the kernel (reducing the memory to store the kernel) and an associated fast and memory-efficient convolution operation of the BIE. 
We here show the approximation of the kernel, or more precisely, the approximation of the BIE. 
The associated key formulas are summarized in 
Table~\ref{KeyFormulas1}. 

Our main concern in this section is to approximate 
the following BIE convolved over Domain F (=Fp, Fs):
\begin{equation}
T^F_i (t):=\sum_j \int^{t_{ij}^+}_{t_{ij}^-} d\tau K^F_{i,j}(\tau) D_j(t-\tau).\nonumber
\end{equation} 
The approximation of this represents the essential part in incorporating H-matrices into the FDPM, and the ART is naturally entailed in it. 
The convolution over Domain F fully involves the above-mentioned singular points along the P- and S-waves, and hence the approximation of this BIE fully comprehends the previously known problem of H-matrices in the wave equation, the main issue of this study. 

In the present study, we do not detail the data-sparse approximations in Domains I and S previously investigated. 
In the algorithm of FDP=H-matrices, 
H-matrices in Domain S are applied to the spatial dependence ${\bf \hat K}^S$ of the kernel, which is exactly the kernel in the spatial BIEM. H-matrices in Domain S are then the same as those of the elastostaic problems. 
H-matrices in Domain I are also applied to the spatial dependence ${\bf \hat K}^I$ of the kernel and work like those of Domain S, given that ${\bf \hat K}^I$ follows the geometrical spreading as ${\bf \hat K}^S$ does (mentioned in \S\ref{FDPH322}). 
Indeed, the LRA of H-matrices has worked successfully in both Domains I and S in the previous studies, for example, in Ref.~\cite{yoshikawa20152}. 

\subsection{Application of H-Matrices to Domain F and Their Accuracy Control in the LRA}
\label{FDPH41}
%Figure 6. âš
\begin{figure}[tbp]
   \begin{center}
  \includegraphics[width=70mm]{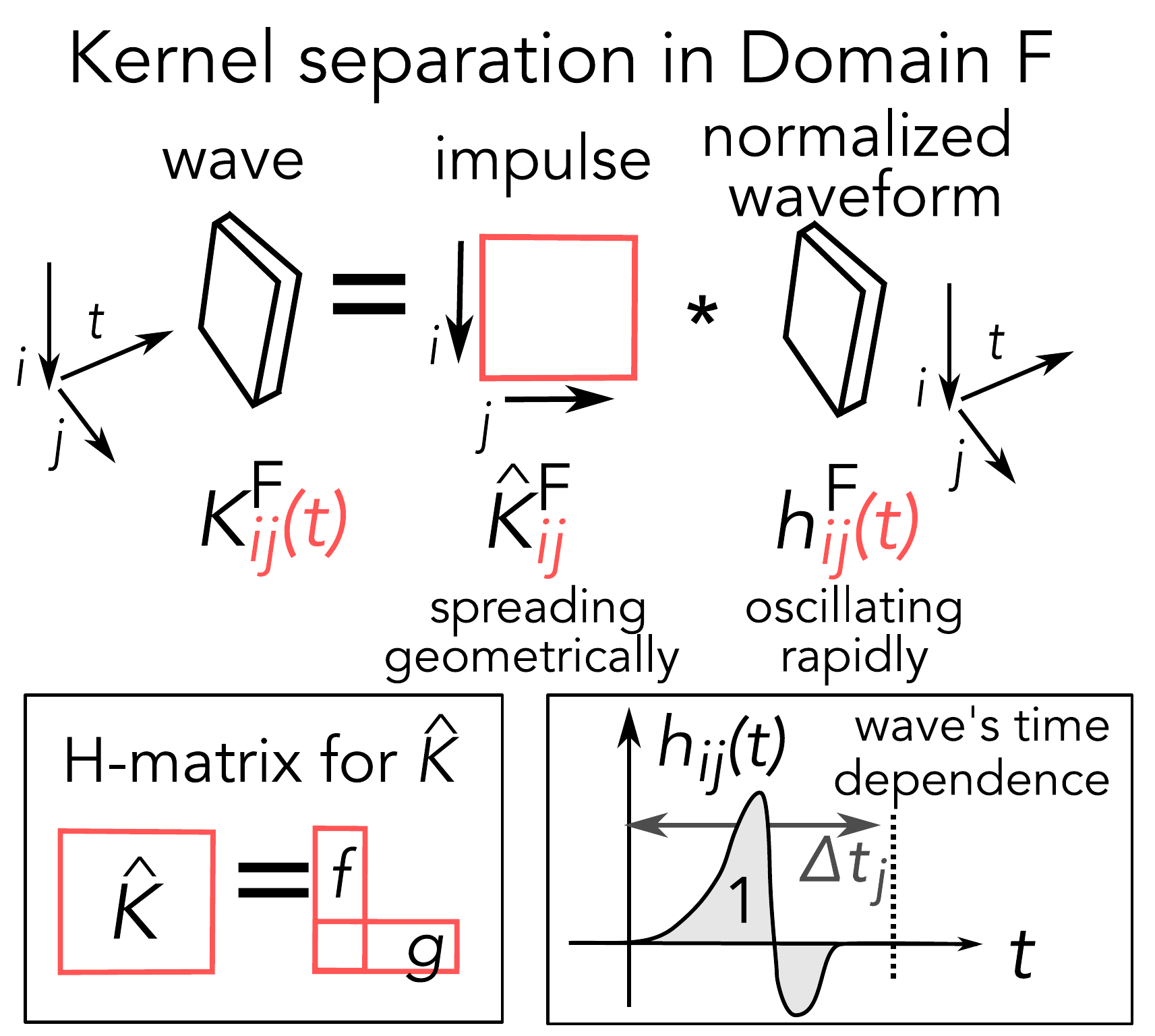}
   \end{center}
\caption{
Separation of the kernel in Domain F.
(Top) The kernel [$K_{ij}^F(t)$] (corresponding to a wave) separated into geometrically spreading $\hat K_{ij}$ (an impulse) and the temporally oscillating $h_{ij}(t)$ (the normalized waveform) for each source $i$ and receiver $j$ over time $t$. (Bottom left) H-matrices applied to $\hat {\bf K}$. (Bottom right) Temporal behaviors of the normalized waveform. Its time integral is set at 1. The time range giving nonzero $h_{ij}(t)$ values is within the time definition range of Domain F, the duration of which is $\Delta t_j$ for each source $j$. %The width of the former for a receiver is bounded by the width of the latter for the receiver...not accurate after the broadening. 
}
\label{FDPHfig:6}
\end{figure}

The singular points of the elasodynamic kernel constitute two spheres (wavefronts) that propagate at the speeds $c(=\alpha,\beta)$ of the P- and S-waves. The coefficients of their delta functions represent the amplitudes of the waves and decay geometrically as power functions of the distance~\citep{ando2016fast,aki2002quantitative}, analogously to the elastostatic kernel. 
The approximation in Domain F then begins with formulating the LRA along the wavefronts as a perturbation series in $diam$ $/$ $(diam+dist)<1/(1+1/\eta)$. 
This formulation is the same as in the H-matrices of the spatial BIEM and thus ensures that H-matrices work along the wavefronts as in the spatial BIEM. 

In roughing out the formulation, we start with the 3D Green's function ${\bf G}^P({\bf x},t)\in\mathbb R^{D_v\times D_v}$ of the P-wave for relative location ${\bf x}$ and time $t$. 
The space-time-dependence of ${\bf G}^P$, given as $G^P_{ab}({\bf x},t):=(4\pi r\rho\alpha^2)^{-1} \gamma_a\gamma_b\delta(t-r/\alpha)$ in a tensorial manner, is expressed by orientation dependence $\gamma_a\gamma_b$, geometrical spreading $r^{-1}$, and the delta function $\delta(t-r/\alpha)$ depending on time. 
The orientation dependence and the geometrical spreading are similar to the static kernel hence favorable, and the remaining delta function is the problematic singularity repeatedly mentioned. 
To eliminate this delta function, we consider the time integral of ${\bf G}^P$, $\int dt {\bf G}^P=(4\pi r\rho\alpha^2)^{-1} \gamma_a\gamma_b$, that is the ``impulse'' of ${\bf G}^P$. 
$\int dt {\bf G}^P$ does not contain the delta function anymore and time-independent, expressing only the orientation-dependent geometrical spreading as the static kernel does. 
Therefore, we can obtain a (fast convergent) Taylor series 
of $\int dt {\bf G}^P({\bf x},t)$ in ${\bf x}$ in the vicinity of the reference value ${\bf x}_0$, given the same logic as in \S\ref{FDPH23} for the static kernels. 
According to the ordinary H-matrices literature~\cite{borm2003hierarchical} mentioned in \S\ref{FDPH23}, such a Taylor series of $\int dt {\bf G}^P$ ensures we obtain its degenerate form: 
\begin{equation}
\int dt {\bf G}^P({\bf x}_{i0}-{\bf x}_{j0},t)
=
\sum_l c_l({\bf x}_i-{\bf x}_{i0})^{p_{1,l}} ({\bf x}_j-{\bf x}_{j0})^{p_{2,l}},
\nonumber
\end{equation}
for receiver $i$ and source $j$ around neighboring locations ${\bf x}_{i0}$ and ${\bf x}_{j0}$, where constants $c_l$, $p_{1,l}$, $p_{2,l}$ at respective effective ranks $l$ are defined in the same manner as the original H-matrices in \S\ref{FDPH23}.
Given this simplicity and the guaranty of the degenerate form, we choose such an impulse form for applying H-matrices, rather than the original form of the Green's function varying over both time and space. 

On the analogy of $\int dt {\bf G}^P$, we introduce the time integral of the kernel ($\hat K_{i,j}\in\mathbb R$, hereafter called an amplitude term) in Domain F (=Fp, Fs):
\begin{equation}
\hat K^F_{i,j}:= \int_0^{\infty} dt K^F_{i,j}(t).
\label{FDPHeq:12}
\end{equation}
We then apply H-matrices to $\hat K^F_{i,j}$ for receiver $i$ and source $j$ as $ij$ entries of matrix ${\bf\hat K}^F$: 
\begin{equation}
{\bf\hat K}^F\simeq \sum_a\sum_l {\bf f}^F_{al} ({\bf g}^F_{al})^T,
\label{FDPHeq:13}
\end{equation}
where ${\bf f}^F_{al}$ and ${\bf g}^F_{al}$ denote
column and row vectors, respectively, associated with the $l$-th largest singular values of ${\bf\hat K}^F$ subdivided for respective admissible leaves $a$, as in the H-matrices of the static problems treated in \S\ref{FDPH23}. 
$\hat K^F_{i,j}$ in Eq.~(\ref{FDPHeq:12}) is exactly a time integral of the kernel over Domain F [$t\in(t_{ij}^-,t_{ij}^+)$, introduced in \S\ref{FDPH22}], as explicitly expressed later as Eq.~(\ref{FDPHeq:28}). 
Recalling the example of ${\bf G}^P$ (or more simply $r^{-1}\delta (t-r/c)$ of the wave equation), we can regard Eq.~(\ref{FDPHeq:13}) as the expansion of geometrically-spreading $\int dt {\bf G}^P$ (the expansion of $1/r$), comparable with that in the PWTD methods~\cite{takahashi2003fast,ergin1998fast} for the elastodynamic and wave-equation problems. 
In summary, the above suite of the definition and the expansion can give the geometrically-spreading kernel and hence its degenerate form~\cite{borm2003hierarchical} in Domain F (the elastodynamical case of which is explicitly shown in Ref.~\cite{takahashi2003fast},  supplemented in \S\ref{FDPHF2}). 
The rank of the low-ranked form of ${\bf \hat K}^F$ is hence $\mathcal O(1)$ for respective Domains Fp and Fs in each admissible leaf $a$, given the existence of the degenerate form of ${\bf\hat K}^F$ as in the case of the elastostatic kernel. 
Compared to the FMM that considers the term-by-term expansion of the kernel, the above equations simply target the numerical numbers taken by the kernel and pass them to H-matrices as a matrix. 
In this manner, the impulsive coefficients of the dynamic kernel corresponding to the singular points, which could have been handled only analytically as in the PWTD method, becomes quite simply compatible with the formulation of H-matrices, executable completely algebraically that is fully numerically. 

Subsequently, we describe the original kernel with $\hat K_{ij}$ by introducing the following normalized kernel $h^F_{ij}(t)(\in\mathbb R)$ to Domain F (=Fp, Fs) for receiver $i$ and source $j$: 
\begin{equation}
h^F_{ij}(t):= K^F_{i,j}(t+t_{ij}^{-})/\hat K^F_{i,j},
\label{FDPHeq:14}
\end{equation}
where the time origin of $h^F_{ij}(t)$ is shifted by $t_{ij}^{-}$ ($:=t_{ij} - \Delta t_j^{-}$, first appearing in \S\ref{FDPH22}) from that of ${\bf K}^F(t)$, for the approximation of the ART shown in \S\ref{FDPH42}. 
Hereafter, we refer to $h^F_{ij}(t)$ as the normalized waveform. The normalized waveform satisfies the normalization condition: $\int dt h^F_{ij}(t)$ $=1$. 
The time range giving nonzero $h_{ij}(t)$ values is fully covered by Domain F, 
and the duration of such a time range is equal to or smaller than the duration $\Delta t_j$ [defined in Eq.~(\ref{FDPHeq:durationofDomF})] of Domain Fp and Fs for each source $j$. 

After the LRA of ${\bf \hat K}^F$ provided by H-matrices is applied, the BIE for $T^F_i(t)$ convolved over Domain F is expressed as 
\begin{equation}
T^F_i (t)\simeq
f^F_i \sum_j g^F _j \int_0^{\Delta t_j}d\tau h^F_{ij}(\tau) D_j (t-t_{ij}^--\tau),
\label{FDPHeq:16}
\end{equation}
where we omitted the rank and leaf number of $f^F_i$ and $g^F_j$ and related summations for brevity. 
The remaining dependence of normalized waveform $h^F_{ij} (t)$ on the pair of receiver $i$ and source $j$ is dealt with by a plane-wave approximation in the next subsection. This is for handling the rapidly oscillating nature of $h^F_{ij} (t)$, which makes itself difficult to be expanded by the LRA techniques (suitable for slowly functions) adopted in H-matrices. 

\subsection{ART}
\label{FDPH42}
In the previous subsection, we referred to that the coefficients of the delta functions in the elastodynamic kernel, representing the wave amplitudes, also follow the geometrical spreading as the static kernel. We then introduced the time integral of the kernel to extract these as matrices to which we apply H-matrices. 
On the other hand, Eq.~(\ref{FDPHeq:16}) contains the travel time $t_{ij}$ and normalized waveform $h^F_{ij}(t)$ [defined in Eqs.~(\ref{FDPHeq:5}) and (\ref{FDPHeq:14}), respectively]. 
They depend on the pair of the receiver $i$ and source $j$ even after $\hat K^F_{i,j}$ is decomposed into $f^F_i$ and $g^F_j$ and require the $\mathcal O(N^2)$ memory, also implying the $\mathcal O(N^2)$ computation time. 
We could tell by the analogy with $r^{-1}\delta(t-r/c)$ that we end up requiring an expansion for $\delta(t-r/c)$ besides the expansion of $1/r$. 
These $t_{ij}$ and $h^F_{ij}(t)$ values depending on $i,j$ pairs can be expressed by the terms that depend on either the receiver $i$ or source $j$ for each admissible leaf [i.e., by the totally $\mathcal O(N\log N)$ components], based on a series of plane wave approximations termed the ART shown below. 

The ART is based on a plane-wave approximation outlined in \S\ref{FDPH420}. 
We then formulate it with the spatial sorting of elements in \S\ref{FDPH421}. 
The ART provides two schemes that have different error bounds described in \S\ref{FDPH422}.

\subsubsection{Overview of the Plane-Wave Approximation}
\label{FDPH420}

%Figure 7_separated_a. âš
\begin{figure*}[tb]
   %\begin{center}
  %\includegraphics[width=80mm]{DegeneratedRay6.png}
  %\includegraphics[width=80mm]{DegeneratedRay8.pdf}
  \includegraphics[width=120mm]{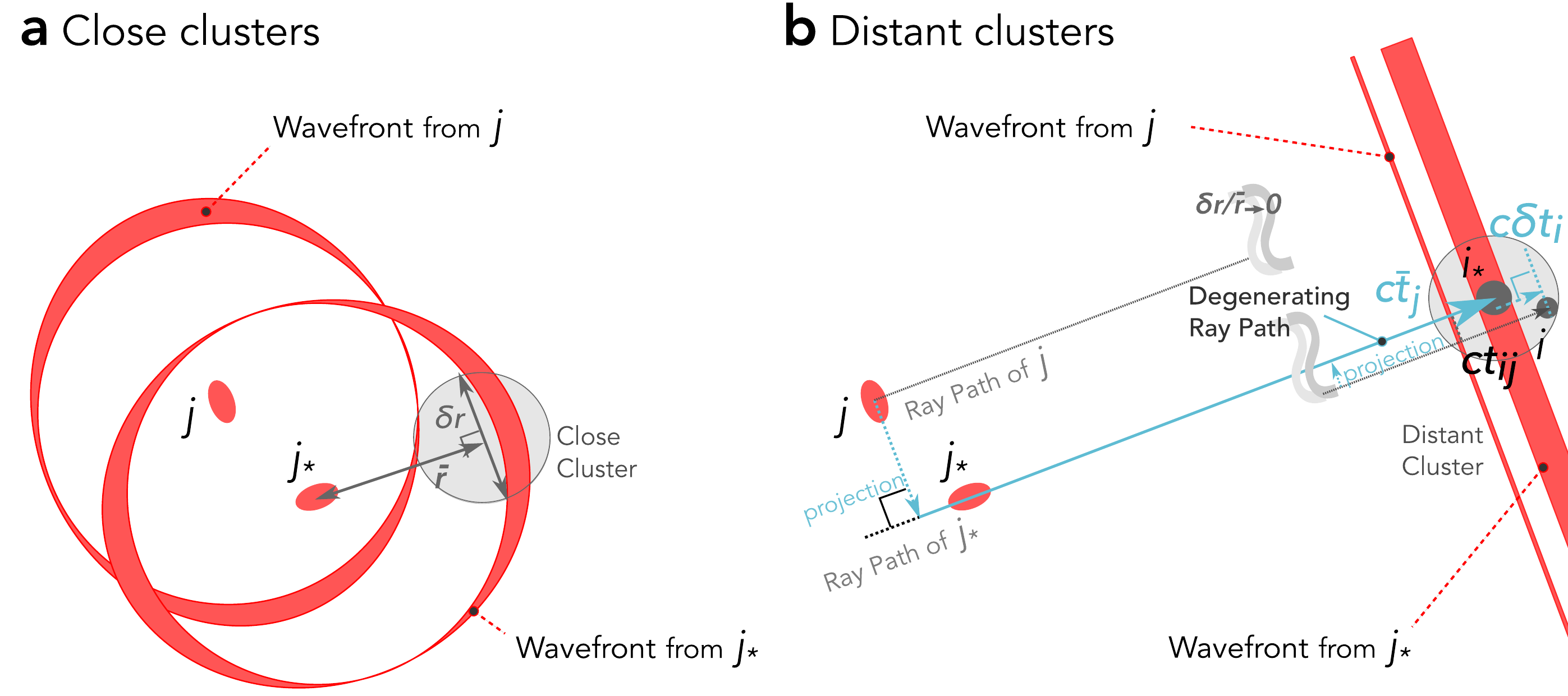}
   %\end{center}
\caption{
Schematic of a plane wave approximation adopted in the ART. 
Source and receiver sets are depicted with the associated wave radiation, expressed by ray paths and wavefronts. 
The distance of receiver $i$ and source $j$ connected by the ray path expresses $t_{ij}$ ($\times c$, where $c$ denotes the wave speed). 
The thickness of the wavefronts does the duration ($\times c$) of the time range giving non-zero values of the normalized waveform $h_{ij}$. 
%Cluster size $\delta r$ vertical to the degenerating ray path indicates the associated perturbation parameter, after divided by distance $\bar r$ between representative receiver $i_*$ and representative source $j_*$, located at the centers of the clusters. 
{\bf a}, 
The case of the close clusters. 
The cluster diameter $\delta r$ is comparable with distance $\bar r$ between the cluster centers $i_*$ and $j_*$.
The panel shows that the thickness of the wavefronts varies significantly 
within the receiver cluster, implying the $ij$ dependence of $h_{ij}$.
%The $ij$ dependence of $t_{ij}$ corresponds to the clear $ij$ dependence of the distance between receiver $i$ and source $j$. 
{\bf b}, 
The case of  the distant clusters. 
The limit of the scale separation $\delta r/\bar r\to 0$ is considered. 
The panel shows that the waves become flat and the variations in the wavefront thickness become negligible within the receiver cluster, implying asymptotic receiver-$i$-independence of $h_{ij}$ [$h_{ij}(t)\approx h_j(t)$]. 
The ray path from $j$ to $i$ is compared with the reference ray path from $j_*$ to $i_*$, termed the degenerating ray path (the DRP). 
The line connecting $i$ and $j$ is projected onto the DRP and the travel time separates asymptotically on the DRP into the travel time $\bar t_j$ from $j$ to $i_*$ and the correction $\delta t_i$ of the travel time between $i_*$ and $i$  ($t_{ij}\approx \delta t_i+\bar t_j$). 
}
\label{FDPHfig:7_a}
\end{figure*}

Fig.~\ref{FDPHfig:7_a} illustrates the basics of the plane-wave approximation and the ART. 
We suppose two clusters gathering neighboring receiver elements ($i$) and source elements ($j$). 
We then set representative receiver $i_*$ virtually at the centers of receivers and representative source $j_*$ likewise. 
We then consider a condition where waves that express the kernel in Domain F are radiated from sources $j$ and are reaching to receivers $i$. 
Fig.~\ref{FDPHfig:7_a} depicts the wave surfaces at fixed time (wavefronts) as well as a part of the source clusters and the receiver clusters. 
Travel time $t_{ij}$ is indicated in the figure by source-receiver distance $r_{ij}$ for a pair of receiver $i$ and source $j$ excluding its normalization factor, wave speed $c$. 
Normalized waveform $h^F_{ij}$ is by the finite thickness of the wavefronts. 
%The handling of these quantities, shown below, are later combined with H-matrices by connecting the clusters with the block cluster of an admissible leaf of H-matrices. 

We see from Fig.~\ref{FDPHfig:7_a} that $i,j$ dependencies of $t_{ij}$ and $h_{ij}$ are affected by the distance between the clusters. 
Receiver($i$)- and source($j$)- dependencies of $h_{ij}$ are related with the varying widths of the circles. Those of $t_{ij}$ are trivially those of $r_{ij}$.
These dependencies are clear particularly when the receivers and sources are relatively close (Fig.~\ref{FDPHfig:7_a}a). 
In contrast, at a distance where the wavefront becomes sufficiently flat, the widths of circles become independent of $i$, i.e., $i$ dependence of $h_{ij}$ cancels  asymptotically (Fig.~\ref{FDPHfig:7_a}b). 
All the rays go through almost the same path there, and the $i$ dependence of distance $r_{ij}$ becomes asymptotically a relative shift from that of the reference $i_*$, i.e., the $i,j$ dependence of $t_{ij}$ separates. 
These are collectively known as the plane-wave approximation~\cite{aki2002quantitative}, which is a perturbation theory concerning the ratios of the cluster diameters to cluster distances of sources and receivers.

In an asymptotic region, as the wavefront becomes flat, normalized waveform $h^F_{ij}$ loses the receiver $i$ dependence and is replaced by that for representative $i_*$ of the receivers in the cluster:
\begin{equation}
h^F_{ij}(\tau)\approx h^F_j (\tau):= h^F_{i_*j}(\tau).
\label{FDPHeq:20}
%\label{FDPHeq:hijdeg}
\end{equation}
We call asymptotic function $h^F_j(t)\in\mathbb R$ the degenerating normalized waveform. 

The asymptotic ray paths for all the pairs of the receivers and sources in the clusters are parallel to a straight line connecting their representatives $i_*$ and $j_*$ (the thick arrow in Fig.~\ref{FDPHfig:7_a}), hereafter called degenerating ray path (the DRP). By projecting the relative locations of the sources and receivers to the DRP, the ART separates the receiver- and source-dependencies of the travel time as
\begin{equation}
t_{ij}\approx \delta t_i+\bar t_j,
\label{FDPHeq:tijsep}
\end{equation}
with 
\begin{flalign}
\delta t_i&= {\bf x}_{ii_*}/c \cdot {\bf x}_{i_*j_*}/r_{i_*j_*} 
\\
\bar t_j&:= r_{i_*j}/c.
\label{FDPHeq:19}
\end{flalign}
Scalar $\bar t_j\in\mathbb R$ describes the travel time from a source $j$ to the representative receiver $i_*$.
Scalar $\delta t_i\in\mathbb R$ describes the effective travel time for the distance of a receiver $i$ from $i_*$ measured along the DRP.
This definition of $\delta t_i$ in Eq.~(\ref{FDPHeq:18}) is hereafter modified to 
\begin{equation}
\delta t_i := (r_{ij_*} - r_{i_*j_*})/c, 
\label{FDPHeq:18}
\end{equation}
for better accuracy [quantified in Eq.~(\ref{FDPHeq:24})].
We call $\bar t_j$ receiver-averaged travel time and $\delta t_i$ receiver-dependent travel-time difference. 
Note that the definitions of $\delta t_i$ and $\bar t_j$ can be further modified slightly by $\mathcal O(\Delta t)$ for the simplification of arithmetics, as explained in \S\ref{FDPH432} and \S\ref{FDPHB25}. 

By substituting Eqs.~(\ref{FDPHeq:20}) and (\ref{FDPHeq:tijsep}) into Eq.~(\ref{FDPHeq:16}), and replacing $t$ with $t + \delta t_i$, we obtain
%Eq.~(\ref{FDPHeq:17}) and Eq.~(\ref{FDPHeq:21}) are substituted into Eq.~(\ref{FDPHeq:16}) (without the higher-order terms) as
\begin{equation}
T^F_i(t+\delta t_i)\approx f^F_i\sum_j g^F_j \int^{\Delta t_j}_0d\tau h^F_j(\tau)  D_j (t-\bar t_{j}^--\tau),
\label{FDPHeq:22}
\end{equation}
where $\bar t_j^-:=\bar t_j-\Delta t_j^-$.
Finally, the source and receiver dependencies fully separate in this convolution.

After seeing the above discussion, one may notice the similarity between the plane-wave approximation and the far-field approximation. The far-field approximation is an asymptotic expansion that takes only the leading term at a distance~\cite{aki2002quantitative,colton2013integral}. For example, it gives $G= ...1/r\delta (t-r/c) +\mathcal O(1/r^3)$ for the Green's function, or equivalently, $G= ...1/r\exp(ik(\omega) r) +\mathcal O(1/r^3)$ in the frequency $\omega$ domain, where $k(\omega)=\omega/c$; in this example, the plane-wave approximation is $\delta(t-r/c)= \delta[t-(\bar r+\delta r)+\mathcal O(\bar r\eta^2)]$~\cite{aki2002quantitative}, or equivalently, $\exp[ik(\omega)r]=\exp[ik(\omega)(\bar r+\delta r+\mathcal O(\bar r\eta^2))]$, where we used ($\bar r,\delta r,\eta$) in the nomenclature of H-matrices. 
Both the far-field and the plane-wave approximations can be regarded as small parameter expansions in $1/r$, and 
the far-field approximation is a term referring to the expansion of the amplitude while the plane-wave approximation referring to that of the phase. 
We used the LRA of H-matrices instead of the far-field approximation (as indeed the kernel in Domain F involves the contribution from the near-field term), and only the phase is the object of the asymptotic expansion in FDP=H-matrices. 
Having said that, one finds that the use of the degenerating normalized waveform tends to involve a sort of far-field approximations in considering the non-impulsive terms of the kernel in Domain F (in the next subsubsection, although that intricacy is supplemented only in \S\ref{FDPHC1}).

\subsubsection{Plane-Wave Approximation for Spatially Sorted Elements}
\label{FDPH421}

%Figure 7_separated_b. âš
\begin{figure}[tbp]
   %\begin{center}
  \includegraphics[width=80mm]{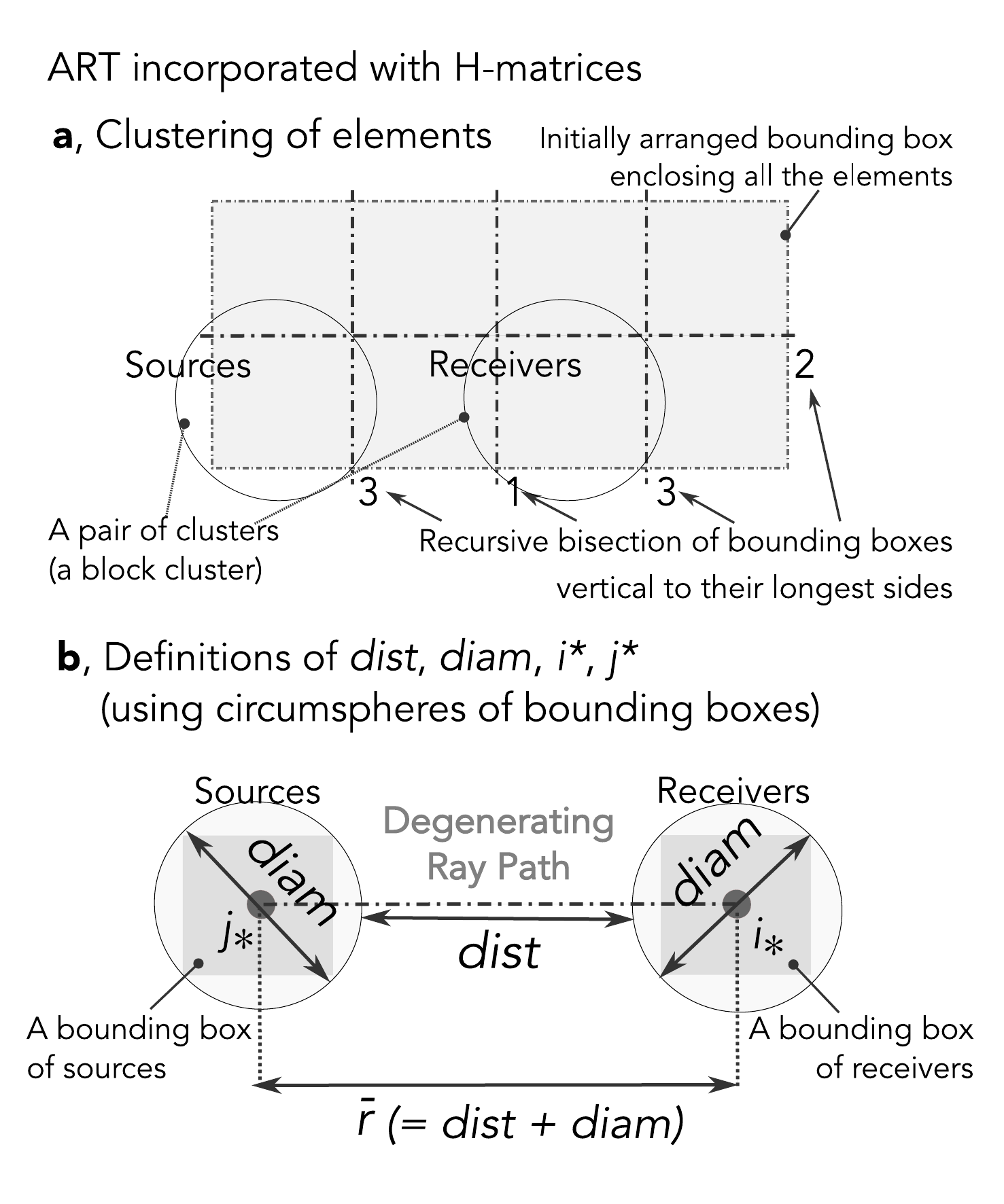}
   %\end{center}
\caption{
Parametrization of the ART using circumspheres of bounding boxes in H-matrices. 
{\bf a}, Clustering of elements. 
H-matrices we employ in FDP=H-matrices arrange a bounding box to enclose all the elements, and create new pairs of bounding boxes (block clusters) recursively such that previous bounding boxes are bisected vertical to their longest sides. 
The level, the number of the division to which a cluster is subjected, is assigned to the related dividing lines in the panel. 
{\bf b}, Definitions of $dist$, $diam$, $i_*$, and $j_*$. 
The representative receiver $i_*$ and source $j_*$ are set at the centers of spheres circumscribing the bounding boxes of the receiver and source clusters in each block cluster. 
The value of $dist$ is given by 
the diameter of the circumspheres (plus the maximum length of the discretized elements enclosed in the two bounding boxes, not illustrated). 
The value of $dist$ is given as $\bar r-diam$. 
}
\label{FDPHfig:7_b}
\end{figure}

The implementation of the ART follows the clustering of elements in H-matrices.
As the admissibility condition of Eq.~(\ref{FDPHeq:8}) is to ensure that source and receiver clusters are distant to certain extent, 
we can introduce the approximation of the ART (for the distant clusters in Fig.~\ref{FDPHfig:7_a}b) to the admissible leaves.
The ART does not apply to the inadmissible leaves (corresponding to the close clusters in Fig.~\ref{FDPHfig:7_a}a). %(See \ref{FDPHE}).

As referred to in \S\ref{FDPH23}, our implementation of H-matrices, adopts the clustering using the bounding boxes (cuboids in the 3D problems and rectangles in the 2D problems) (Fig.~\ref{FDPHfig:7_b}a). 
This implementation first sets an initial bounding box that encloses all the elements.
A related subset of elements (the cluster) is then defined as elements the centers of masses of which are located in a bounding box. 
We bisect the bounding box recursively by equally dividing its largest side, and define the related subsets recursively in the above-mentioned way. 
We also define the block clusters (pairs of the clusters) in a recursive manner that a parental block cluster generates four children with bisecting the two bounding boxes of source and receiver clusters constituting the parental block cluster. 

We introduce $i_*$, $j_*$, $diam$, and $dist$ to each admissible leaf in the following manner (Fig.~\ref{FDPHfig:7_b}b). 
%We outline this clustering process below to supplement the explanation in \S\ref{FDPH23}. Please refer to Ref.~\cite{borm2003hierarchical} for details. A bounding box is first prepared to enclose all the centers of elements, and new boxes are created recursively as previous boxes bisected vertical to their longest sides (Fig.~\ref{FDPHfig:7_b}). According to the bisection, we introduce and compute a subset of elements enclosed in a bounding box recursively. This clustering of elements allows us to cluster the source-receiver pairs in the following manner. First, a pair of source and receiver clusters (a block cluster) is uniquely given at the initial level (the number of the division to which a cluster is subjected), as the largest bounding box is the only one box to give the receiver or source clusters at the initial level. Then, we examine whether a block cluster satisfies the admissibility or inadmissibility conditions. If so, it is adopted as admissible or inadmissible leaves, respectively. If not, associated source and receiver clusters are bisected individually, and four pairs of bisected sources and receiver clusters are obtained as block clusters at the next level. We continue this test recursively in each level, as long as the block clusters exist at that level. 
The representatives $i_*$ and $j_*$ are set at the centers of cuboids for the receivers and sources, respectively (shown in Fig.~\ref{FDPHfig:7_b}). 
The value of $diam$ in H-matrices is given as the maximum diagonal length of bounding boxes plus the maximum length of the boundary elements enclosed in the boxes. 
The maximum diagonal lengths of cuboids take the same value for the receiver and source clusters in the above-mentioned implementation (shown in Fig.~\ref{FDPHfig:7_b}), as they necessarily belong to the same level (and then have the same shape and size in the above-mentioned implementation). 
The value of $dist$ is given as distance $\bar r$ between $i_*$ and $j_*$ (distance between the centers for the source and receiver cuboids) minus $diam$ ($dist=\bar r-diam$).
%Hereafter, for notational simplicity, we give the $diam$ value to a block cluster, as the average of the $diam$ values of corresponding source and receiver clusters; $diam$ later appeared in this paper represents a $diam$ value given to the block cluster. 
%Variables $dist$ and $diam$ are calculated to respective block clusters in the clustering process of H-matrices, and their calculation results in calculating the locations of $i_*$ and $j_*$ together. 
%These definitions of $dist$, $diam$, $i_*$, and $j_*$ is helpful for incorporating the aforementioned approximations of the ART with H-matrices.

The error of the ART is associated with the element configuration in the admissible leaves. 
In particular, the following error bound of the travel time is determined mostly by just the configuration of the bounding boxes. 
The bound comes from the admissibility condition $diam/dist < \eta$ [Eq.~(\ref{FDPHeq:8})] all the admissible leaves obey. 
We can rewrite the above admissibility condition as $diam/r_{i_*j_*}<(1 + \eta^{-1})^{-1}$ by using $\bar r=dist+diam$ and $r_{i_*j_*}=\bar r$, 
which are deduced from the aforementioned definitions of ($diam$, $dist$) and those of ($i_*$, $j_*$), respectively. 
Using this rewritten admissibility condition and further utilizing that $diam$ in our definition bounds the diameters of the circumspheres of the bounding boxes, 
we obtain the following perturbation series of the travel time in $(r_{ij}-r_{i_*j_*})/r_{i_*j_*}$:
\begin{equation}
t_{ij} = \delta t_i+ \bar t_j + \mathcal O\left [(1 + 1/\eta)^{-2} dist \right].
\label{FDPHeq:17}
\end{equation}
This shows the approximation of the travel time in Eq.~(\ref{FDPHeq:tijsep}) including its error terms.
The ART neglects the higher-order term in Eq.~(\ref{FDPHeq:17}) as $t_{ij}\approx \delta t_i+\bar t_j$.

We further estimate the error due to the approximation of Eq.~(\ref{FDPHeq:20}) that drops the receiver dependence of the normalized waveform. The associated error of the BIE fully comes from Eq.~(\ref{FDPHeq:16}) that convolves the normalized waveform temporally, and then it is enough to consider Eq.~(\ref{FDPHeq:16}) for the error estimates of the approximation of Eq.~(\ref{FDPHeq:20}) (as far as we consider the error estimates of the BIE). 
On one hand, the error is estimated to be of order 1) the variation of the azimuthal angle, being $\mathcal O[1/(1+1/\eta)]$ for an admissible leaf; it can also be of order 2) the source-receiver distance, also $\mathcal O[1/(1+1/\eta)]$
(Please refer to \S\ref{FDPHC1} for details). 
On the other hand, 
the associated error does not emerge when 
$D_j$ is constant within Domain F given the normalization condition Eq.~(\ref{FDPHeq:14}) of the normalized waveform: $\int dt h_{ij}^F(t)= \int dt h_{j}^F(t)=1$. 
That is,
the associated error is also of order the variation in $D_j$ within Domain F, $\mathcal O(\Delta t_j \partial_t D_j )$. 
Through the multiplication of these two, the error resulting from the convolution is  estimated as
\begin{flalign}
&\int_0^{\Delta t_j}d\tau h^F_{ij}(\tau)D_j(t-\tau-t^-_{ij})
\nonumber
\\=&\int_0^{\Delta t_j}d\tau h^F_{j}(\tau)D_j(t-\tau-t^-_{ij})
\nonumber
\\&
+\mathcal O[(1+\eta^{-1})^{-1} \Delta t_j\partial_tD_j].
\label{FDPHeq:21}
\end{flalign}
We note that this estimate is for the kernel being independent of the receiver orientation, such as the displacement nucleus and stress nucleus we consider. 
The projection of the stress to the traction, using the normal vector of the receiver element, then requires some caution. The error increases to $\mathcal O(\Delta t_j\partial_t D_j)$ when the normalized waveform is calculated carelessly to the kernel of the traction due to its receiver-orientation dependence (supplemented in \S\ref{FDPHC1}).

We also add that more precisely, the error (the third term) of the travel time in Eq.~(\ref{FDPHeq:17}) comes from the perturbation in the ratio $\delta r/\bar r$ of 1) cluster diameter $\delta r$ projected onto the DRP 
to 2) distance $\bar r$ between cluster centers, rather than from that in $diam/\bar r$. 
This results in that the error order is $\mathcal O[(\delta r)^2/\bar r)]$. 
Indeed, when all the sources and receivers are exactly on the DRP, 
the travel time exactly separates without any errors ($t_{ij}=\delta t_i +\bar t_j$).
This is a direct consequence of the triangle inequality of vectors, 
by considering that $t_{ij}$, $\delta t_i$, $\bar t_j$ are associated with the distances between $i$ and $j$, between $i$ and $i_*$ (along the DRP), and $i_*$ and $j$, respectively (See Fig.~\ref{FDPHfig:7_a}). 
A more detailed discussion can be found in Ref.~\cite{aki2002quantitative} although their nomenclature is different from ours. 

%Sometimes in other studies, to avoid using the center of mass of an element, $diam$ is defined as the longest distance of two points on the boundaries enclosed in each cuboids, and $dist$ as the shortest distance of a point on the boundaries enclosed in the source cuboids and that in the receiver cuboids~\cite{borm2003hierarchical}. Our definitions of $diam$ and $dist$ are respectively the upper bound of $diam$ and lower bound of $dist$ given by such definitions for the undiscretized boundary.
%%Cluster divisions are repeated until $diam$ becomes smaller than cutoff length $l_{min}$ being a parameter of H-matrices (mentioned in \S\ref{FDPH23}). 
%Therefore, the admissibility condition for such undiscretizing boundaries is satisfied when the admissibility condition we provide holds. 

\subsubsection{Two Admissibility Conditions of H-matrices in Regulating the Error Due to the Travel-Time Approximation}
\label{FDPH422}
%Figure 8. âš
\begin{figure*}[tb]
   \begin{center}
  \includegraphics[width=130mm]{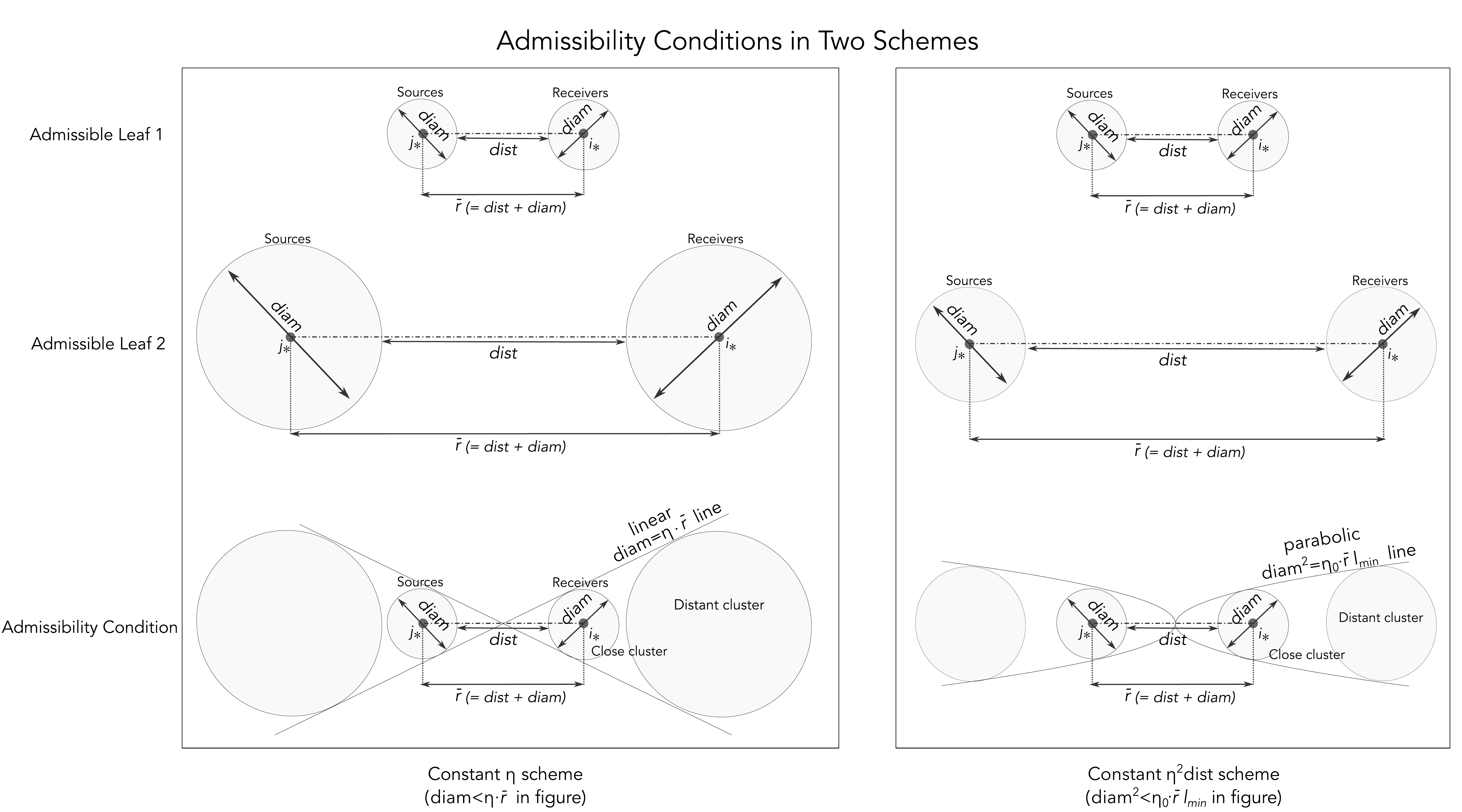}
   \end{center}
\caption{
Schematic of the two schemes in the ART varying $\eta$ of the admissibility condition $diam<\eta\cdot dist$, using two admissible block clusters (sufficiently distant two pairs of clusters) of different distances. The $dist$ value in the admissibility condition is replaced with the value of $\bar r=diam+dist$ in the figure for brevity. 
(Left) Constant $\eta$ scheme, assuming the lower bound of $dist$ as a linear function of $diam$. 
(Right) Constant $\eta^2dist$ scheme, assuming the lower bound of $dist$ as a parabolic function of $diam$.
}
\label{FDPHfig:8}
\end{figure*}

The error of $t_{ij}$ in Eq.~(\ref{FDPHeq:17}) is $\mathcal O[(1 + \eta^{-1})^{-2}dist]$ and diverges when $dist\to \infty$ for the cases of constant $\eta$ while the error in Eq.~(\ref{FDPHeq:21}) associated with $h^F_{ij}$ is regulated within a finite value with constant $\eta$. 
The handling of this error in Eq.~(\ref{FDPHeq:17}) gives us two schemes to incorporate the ART with H-matrices (illustrated in Fig.~\ref{FDPHfig:8}). 
Both are expressed by the admissibility conditions and are given by the distance ($dist$) dependence of the $\eta$ value. We call them constant $\eta$ scheme and constant $\eta^2dist$ scheme. They differ in accuracy and are comparable with the multi-level and two-level schemes in the PWTD method~\cite{ergin1998fast}, respectively. (The latter may be more similar to the single-level FMM in the frequency domain~\cite{chaillat2008multi}.) We note that all the estimates of the costs and accuracy in the paper are for the constant $\eta$ scheme unless we specify the other.

\paragraph{Constant $\eta$ Scheme}
The constant $\eta$ scheme assumes a constant $\eta$ value, which corresponds to the admissibility condition usually adopted in H-matrices~\cite{borm2003hierarchical}. 
This scheme achieves the $\mathcal O(N\log N)$ costs, as later discussed in \S\ref{FDPH6}. 
The constant $\eta$ scheme keeps the $diam/dist$ value $\mathcal O(\eta)$ regardless of the $dist$ value [Fig.~\ref{FDPHfig:8} (left)].

%In this scheme, the travel time separation [Eq.~(\ref{FDPHeq:17})] is regarded as an approximation of the (P and S) wave speeds as follows. In t
In the constant $\eta$ scheme, the error associated with the use of Eq.~(\ref{FDPHeq:17}) can be simply estimated for each pair of receiver $i$ and source $j$ by using a following quantity: 
\begin{equation}
c_{ij} :=r_{ij}/(\delta t_i +\bar t_j).
\label{FDPHeq:23}
\end{equation}
We call it effective wave speed. 
The error of effective wave speed $c_{ij}$ is expressed as
\begin{equation}
|c_{ij}/c - 1| < \frac 1 4 (1 + \eta^{-1})^{-2} + \mathcal O[(1 + \eta^{-1})^{-3}], 
\label{FDPHeq:24}
\end{equation}
by using original wave speed $c$. 
Eq.~(\ref{FDPHeq:24}) is obtained from the comparison between Eq.~(\ref{FDPHeq:17}) and the summation of Eqs.~(\ref{FDPHeq:19}) and (\ref{FDPHeq:18}) in a perturbative manner treating $1/(1+1/\eta)$ as a small parameter. 
Eq.~(\ref{FDPHeq:24}) shows that the error of an effective wave speed [of $\mathcal O((1 + \eta^{-1})^{-2})$] is kept finite without divergence at a distance even supposing the constant $\eta$ value while the error in the approximated travel time can be unbounded as mentioned earlier. 
Eq.~(\ref{FDPHeq:24}) enables us to regard the use of Eq.~(\ref{FDPHeq:17}) in the constant $\eta$ scheme as an approximation of the wave-speed of the $\mathcal O [(1 + \eta^{-1})^{-2}]$ accuracy. 

It may be an additional appeal that this scheme does not induce any numerical dispersity (the artificial wavelength-dependencies of the effective wave speed).
The wave-speed approximation has been verified well for the volume-based methods of the elastodynamic problems~\cite{day2005comparison,pelties2012three} while their simulated acoustic speed is dispersive~\cite{andrews1976rupture}. 
In the constant $\eta$ scheme of FDP=H-matrices, the wave-speed error shown in Eq.~(\ref{FDPHeq:24}) depends on the $\eta$ value and is independent of $dist$. This expresses negligible dispersity, which is examined in \S\ref{FDPH632}.

\paragraph{Constant $\eta^2dist$ Scheme}
The constant $\eta^2dist$ scheme assumes a constant $\eta^2 dist$ value, which is given by the following admissibility condition:
\begin{equation}
diam <\sqrt{\eta_0 l_{min} dist},
\label{FDPHeq:25}
\end{equation}
where $\eta_0$ is the maximum value of $\eta$ bounding the ratio $diam/dist$ ($diam/dist<\eta:=\sqrt{\eta_0 l_{min}/dist}$). 
Eq.~(\ref{FDPHeq:25}) geometrically implies that $dist$ is asymptotically proportional to the square of $diam$ [Fig.~\ref{FDPHfig:8} (right)].
The value of $\eta$ varies in this scheme, and is maximized (as $\eta=\eta_0,$ $diam<\eta_0 dist$) when $diam$ takes its minimum value $diam=l_{min}$ for the admissible leaves. The total computation cost of the constant $\eta^2dist$ scheme is estimated to be almost $\mathcal O(N^{3/2})$, numerically in Fig.~\ref{FDPHfig:9_b} and analytically in \S\ref{FDPHC2}.

This scheme [Eq.~(\ref{FDPHeq:25})] regulates the travel-time error of $\mathcal O[(1 + 1/\eta)^{-2}dist]$ in Eq.~(\ref{FDPHeq:17}) within a constant value as $\eta$ decreases in inverse proportion to the square root of $dist$ ($\eta \propto 1/\sqrt{dist}$). The travel-time error in the constant $\eta^2dist$ scheme
is evaluated as
\begin{equation}
|\delta t_i+\bar t_j- t_{ij}|<\frac 1 4 \eta_0 l_{min}/c + \cdots.
\label{FDPHeq:26}
\end{equation} 
We obtain this by substituting $\eta=\sqrt{\eta_0l_{min}/dist} $ into the inequality in Eq.~(\ref{FDPHeq:24}) for the constant $\eta$ scheme. The higher-order term in Eq.~(\ref{FDPHeq:26}) is of $\mathcal O [(1+1/\eta)^{-3}]$ as in Eq.~(\ref{FDPHeq:24}). Eq.~(\ref{FDPHeq:26}) shows the asymptotic independence of the travel-time error from $dist$ in the leading order term. 

By substituting Eq.~(\ref{FDPHeq:26}) in Eq.~(\ref{FDPHeq:21}), we estimate the error of the travel-time approximation as $\mathcal O(\partial_tD$ $\delta x/c)$ with defining a characteristic length $\delta x :=\eta_0 l_{min}/4$.
It then allows us to treat the travel-time approximation as 
an approximate time shift by $\delta x/c$ of the temporally convolved slip- and opening-rates $D$ in the constant $\eta^ 2dist$ scheme. 
Meanwhile, the other approximation error of the ART, that of $h^F$ in Eq.~(\ref{FDPHeq:21}), becomes $\mathcal O(1/\sqrt{dist})$ and vanishes asymptotically at a distance in this scheme. 

\subsection{Temporal Discretization of a BIE Convolved over Domain F}
\label{FDPH43}
Last, we obtain a temporally discrete form of Eq.~(\ref{FDPHeq:22}).
We consider the temporal collocation in \S\ref{FDPH432} by treating $\delta t_i$ in Eq.~(\ref{FDPHeq:22}) as a correction factor. 
We then discretize the time integral of Eq.~(\ref{FDPHeq:22}) in \S\ref{FDPH431}.

For brevity, we here suppose $\epsilon_t=1$ in Eq.~(\ref{FDPHeq:Tcollocation}) 
without loss of generality [i.e., we can consider $t=(n+1)\Delta t$ in Eq.~(\ref{FDPHeq:22}) by regarding $\delta t_i+(1-\epsilon_t)\Delta t$ as a redefined $\delta t_i$ value]. 
We use the piecewise-constant temporal interpolation [Eq.~(\ref{FDPHeq:Dinterpolation})] of the slip- and opening-rates for the discretization.

\subsubsection{Time Shifts of the Collocation Points for Evaluating a BIE Convolved over Domain F}
\label{FDPH432}
The receiver-dependent travel-time difference $\delta t_i$ shifts the collocation time in the left hand side of Eq.~(\ref{FDPHeq:22}). Meanwhile, 
since differences between possible values of $\delta t_i$ for different receivers $i$ are not necessarily the integer multiples of $\Delta t$, 
there is not generally a special choice $t$ such that $t+\delta t_i$ coincides with the collocation times of all the receivers $i$ in all the block clusters of the admissible leaves. We then need certain consideration on it examined below. 

A simple way to relate continuous $t+\delta t_i$ to the collocation time is to use an appropriate discrete value $\delta m_i\in\mathbb Z$ (called receiver-dependent travel-time-step difference) as
\begin{equation}
\delta t_i= \delta m_i\Delta t,
\label{eq:discretedelti}
\end{equation}
instead of the continuous value of $\delta t_i$ given by Eq.~(\ref{FDPHeq:18}). Eq.~(\ref{eq:discretedelti}) adjusts $t+\delta t_i$ to a collocated time for a time step $n+\delta m_i$ and gives 
$T_i(t+\delta t_i)=T_{i,n+\delta m_i}$ for the case of $t=(n+1)\Delta t$.
Eq.~(\ref{eq:discretedelti}) can be the rounding-down of Eq.~(\ref{FDPHeq:18}), that is 
\begin{equation}
\delta m_i=
\left\lfloor 
\frac{r_{ij_*}-r_{i_*j_*}}{c\Delta t}
\right\rfloor,
\label{FDPHeq:30}
\end{equation}
%The approximation $\delta t_i/\Delta t \approx \delta m_i$ is rough yet sufficient for the constant $\eta$ scheme, since t
as well as the rounding-up, rounding-off, or other variations,
where $\lfloor\rfloor$ denotes the floor function. 
The neglected $\mathcal O(\Delta t)$ part due to replacing Eq.~(\ref{FDPHeq:18}) with Eq.~(\ref{eq:discretedelti}) is regarded as a small fraction in the travel-time approximation of the ART shown in Eq.~(\ref{FDPHeq:17}). 
The use of Eq.~(\ref{eq:discretedelti}) will be satisfactory for the constant $\eta$ scheme, since such an $\mathcal O(\Delta t)$ change in $\delta t_i$ just gives negligible $\mathcal O(c\Delta t/dist)$ error in the effective wave speed evaluated in Eq.~(\ref{FDPHeq:24}). 

When using Eq.~(\ref{eq:discretedelti}) in Eq.~(\ref{FDPHeq:22}), we obtain
\begin{equation}
T^F_{i,n+\delta m_i}
\approx f_i^F\sum_j g^F_j
\int^{\Delta t_j}_0d\tau h^F_j(\tau) D_j (t-\bar t_{j}^--\tau).
\label{FDPHeq:27}
\end{equation}
The discrete choice, Eq.~(\ref{eq:discretedelti}), of $\delta t_i$ is intrinsically a temporal interpolation of $T_{i,n}^F$. 
Although we adopted the rounding-down in Eq.~(\ref{FDPHeq:30}) in this study for keeping the causality, rounding-off may help to avoid the systematic errors in the approximation of the travel time. 
We can also consider the higher order interpolations.
%Corresponding higher order interpolations can be obtained in a similar way as well. 
%More accurate interpolation is formally constructible, such as $T_i((n+1+\delta m_i)\Delta t)\approx T_i((n+1)\Delta t+\delta t_i)(1-\delta t^\prime_i/\Delta t)+ T_i(n\Delta t+\delta t_i)\delta t^\prime_i/\Delta t$ of $\mathcal O(\partial_t^2 T_i(\Delta t)^2)$ errors, where $\delta t^\prime_i/\Delta t$ is the decimal part of $\delta t_i/\Delta t$. 

\subsubsection{Temporal Discretization of the Kernel After Applying the ART in Continuous Time}
\label{FDPH431}
When we formally suppose $i=i_*$ and $h^F_j=K^F_{ij}$,
the integrand of Eq.~(\ref{FDPHeq:27}) is identified with that of the original FDPM in Domain F. 
Then supposing the case of $i=i_*$, we can map the discretization of $h^F_j$ in Eq.~(\ref{FDPHeq:27}) to that of $K^F_{ij}$ in the original FDPM (shown in Fig.~\ref{FDPHfig:3}a). 

For discretizing $\bar t_j^-$ and $\Delta t_j$, 
we introduce two integers, $\bar m_j^-\in\mathbb Z$ (hereafter called receiver-averaged travel time step) and $\Delta m_j\in\mathbb Z$. 
They are defined as $m_{ij}^-$ and $m_{ij}^+-m_{ij}^-$, respectively, of the original FDPM for $i=i_*$. Integers $m_{ij}^\pm$ defined in \S\ref{FDPH22} are illustrated in Fig.~\ref{FDPHfig:3}a.
The explicit values of them are given as 
%To discretize the time integral, as $m_{ij}^\pm$ in the FDPM, 
%%$\bar m_j^-$ and $\bar m_j^+$ 
%$\bar m_j^{-}:=\lfloor \bar t_j^{-}/\Delta t\rfloor$ 
%and 
%$\bar m_j^{+}:=\lceil \bar t_j^{+}/\Delta t\rceil$
\begin{flalign}
\bar m_j^{-}&:= 
\left\lceil 
\frac{r_{i_*j} -\Delta x_j/2}{c\Delta t}-\delta C^{c-}_j
\right\rceil
\label{eq:origdefofbarmj}
\\
\Delta m_j&:= 
\left\lfloor 
\frac{r_{i_*j} +\Delta x_j/2}{c\Delta t}+\delta C^{c+}_j
\right\rfloor -\bar m_j^{-}
\label{eq:origdefofDelmj}
, 
\end{flalign}
as in the original FDPM \cite{ando2016fast} (for the case of $i=i_*$), where $\lceil\rceil$ denotes the ceiling function. 
See \S\ref{FDPHC3} for details. 
%The integer $\bar m_j^-$ is the time step ($m_{i_*j}^-$) experiencing the wave arrival for the source $j$ and receiver $i_*$. The integer $\Delta m_j$ is the elapsed time step from $\bar m_j^-$  until the time step ($m_{i_*j}^+$) experiencing the completion of the wave passage. 

To obtain a simple discrete convolution,
we further subtly modify the continuous value of $\bar t_j^-$ to a discrete one, 
\begin{equation}
\bar t_j^-= \bar m_j^-\Delta t,
\label{eq:discretizationofbartj}
\end{equation} 
for each source $j$. That is, we adopt $\bar t_j:=\bar m_j^-\Delta t+\Delta t_j^-$ instead of Eq.~(\ref{FDPHeq:19}).
The above-mentioned integer $\bar m_j^-$ here appears. 
This adoption of Eq.~(\ref{eq:discretizationofbartj}) instead of Eq.~(\ref{FDPHeq:19}) will satisfactory for the constant $\eta$ scheme for the same reason as the use of Eq.~(\ref{eq:discretedelti}) for $\delta t_i$ in \S\ref{FDPH432}. 
Further, we take $\Delta t_j$ as an integer-multiple of $\Delta t$ by adjusting safe coefficients $\delta C_j^{c\pm}$ in
the definition of $\Delta t_j^\pm$ [Eqs.~(\ref{FDPHeq:6}) and (\ref{FDPHeq:7})] under the following rule, 
\begin{equation}
\delta C_j^{c+}+\delta C_j^{c-}=
\left\lceil 
\frac{\Delta x_j}{c\Delta t}
\right\rceil 
-
\frac{\Delta x_j}{c\Delta t}.
\label{eq:safecoeffordiscDeltj}
\end{equation}
When $\bar t_j^-$ and $\Delta t_j$ are discretely adjusted as Eqs.~(\ref{eq:discretizationofbartj}) and (\ref{eq:safecoeffordiscDeltj}),
the ceiling and floor functions in the right hand sides of Eqs.~(\ref{eq:origdefofbarmj}) and (\ref{eq:origdefofDelmj}) (corresponding to $\lceil\bar t_j^-\rceil$ and $\lfloor \bar t_j+\Delta t_j\rfloor-\lceil\bar t_j^-\rceil$) are dropped, and thus
the ratio of $\Delta t_j/\Delta t$ coincides with $\Delta m_j$, that is, 
\begin{equation}
\Delta t_j=\Delta m_j\Delta t.
\label{eq:discretizationofDeltj}
\end{equation}
%We used $\bar t_j^\pm =\bar t_j\pm \Delta t_j^\pm$ and $\Delta t_j=\Delta t^+_j+\Delta t^-_j$ to obtain this relation from Eqs.~(\ref{eq:origdefofbarmj}) and (\ref{eq:origdefofDelmj}).  
Note that unlike the analogy of $\delta t_i$,  Eq.~(\ref{eq:discretizationofbartj}) can also be implemented as an error-free tuning of the $\Delta t^\pm_j$ values (\ref{FDPHC32}), instead of as the aforementioned discretization process of the continuous $\bar t_j$ values inducing the $\mathcal O(c\Delta t/dist)$ error, although their difference is quite subtle; for that case, we retain Eq.~(\ref{FDPHeq:19}) and tune one degree of freedom remaining in the set of the paired ($\delta C^{c+}, \delta C^{c-}$) values [or equivalently, in ($\Delta t_j^+,\Delta t_j^-$)] as a leaf-dependent parameter after the condition of Eq.~(\ref{eq:safecoeffordiscDeltj}) erases their one degree of freedom. 

Given Eqs.~(\ref{eq:discretizationofbartj}) and (\ref{eq:discretizationofDeltj}), and
by substituting $t=(n+1)\Delta t$ and $D_{j}(t)=$ $ \sum_m D_{j,m}[H(m\Delta t)-H((m+1)\Delta t)]$ into Eq.~(\ref{FDPHeq:27}), 
we obtain the following fully discretized BIE (\S\ref{FDPHC3}):
\begin{equation}
T^F_{i,n+\delta m_i}
\approx f^F_i\sum_jg^F_j
\sum_{m=0}^{\Delta m_j-1} h^F_{j,m}D_{j,n-(m+\bar m^{-}_j)},
\label{FDPHeq:31}
\end{equation}
where $h^F_{j,m}$ is the temporally discretized form of the normalized waveform given by Eq.~(\ref{FDPHeq:13}): 
\begin{equation}
h^F_{j,m}:=\frac 1 {\hat K^F_{i_*,j}} 
\int_{m\Delta t+t_{i_*j}^-}
^{(m+1)\Delta t+t_{i_*j}^-}
d\tau K_{i_*,j} (\tau).
\label{FDPHeq:29}
\end{equation}
We note that the expression of $h^F_{j,m}$ is altered to another lengthy form when Eqs.~(\ref{eq:discretizationofbartj}) and (\ref{eq:discretizationofDeltj}) are not adopted (also supplemented in \S\ref{FDPHC3}). 
$\hat K^F_{i_*,j}$ in Eq.~(\ref{FDPHeq:29}) is obtained as the amplitude term $\hat K^F_{i,j}$ [defined by Eq.~(\ref{FDPHeq:12})] of $i=i_*$. 
Eq.~(\ref{FDPHeq:12}) assigns 
the numerical value of $\hat K^F_{i,j}$ to an arbitrary pair of receiver $i$ and source $j$ as 
\begin{equation}
\hat K^F_{i,j}=\int^{t^{+}_{ij}}_{t^{-}_{ij}} d\tau K_{i,j} (\tau),
\label{FDPHeq:28}
\end{equation}
that is a time integral of the kernel over Domain F [$\tau\in(t_{ij}^-,t_{ij}^+)$]. 
These integral forms of Eqs.~(\ref{FDPHeq:29}) and (\ref{FDPHeq:28}) exactly coincide with the original kernel discretized by the temporally-piecewise-constant slip- and opening-rate, while the integral intervals are $\Delta t$ in  Eq.~(\ref{FDPHeq:29}) as in the original ST-BIEM and are $\Delta t_j$ in Eq.~(\ref{FDPHeq:28}).  
This coincidence allows us to calculate Eqs.~(\ref{FDPHeq:29}) and (\ref{FDPHeq:28}) from the analytical expressions of the discrete kernel in the original ST-BIEM, the double-layer expressions of which for the piecewise-constant time interpolation are found both in the 2D~\cite{tada2001dynamic} and 3D~\cite{tada2006stress} settings. 
%by using the constituent functions, which describe the solutions of the integral equation when the temporal behavior of the slip- and opening-rate is given as a step function. 

%Only here, it is sensitive for the accuracy to distinguish the stress and traction. The kernel needs to describe the stress at receivers [as in Eq.~(\ref{FDPHeq:1})] to ensure the accuracy of the degenerating normalized waveform; if the kernel is for describing the traction of receivers, the accuracy degrades from Eq.~(\ref{FDPHeq:21}) (mentioned in \S\ref{FDPHC1}). 

Besides, in the 2D problems, we will increase $\bar t_j^+$ [i.e., increase $\Delta m_j$ and $\delta C_j^{c+}$ from those of Eqs.~(\ref{eq:origdefofDelmj}) and (\ref{eq:safecoeffordiscDeltj}) by positive integer number $n_c$] for the 2D-specific error handling of the FDPM (\S\ref{FDPH22}), as supplemented in \ref{FDPHH}.

\section{Arithmetic of FDP=H-Matrices in Domain F}
\label{FDPH5}
Based on the data-sparse approximation developed in the previous section, this section treats of the operations of FDP=H-matrices that accomplish the $\mathcal O(N\log N)$ total memory consumption and $\mathcal O(N\log N)$ computation time per time step. 
As in the previous section, our main focus is Domain F. The starting point of the operation development is the fully reduced BIE Eq.~(\ref{FDPHeq:31}) for Domain F. 
We decompose Eq.~(\ref{FDPHeq:31}) into three formulae in \S\ref{FDPH51} and obtain an arithmetic for Domain F in \S\ref{FDPH52}. 
Arithmetics for Domains I and S are constructed in similar manners (Please refer to \ref{FDPHB}). 
The derived key formulas for the arithmetic in Domain F will be summarized in Table~\ref{KeyFormulas2}. 

\subsection{Three Formulae for Evaluating the Discretized BIE in Domain F with FDP=H-Matrices}
\label{FDPH51}

Eq.~(\ref{FDPHeq:31}) evaluates a three-rank tensor and expresses a summation over the time steps $m$ and sources $j$ for all the receivers $i$. 
The reduced form of Eq.~(\ref{FDPHeq:31}) allows us to separate this set of operations involving $m$, $j$, and $i$ into three formulae.

%utilize the receiver $i$ independence of the temporal variables $\Delta m_j$ and $\bar m_j^-$. We define 
The convolution over the time step $m$ in Eq.~(\ref{FDPHeq:31}) 
gives a temporally evolving variable of the source: 
\begin{equation}
\hat D^F_{j,n} := \sum_{m=0}^{\Delta m_j-1} h^F_{j,m} D_{j,n-m}.
\label{FDPHeq:32}
\end{equation}
This is the first formula of FDP=H-matrices, converting $D$ to $\hat D^F$ in a receiver-$i$-independent manner.
$\hat D^F$ simplifies Eq.~(\ref{FDPHeq:31}) to
\begin{equation}
T^F_{i,n+\delta m_i}\approx f^F_i\sum_j g^F_j \hat D^F_{j, n-\bar m^-_j}.
\label{FDPHeq:33}
\end{equation}
Hereafter for explanatory simplicity, we consider one rank and one admissible leaf and omit the summation over the ranks and leaves as Eq.~(\ref{FDPHeq:33}) does.

Eq.~(\ref{FDPHeq:33}) can be comparable to the formula, ${\bf T}=K{\bf E}\approx {\bf f}[{\bf g}\cdot {\bf E}]$, of H-matrices in the static problems (Fig.~\ref{FDPHfig:11}a), which 
%The computation from $D$ to $T$ by FDP=H-matrices is decomposed into two computations converting $D$ to $\hat D$ and $\hat D$ to $T$. 
%The computation ${\bf T}\approx {\bf f}[{\bf g}\cdot {\bf E}]$ of H-matrices 
separates into a 
receiver-independent product $\bar T:=[{\bf g}\cdot {\bf E}]$ and source-independent product ${\bf T}\approx {\bf f}\bar T$. 
We can identify the computation of convolution in Eq.~(\ref{FDPHeq:33}) with that of H-matrices, 
excluding the time shift of $\hat D$ by a scalar $\bar m_j^-$ (Fig.~\ref{FDPHfig:11}b).
Such a time shift of making unique difference of them operates to extract $\hat D_{j,\bar m_j^-}$ from the entire history of $\hat D_{j,m}$ in accord with relation $m=\bar m^-_j$. 
The value of $\bar m_j^-$ represents a finite time step taken for the wave propagation from source $j$ to representative receiver $i_*$ in admissible leaf $a$. 
Relation $m=\bar m^-_j$ constitutes line $m=j \Delta x/(c\Delta t)+const.$ on a submatrix for the case of the 2D planar fault and depicts the role of $\bar m_j^-$ as a wave propagation time (Fig.~\ref{FDPHfig:11}b). 

Scalar $\bar T$ of H-matrices may correspond to the stress at the representative receiver position. %denoted by $i_*$. 
We introduce its time-step-($m$-)dependent value $\bar T_m$ into FDP=H-matrices; 
\begin{equation}
\bar T_m :=\sum_j g_j \hat D_{j,m-\bar m ^-_j},
\label{FDPHeq:34}
\end{equation}
where $\bar T_m$ is defined for arbitrary $m$ independent of the current time step $n$. This is the second formula of FDP=H-matrices, converting $\hat D$ to $\bar T$ (Fig.~\ref{FDPHfig:11}b). 
Hereafter, superscript $F$ in this section is omitted in equations for notational simplicity.
We refer to $\bar T$ as the representative stress. 
The history of $\bar T$ is stored as a vector in FDP=H-matrices while $\bar T$ is a scalar in H-matrices.
The required vector length for the history of $\bar T_m$ is of order $(\delta m_i+\bar m_j^-)$, the approximated travel time step, as detailed in \S\ref{FDPH52} and \S\ref{FDPH523}. 
$\bar T$ is given for each rank and each admissible leaf as in H-matrices. 
The representative stress $\bar T$ gives a simple expression of the stress at current time step $n$ with the time shift by $\delta m_i$: 
\begin{equation}
T_{i,n}=f_i\bar T_{n-\delta m_i}.
\label{FDPHeq:35}
\end{equation}
This is the third formula of FDP=H-matrices, converting $\bar T$ to $T$.
%The computation converting $\hat D$ to $T$ by FDP=H-matrices is decomposed into two computations converting $\hat D$ to $\bar T$ and $\bar T$ to $T$. 

The conversions from $\hat D$ to $\bar T$ [Eq.~(\ref{FDPHeq:34})] and $\bar T$ to $T$ [Eq.~(\ref{FDPHeq:35})]  define a different arithmetic of FDP=H-matrices from that of H-matrices because of the time shifts by $\delta m_i$ and $\bar m^-_j$. 
$\bar T_m$ at time step $m$ in Eq.~(\ref{FDPHeq:34}) is contributed from the motion of the source ($j$) in the past by $\bar m_j^-$ (the receiver-averaged travel time step). 
The delay of the interaction in FDP=H-matrices is caused by the wave propagation, or intrinsically by the causality, contrasting to the original H-matrices in the static problems formally assuming the instantaneous action. 
Eq.~(\ref{FDPHeq:35}) uses the representative stress of the past by $\delta m_i$ (the receiver-dependent travel-time-step difference) for computing the stress $T_{i,n}$, and such time shift is due to the difference in the travel times between individual receivers.

To implement these time shifts in the arithmetic, 
it is useful to define the following sparse matrices. 
The receiver-averaged travel time step 
$\bar m_j^-$ allows us to define time-shift matrix 
${\bf S}^{source}$ ($\in\mathbb R^{[\max_j(\bar m^-_{a,j})-\min_j(\bar m^-_{a,j})]\times N_{s,a}}$)
 for sources ($j$) in a tensorial manner:
\begin{equation}
S_{m,j}^{source}:= \delta_{m,-\bar m_j^-},
\label{FDPHeq:36}
\end{equation}
where $N_{s,a}$ denotes the number of sources in admissible leaf $a$, and we signalize the $a$-dependence of $\bar m_j^-$ only here for showing the dimension of ${\bf S}^{source}$. Integer [$\max_j(\bar m^-_{a,j})-\min_j(\bar m^-_{a,j})$] is noticed to be $\mathcal O[diam/(c\Delta t)]$ given that the variance of $\bar m^-_{a,j}$ [Eq.~(\ref{eq:origdefofbarmj})] is due to the variation of source locations within a sphere of diameter $diam$.
Receiver-averaged travel time step $\bar m_j^-$ represents the number of time steps elapsing during the wave propagation from source $j$ to representative receiver $i_*$. 
Similarly, 
we define 
time-shift matrix ${\bf S}^{receiver}$ ($\in\mathbb R^{N_{r,a}\times[\max_i(\delta m_{a,i})-\min_i(\delta m_{a,i})]}$)
for receivers ($i$) as
\begin{equation}
S_{m,i}^{receiver}:= \delta_{m,\delta m_i},
\label{FDPHeq:37}
\end{equation}
with receiver-dependent travel-time-step difference $\delta m_i$,
where $N_{r,a}$ denotes the number of receivers in admissible leaf $a$. We signalize the $a$-dependence of $\delta m_i$ only here for showing the dimension of ${\bf S}^{receiver}$. Integer [$\max_{i}(\delta m_{a,i})-\min_i(\delta m_{a,i})$] is estimated to be $\mathcal O[diam/(c\Delta t)]$ given the definitional identity of $\delta m_{a,i}$, Eq.~(\ref{FDPHeq:30}).
Scalar $\delta m_i$ represents the difference in the discretized wave-propagation time between receiver $i$ and representative receiver $i_*$. 
%The matrices ${\bf S}^{source}$ and ${\bf S}^{receiver}$ are sparse, and 
The numbers of nonzero components in ${\bf S}^{source}$ and ${\bf S}^{receiver}$ are respectively equal to $N_{s,a}$ and $N_{r,a}$ in admissible leaf $a$ because 
every source $j$ and receiver $i$ have its own single value of $\bar m^-_j$ and that of $\delta m_i$ in $a$, respectively. 

\subsection{Operations of FDP=H-matrices in Domain F with Sparse Matrices}
\label{FDPH52}
%Figure 11. âš
\begin{figure*}[tbp]
   \begin{center}
  \includegraphics[width=120mm]{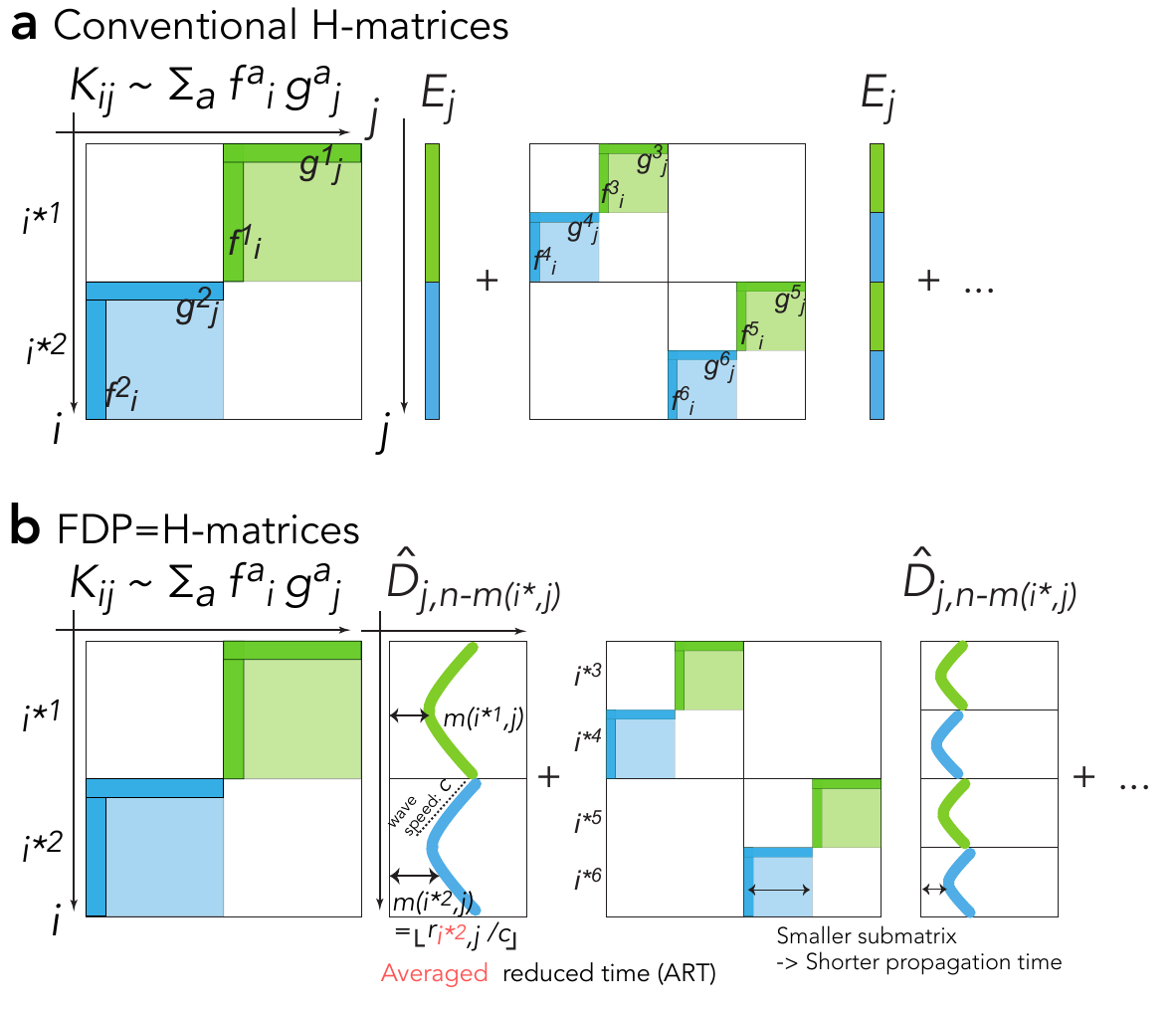}
   \end{center}
\caption{
Schematic of computations convolving the kernel and boundary variables in H-matrices and FDP=H-matrices [the right hand side of Eq.~(\ref{FDPHeq:33})]. Among separated convolutions of different levels (the number of the division the cluster is subjected to), the computation of the levels 0 and 1 are illustrated explicitly. 
Submatrices of the kernel and corresponding convolved components of the boundary variables are painted by green and blue when leaf number $a$ is respectively odd and even, respectively. 
The rank dependence of the kernel is omitted here for brevity. 
{\bf a}, Convolution in H-matrices, summing the product of kernel $K_{ij}$ and slip and opening $E_j$ over source $j$ for each receiver $i$. 
The submatrix of each admissible leaf $a$ is reduced to receiver dependence $f_i^a$ and source dependence $g_j^a$. 
{\bf b}, Convolution in Domain F of FDP=H-matrices, summing the product of kernel $\hat K_{ij}$ and $\hat D_{j,n-\bar m_j^-}$, a temporal convolution of the slip- and opening-rate and the normalized waveform, over source $j$ for each $i$. Components of $\hat D$ located at 
receiver averaged travel time $\bar m_j^-$ is selected among the history of $\hat D_{j,n-m}$ over time step $m$ and gets convolved with the kernel. 
For explanatory simplicity, 
the notation and definition of $\bar m_j^-$ is modified as $m(i_{*a},j)$ with omitting $\Delta t_j^-$ in the figure, to indicate that $\bar m_j^-$ given by the ART is intrinsically the discrete value of travel time $t_{i_{*a}j}=r_{i_{*a}j}/c-\Delta t_j^-$ between the representative receiver $i_{*a}$ in admissible leaf $a$ and source $j$. 
}
\label{FDPHfig:11}
\end{figure*}

Each of the three formulae obtained in \S\ref{FDPH51} 
represents any of the following three kinds of the variable conversions: 1) from slip- and opening-rate $D$ to $\hat D$ convolving $D$ and the normalized waveform ($D\to \hat D$), 2) from $\hat D$ to representative stress $\bar T$ 
($\hat D\to \bar T$), and 3) from representative stress $\bar T$ to stress $T$ ($\bar T\to T$). 
Below, we construct from these the operations of FDP=H-matrices in Domain F. 

The definitional identity of $\hat D_{j,n}$, Eq.~(\ref{FDPHeq:32}) 
gives the conversion $D\to \hat D$ straightforwardly. We compute $\hat D^F_{j,n}$ for all the sources $j$ contained in respective admissible leaves from ${\bf D}_n$ in each time step $n$ with Eq.~(\ref{FDPHeq:32}).

%We then focus on the conversions ($\hat D\to \bar T$) and ($\bar T\to T$) [that respectively correspond to Eqs.~(\ref{FDPHeq:34}) and (\ref{FDPHeq:35})]. 

\begin{figure}[tb]
   %\begin{center}
  \includegraphics[width=100mm]{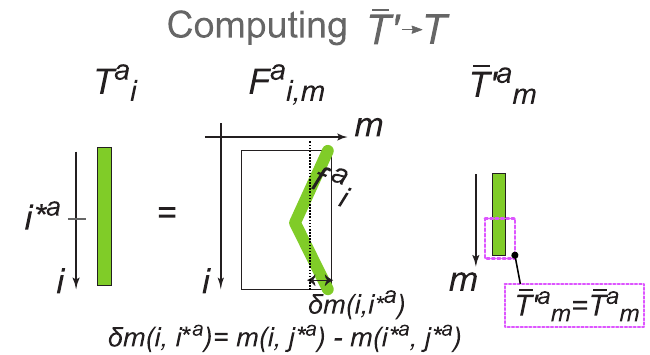}
  %\includegraphics[width=100mm]{h-art_arithm_separated_a.pdf}
  %\includegraphics[width=80mm]{h-art_arithm_separated_a.pdf}
   %\end{center}
\caption{
Schematic of the arithmetic in Domain F, with regard to the computational procedure of $\bar T^\prime\to T$~[Eq.~(\ref{FDPHeq:39_tensorial_prime}), obtained from Eq.~(\ref{FDPHeq:39_tensorial}) via Eq.~(\ref{eq:discretizedcausality})]. 
The illustration method and settings are the same as those in Fig.~\ref{FDPHfig:11}, and the trivial dependence on current time step $n$ is omitted here for brevity. 
Stress $T_i^a$ at receiver $i$, in each admissible leaf $a$, is calculated as a product of $f_i^a$ and a component of conditionally predicted representative stress $\bar T^{\prime a}_m$ located at the associated receiver-dependent travel-time-step difference $\delta m_i=m$. 
The origin $m=0$ of the time step is indicated in the figure by the dotted line in $F^a_{i,m}$. 
The notation of $\delta m_i$ is changed to $\delta m(i,i_{*a})$ in the figure to intuitively indicate that $\delta m_i$ given by the ART represents the discrete value of the travel time difference between receiver $i$ and representative receiver $i_*$ for the wave radiated from representative source $j_*$; the definition of $\delta m_i$, shown in \S\ref{FDPH43}, is subtly modified in the figure for explanatory simplicity. 
The sets of the $m$-th components, $\bar T^{\prime a}_m$, required in this computation are surrounded by a purple square in respective admissible leaves $a$, where $\bar T^{\prime a}=\bar T_{n-m}^a$ (shortened as $\bar T_m^a$ to omit the trivial $n$-dependence) holds. 
A second-rank tensor, $F^a_{i,m}$ (being a sparse matrix), represents all the above computational procedure of the $\bar T^\prime\to T$ computation. 
}
\label{FDPHfig:11p5_a}
\end{figure}

We rewrite Eq.~(\ref{FDPHeq:35}) in the following way to
convert the representative stress to the stress efficiently ($\bar T\to T$):
\begin{equation}
T_{i,n}=\sum_m F_{i,m}  \bar T_{n-m},
\label{FDPHeq:39_tensorial}
\end{equation}
with the product of time-shift matrix $S^{receiver}$ and $f$:
\begin{equation}
F_{i,m}:= f_i S_{m,i}^{receiver}.
\label{FDPHeq:38}
\end{equation}
We then obtain $T_{i,n}$ of all the receivers $i$ at each time step $n$ from $\bar T_m$ by using Eq.~(\ref{FDPHeq:39_tensorial}) once. 
Fig.~\ref{FDPHfig:11p5_a} shows that 
Eq.~(\ref{FDPHeq:39_tensorial}) serves both the time shift by $\delta m_i$ and the multiplication of $\bar T_{n-m}$ by $f_i$. 
Note that $\delta m_i$ is expressed as $\delta m(i,i_{*a})$ in the figure to indicate that $\delta m_i$ depends on receiver $i$ and representative receiver $i_*$ of admissible leaf $a$. Besides, $\bar T$ and $\bar T^\prime$ after-mentioned are identified in the figure for brevity.

%Figure 11.5_separated_a âš
%\begin{figure}[H] %% usepackage{here}

Conversion $\hat D\to \bar T$ is obtained from the definitional identity, Eq.~(\ref{FDPHeq:34}), of representative stress $\bar T_m$. 
Its simple implementation is using a ``divide-and-conquer'' algorithm (detailed in \S\ref{FDPH523}). There we compute $\bar T_m$ of each time step $m$ successively and certainly the time complexity becomes of $\mathcal O(N\log N)$ per time step. 
However, direct computation of Eq.~(\ref{FDPHeq:34}) requires to store the history of $\hat D_{j,n-m}$ ranging $0\leq m < \bar m_j^-$ (Fig.~\ref{FDPHfig:11}b) (or ${\bf D}_{n-m}$ at $0\leq m < \bar m_j^-+\Delta m_j$). 
It results in the $\mathcal O[NL/(\beta\Delta t)]$ memory requirements of this implementation, which are mostly due to the large block clusters of $dist=\mathcal O(L)$ that give $\bar m_j=\mathcal O[L/(c\Delta t)]$ and $N_{r,a}, N_{s,a}=\mathcal O(N)$ (detailed in \S\ref{FDPH523} as well). 

To obviate such $\mathcal O[NL/(\beta \Delta t)]$ history of the boundary variables, 
we evaluate $\bar T_m$ in an equivalent yet recursive (so-called ``dynamic programming'') manner instead. 
We first define tensor $G_{m,j}$ in an analogous form with $F_{i,m}$,
\begin{equation}
G_{m,j}:=g_j S^{source}_{m,j},
\label{FDPHeq:41}
\end{equation}
by using vector $g_j$ and sparse matrix $S^{source}_{m,j}$.
Next, along the line of an analogy of $F_{i,m}\bar T_{n-m}$ in Eq.~(\ref{FDPHeq:39_tensorial}), we aim to construct $\bar T_m$ [Eq.~(\ref{FDPHeq:34})] from $G_{m,j} \hat D_{j,n}$. 
For that purpose, we rewrite Eq.~(\ref{FDPHeq:34}) and express 
the involved time shift of $\hat D$ as an delta-functional extraction of $\hat D$ from the history space: 
\begin{equation}
\bar T_m =
\sum_{m^\prime=-\infty}^\infty\sum_j g_j \hat D_{j,m+m^\prime}\delta _{m^\prime,-\bar m_j^-}.
\label{FDPHeq:barTdcomptointersection}
\end{equation}
Here we used $\sum_a f_{a}\delta_{a,b}=f_b$ for arbitrary function $f_a$ and subscripts $a$ and $b$. 
Comparing Eq.~(\ref{FDPHeq:barTdcomptointersection}) with the definitional identity Eq.~(\ref{FDPHeq:41}) of $G_{m,j}$ [and  Eq.~(\ref{FDPHeq:36}) of $S^{source}_{m,j}$], 
we find that the $n$ value such that $n=m+m^\prime$ yields the desired sparse-matrix-vector product $G_{m^\prime,j}\hat D_{j,n}$ as $g_j \hat D_{j,m+m^\prime}\delta _{m^\prime,-\bar m_j^-}=G_{m^\prime,j}\hat D_{j,n}$. 
As illustrated in Fig.~\ref{FDPHfig:11}b, 
summation $\sum_j G_{m^\prime,j}\hat D_{j,n}$ at $n=m+m^\prime$ is an operation that searches the $m^\prime$ space for the intersection ($-\bar m^-_j=n-m$) of lines (causal cones) $m^\prime=-\bar m^-_j$ [$m=j \Delta x/(c\Delta t)+const.$] and $m^\prime=n-m$; the former line expresses the time shift due to the wave propagation and the latter specifies the certain value of relative time step $n-m$. 
As $n$ increases, the associated $m^\prime=n-m$ value and the intersection also move, 
and then $\sum_j G_{m^\prime,j}\hat D_{j,n}$ cumulatively computes the $\bar T_m$ value through Eq.~(\ref{FDPHeq:barTdcomptointersection}) for each $m$. 
That is, Eq.~(\ref{FDPHeq:barTdcomptointersection}) represents an operation procedure for cumulatively constructing $\bar T_m$ by summing up $G_{n-m,j}\hat D_{j,n}$ over sources $j$ in each time step $n=0,1,...$ as $[\sum_jG_{-m,j}\hat D_{j,0}+\sum_jG_{1-m,j}\hat D_{j,1}+...]$. 
This cumulative nature of the computation is attributable to the independence of the original kernel in Eq.~(\ref{FDPHeq:3}) from the absolute time $t$ and $\tau$, as the kernel depends on only relative time $t-\tau$ [associated with $m^\prime$ in Eq.~(\ref{FDPHeq:barTdcomptointersection})], which is intrinsically the temporal translational symmetry of the Green's function. 

\begin{figure}[tb]
   \begin{center}
  \includegraphics[width=100mm]{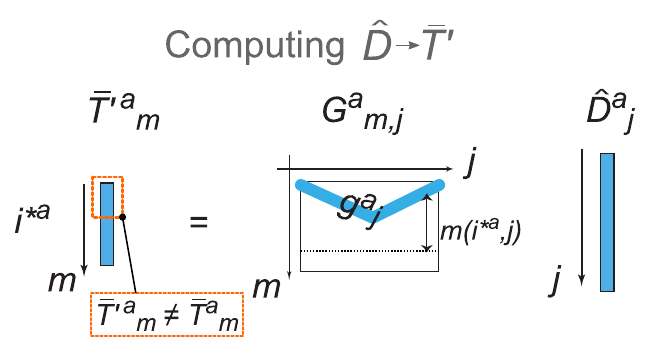}
   \end{center}
\caption{
Schematic of the arithmetic in Domain F, with regard to the computational procedure of $\hat D\to \bar T^\prime$ [Eq.~(\ref{FDPHeq:42_tensorial})]. 
The illustration method and settings are the same as those in Fig.~\ref{FDPHfig:11}, and the trivial dependence on current time step $n$ is omitted here for brevity. 
Trivial time shift of $T^\prime$ (expressed as the multiplication of $\mathcal M$ to $T^\prime$) is omitted here for brevity.
Conditionally predicted representative stress $\bar T^{\prime a}_m$ at time step $m$, in each admissible leaf $a$, is calculated as a summation of the product of $g_j^a$ and a component of $\hat D_j$ over the sources $j$ that have the associated receiver-averaged travel time step, $\bar m_j^-=-m$. 
The origin $m=0$ of the time step is indicated by dotted lines in $G^a_{m,j}$. 
Sets of the $m$-th components, $\bar T^{\prime a}_m$, incremented in this computation are surrounded by an orange square in respective admissible leaves $a$, where $\bar T^{\prime a}=\bar T_{n-m}^a$ does not hold in contrast to in the purple box in Fig.~\ref{FDPHfig:11p5_a}. 
A second-rank tensor, $G^a_{m,j}$ (being a sparse matrix), represents these time-shifted products of the $\hat D\to \bar T^\prime$ computation. 
}
\label{FDPHfig:11p5_b}
\end{figure}

The sparse-matrix-vector product $G_{m^\prime,j}\hat D_{j,n}$ computing $\bar T$ can be illustrated as Fig.~\ref{FDPHfig:11p5_b}. 
Similarly to $T_{i,n}=F_{i,m}\bar T_{n-m}$ of computing $F_{i,m}\bar T_{n-m}$ (Fig.~\ref{FDPHfig:11p5_a}), 
$\hat D_{j,n}$ is multiplied by a vector ($g_j$) and contributes to $G_{m^\prime,j}\hat D_{j,n}$ of $m^\prime=-\bar m_j$ at each time step $n$. 
In the figure, the notation of $\bar m_j^-$ is modified as $m(i_{*a},j)$ to indicate that $\bar m_j^-$ depends on representative receiver $i_{*}$ of admissible leaf $a$ and source $j$.
By considering that the computation of $\bar T$ is originally intended to evaluate $\bar T_{n-m}$ in Eq.~(\ref{FDPHeq:39_tensorial}) for obtaining $T_{i,n}$, we replace $m$ with $n-m$ in Eq.~(\ref{FDPHeq:barTdcomptointersection}) as follows like the moving coordinate in the figure: 
\begin{equation}
\bar T_{n-m} =
\sum_{m^\prime=-\infty}^\infty\sum_j 
G_{m^\prime,j}
\hat D_{j,n-m+m^\prime}.
\label{eq:whatwantedoriginallyforthethridformula}
\end{equation}
Note $g_j \delta _{m^\prime,-\bar m_j^-}=G_{m^\prime,j}$.

%Figure 11.5_separated_b âš
%\begin{figure}[H] %% usepackage{here}

As above, accumulating $\sum_jG_{m^\prime,j}\hat D_{j,n}$ at each time step $n$, we can obtain 
the representative stress $\bar T_m$ of given time step $m$ from Eq.~(\ref{eq:whatwantedoriginallyforthethridformula}). On the other hand, 
accumulated $\sum_jG_{m^\prime,j}\hat D_{j,n}$ is a partial sum of Eq.~(\ref{eq:whatwantedoriginallyforthethridformula}) originally summed over $m^\prime=-\infty,...,\infty$, and then we need additional consideration to relate the former to the latter defined in the limit. We deal with it by defining a substitute, for $\bar T$, available from the accumulation of $\sum_jG_{m^\prime,j}\hat D_{j,n}$. 
Such a substitute is found in the above expression Eq.~(\ref{eq:whatwantedoriginallyforthethridformula}) of $\bar T$, as
a part evaluable with only the history of $\hat D_{j,n-m+m^\prime}$ within the time steps $n-m+m^\prime<n$ before the current time step $n$: 
\begin{equation}
\bar T_{n-m} =
\left[
\sum_{m^\prime=-\infty}^{m-1}
+
\sum_{m^\prime=m}^\infty
\right]
\sum_j G_{m^\prime,j} \hat D_{j,n-m+m^\prime}.
\label{FDPHeq:barTdcomptointersection_pastfuture}
\end{equation} 
We arranged the decomposition in Eq.~(\ref{FDPHeq:barTdcomptointersection_pastfuture}) such that
$\hat D_{j,n-m+m^\prime}$ in the first summation within $m^\prime<m$ covers the history of $\hat D_{j,m}$ exactly ranging over $m<n$, the time step $m$ before the current time step $n$. 
In that manner, we isolate the first term in Eq.~(\ref{FDPHeq:barTdcomptointersection_pastfuture}), defined as
\begin{equation}
\bar T^\prime_{n,m}:=
\sum_{m^\prime=-\infty}^{m-1}\sum_j 
G_{m^\prime,j}\hat D_{j,n-m+m^\prime},
\label{eq:defofpredrep}
\end{equation} 
which represents the conditional summation of $\sum_jG_{m^\prime,j}\hat D_{j,m}$ over time steps $m<n$, 
from the other part of the summation; 
the other is represented by the second term in Eq.~(\ref{FDPHeq:barTdcomptointersection_pastfuture}), which is associated with $\hat D_{j,m}$ at current time step $m=n$ and future time steps $m>n$.
Eq.~(\ref{eq:defofpredrep}) corresponds to the above-mentioned incremental temporal summation $[\sum_jG_{-m,j}\hat D_{j,0}+\sum_jG_{1-m,j}\hat D_{j,1}+...]$ for $\bar T$.

The difference between $\bar T^\prime_{n+1,m+1}$ 
and 
$\bar T^\prime_{n,m}$ constitutes the increment of $\bar T_{n-m}$ due to $\hat D_{j,n}$
as both of these $\bar T^\prime$ components correspond to $\bar T_{n-m}$; 
\begin{flalign}
\bar T^\prime_{n+1,m+1}-\bar T^\prime_{n,m}
=&
\sum_{m^\prime=-\infty}^{m}\sum_j 
G_{m^\prime,j}\hat D_{j,n-m+m^\prime}.
\nonumber\\
&
-
\sum_{m^\prime=-\infty}^{m-1}\sum_j 
G_{m^\prime,j}\hat D_{j,n-m+m^\prime}.
\label{eq:recursivetbarprimederiv1}
\\=&
\sum_j G_{m,j}\hat D_{j,n},
\label{eq:recursivetbarprimederiv2}
\end{flalign}
We obtain Eq.~(\ref{eq:recursivetbarprimederiv2}) from Eq.~(\ref{eq:recursivetbarprimederiv1}) by considering that the difference between summation ranges $m^\prime\in(-\infty,m]$ and $m^\prime\in(-\infty,m-1]$ is equal to $m^\prime=m$. 
The term in Eq.~(\ref{eq:recursivetbarprimederiv2}) is exactly above-mentioned $\sum_j G_{m,j}\hat D_{j,n}$.
By replacing $m+1$ with $m$ in the above result (as $\bar T^\prime_{n+1,m}-\bar T^\prime_{n,m-1}=\sum G_{m-1,j}\hat D_{j,n}$), we derive its symbolic form:
\begin{equation}
\bar T^\prime_{n+1,m}=
\sum_{m^\prime}
\mathcal M_{m,m^\prime}
\left[\bar T^\prime_{n,m^\prime}
+\sum_j G_{m^\prime,j}\hat D_{j,n}\right],
\label{FDPHeq:42_tensorial}
\end{equation}
with
\begin{equation}
\mathcal M_{m,m^\prime} :=\delta_{m,m^\prime+1}
\label{FDPHeq:40}
\end{equation}
to express the shift of time step $m$ by $1$. 
As above, $\bar T^\prime_{n,m}$ gives a recursive key relation to compute $\bar T_{n-m}$ from $\sum_j G_{m,j}\hat D_{j,n}$.
We term $\bar T^\prime_{n,m}$ the conditionally predicted representative stress, given its characteristic conditional summation for forecasting the representative stress. 

The recursive summation (the second term) in Eq.~(\ref{FDPHeq:42_tensorial}) accumulating a part of $\bar T_{n-m}$ that stems from $\hat D_{j,n}$ gets completed when
$\sum_{m^\prime=-\infty}^{m-1}\sum_j G_{m^\prime,j}$ becomes identical to $\sum_{m^\prime=-\infty}^\infty\sum_j G_{m^\prime,j}$. 
Further, the variations in the $m$-th components $\bar T_{n,m}^\prime$ raised by $G_{m,j}\hat D_{j,n}$ at time step $n$ are within $m\leq -\min\bar m_j^-$ in Eq.~(\ref{FDPHeq:42_tensorial}) (surrounded by orange boxes in Fig.~\ref{FDPHfig:11p5_b}). Given these, we find that 
$\bar T_{n,m}^\prime$ converges to $\bar T_{n-m}$  
when
$m>-\min\bar m^-_j$: 
\begin{equation}
%^\forall 
m>-\min_j \bar m^-_j,\hspace{5pt} \bar T^\prime_{n,m}=\bar T_{n-m},
\label{FDPHeq:43}
\end{equation}
where $\min_j\bar m^-_j$ expresses the minimum of $\bar m^-_j$ in an admissible leaf.
This indicates that $\bar T_{n,m}^\prime$ substitutes for the component $\bar T_{n-m}$ of the representative stress at $m>-\min_j \bar m^-_j$. 

$\bar T^\prime_{n,m}$ computed in the above manner is then employed as $\bar T_{n-m}$ to compute $T_{i,n}$ by Eq.~(\ref{FDPHeq:39_tensorial}). 
$\bar T_{n-m}$ required for evaluating $T_{i,n}$ in Eq.~(\ref{FDPHeq:39_tensorial}) is localized in the range $m\geq \min\delta m_i$ (surrounded by purple boxes in Fig.~\ref{FDPHfig:11p5_a}, where $\bar T_{n-m}$ is described as $\bar T_m$ for brevity). We then need the increments due to $\hat D$ completing there before current time step $n$, to guarantee the equality $\bar T^\prime_{n,m}=\bar T_{n-m}$ (as the purple box in Fig.~\ref{FDPHfig:11p5_a} do not overlap with the orange box in Fig.~\ref{FDPHfig:11p5_b}, only where $\bar T^{\prime}_{n,m}\neq \bar T_{n-m}$). 
This requirement is satisfied if and only if the following discretized causality holds in each admissible leaf: 
\begin{equation}
\min(\delta m_i+\bar m_j^-)>0, 
\label{eq:discretizedcausality}
\end{equation}
where $\min(.)$ expresses the minimum in the concerned admissible leaf. 
As far as Eq.~(\ref{eq:discretizedcausality}) holds,
Eq.~(\ref{FDPHeq:43}) allows us to substitute $\bar T^\prime_{n,m}$ for $\bar T_{n-m}$ in Eq.~(\ref{FDPHeq:39_tensorial}) as
\begin{equation}
T_{i,n}=F_{i,m}  \bar T^\prime_{n,m}.
\label{FDPHeq:39_tensorial_prime}
\end{equation}
The condition to satisfy Eq.~(\ref{eq:discretizedcausality}) depends on the definition  
of $\delta m_i$ and $\bar m_j^-$ (intrinsically the method in approximating $t_{ij}$), and we supplement its explicit expression in \S\ref{FDPHdisccausality} considering the setting of our implementation shown in \S\ref{FDPH421} and \S\ref{FDPH43}. 

We have obviated the above-mentioned $\mathcal O(NL)$ history of $\hat D_{j,n}$ 
by using
Eq.~(\ref{FDPHeq:42_tensorial}) ($\hat D\to\bar T^\prime$) 
and 
Eq.~(\ref{FDPHeq:39_tensorial_prime}) ($\bar T^\prime\to T$) requiring $\hat D_{j,n}$ only at the current time step $n$. 
The required history of ${\bf D}_{n-m}$ (for evaluating $\hat D_{j,n}$) now ranges within $0\leq n\leq \max\Delta m_j$ only. 
We also require to store the non-zero components of $\bar T^\prime_{n,m}$ only within $\max_i \delta m_i \geq m > -\max_j \bar m_j^-$ for computing
Eq.~(\ref{FDPHeq:42_tensorial}) ($\hat D\to\bar T^\prime$) and  Eq.~(\ref{FDPHeq:39_tensorial_prime}) ($\bar T^\prime\to T$), and $\bar T^\prime_{n,m}$ is always zero within $m\leq-\max_j\bar m_j^-$, where the maximum ($\max$) is evaluated in an admissible leaf. 

As above, we compute ${\bf T}_{n+1}$ from ${\bf D}_{n-m}$ ($0\leq n\leq \max\Delta m_j$) with the additional variables $(\bar T^\prime_{m,n}, \hat D_{j,n})$ at each time step $n$. 
The required quantities are these $n$-dependent variables and ($\delta m_i,\bar m_j^-,\Delta m_j$, $f_i,g_j,h_{j,m}$), the memory to store all of which is scaled by the number of elements in the associated block clusters, i.e., $\mathcal O(N\log N)$ in total (supplemented in \ref{FDPHB5}). Although $h_{j,m}$ has two subscripts, its $m$-range is $0\leq n\leq \max\Delta m_j$ [$=\mathcal O(1)$] as for ${\bf D}_{n-m}$, and then the associated costs are $\mathcal O(N\log N)$. 
Our implementation is intrinsically a sparse-matrix arithmetic using $F_{i,m}$, $G_{m,j}$, and $\mathcal M_{m,m^\prime}$, in contrast to the vector operations in the ordinary H-matrices (also supplemented in \ref{FDPHB5}).

\subsection{A Simple Procedure for Computing $\hat D \to \bar T$}
\label{FDPH523}
In the arithmetic of Domain F, $\bar T$ can be computed simply with its definitional identity, Eq.~(\ref{FDPHeq:34}), instead of incrementing $\bar T$ through Eq.~(\ref{FDPHeq:42_tensorial}). 
Indeed, this is exactly what is executed in the PWTD method in its respective spatiotemporal clusters~\cite{ergin1999plane} although variable $\bar T$ is not explicitly defined in the PWTD method.
In this alternative procedure, 
it is enough to compute $\bar T_{n-m}$ [Eq.~(\ref{FDPHeq:34})] only for $n-m=n-\min\delta m_i$, that is, the largest $n-m$ in Eq.~(\ref{FDPHeq:39_tensorial}) [Eq.~(\ref{FDPHeq:39_tensorial}) requires $\bar T_{n-m}$ of $n-\max\delta m_i\leq n-m\leq n-\min\delta m_i$ to evaluate ${\bf T}_n$].
The other components of $\bar T_{n-m}$ can be stored beforehand as they correspond to
$\bar T_{(n-m^\prime)-\max\delta m_i}$ computed at the past time steps of $n-m^\prime=n-1,n-2,...$. 

The time complexity of the arithmetic using this $\bar T$ computation is $\mathcal O(N\log N)$ per time step, as that of the arithmetic using Eq.~(\ref{FDPHeq:42_tensorial}). 
Meanwhile, 
Eq.~(\ref{FDPHeq:34}) requires the history of $\hat D_{j,n-m}$ (or $D_{j,n-m}$ eventually) for $\min_j m_j^{-}\leq m<\max_j m_j^{-}$ for respective sources $j$ in each admissible leaf. 
The memory usage for storing it is of 
$\sum_jm_j^{-}$ and then amounts to $\mathcal O(NL)$ for the leaves of the maximum size, as in the PWTD method. 
%There may be the case where reducing $\mathcal O(NL)$ memory of the slip- and opening-rate is secondary, for example the case to store the slip- and opening-rate without thinning until the end of the simulation for reducing the output cost. 

\section{Numerical Experiments}
\label{FDPH6}
We have developed the data-sparse approximations and operations of FDP=H-matrices. In this section, we detail and confirm the properties of FDP=H-matrices with our numerical implementation of the algorithm. 

We solve 2D anti-plane problems as the simplest applications of FDP=H-matrices. In the 2D problems, the numerical cost is low, the kernel becomes simple, and these make it possible to compare the implementation of FDP=H-matrices thoroughly with the original BIE implementation. Although Domain I does not exist in the anti-plane problem, we can examine the accuracy and cost of Quantization by using it in Domain S in the 2D problems (shown in \S\ref{FDPH721} and \S\ref{FDPHB3}). In \ref{FDPHH}, we supplement the additional handling of truncation errors specific to the 2D cases due to the replacement of the kernel in Domain S by the static form. Such an error handling does not exist in the 3D cases~\cite{ando2016fast} being the primarily intended application of FDP=H-matrices. 

We normalize the stress by the self interaction ($K_{i,j,m}$ of $i=j$ at $m=0$, i.e., the radiation damping term), and adopt $\Delta t = \beta = 1$ with the Courant-Friedrichs-Lewy (CFL) parameter set at $\beta\Delta t/\Delta x = 1/2$. 

This section is organized as follows. 
In \S\ref{FDPHschemedepcosts}, 
we confirm the scheme dependence of the numerical costs (considering the constant $\eta$ and constant $\eta^2dist$ schemes). 
In \S\ref{FDPH61}, we separately examine the accuracy of each approximation detailed in \S\ref{FDPH4}. In \S\ref{FDPH62}, we demonstrate the accuracy and cost of FDP=H-matrices combining whole approximations by simulating dynamic rupture problems. In \S\ref{FDPH63}, we investigate how the simulated solution is affected by the chosen values of the approximation parameters associated with the operations in Domain F. 

\subsection{Typical Costs of Two Schemes}
\label{FDPHschemedepcosts}
In the calculation of the ST-BIEM using FDP=H-matrices, the boundary shape is first set as in the original ST-BIEM. Second, the discrete elements are clustered and the LRA of the kernel is performed by following that clustering (they constitute the data-sparse approximation of FDP=H-matrices). Third, the low-rank approximated kernel is used to simulate the given initial boundary value problem (the operation part of FDP=H-matrices). The associated clustering of elements by H-matrices are independent of the initial and boundary conditions of the problems (as it is simply the approximation of the BIE) and is uniquely determined when the discrete boundary shape is set. The structure of the block clusters (information of the level and the number of elements in each block cluster both for the admissible and inadmissible leaves) determines the cost-size scaling expected to the algorithm, and the explicit form of the kernel is not affecting the scaling, as far as the ranks of the approximated submatrices are $\mathcal O(1)$ in the respective admissible leaves. 
This property is the same as that of the original H-matrices, 
and then such a typical cost scaling FDP=H-matrices should achieve can be evaluable without specifying the specific kernel, as in the case of H-matrices~\cite{bebendorf2003adaptive}. 
We here numerically check the $N$ dependencies of such typical numerical cost orders. 

FDP=H-matrices have two schemes, namely the constant $\eta$ and constant $\eta^2dist$ schemes, and here we investigate the costs of both cases. 
We focus on the costs of the admissible leaves and do not consider the costs associated with the inadmissible leaves here because those of the latter are strictly $\mathcal O(N)$ as far as we choose finite $l_{min}$ in the inadmissibility condition Eq.~(\ref{FDPHeq:9}) [detailed in \ref{FDPHE}]. 
The rank and accuracy are not referred to below and are investigated in \S\ref{FDPH611} and \S\ref{FDPH63} in the actual elastodynamic simulations. 
We will here use $diam<\eta \bar r$ and $diam < \sqrt{\eta_0 l_{min}\bar r}$ (where $\bar r = dist + diam$) instead of Eqs.~(\ref{FDPHeq:8}) and (\ref{FDPHeq:25}) as tractable alternatives of the constant $\eta$ scheme and constant $\eta^2 dist$ scheme, respectively. 
These subtle modifications of the schemes do not affect their cost orders and are simply for checking the asymptotic size scaling quickly. 

\begin{figure*}[tbp]
   \begin{center}
  \includegraphics[width=110mm]{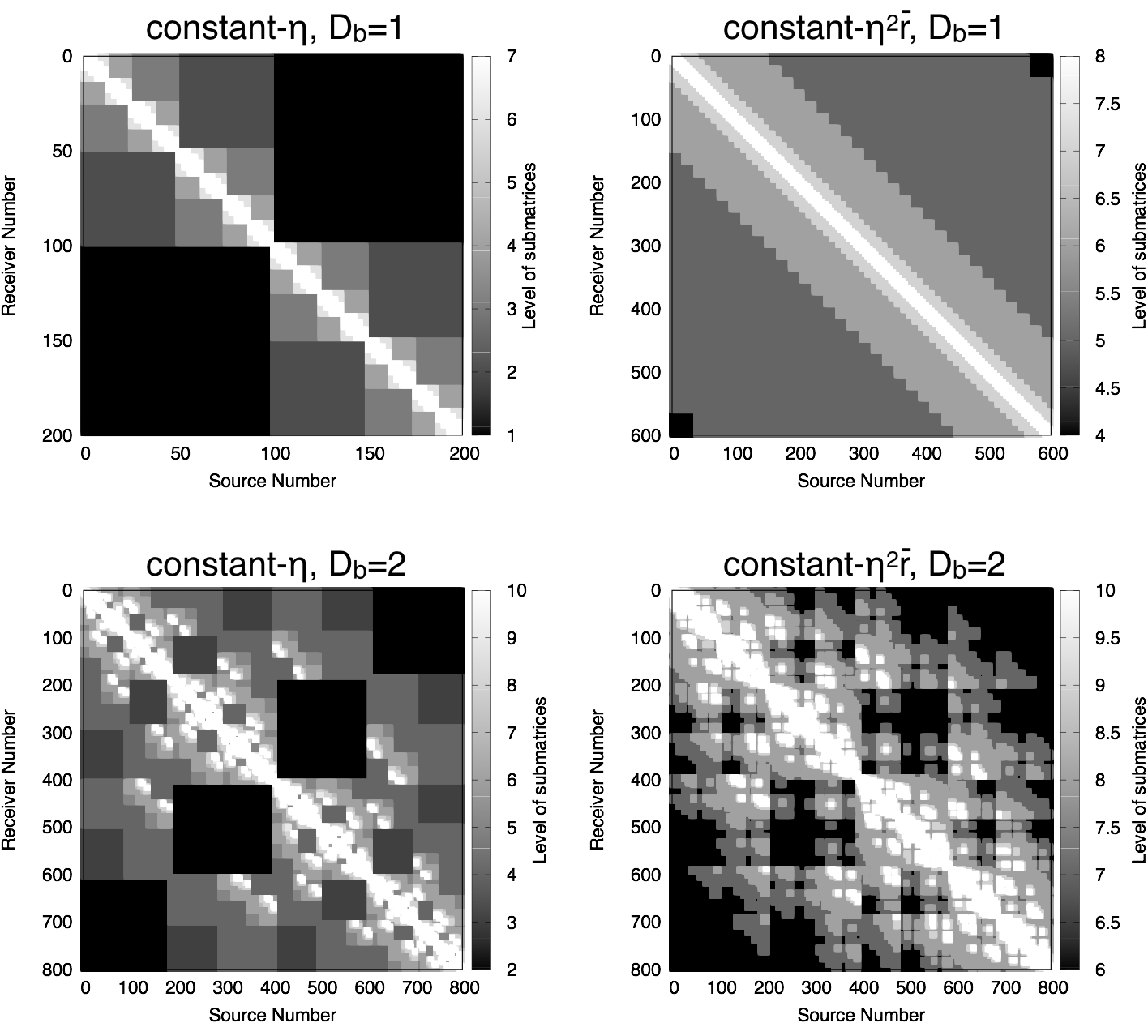}
  %\hspace{5pt}
  %\includegraphics[width=70mm]{fig_causalitycost_multi.eps}
  %\includegraphics[width=140mm]{fig9sum.png}
   \end{center}
\caption{
Submatrix distributions of H-matrices for 1D and 2D boundary configurations with the two schemes of the ART. 
The assumed geometries follow the ones introduced in \S\ref{FDPHschemedepcosts}, and the boundary dimension ($D_b$) and adopted scheme are shown at the top of each panel. 
The axes and color bar indicate the element numbers and the levels (the numbers of cluster splittings to get the corresponding submatrices) of the block clusters, respectively. Parameter values are set at $(\eta_0, l_{min}/\Delta x) = (1, 5)$ for the constant $\eta$ scheme of $D_b=1$ and at $(0.85, 2.5)$ for the others. 
}
\label{FDPHfig:9_a}
\end{figure*}

The example boundaries are shaped as follows. 
As seen below, the effective dimension $D_b$ of the boundary $\Gamma$ affects the cost scaling of the constant $\eta^2dist$ scheme, and then we consider two example cases, 
where 
$D_b$ is defined such that $[L/(\beta\Delta t)]^{D_b}=N$ for the characteristic size $L$ of the system, that is as $D_b:=\log N/\log [L/(\beta\Delta t)]$. $D_b$ can be larger (or smaller) than the primitive estimate, $D_v-1$. 
As a one-dimensional (1D) geometry of $D_b=1$, 
we consider the set of linearly aligned elements of length $L$; that gives $N=L/\Delta x$ with constant element length $\Delta x$ and discretizes $x\in(0,L)$ with the elements $\Gamma_i$ that cover $x\in(i\Delta x,(i+1)\Delta x)$ of the $x$-axis. 
For $D_b=2$, we consider a set of elements randomly and uniformly dispersed in a square the side length of which takes $L$; the number density per area $N/[L/(\Delta x/2)]^2$ is fixed at $0.08$ as a specific example. 
Other adopted parameter values are listed in the caption of Fig.~\ref{FDPHfig:9_a}. 
We note that the elements are sorted by the clustering procedure of H-matrices mentioned in \S\ref{FDPH421} as the elements of small $x$ or $y$ values take smaller numbers $i$ and $j$. 

Both in terms of the total memory and time complexity per time step,
the reduced costs of the convolution are of order the associated spatial or temporal integration lengths for each admissible leaf (shown in \S\ref{FDPH52} and in \ref{FDPHB5}). 
The spatial integration size corresponds to sum $\sum_a(N_s^a+N_r^a)$ of the number of boundary elements ($N_s^a$ sources and $N_r^a$ receivers) contained in admissible leaves $a$. 
The temporal one corresponds to sum $\sum_a \bar r^a$ of $\bar r^a$ over admissible leaves, which is $\mathcal O(\sum_a dist^a)$; normalization by $c\Delta t$ is omitted here for brevity. The temporal integration size in each domain are bounded by $\bar r^a/(c\Delta t)$ even for the large number ($M$) of time steps. 
We then evaluate these sums $\sum_a(N_s^a+N_r^a)$ and $\sum_a \bar r^a$ as indicators of the numerical cost orders. 
Hereafter, 
the index to express leaf number $a$ is omitted from $N_s$, $N_r,\bar r$, and $dist$ for brevity.

We first construct the block cluster tree (the structure to divide the kernel matrix, detailed in \S\ref{FDPH23}) in association with the cost investigation (Fig.~\ref{FDPHfig:9_a}). 
Fig.~\ref{FDPHfig:9_a} shows the obtained submatrix distribution in the block-cluster tree, visualized by the color map of the levels of submatrices.
The block-cluster distribution of the constant $\eta$ scheme shows simple fractal sieves in both dimensions. That for the constant $\eta^2 dist$ scheme shows a linear form in the 1D configuration of boundary elements while quite scattered in the 2D boundary configuration. 

%Figure 9a. âš

\begin{figure*}[tbp]
   \begin{center}
   \includegraphics[width=100mm]{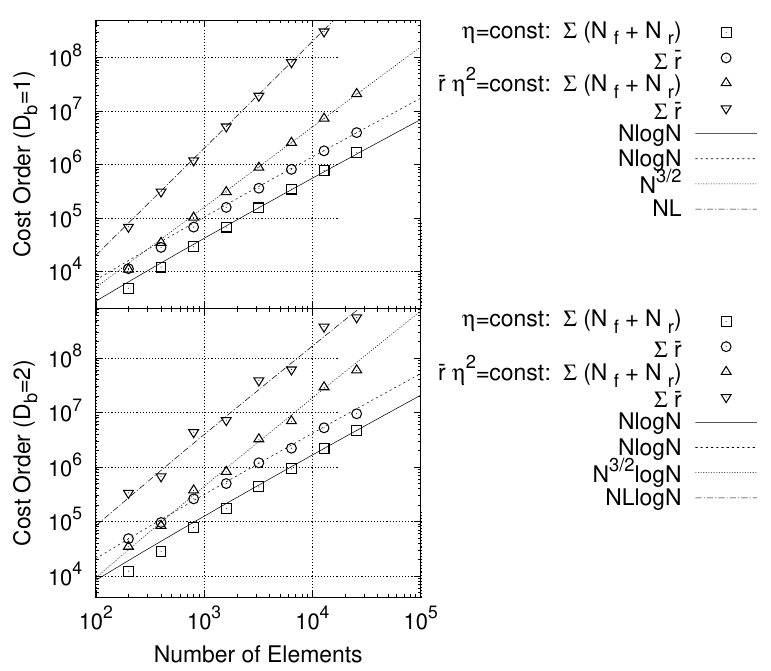}
   \end{center}
\caption{
Typical numerical costs of the admissible leaves for the two schemes of the ART, evaluated by using the 1D and 2D boundary configurations introduced in \S\ref{FDPHschemedepcosts}. 
The problem settings are the same as in Fig.~\ref{FDPHfig:9_b} except the parameters of H-matrices set at $(\eta_0, l_{min}/\Delta x) = (0.85, 5)$. 
The top and bottom panels show the results of $D_b=1$ and $D_b=2$, respectively.
}
\label{FDPHfig:9_b}
\end{figure*}

The $N$ dependencies of $\sum(N_s + N_r)$ and $\sum\bar r$ are shown in Fig.~\ref{FDPHfig:9_b}. 
They are scaled by $\mathcal O(N\log N)$ in the case of the constant $\eta$ scheme. 
As $\sum(N_s + N_r)$, being the cost order of the spatial integration for the admissible leaves of FDP=H-matrices, is also the cost order of the admissible leaves of H-matrices in the spatial BIEM, its $\mathcal O(N\log N)$ order for the case of supposing a constant $\eta$ value is evident. 
The $\mathcal O(N\log N)$ scaling of
$\sum\bar r$, the cost indicator of the temporal integration, 
is also natural under the constant $\eta$ condition, given the order estimate $dist\sim diam/\eta \sim [N_r^{1/D_b}+ N_s^{1/D_b}]/\eta\leq [N_r+ N_s]/\eta$ for $D_b\geq 1$; note $diam/dist=\mathcal O(\eta)$. 
In the constant $\eta^2 dist$ scheme cases, $\sum(N_s + N_r)$ and $\sum\bar r$ are respectively fitted well by the scaling lines of almost $\mathcal O(N^{3/2})$ [i.e., $\mathcal O(N^{3/2})$ with log factors] and of almost $\mathcal O(NL)$ using characteristic length $L$ of the system mentioned earlier.
As $D_b=1$ is special where the separation of the travel time is exactly met ($t_{ij}=\delta t_i +\bar t_j$, mentioned in \S\ref{FDPH421}), the $\eta^2 dist$ scheme is unnecessary at $D_b=1$ [where $\mathcal O(N^{3/2})+\mathcal O(NL)=\mathcal O(NL)=\mathcal O(N^2)$], and so the $\eta^2dist$ scheme will substantially be regarded as the scheme of the almost $\mathcal O(N^{3/2})$ costs [$\mathcal O(N^{3/2})$ memory and time complexity per time step] for its main coverage of $D_b=2,3$ [where $\mathcal O(N^{3/2})+\mathcal O(NL)=\mathcal O(N^{3/2})$].
All the numerical results are consistent with the scale analysis shown in \S\ref{FDPHC2}, the scale analysis of which further predicts that these scalings hold also in the 3D problems.

%Figure 9b. âš

As above, we obtain the cost estimates of FDP=H-matrices shown in \S\ref{FDPH32}. 
The $L$ factor is excluded from the cost estimates in typical geometries by the aforementioned logic, and it holds in closely spaced boundaries [$D_b\geq D_v-1$ giving $L/(\beta \Delta t)\leq N$] which is the main focus of the algorithm. 
We supplement the cost estimates containing $L$ factors in \ref{FDPHB5} after the arithmetics of Domains I and S are developed as that of Domain F. 

\subsection{Numerical Evaluation of Error Control and Cost Reduction in Domain F}
\label{FDPH61}
Below, we evaluate the cost and accuracy of H-matrices applied to each domain in \S\ref{FDPH611} and those of the ART in \S\ref{FDPH612}.
\subsubsection{H-matrices along Wavefronts in Domain F}
\label{FDPH611}
%Figure 12_1. âš
\begin{figure*}[tb]
   \begin{center}
  \includegraphics[width=120mm]{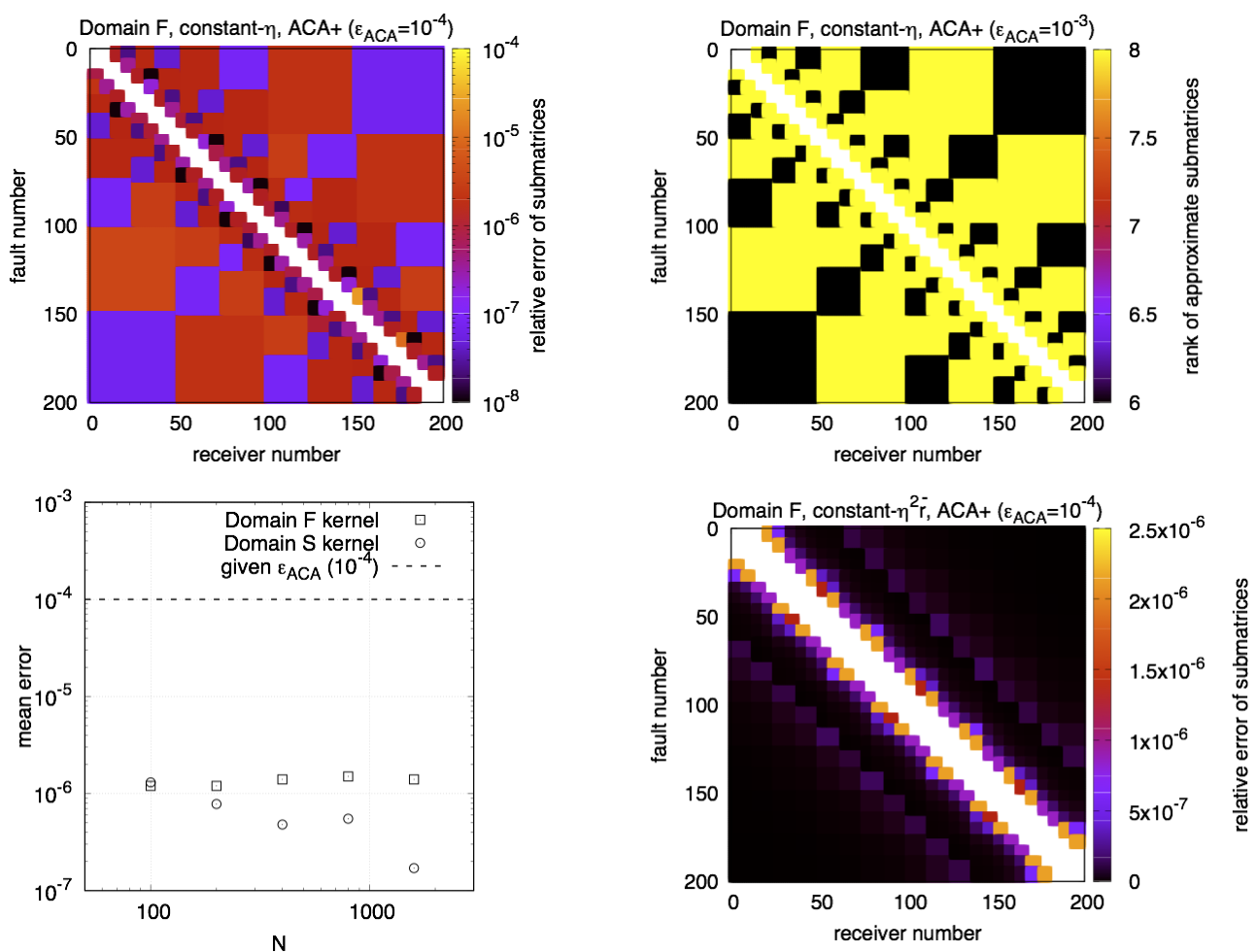}
   \end{center}
\caption{
Error and rank distributions of the LRA using ACA+, explained in \S\ref{FDPH611}.
The axes express the element numbers, and color bars indicate the relative errors or ranks of approximated submatrices.  
Parameters are set at $\eta_0 = 2, l_{min} = 14$ $(7 \Delta x)$ and Domain F is broadened by 3$\Delta x/\beta$ (detailed in \ref{FDPHH}). Required error bound $\epsilon_{ACA}$ is indicated in the panels. 
(Top left) Error distribution in $\hat K^F$ for constant $\eta$. 
%The errors of the approximate matrix are suppressed below $\epsilon_{ACA}$. 
(Top right) Rank distribution in $\hat K^F$ for constant $\eta$. 
(Bottom left) Mean error (over submatrices, defined in \S\ref{FDPH612}) versus the number of element $N$.
%The ranks of submatrices are independent from the number of the submatrix components.
%The errors become smaller as $dist$ increases due to the decreases of $\eta$.
(Bottom right) Errors in $\hat K^F$ for constant $\eta^2 dist$. 
}
\label{FDPHfig:12}
\end{figure*}

The following numerically tests the accuracy and cost of H-matrices applied along Domain F, introduced in \S\ref{FDPH41}. 
We here choose a planar fault as a simple application example, in the same unit and discretization as the $D_b=1$ case in \S\ref{FDPHschemedepcosts}). 
Now the kernel is explicitly computed, and the units and CFL parameter value are the already mentioned ones $K_{0,0,0}=\Delta t = \beta = 1$ and $\beta\Delta t/\Delta x = 1/2$. Unspecified adopted parameter values are listed in Fig.~\ref{FDPHfig:12}. The constant $\eta$ scheme being our main proposal is investigated basically, and the results of the constant $\eta^ 2dist$ scheme are briefly mentioned. 

As mentioned in \S\ref{FDPH23}, H-matrices approximate the submatrix ${\bf K}_a$ of admissible leaf $a$ to a low-ranked one, denoted by ${\bf K}_{a,LRA}$. The error criterion is set as $|{\bf K}_a-{\bf K}_{a,LRA}|<\epsilon_H|{\bf K}_a|$ with regard to the Frobenius norm $|\cdot|$ by using given constant $\epsilon_H$. 
This criterion gives the candidates (denoted by ${\bf K}_{a,LRA,l}$ for rank $l$), and the minimum-rank candidate is adopted as ${\bf K}_{a,LRA}$. 
A fast approximation technique of the $\mathcal O(N\log N)$ complexity and memory is typically utilized to amend the $\mathcal O(N^3)$ computational time and $\mathcal O(N^2)$ memory capacity of the exact LRA~\cite{borm2003hierarchical}. 
A common basis-selection method is the ACA~\cite{bebendorf2003adaptive} of partial pivoting. 
The error criterion of the ACA~\cite{borm2003hierarchical} is
$|{\bf K}_{a,LRA,l}-{\bf K}_{a,LRA,l+1}|/$ $|{\bf K}_{a,LRA,l+1}|$ $<\epsilon_{ACA}$, where ${\bf K}_{a,LRA, l+1}$ of the 1 higher rank replaces original submatrix ${\bf K}_a$ in the original criterion for ${\bf K}_{a,LRA}={\bf K}_{a,LRA,l}$ besides the subtle modification of the bounding parameters: $\epsilon_H\to\epsilon_{ACA}$. 
This altered error criterion exactly observes the original one (with $\epsilon_H=\epsilon_{ACA}$) for the complete pivoting, and the partially pivoting ACA executes the LRA in an approximate yet fast manner of the partial pivoting, expecting $\epsilon_H\approx \epsilon_{ACA}$~\cite{bebendorf2003adaptive,borm2003hierarchical}. 
A relation $|{\bf K}_a-{\bf K}_{a,LRA}|\lesssim \epsilon_{ACA}|{\bf K}_a|$ holds if the ACA works successfully. 
Although the above criterion of the ACA is originally for ${\bf K}_{a,LRA,l}$, we adopted ${\bf K}_{a,LRA,l+1}$ as the low-ranked kernel in this study when the above criterion is satisfied. 

This study uses ACA+~\cite{grasedyck2005adaptive}, which improves the accuracy of the partially pivoting ACA by using a randomly selected point as an additional candidate of the pivoting point in the pivoting process. 
In our investigation, the partially pivoting ACA was sometimes erroneous even in the spatial BIEM (\ref{FDPH_ACAcheckapp}). 

Regarding the numerical accuracy, 
we evaluate whether each low-ranked submatrix satisfies the expected accuracy $\epsilon_H\lesssim\epsilon_{ACA}$. For this accuracy evaluation, if the LRA does not converge as sometimes occurring in the partially-pivoting ACA cases of our investigation, we terminate the LRA when the rank exceeds the original rank of each submatrix. To clarify the degree of the convergence, we do not employ any exception handling for the approximated matrices obtained through the LRA in \S\ref{FDPH611}, and also in \ref{FDPH_ACAcheckapp}. 

Regarding the numerical costs, we measure the rank of each submatrix. If the approximation works well, the rank of an approximate matrix is expected to be $\mathcal O(1)$ and is independent of the number of submatrix components. These are crucial to achieve $N\log N$ costs by FDP=H-matrices, and their confirmation is the test of our statement that H-matrices work successfully along the singular wavefronts.

\paragraph{Constant $\eta$ Scheme}
The result of the constant $\eta$ scheme is described below. 
%We examined what kind of low-rank approximation methods can achieve an almost $\mathcal O (N)$ cost with high accuracy.

With ACA+, the accuracy is satisfactory in Domain F [Fig.~\ref{FDPHfig:12} (top left)], as for the static kernel of the spatial BIEM~\cite{borm2003hierarchical} corresponding to the kernel in Domain S. As later shown, ACA+ worked for all the matrices expressing the spatial dependence of the kernels implemented in this paper (shown in \S\ref{FDPH631} and Table~\ref{FDPHtab:1}). 
The norm of the relative error due to the LRA is approximately $10^{-2}$ times smaller than $\epsilon_{ACA}$ in most submatrices. This smallness may be due to the aforementioned alteration of the error criterion that we adopt ${\bf K}_{a,LRA,l+1}$ (more accurate one) when the error criterion for ${\bf K}_{a,LRA,l}$ is satisfied. Aside from that detail, the error regulation $|{\bf K}_a-{\bf K}_{a,LRA}|\lesssim \epsilon_{ACA}|{\bf K}_a|$ is satisfied in all the submatrices as expected.

The rank of an approximated submatrix is independent from the number of elements in the submatrix [Fig.~\ref{FDPHfig:12} (top right)]. The ranks are almost constant and of $\mathcal O(1)$. This will be the first numerical confirmation that H-matrices work in Domain F, namely along wavefronts of the elastodynamic kernel. 

Additionally, we notice the fractal patterns of
the accuracy and rank distributions appearing along the direction from the center to the top right or bottom left end in all the panels of Fig.~\ref{FDPHfig:12} for the constant $\eta$ scheme. 
Such an oscillatory behavior corresponds to the (hierarchically repeating) variations in the values of $diam / dist$ within $\eta / 2 <diam / dist <\eta$ occurring between block clusters at each level. 
This behavior is consistent with the expected nature of the LRA applied to the kernel in Domain F (i.e., along the wavefronts) that the LRA is there substantially an expansion about $diam / (diam+dist)$ as for the static kernel of Domain S, as formulated in \S\ref{FDPH41}. 
These vibrations would not matter as the error is always much lower than $\epsilon_{ACA}$ and the rank is $\mathcal O (1)$. 

Fig.~\ref{FDPHfig:12} (bottom left) shows the $N$-dependence of the error in the LRA. 
The selected parameter values are unchanged from those in the above experiments except the $N$ values. 
We measured the accuracy of the LRA by using the average of the relative error norm $|{\bf K}_a - {\bf K}_{a,LRA}|/|{\bf K}_a|$ of submatrices weighted by the numbers of the submatrix entries, 
called a mean error here. 
It represents the effective relative error expected in each matrix entry. 
The mean errors of the kernels are shown to be smaller than the specified $\epsilon_{ACA}$ value ($\epsilon_{ACA}=10^{-4}$) in the studied $N$ range. 
The error of the asymptotic Domain S kernel (corresponding to the spatial BIEM kernel) tends to decrease as $N$ increases. The error of the kernel of Domain F is roughly independent of $N$. 
As above, the difference exists in the size dependence between Domains F and S kernels. This will be ascribed to the difference in the attenuating natures, a possibly unique difference of these two kernels in this setting. 
Although the investigated size range is here not so large for the application of H-matrices, 
by considering that the studied fault size ($N\geq100\Delta x$) is much larger than $l_{min}(=7\Delta x)$ and $\Delta x$ in Fig.~\ref{FDPHfig:12} (bottom left), these observed tendencies are expected to be within the asymptotic region, maintained even at larger $N$ values.

\paragraph{Constant $\eta^2 dist$ Scheme} 
ACA+ worked successfully in the case of the constant $\eta ^ 2 dist$ scheme as in the case of the constant $\eta $ scheme [Fig.~\ref{FDPHfig:12} (bottom right)]. 
Besides, the accuracy improvement was observed for the constant $\eta^ 2 dist$ scheme as $dist$ increases. 
It implies that the LRA applies more safely with the constant $\eta^ 2 dist$ scheme than with the constant $\eta$ scheme.
This could be interpreted as the fast convergence provided by the nature of this scheme that the ratio $diam / (diam+dist)$ [$=\mathcal O(\eta)$], i.e. the perturbation parameter in the LRA, gets smaller as $dist$ increases. 
It would be another support for that the LRA in Domain F of our implementation was successfully an expansion about $diam / (diam+dist)$. 

\subsubsection{ART}
\label{FDPH612}
%Figure 13. âš
\begin{figure*}[tbp]
   \begin{center}
  \includegraphics[width=120mm]{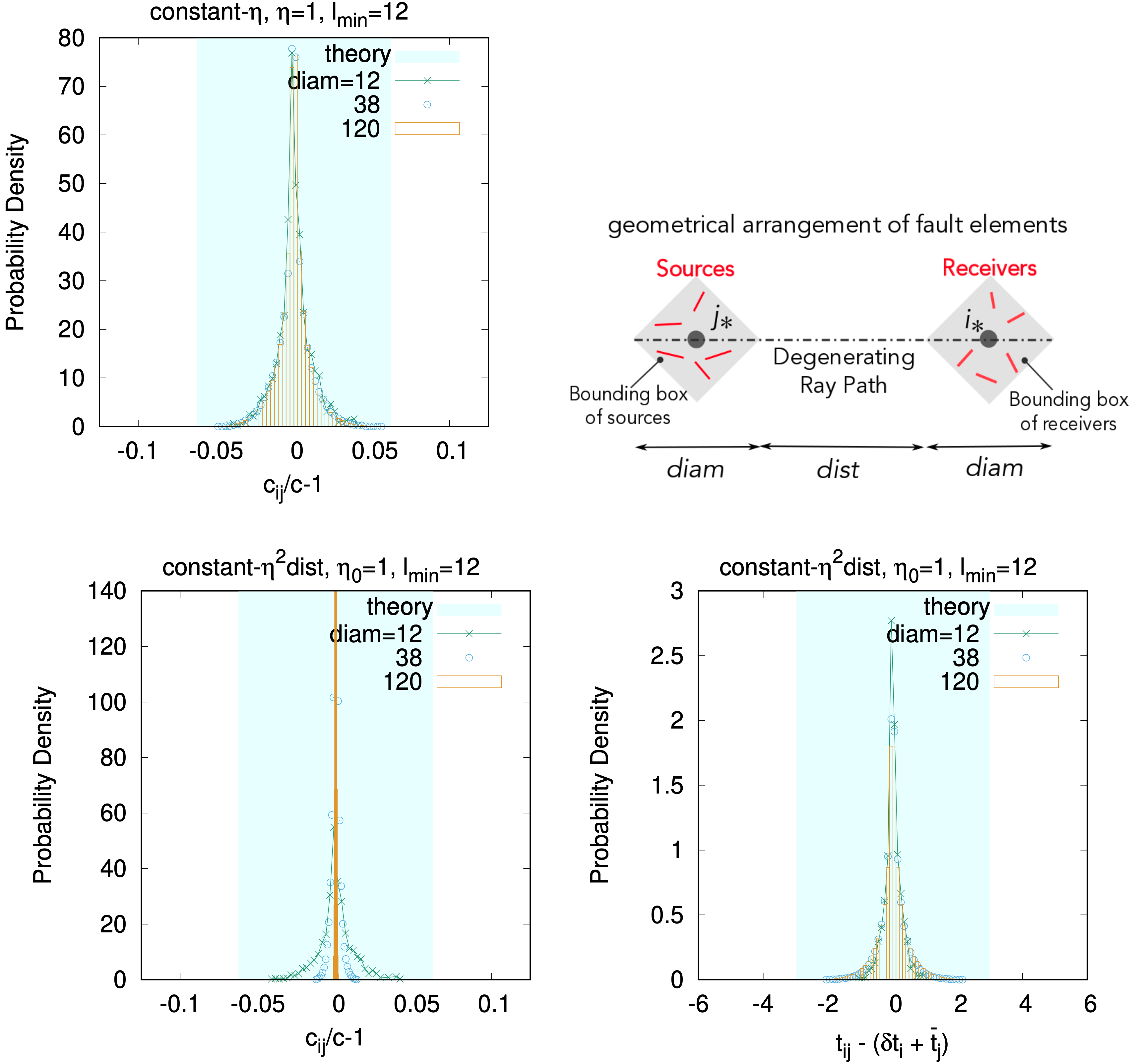}
  %\includegraphics[width=55mm]{pict_cij_consteta.eps}
  %\includegraphics[width=62mm]{distributedfaultiinadmissiblleleaves.png}
  %\\
  %\includegraphics[width=55mm]{pict_cij_constetasquaredist.eps}
  %\includegraphics[width=55mm]{pict_tij_constetasquaredist.eps}
   \end{center}
\caption{
Error distributions of the effective wave speeds and travel times generated by the ART. The problem setting is detailed in \S\ref{FDPH612}. 
The theoretical upper bounds [Eqs. (\ref{FDPHeq:24}) and (\ref{FDPHeq:26})] of their errors are also indicated (colored regions termed theory) in the panels. 
(Top right) Simulated geometry. The degenerating ray path overlaps diagonal lines of bounding boxes and realizes $diam$ = $\eta dist$. 
(Top left) Error distributions of the effective wave speeds 
in the constant $\eta$ scheme, independent of $diam$. 
(Bottom left) Error distributions of the effective wave speeds 
in the constant $\eta^2dist$ scheme, becoming impulsive as $diam$ increases. 
(Bottom right) 
Error distributions of the effective travel time 
in the constant $\eta^2dist$ scheme, independent of $diam$. 
}
\label{FDPHfig:13}
\end{figure*}

The ART provides its two schemes, namely the constant $\eta$ scheme and the constant $\eta^2dist$ scheme. 
The constant $\eta$ scheme regards the separation of the travel time [Eq.~(\ref{FDPHeq:17})] as an approximation regulating the error of the wave speeds [defined in Eq.~(\ref{FDPHeq:23})], and the associated error bound is given by Eq.~(\ref{FDPHeq:24}). The constant $\eta^2dist$ scheme straightforwardly regards Eq.~(\ref{FDPHeq:17}) as an approximation regulating the error of the travel time, bounded by Eq.~(\ref{FDPHeq:26}). 
These approximations and bounds are investigated below. 
The accuracy of the other approximation in the ART is related to the normalized waveform [Eq.~(\ref{FDPHeq:21})] and is affected by the temporal change rate in the slip- and opening-rate that depends on the given problem, and then we evaluate it later in the dynamic rupture simulation (in~\S\ref{FDPH62}). 

Fig.~\ref{FDPHfig:13} (top right) shows a configuration supposed in the following accuracy test. 
The fault elements of constant length $\Delta x$ are distributed uniformly within a pair of 2D bounding boxes (with the number density $= 1/4$, as an example). 
There, ratio $diam/dist$ can take the maximum value $\eta(-0)$ and the degenerating ray path overlaps with some diagonal lines of the source and receiver bounding boxes. 
It is one of the demanding cluster configurations for using the approximation of Eq.~(\ref{FDPHeq:17}) among the available choices under a given admissibility condition. 
We did not study the linearly aligned faults despite their geometrical simplicity because the travel-time approximation Eq.~(\ref{FDPHeq:17}) yields no errors on a straight line, as mentioned in \S\ref{FDPH421}. 
We varied $diam/dist$ to study the accuracy of the travel-time approximation, by considering the prediction of Eq.~(\ref{FDPHeq:17}) that the error bound of the travel-time approximation is scaled by $diam/dist$. As the parameter values of $\eta_0$ and $l_{min}$ do not influence the approximation of the ART qualitatively, we investigate only one parameter set ($\eta_0 = 1$, $l_{min} = 12$) with respect to them. The travel-time approximation Eq.~(\ref{FDPHeq:17}) is fully described by the spatial configuration without the information of the temporal discretization and the kernel components, and we do not specify them here. 

Fig.~\ref{FDPHfig:13} (top left) shows the errors of the effective wave speeds [Eq.~(\ref{FDPHeq:23})] 
in the constant $\eta$ scheme. The value of $diam$ is initially set at $\eta_0l_{min}$ and changes $\sqrt{10}$-fold and $10$-fold in the figure. 
The errors of the effective wave speeds in these cases obeyed almost the same distribution independent of $dist$. It is consistent with the expected non-dispersive nature of the constant $\eta$ scheme described in \S\ref{FDPH422}. 
Besides, most of the errors were within (and moreover, much smaller than) the theoretical approximate upper bound $(1+\eta^{-1})^{-2}/4$ given by Eq.~(\ref{FDPHeq:24}), represented by the bluish-green frame in Fig.~\ref{FDPHfig:13}. 

Fig.~\ref{FDPHfig:13} (bottom left) shows
the distribution of the effective wave speeds in the constant $\eta^2 dist$ scheme. 
As expected, it becomes delta functional as the $dist$ value increases, and the errors almost disappear. 
Fig.~\ref{FDPHfig:13} (bottom right) further confirms that 
the errors of the travel times are regulated within the approximate upper bound $\eta_0 l_{min} / (4c)$ given by Eq.~(\ref{FDPHeq:26}) and are finite even at a distance. 

In summary, the error upper bounds of the ART were shown well evaluated by the analytic Eqs.~(\ref{FDPHeq:24}) and (\ref{FDPHeq:26}). 
The measured error distribution also showed that 
the error values were much smaller than these analytical bounds in most cases. 
They suggest that FDP=H-matrices can be highly accurate even on nonplanar boundaries, a demanding example of which is the distributed boundary elements analyzed in this subsection.

\subsection{Dynamic Rupture Simulations}
\label{FDPH62}

We get into the investigation of the cost and accuracy of FDP=H-matrices with actual numerical simulations. The cost investigation is shown in \S\ref{FDPH621}, and the accuracy in \S\ref{FDPH622}. 

In this subsection, we treat the dynamic rupture problem as an example of the elastodynamic simulation. 
The dynamic rupture problem is an initial boundary value problem; its problem setting comprises the boundary geometry, the boundary condition, and the initial condition. 
The geometry and the initial condition will be detailed in \S\ref{FDPH622} where the physical setting becomes relevant. The adopted parameter values are listed in the figures for reproducibility; by association, we will show the values of the parameters concerning the 2D specific approximations, defined in \ref{FDPHH}. The figures of dynamic rupture solutions are thinned out for visibility. 

The boundary condition of the dynamic rupture problem is ordinarily a mixed boundary condition that takes the displacement-discontinuity condition on the unruptured area and the traction boundary on the fractured area of the crack surface. 
On the unruptured area, we assume the anti-plane shear displacement-discontinuity $\Delta u({\bf x},t)$ is time-independent: 
\begin{equation}
\Delta \dot u({\bf x},t)=0.
\nonumber
\end{equation}
This is an example of ${\bf f}_{\Delta u}({\bf x},t)$ (in a temporally differentiated form) mentioned in \S\ref{FDPH21}. 
On the ruptured area, we assume the exponential slip weakening law for the shear traction $T_{shear}$ at location ${\bf x}$: 
\begin{equation}
T_{shear} = (T_{th} -T_{dy})\exp(-\Delta u/D_c)+T_{dy},
\nonumber
\end{equation}
where $T_{th}$ denotes the yielding value of the traction, $T_{dy}$ the shear traction in the fully fractured zone, and $D_c$ a characteristic slip-weakening distance. 
This is an example of ${\bf f}_{T}({\bf x},t)$ mentioned in \S\ref{FDPH21}. 
Besides, we assume that the unruptured area transitions to the ruptured one when the traction value $T_{shear}$ on it reaches to the threshold $T_{th}$. 
The appearing parameters $T_{th}$, $T_{dy}$, and $D_c$ of the above boundary condition are assumed to be spatially homogeneous in this study. 

Hereafter, we modify the implementation of ACA+ from the test of the LRA executed in \S\ref{FDPH631}.
We replace the approximate submatrix with the original submatrix 
when the rank of the approximated submatrix exceeds that of the original submatrix. 
We required such exception handling occasionally in the neighboring clusters of originally small ranks even with ACA+, for the cases of nonplanar faults. 

\subsubsection{Cost Scaling}
\label{FDPH621}

We here measure the numerical costs (the total memory consumption and time complexity) of a dynamic rupture simulation with a simple planar boundary geometry same as that in \S\ref{FDPH61}. 
The boundary and initial conditions follow those of the later-mentioned planar problems in \S\ref{FDPH622}; the initial and boundary conditions do not affect the leading orders of the time complexity and memory, which is for evaluating the BIE by FDP=H-matrices, and then the following would not be the condition-specific result. 
To focus on the geometry-independent aspects of the cost scaling, we evaluate the numerical costs of the original ST-BIEM without any reduction assuming the translational symmetry of the boundary that holds only in the planar boundary cases of structured elements. 
The time complexity is measured without any parallelization on a laptop (MacBook Pro MF839). 
The time complexity per time step is quantified as the ratio of the wall-clock time (taken by the whole simulation) to the number of time steps, which is below referred to as the computation time per time step. 

%Figure 14. âš
\begin{figure}[tbp]
%\begin{figure}[H] %% usepackage{here}
   \begin{center}
  \includegraphics[width=75mm]{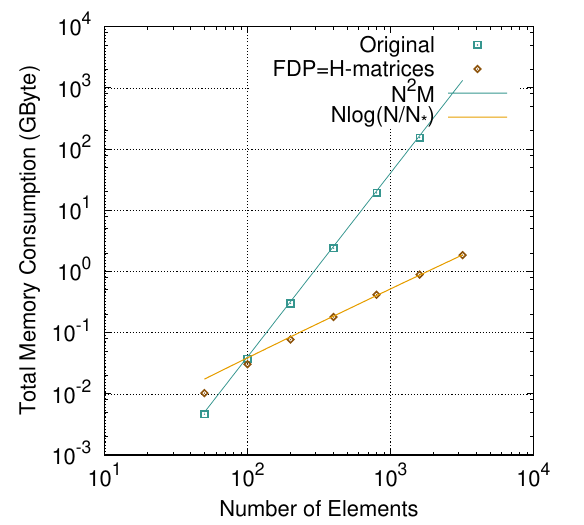}
\\
  \includegraphics[width=75mm]{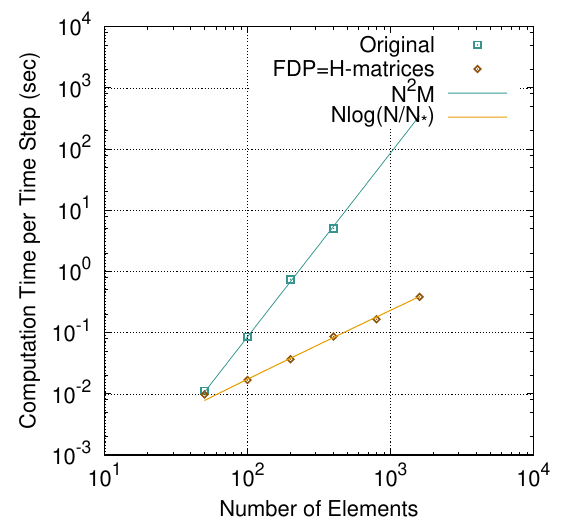}
   \end{center}
\caption{
Measured costs of FDP=H-matrices, compared with the original ST-BIEM ones. Plotted results are of a planar problem detailed in \S\ref{FDPH622}, 
for parameter values $M = 5N$, $\epsilon_{Q} =\epsilon_{st} =\epsilon_{ACA}= 10^{-2}$, $l_{min}/\Delta x=5$, and $\eta = 2$, with Domain F broadened by $3\Delta x /\beta$. 
Asymptotes of $\mathcal O(N\log N/N_*)$ and of $\mathcal O(N^2M)$ are indicated by lines with a constant $N_*$.
%The results demonstrate the almost $\mathcal O(N)$ cost scaling of FDP=H-matrices. âš
(Top) Total memory consumption. (Bottom) The computation time per time step (the total computation time to complete the whole simulation, divided by the number of the total time steps). 
}
\label{FDPHfig:14}
\end{figure}

Fig.~\ref{FDPHfig:14} compares the numerical costs of FDP=H-matrices with those of the original ST-BIEM. %The adopted values of approximation parameters are described in the caption. 
Both the total memory consumption and computation time per time step are $\mathcal O(N^2M)$ in the original ST-BIEM. 
As expected, both show the almost $\mathcal O(N)$ scaling in the results of FDP=H-matrices. 

More precisely, the costs of FDP=H-matrices are well fitted to the $N\log(N/N_*)$ scaling with constant $N_*$, indicating $N\log N$ at $N \gg N_*$. 
This is natural as FDP=H-matrices have $\mathcal O(N)$ costs of inadmissible leaves and $\mathcal O(N\log N)$ costs of admissible leaves. 
In the figure, $N_*\sim 10$ is obtained and we investigate the parameter dependence of $N_*$ in \S\ref{FDPH632}. 
$N\log(N/N_*)$ yields the expected $N\log N$ asymptote and confirms that FDP=H-matrices achieve $N\log N$ scaling in the elastodynamic problem.

\subsubsection{Spatiotemporal Patterns of Solution Accuracy}
\label{FDPH622}
Here, we simulate two examples of a planar boundary and a nonplanar one with the constant $\eta$ scheme. The value of $\eta$ used here is near 1, the typical order of $\eta$ values in H-matrices; for example, $\eta = 2$ is used in a previous study~\cite{ohtani2011fast} of a 3D elastostatic problem.

\paragraph{Accuracy in Planar Problems}
%Figure 15. âš
\begin{figure*}[tbp]
   \begin{center}
  %\includegraphics[width=70mm]{fig_Dxt_planar_original_thinned6.eps}
  %\includegraphics[width=70mm]{fig_Dxt_planar_eta1_thinned6.eps}
	%\\
  % \includegraphics[width=70mm]{PlanarFaultExample_Setting.pdf}
  %\includegraphics[width=70mm]{fig_Dx_snapshot_planar_time.eps}
  \includegraphics[width=120mm]{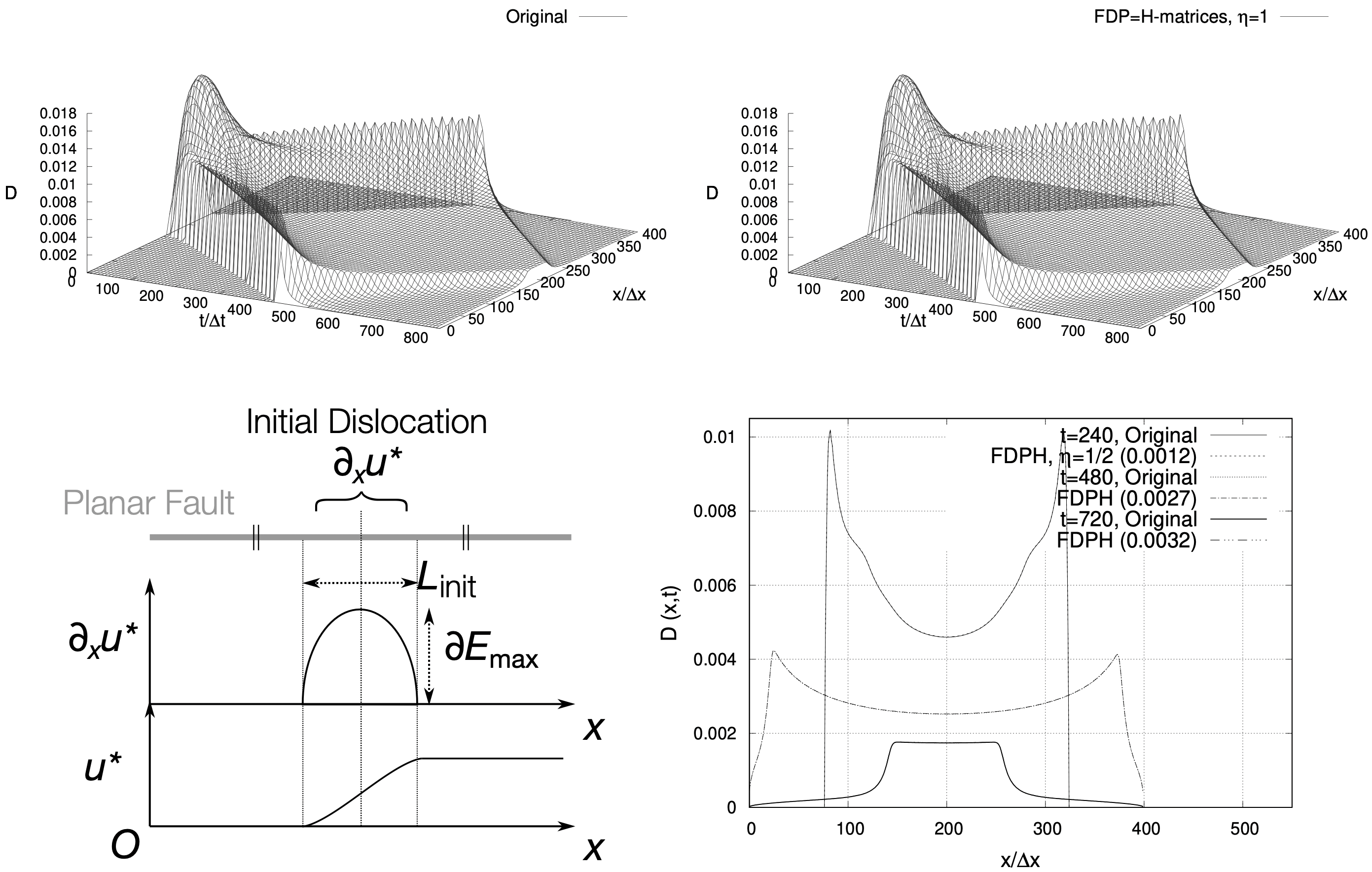}
   \end{center}
\caption{
Simulated dynamic rupture on a planar fault. 
Parameter values of the given problem are set at $(T_{th},T_{bg}/T_{th},T_{dy} /T_{th},D_c/(\Delta x/2),$ $L_{init} /\Delta x)$ $= (10^{-2},0.35,0,0.1,50)$. The parameters of FDP=H-matrices are set at $(l_{min} /\Delta x, \epsilon_{ACA}, \epsilon_{Q}, \epsilon_{st}) = (5,10^{-3},10^{-3},10^{-6})$, and Domain F is broadened by $10\Delta x/\beta$. 
(Top left) The original solution of slip rate $D$ evolving over space $x$ and time $t$. (Top right) $D$ simulated by FDP=H-matrices evolving over space $x$ and time $t$. (Bottom left) Simulated planar geometry and excitation, detailed in \S\ref{FDPH622}. The elliptic dislocation $\partial_x u_*(\leq \partial E_{max})$ of dimension $L_{init}$ increments the initial stress field on the fault lying along the $x$-axis. (Bottom right) Snapshots of the top panels at given time $t$. FDP=H-matrices are abbreviated to FDPH. Bracketed numbers at the ends of the legends of FDP=H-matrices indicate the errors from the original solution in respective snapshots, evaluated by the ratios of the Euclidean norms of the residuals to the Euclidean norms of the original solution at the given time steps. 
}
\label{FDPHfig:15}
\end{figure*}

First, we consider a planar fault as the simplest geometry case. 
The boundary is the same as that in \S\ref{FDPH61}, where the elements $i=0,...,N-1$ is located along $x$-axis (i.e. $y=0$) and covers $x\in(i\Delta x,(i+1)\Delta x)$ of length $\Delta x$. The fault dimension is here denoted by $L$, which satisfies $L=N\Delta x$. 

The initial condition in the following planar boundary problem is shown in Fig.~\ref{FDPHfig:15}. 
The initial traction is 
the sum of a constant background value $T_{bg}$ and the quasistatic traction field incurred by the elliptically distributed dislocation $\partial_1 u^*$, the length of which is here denoted by $L_{\rm{init}}$, and the maximum value of which is $\partial E_{\max}$ ($T_{bg}+K^{stat}*\partial u^*$, where $*$ denotes the spatial convolution and $K^{stat}$ denotes the aforementioned elastostatic kernel); the center of the dislocation is set to coincide with that of the fault line. 
The ruptured area is initially identified with the area giving nonzero initial dislocation values while the slip (the shear components of the displacement discontinuity) is initially set at zero homogeneously over the entire boundary. Below, $\partial E_{\max}$ is set just at the threshold value such that $\max(T_{bg}+K^{stat}*\partial u^*)=T_{th}$, giving an yielding point on the boundary. 

Fig.~\ref{FDPHfig:15} (top left) shows the spatiotemporal evolution of the slip rate in the original ST-BIEM of the adopted parameter values. The rupture propagates over the fault starting from the initially ruptured area. Fig.~\ref{FDPHfig:15} (top right) shows the solution obtained by using FDP=H-matrices. The solution of FDP=H-matrices is shown to reproduce the original solution well. 

Fig.~\ref{FDPHfig:15} (bottom right) shows the snapshots of the solutions at given time steps, indicating the detail of the error distribution. The error of the solution ($D ^{FDPH}_{i,n}$) of FDP=H-matrices distributed over elements $i$ at each time step $n$ is quantified with the relative absolute error, $[\sum_i (D^{FDPH}_{i,n} - D^{orig}_{i,n})^2/$ $\sum_i (D^{orig}_{i,n})^ 2]^{1/2}$, from original solution $D ^{orig}_{i,n}$. 
The values of this error are shown in brackets at the end of the legend of FDP=H-matrices. 
The errors were below $0.4\%$ at $\eta=1/2$ even after the roughly 500 steps [Fig.~\ref{FDPHfig:15} (bottom right)]. 
Also remarkably, there were no observable errors in the rupture propagation speed that is extensively investigated in the fracture mechanical literature (e.g., Ref.~\cite{andrews1976rupture}).
These observations imply that the accuracy of FDP=H-matrices will be sufficient in many cases despite its cumulative property. Indeed, $0.4\%$ is approximately 0.1 times smaller than the cumulative short-wavelength numerical oscillations frequently observed owing to given numerical conditions and rounding errors of the kernel evaluation \cite{day2005comparison,tada2001dynamic}, Fig.~3 in Ref.~\cite{tada2001dynamic}. 

\paragraph{Accuracy in Nonplanar Problems}
%Figure 16. âš
\begin{figure*}[tbp]
   \begin{center}
  %\includegraphics[width=75mm]{fig_Dxt_nonplanar_original_thinned.eps}
  %\includegraphics[width=75mm]{fig_Dxt_nonplanar_eta05_thinned.eps}
	%\\
  %\includegraphics[width=75mm]{simulatedgeometry.png}
  %\includegraphics[width=75mm]{fig_Dx_snapshot_nonplanar.eps}
  %\includegraphics[width=150mm]{Fig16sum.pdf}
  %\includegraphics[width=160mm]{Fig16sumpng2.png}
  \includegraphics[width=125mm]{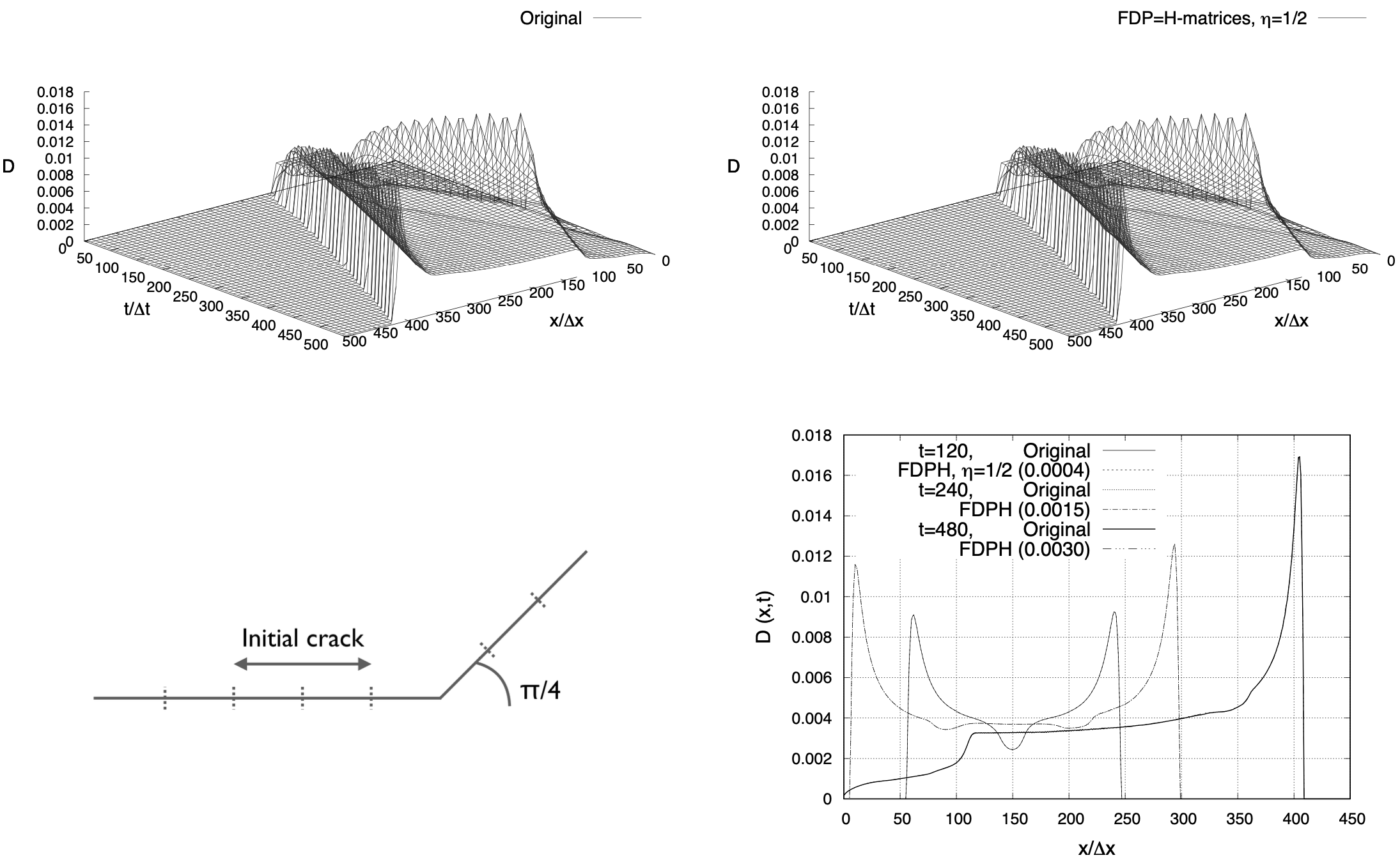}
   \end{center}
\caption{
Simulated dynamic rupture on a nonplanar fault. 
The parameter values are set as $(T_{th},T_{bg}/T_{th},T_{dy} /T_{th},D_c/(\Delta x/2)) = (10^{-2},0.35,0,0.1)$, $(L_{init} /\Delta x, E_{init} /(\Delta x/2)) = (50,0.02)$, and $(l_{min}/\Delta x, \epsilon_{ACA}, \epsilon_{Q}, \epsilon_{st}) = (5,10^{-3}, 10^{-3}, 10^{-6})$, and Domain F is broadened by $10\Delta x/\beta$. 
(Top left) 
Original solution of slip rate $D$ evolving over space $x$ and time $t$. 
(Top right) $D$ simulated by FDP=H-matrices evolving over space $x$ and time $t$. 
(Bottom left) Simulated geometry and excitation, detailed in \S\ref{FDPH622}. The kink (by angle $\pi/4$) is located between $x/\Delta x = 249$ and $250$. 
The elliptic crack increments the initial stress field on the fault. 
(Bottom right) Snapshots of the top panels at given times $t$. 
FDP=H-matrices are abbreviated to FDPH. Bracketed numbers in the legends indicate the relative errors of FDP=H-matrices in the same manner as in Fig.~\ref{FDPHfig:15}.
}
\label{FDPHfig:16}
\end{figure*}

Fig.~\ref{FDPHfig:16} (bottom left) shows a simulated example of a nonplanar boundary geometry, which is a line fault (corresponding to the previous planar example) kinked at 5/8 length by $\pi /4$. 
Initially, we suppose that the shear traction is at a constant value $T_{bg}$. 
An elliptical slip of radius $L_{init}$ is next quasistatically imposed at time $t=0$ such that the maximum slip is equal to $E_{init}$, and we solve the consequential dynamic rupture propagation. The initially ruptured area is exactly that of the quasistatically imposed elliptical slip. 

Fig.~\ref{FDPHfig:16} (top left) and Fig.~\ref{FDPHfig:16} (top right) respectively show the spatiotemporal evolution of the slip rates simulated by the original ST-BIEM and FDP=H-matrices. 
In the original result, the rupture first propagates over a plane before the time step $t/\Delta t \sim 100$. The rupture subsequently extends to the whole fault area beyond the kink (located between elements $i = 249$ and $250$). The result of FDP=H-matrices reproduces the original solution well.

The snapshots [Fig.~\ref{FDPHfig:16} (bottom right)] show that FDP=H-matrices accurately reproduced the original solution even in this nonplanar fault geometry. The error is shown temporally cumulative yet satisfactorily small. These are the same as in the planar problem. The magnitude of the error can be roughly the same as in the planar problem. 

\subsection{Parameter Dependence of Costs and Accuracy}
\label{FDPH63}
We end the numerical experiments by investigating the dependencies of the cost and accuracy on the parameters of FDP=H-matrices that control the characteristic approximations in Domain F, described in \S\ref{FDPH4}. First, we study the influence of $\epsilon_{ACA}$ (the approximate error bound of the LRA in H-matrices). Second, we study the influence of $\eta$ (upper bound of $diam / dist$) determining the approximation accuracy of the ART. Other parameters for handling the 2D specific errors are detailed in \ref{FDPHH}.

In the following text, we focus on the constant $\eta$ scheme which can achieve the $\mathcal O (N\log N)$ cost scaling. 

\subsubsection{$\epsilon_{ACA}$ Dependence}
\label{FDPH631}

We summarized the influence of $\epsilon_{ACA}$ on the cost and accuracy in Table~\ref{FDPHtab:1}. 
We first measured the direct effect of $\epsilon_{ACA}$ on the solution by the error in the solution (quantified in the same way as that in \S\ref{FDPH622}), but it was mostly independent of the exponential variations in the $\epsilon_{ACA}$ values (Table~\ref{FDPHtab:1}, errors in solutions, abbreviated to soln). 
This suggests that H-matrices in FDP=H-matrices can provide sufficient accuracy within the range of the $\epsilon_{ACA}$ values investigated in this study, which is near those of the conventional H-matrices in the previous studies (for example, $\epsilon_{ACA}=10^{-4}\sim10^{-6}$ in Refs.~\cite{borm2003hierarchical,ohtani2011fast}). 
It is quite affirmative result but also inhibits us from accessing the detailed evaluations of the influence of $\epsilon_{ACA}$ just by seeing the simulated solution of the elastodynamic problem. 
In the following, we then investigate the detail of the influence of $\epsilon_{ACA}$ by investigating the accuracy and cost of the LRA that are directly affected by $\epsilon_{ACA}$, as done in the previous studies of H-matrices~\cite{borm2003hierarchical}. 

The accuracy and cost of the LRA are respectively measured with using the weighted mean of the relative error norm (the mean error, introduced in \S\ref{FDPH611}) and the rank (called mean rank). The weight coefficients of these means are set at the numbers of the included submatrix entries, and these means express the effective relative error and rank expected in each matrix entry. 
We did not consider the variations of the accuracy and rank from the consideration as they are relatively small, as shown in Fig.~\ref{FDPHfig:12}. 

The values of the mean error and average rank for several $\epsilon_{ACA}$ values are shown in Table~\ref{FDPHtab:1}. 
Indices (F, S, and tr) correspond to the Domain F kernel, (asymptotic) Domain S kernel, and transient kernel in Domain S (introduced in \S\ref{FDPH721}), respectively. The involved parameters are set at the same values as those in Fig.~\ref{FDPHfig:12}. 

The mean error was $10^{-2}$ times smaller than $\epsilon_{ACA}$ in the range of $\epsilon_{ACA} = 10 ^{-2} - 10 ^{-5}$ (mean error in Table~\ref{FDPHtab:1}). It is consistent with the error distribution in Fig.~\ref{FDPHfig:12}, and will be ascribed to the error criterion we adopted, as mentioned is \S\ref{FDPH611}. In addition, the mean error was roughly in proportion to $\epsilon_{ACA}$. 

The mean rank increased in proportion to $\log \epsilon_{ACA}$ (mean rank in Table~\ref{FDPHtab:1}). This $\epsilon_{ACA}$ dependence of the rank is consistent with the theoretical cost estimates of the ACA~\cite{bebendorf2003adaptive}. 
Considering that the change in the rank is $\mathcal O(1)$, even when $\epsilon_{ACA}$ increases 1000-fold as in Table~\ref{FDPHtab:1}, $\epsilon_{ACA}$ seems to have little impact on the numerical costs after the kernel matrices are approximated. 

\begin{table}[htb]
\begin{center}
\caption{
Mean error and mean rank (and the solution error), introduced in \S\ref{FDPH631}, versus $\epsilon_{ACA}$. 
Indices F, S, and tr correspond to the Domain F kernel, Domain S asymptotic kernel, and transient kernel in Domain S (defined in \S\ref{FDPH721}), respectively. 
The solution error (soln) is also listed, which is evaluated at $t=480$ under the same definition with 
the same parameter values as those in Fig.~\ref{FDPHfig:16} except the specified $\epsilon_{ACA}$ values. 
}
  \begin{tabular}{|c|c|c|c|c|c|c|c|}
\hline
\multicolumn{1}{|c|}{$\epsilon_{ACA}$}
&
\multicolumn{3}{|c|}{mean error}
&
\multicolumn{3}{|c|}{mean rank}
&
\multicolumn{1}{|c|}{error}
\\\hline
&F&S&tr&F&S&tr&soln
\\\hline
$10^{-2}$&$3{\small\times}10^{-5}$&$6{\small\times}10^{-5}$&$2{\small\times}10^{-5}$&6&6&6&0.003
\\
$10^{-3}$&$2{\small\times}10^{-6}$&$2{\small\times}10^{-6}$&$2{\small\times}10^{-6}$&7&7&7&0.003
\\
$10^{-4}$&$1{\small\times}10^{-6}$&$7{\small\times}10^{-7}$&$5{\small\times}10^{-7}$&8&8&8&0.003
\\
$10^{-5}$&$6{\small\times}10^{-8}$&$3{\small\times}10^{-7}$&$3{\small\times}10^{-8}$&9&9&9&0.003
\\\hline
  \end{tabular}
\end{center}
\label{FDPHtab:1}
\end{table}

\subsubsection{$\eta$ Dependence}
\label{FDPH632}
%Figure 17. âš
\begin{figure*}[tbp]
\centering
  \includegraphics[width=120mm]{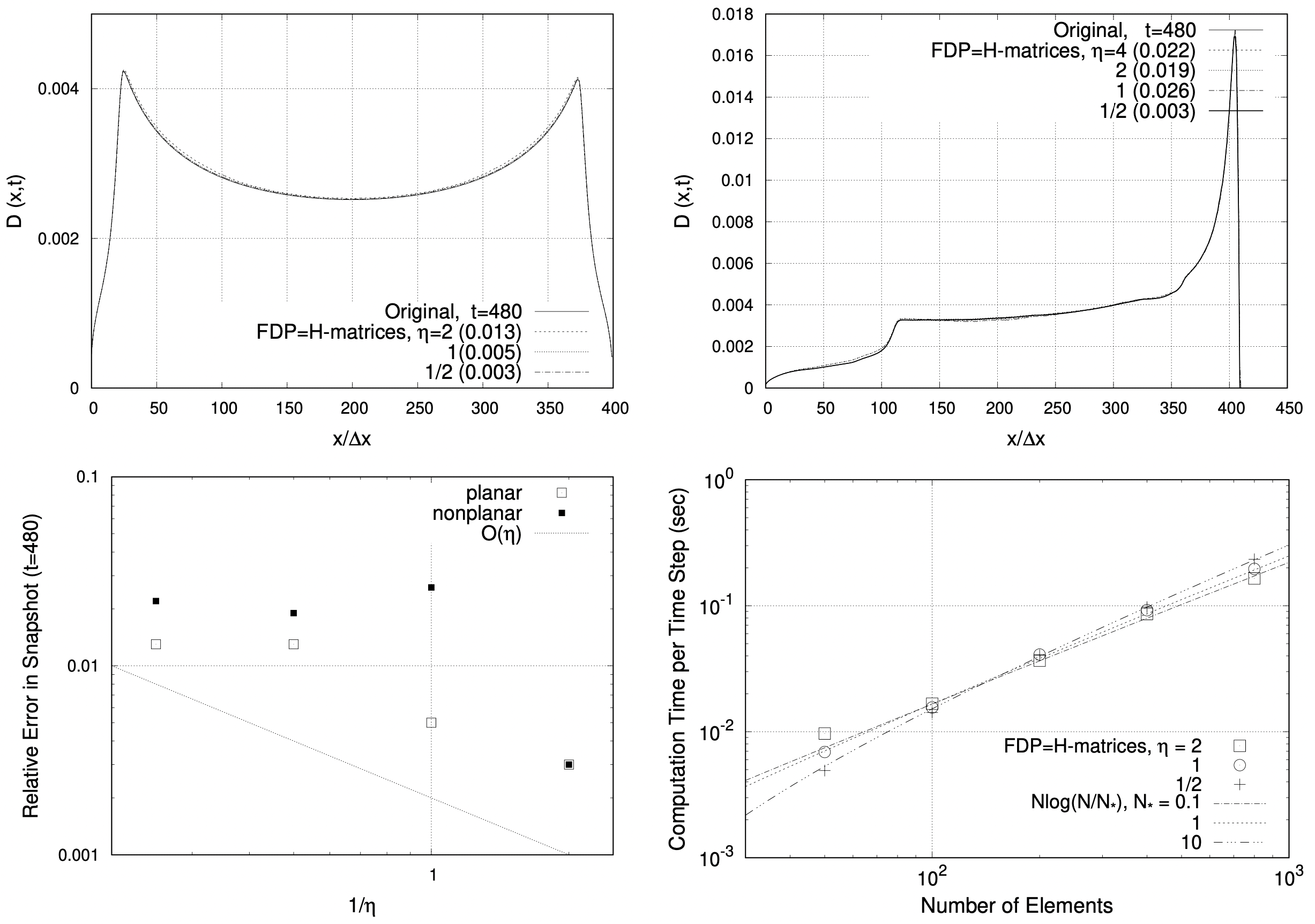}
%\begin{minipage}{1.0\hsize}
%\centering
%  \includegraphics[width=60mm]{fig_Dx_snapshot_planar_eta.pdf}
%  \includegraphics[width=60mm]{fig_Dx_snapshot_nonplanar_eta.pdf}
%\end{minipage}
%\\
%\centering
%\begin{minipage}{1.0\hsize}
%\centering
%  \includegraphics[width=60mm]{etadependenceofsnapshoterror.pdf}
%  \includegraphics[width=60mm]{fig_timeconsumption_mac_etadep.pdf}
%\end{minipage}
\caption{
Error and cost versus $\eta$. Bracketed numbers in the legends of the top left and top right panels indicate the relative errors of FDP=H-matrices in the same manner as in Fig.~\ref{FDPHfig:15}.
(Top left) Snapshots of slip rate $D (x, t)$ at $t = 480$ on a planar fault. Parameter values are the same as in Fig.~\ref{FDPHfig:15} except the $\eta$ value. 
(Top right) Snapshots of slip rate $D (x, t)$ at $t = 480$ on a nonplanar fault. Parameter values are the same as in Fig.~\ref{FDPHfig:16} except the $\eta$ value. 
(Bottom left) Relative error versus $\eta$. 
The $\mathcal O(\eta)$ asymptote is shown by a line for the planar fault case. 
The settings are the same as in the top panels. 
(Bottom right) The computation time per time step measured in \S\ref{FDPH62} versus $\eta$. 
Fitted curves of $N\log N/N_*$ scaling are shown by lines. 
Measurements are made on a planar fault with the same setting as the case in \S\ref{FDPH621} except the adopted $\eta$ values. 
}
\label{FDPHfig:17}
\end{figure*}

Fig.~\ref{FDPHfig:17} shows the $\eta$ dependence of the solution errors in the dynamic rupture problems, simulated in \S\ref{FDPH622}.
It indicates that the solution with FDP=H-matrices converges to the original solution as $\eta$ decreases both in the planar and nonplanar cases. Especially in the planar case, when $\eta$ is small, the error is approximately proportional to $\eta$. This $\eta$ dependence is ascribable to the error of $\mathcal O [1 / (1 /\eta+ 1)]$ concerning the degenerating normalized waveform [Eq.~(\ref{FDPHeq:21})], 
given that the travel-time (or wave-speed) approximation, that is the other possible cause of the error depending on $\eta$, 
becomes exact in a 2D planar fault case (mentioned in \S\ref{FDPH612}). 
The nonplanar fault case shows larger errors than those of the planar case at $\eta\geq1$, probably because an approximation error of the effective wave speed is also contained in nonplanar fault geometries. On the other hand, such increased error in the nonplanar problem safely reduces to the same level as that in the planar problem at relatively small $\eta = 1/2$.

Fig.~\ref{FDPHfig:17} (bottom right) shows the $\eta$ dependence of the cost, fitted by $N$ $\log N / N_*$ with $\eta$-dependent $N_*$. 
The cost is measured in the dynamic rupture problem on a planar fault under the same setting as in the case described in \S\ref{FDPH621}, except for the $\eta$ values. 
Here, we show only the computation time per time step for brevity, 
given that the total memory consumption and computation time per time step have showed the same size dependence in \S\ref{FDPH621}. 
The cost of FDP=H-matrices is shown to retain the scaling of $\mathcal O (N \log N/N_*)$ even when $\eta$ varies. 
In our measurement, $N_*$ was proportional to $1/\eta$.
It would be ascribable to that $N_*$ [that balances $\mathcal O(N\log N)$ costs of the admissible leaves and $\mathcal O(N)$ costs of the inadmissible leaves] correlates with the minimum size of the admissible leaves, being on the order of the minimum value $l_{min}/\eta$ of $dist$.

\section{Discussion}
\label{FDPH7}
We have developed the data-sparse approximations and operations of FDP=H-matrices and investigated their detail through their numerical implementation in the 2D anti-plane problems. We summarize their error and cost controls in \S\ref{FDPH71} for the algorithm tuning in the prospective use. 
We also refer to some associated works in \S\ref{FDPH73}.

\subsection{Summary of Error and Cost Controls in FDP=H-Matrices}
\label{FDPH71}
We overview the dependence of the error and cost on the main error-control parameters $(\epsilon_{ACA}, \eta, l_{min})$ of FDP=H-matrices, which have been evaluated analytically in \S\ref{FDPH4} and numerically in \S\ref{FDPH6}. 
The associated dependence on the schemes (the constant $\eta$ and constant $\eta^2dist$ schemes) are also included in them here.
The left error and cost controls---their dependence on $\epsilon_Q$ of Quantization, investigated in \S\ref{FDPHA3}, and the 2D specific error handling, detailed in \ref{FDPHH}--are also summarized, and this section serves the full summary of the error and cost controls of FDP=H-matrices.

The cost and accuracy of the LRA in H-matrices are affected by the selected method of the LRA and the adopted $\epsilon_{ACA}$ values. 
ACA+ worked with satisfactory accuracy in most cases while the partially pivoting ACA sometimes did erroneously in our investigation. 
On the other hand, even with ACA+, we required exception handling occasionally in the neighboring clusters, especially for the cases of the nonplanar boundaries. We substituted the original submatrix for the approximate one when the rank of the nominally low-ranked submatrix exceeds the original one in \S\ref{FDPH62} and \S\ref{FDPH63}. 
ACA+ achieved substantial error bound ($\epsilon_ H$) of order (or smaller than) that we specified ($\epsilon_ {ACA}$) (Fig.~\ref{FDPHfig:12} and Table~\ref{FDPHtab:1}). 
Table~\ref{FDPHtab:1} implies that ACA+ with $\epsilon_{ACA} = 10 ^ {-2}\sim 10^{-5}$ seems to guarantee the same accuracy as that of the dynamic rupture problems (of $\epsilon_{ACA} = 10 ^ {-4}$) of this study. 

The errors of the travel time and normalized waveform are controlled by two constants $\eta$ and $l_{min}$ in the approximation of the ART. 
The constant $\eta$ scheme suppresses the error of the wave speed approximately below $4^{-1} / (1 + 1 / \eta) ^ 2$ [Eq.~(\ref{FDPHeq:24})], i.e., in a non-dispersive manner. The bound is independent of $l_{min}$ besides. Eq.~(\ref{FDPHeq:24}) shows that the error decays rapidly in the inverse-square proportion to the $\eta$ value and for example gives less-than-about $6\%$ wave-speed error at $\eta = 1$. The error of the normalized waveform $h_j$ shown in Eq.~(\ref{FDPHeq:21}) is of $\mathcal O[1/(1 + 1 / \eta)]$ and moreover of order duration $\Delta t_j$ of Domain F [$\mathcal O(\Delta t_j)$], which is also of order the originally discretized time interval $\Delta t$; as the approximation of $h_j$ is intrinsically the temporal interpolation of the kernel, the error order of Eq.~(\ref{FDPHeq:21}) may be improved to $\mathcal O[(\Delta t_j)^2/(1 + 1 / \eta)]$ in some sort of time-marching schemes of the original ST-BIEM~\cite{noda2020comparison} achieving $\mathcal O[(\Delta t)^2]$ about it. 
As far as we examined, the solution of $\eta = 1/2$ converged to the original solution within about $0.3\%$ relative error (\S\ref{FDPH622}), which is near 10 times smaller than the error frequently occurring due to the spatiotemporal discretization~\cite{day2005comparison,tada2001dynamic}. In the constant $\eta^2 dist$ scheme, $\eta$ is a function of space $\eta\propto1/\sqrt{dist}$, and both of $\eta$ and $l_{min}$ contribute to the accuracy as described in Eq.~(\ref{FDPHeq:26}). There, the error of the ART can be negligible as it can become smaller than the original discretization error of the boundary elements.

The solution of FDP=H-matrices is not observably affected by the variations in $\epsilon_Q$ (\S\ref{FDPHA3} and \S\ref{FDPH723}) of Quantization; 
as mentioned in the opening of \S\ref{FDPH6}, we applied Quantization to Domain S kernel of the 2D cases to check its property.
The solution error in our evaluation was unchanged from $0.3\%$ relative error in the range of $\epsilon_Q = 10^{-3} \sim 10^{-1}$ (\S\ref{FDPHA3}) as far as the absolute-error bound ($\epsilon_{st}$) is set at $10 ^ {-6}$ (Fig.~\ref{FDPHfig:19}); $\epsilon_{st}$ required much small values to deal with the 2D specific errors (detailed in \ref{FDPHH} and summarized below) and secondarily $\epsilon_Q$ became irrelevant to the accuracy. 
Regarding the cost, the value $\epsilon_{st}$ of the allowable absolute error (\S\ref{FDPH723}) was less relevant than the relative allowable error value $\epsilon_Q$ (\S\ref{FDPHA3}); the cost change was proportional to $\ln\epsilon_{st}$ and $1/\epsilon_Q$ although their proportionality factors were both quite small. 
Given these, even considering the 3D setting, the additional absolute error condition may be preferable to be introduced for reducing the cost in retaining the accuracy. 

Additional errors of the FDPM exist in the 2D cases due to the approximate spatiotemporal separation of the kernel. Its primary handling was enlarging the width of Domain F (detailed in \ref{FDPHH}) as in the conventional 2D implementation of the FDPM~\cite{ando2007efficient}. 
We further improved the accuracy in the admissible leaves by adding the LRA of the third-order tensor [detailed in \S\ref{FDPH721} with Fig.~\ref{FDPHfig:19} (top), referred to as TCA]. By setting the allowable absolute error ($\epsilon_{st}$) at about $10^{-6}$ and the additional width of Domain F at about $10\beta\Delta x$, we suppressed the solution error below about $0.3\%$ (Fig.~\ref{FDPHfig:19}). These modifications did not change the cost largely (Fig.~\ref{FDPHfig:19}). Our investigation indicated that the 2D specific errors are predominant accuracy-controlling factors in our implementation for the 2D cases. This implies that the inherent errors existing in both the 2D and 3D cases of FDP=H-matrices are satisfactorily small.

Last, we emphasize that the cost scaling of $N\log N$ is kept throughout the aforementioned parameter tuning to reduce the errors. 
As indicated both numerically and analytically, the parameter dependence of the cost is basically represented by the prefactors of the scaling. 
In the actual use, these parameter dependencies of the accuracy will be automatically checked through the robustness check of the results against these parameters like against discretization length $\Delta x_j$ of receiver $j$ and time step $\Delta t$. 
Even considering the cost of such robustness check, FDP=H-matrices will be sufficiently faster than the original implementation. 

\subsection{Applicability, Extensions, and Parallel Computations of FDP=H-Matrices}
\label{FDPH73}
\begin{figure}[tbp]
   \begin{center}
  \includegraphics[width=68mm]{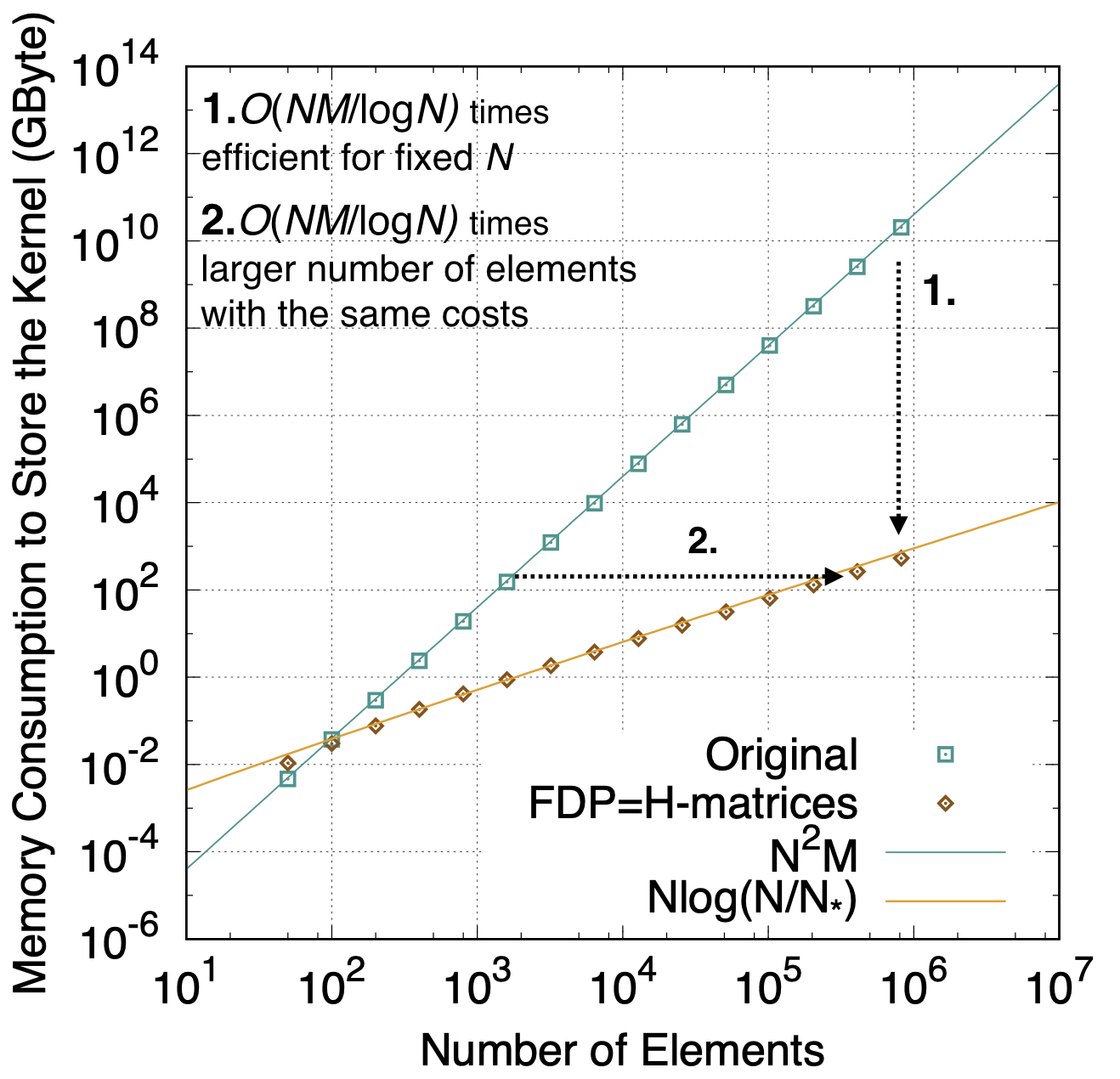}
   \end{center}
\caption{
Memory consumptions to store the kernel for the cases of the original ST-BIEM and of FDP=H-matrices, in the size range $10 < N < 10^7$. Numbered arrows represent the cost comparison between the original ST-BIEM and FDP=H-matrices. Parameter values and notations are the same as Fig.~\ref{FDPHfig:14}. Some acceleration techniques, not used in Fig.~\ref{FDPHfig:14} (introduced for the parameter studies shown in the appendix) are used, and so the cost of FDP=H-matrices changes within a factor.
}
\label{FDPHfig:addedmemoryevaluation}
\end{figure}

We obtained an algorithm for simulating the elastodynamic BIEM with the $\mathcal O (N\log N)$ memory and the $\mathcal O(N\log N)$ time complexity per time step [that is the $\mathcal O(NM\log N)$ complexity in total]. 
To our knowledge, the algorithm based on FDP=H-matrices is the first versatile one that serves both the $\mathcal O(N\log N)$ whole memory and $\mathcal O(N\log N)$ time complexity per time step in executing the transient elastodynamic (more generally, hyperbolic-equational) boundary analyses. 
These cost reductions allow the ST-BIEM to simulate the same-sized problem with $NM/\log N$ times smaller computational resources, and $NM/\log N$ times larger problems with the same costs, as illustrated in Fig.~\ref{FDPHfig:addedmemoryevaluation}. 
FDP=H-matrices will have wide applications in realistic (particularly elastodynamic) problems, where the memory storage is the bottleneck of the modeling~\cite{ando2016fast}. 
Please refer to \ref{FDPHB5} for cost estimate details. 

%Fig âš

The algorithmic progress provided by FDP=H-matrices separates into that of the data-sparse (kernel-low-rank) approximations and that of the associated operations (arithmetics). 
As stated in the introduction, our initial motivation has been to solve a known problem of H-matrices in approximating the kernel function of the wave problems. 
We solved it by applying H-matrices along Domain F of the FDPM fully involving the wavefronts. 
This technique is purely for dealing with the singularity distributed along the causal cones, and hence other classes of hyperbolic partial differential equations, such as the wave equation, suffering from the same problem can also be within the realm of the applicability of FDP=H-matrices. 
We in this paper employed the time integral of the kernel over Domain F (the amplitude term) as one of implementations of such {\it an LRA along wavefronts}, with the analogy of the impulse of the pulse force (Green's function). 
Obviously, there can be other ways to apply H-matrices along Domain F, such as applying the LRA to the kernel submatrix sliced along respective reduced-time steps in Domain F, as originally suggested in Ref.~\cite{ando2016fast}. We need further investigation about the implementation of the LRA along wavefronts. 
Meanwhile, we also found the associated arithmetics unexpectedly and erased the $\mathcal O(NM)$ memory to store the history of the boundary elements in the evaluation of the BIEs. 
This arithmetic developed in this research can be combined also with the analytically expanded kernel in the PWTD method just with the replacement of the LRA in H-matrices with the kernel expansion in the PWTD method, e.g., of Ref.~\cite{takahashi2003fast}. Such a derivative implementation may be called the FDP-PWTD method.
Besides, $\mathcal H^2$-matrices~\cite{hackbusch2002data} of $\mathcal O(N)$ costs in the spatial BIEM may further allow FDP=H-matrices to erase a logarithmic factor of their numerical costs.

As all the homogeneous elastodynamic kernels (both the single- and double-layer potentials~\cite{bonnet1999boundary}) comprise the integrodifferential forms of the Green's function, being expandable along the wavefronts as mentioned earlier, FDP=H-matrices can offer various extensional usages by simply replacing the explicit functional form of the kernel. 
We focused on crack problems evaluating the double-layer potentials in the simulation. 
Their application to the other problems using the single-layer potential as diffractive and scattering  problems~\cite{maruyama2016transient,desiderio2017h} will be done in other places. 
Similarly, their application to elastic heterogeneity is also possible with a multi-regional approach~\cite{bonnet1999boundary,kame2012proposal} subdividing the heterogeneous media into the homogeneous ones. 
Although we considered the piecewise-constant interpolation in space, we can apply FDP=H-matrices to other spatial local basis functions, e.g., the spline basis~\cite{nishimura1989regularized}, without any modifications of the algorithm, even with unstructured meshes. 
Temporal basis functions other than piecewise-constant ones are also available, as far as they possess the equally-spaced nature, which is indispensable for obtaining temporally translationally symmetric discretized kernel $K_{i,j,m}$ assumed in FDP=H-matrices; 
some (adaptive-)hierarchical-time-stepping implementation~\cite{lapusta2000elastodynamic}, using the kernel of the equally-spaced basis, will be also within the range of the application. 
The application to other methods of weighted mean residuals than collocation methods, such as Galerkin methods~\cite{bonnet1998symmetric,fischer2004multipole} may give us another perspective.

For investigating the detail of the data-sparse approximation and the algorithms of FDP=H-matrices, our numerical experiments have been limited to the 2D example. 
On the other hand, the case most requiring the fast computation is the 3D case originally much heavier than the 2D case. 
We can expect that FDP=H-matrices well approximate the impulsive kernel in Domain F since the geometrical nature of that kernel is common in both the 2D and 3D cases. 
We will treat of those 3D examples in the upcoming reports; there we will see that as suggested from the geometrical spreading nature of the 3D kernel in Domain F (Ando, 2016), H-matrices 
work efficiently and the $\mathcal O(N\log N)$ scaling hold for those cases. The application for the wave equation may also be discussed there.

The aim of this study has been regarding the method proposal and its numerical precise investigation. 
Because of that nature, while the investigated system size was enough large for the investigation of the asymptotic cost orders, 
the computed size scale in this study has been intentionally set at relatively small ranges, even in Fig.~\ref{FDPHfig:addedmemoryevaluation}. 
The application of FDP=H-matrices to large $N$ problems and the associated efficient parallel computations should be regarded as upcoming key issues. 
The efficiency of the parallelization will depend on task assignment 
as in H-matrices~\cite{bebendorf2009}, the FMM, and the PWTD method due to the common circumstance that the sizes of the computed vectors ranging from $\mathcal O(1)$ to $\mathcal O(N)$, or intrinsically due to the hierarchical division of the BIE. 
As the root of the difficulty is the same, 
it is a desirable collaboration to combine FDP=H-matrices with a highly efficient parallel computation library of H-matrices, such as HACApK~\cite{ida2014parallel}. 
Meanwhile, as the scaling merit remains even at large $N$ as in Fig.~\ref{FDPHfig:addedmemoryevaluation}, 
there will be certain cases where the original implementation has required large parallelization yet FDP=H-matrices run enough quickly with simple open MP implementation, as in some parallel-implementation reports~\cite{otani2007fast} of the PWTD-method.

\section{Conclusion}
We have developed FDP=H-matrices for solving transient elastodynamic problems in a fast and memory-efficient manner, by combining the FDPM and H-matrices with newly developed modules named Quantization and the ART. 
FDP=H-matrices reduce 
both
the time complexity of the spatiotemporal convolution of a given BIE per time step and whole memory consumption required in the repetitive evaluations of the BIE, that have both been 
$\mathcal O(N^2M)$ in the original ST-BIEM,
to $\mathcal O(N\log N)$ for $N$-element and $M$-time-step problems. 
%This was an unprecedented level of the cost reduction, even compared with those of the previous methods such as $\mathcal O(N^2)$ of H-matrices or the FDPM and $\mathcal O(N\log^2N)$ (complexity per time step) and $\mathcal (NM)$ (total memory) of the PWTD method. 
First, 
by introducing the approximations along the wavefronts, we constructed arithmetics of FDP=H-matrices for both the 2D and 3D problems.
We next implemented FDP=H-matrices in the 2-D anti-plane problems to investigate the detail of the cost reduction and accuracy. 
The present numerical experiments demonstrated that FDP=H-matrices achieve the log-linear [$\mathcal O(N\log N)$] cost order with retaining the high accuracy of the original ST-BIEM.

%\section{Section title}
%\label{sec:1}
%%Text with citations \cite{RefB} and \cite{RefJ}.
%\subsection{Subsection title}
%\label{sec:2}
%as required. Don't forget to give each section
%and subsection a unique label (see Sect.~\ref{sec:1}).
%\paragraph{Paragraph headings} Use paragraph headings as needed.
%\begin{equation}
%\end{equation}

%% For one-column wide figures use
%\begin{figure}
%% Use the relevant command to insert your figure file.
%% For example, with the graphicx package use
%  \includegraphics{example.eps}
%\end{figure}
%%
%% For two-column wide figures use
%\begin{figure*}
%% Use the relevant command to insert your figure file.
%% For example, with the graphicx package use
%  \includegraphics[width=0.75\textwidth]{example.eps}
%\end{figure*}
%\begin{table}
%\begin{tabular}{lll}
%\hline\noalign{\smallskip}
%\end{tabular}
%\end{table}
%

\section*{Acknowledgements}
We would first like to express our deepest gratitude to Dr. Marc Bonnet for his generous and patient help in thoroughly improving the manuscript. 
We also acknowledge helpful discussions with A. Ida, N. Kame, M. Ohtani, and P. Romanet. 
This work was supported by JSPS KAKENHI Grant Numbers JP25800253 and MEXT KAKENHI Grant Numbers JP26109007, and by the ``Joint Usage/Research Center for
Interdisciplinary Large-scale
Information Infrastructures'' and ``High
Performance Computing Infrastructure'' in Japan (Project ID: jh180043-NAH).

%\section*{References}
%\bibliography{reference_FDPH}   % name your BibTeX data base
%\bibliography{thebibliography}   % name your BibTeX data base

\begin{thebibliography}{10}
\expandafter\ifx\csname url\endcsname\relax
  \def\url#1{\texttt{#1}}\fi
\expandafter\ifx\csname urlprefix\endcsname\relax\def\urlprefix{URL }\fi
\expandafter\ifx\csname href\endcsname\relax
  \def\href#1#2{#2} \def\path#1{#1}\fi

\bibitem{wannamaker1984electromagnetic}
P.~E. Wannamaker, G.~W. Hohmann, W.~A. SanFilipo, Electromagnetic modeling of
  three-dimensional bodies in layered earths using integral equations,
  Geophysics 49~(1) (1984) 60--74.

\bibitem{jones1974integral}
D.~Jones, Integral equations for the exterior acoustic problem, The Quarterly
  Journal of Mechanics and Applied Mathematics 27~(1) (1974) 129--142.

\bibitem{schanz1999boundary}
M.~Schanz, A boundary element formulation in time domain for viscoelastic
  solids, Communications in Numerical Methods in Engineering 15~(11) (1999)
  799--809.

\bibitem{rice1993spatio}
J.~R. Rice, Spatio-temporal complexity of slip on a fault, Journal of
  Geophysical Research: Solid Earth 98~(B6) (1993) 9885--9907.

\bibitem{ando2018dynamic}
R.~Ando, Y.~Kaneko, Dynamic rupture simulation reproduces spontaneous
  multifault rupture and arrest during the 2016 mw 7.9 kaikoura earthquake,
  Geophysical Research Letters 45~(23) (2018) 12--875.

\bibitem{nishimura1989regularized}
N.~Nishimura, S.~Kobayashi, A regularized boundary integral equation method for
  elastodynamic crack problems, Computational mechanics 4~(4) (1989) 319--328.

\bibitem{beskos1987boundary}
D.~E. Beskos, Boundary element methods in dynamic analysis.

\bibitem{bonnet1999boundary}
M.~Bonnet, Boundary integral equation methods for solids and fluids, Meccanica
  34~(4) (1999) 301--302.

\bibitem{aliabadi2002boundary}
M.~H. Aliabadi, The boundary element method, applications in solids and
  structures, Vol.~2, John Wiley \& Sons, 2002.

\bibitem{zhang1991novel}
C.~Zhang, A novel derivation of non-hypersingular time-domain bies for
  transient elastodynamic crack analysis, International Journal of Solids and
  Structures 28~(3) (1991) 267--281.

\bibitem{nishimura2002fast}
N.~Nishimura, Fast multipole accelerated boundary integral equation methods,
  Applied mechanics reviews 55~(4) (2002) 299--324.

\bibitem{day2005comparison}
S.~M. Day, L.~A. Dalguer, N.~Lapusta, Y.~Liu, Comparison of finite difference
  and boundary integral solutions to three-dimensional spontaneous rupture,
  Journal of Geophysical Research: Solid Earth 110~(B12).

\bibitem{tada1997non}
T.~Tada, T.~Yamashita, Non-hypersingular boundary integral equations for
  two-dimensional non-planar crack analysis, Geophysical Journal International
  130~(2) (1997) 269--282.

\bibitem{takahashi2003fast}
T.~Takahashi, N.~Nishimura, S.~Kobayashi, A fast biem for three-dimensional
  elastodynamics in time domain, Engineering analysis with boundary elements
  27~(5) (2003) 491--506.

\bibitem{ergin1999plane}
A.~A. Ergin, B.~Shanker, E.~Michielssen, The plane-wave time-domain algorithm
  for the fast analysis of transient wave phenomena, IEEE Antennas and
  Propagation Magazine 41~(4) (1999) 39--52.

\bibitem{rokhlin1985rapid}
V.~Rokhlin, Rapid solution of integral equations of classical potential theory,
  Journal of computational physics 60~(2) (1985) 187--207.

\bibitem{mavaleix2020fast}
D.~Mavaleix-Marchessoux, M.~Bonnet, S.~Chaillat, B.~Lebl{\'e}, A fast boundary
  element method using the z-transform and high-frequency approximations for
  large-scale three-dimensional transient wave problems, International Journal
  for Numerical Methods in Engineering 121~(21) (2020) 4734--4767.

\bibitem{lubich1988convolutionI}
C.~Lubich, Convolution quadrature and discretized operational calculus. i,
  Numerische Mathematik 52~(2) (1988) 129--145.

\bibitem{lubich1988convolutionII}
C.~Lubich, Convolution quadrature and discretized operational calculus. ii,
  Numerische Mathematik 52~(4) (1988) 413--425.

\bibitem{banjai2009rapid}
L.~Banjai, S.~Sauter, Rapid solution of the wave equation in unbounded domains,
  SIAM Journal on Numerical Analysis 47~(1) (2009) 227--249.

\bibitem{chaillat2017fast}
S.~Chaillat, M.~Darbas, F.~Le~Lou{\"e}r, Fast iterative boundary element
  methods for high-frequency scattering problems in 3d elastodynamics, Journal
  of Computational Physics 341 (2017) 429--446.

\bibitem{maruyama2016transient}
T.~Maruyama, T.~Saitoh, T.~Bui, S.~Hirose, Transient elastic wave analysis of
  3-d large-scale cavities by fast multipole bem using implicit runge--kutta
  convolution quadrature, Computer Methods in Applied Mechanics and Engineering
  303 (2016) 231--259.

\bibitem{ando2007efficient}
R.~Ando, N.~Kame, T.~Yamashita, An efficient boundary integral equation method
  applicable to the analysis of non-planar fault dynamics, Earth, planets and
  space 59~(5) (2007) 363--373.

\bibitem{ando2016fast}
R.~Ando, Fast domain partitioning method for dynamic boundary integral
  equations applicable to non-planar faults dipping in 3-d elastic half-space,
  Geophysical Supplements to the Monthly Notices of the Royal Astronomical
  Society 207~(2) (2016) 833--847.

\bibitem{hackbusch1999sparse}
W.~Hackbusch, A sparse matrix arithmetic based on-matrices. part i:
  Introduction to-matrices, Computing 62~(2) (1999) 89--108.

\bibitem{lapusta2000elastodynamic}
N.~Lapusta, J.~R. Rice, Y.~Ben-Zion, G.~Zheng, Elastodynamic analysis for slow
  tectonic loading with spontaneous rupture episodes on faults with rate-and
  state-dependent friction, Journal of Geophysical Research: Solid Earth
  105~(B10) (2000) 23765--23789.

\bibitem{chaillat2008multi}
S.~Chaillat, M.~Bonnet, J.-F. Semblat, A multi-level fast multipole bem for 3-d
  elastodynamics in the frequency domain, Computer Methods in Applied Mechanics
  and Engineering 197~(49-50) (2008) 4233--4249.

\bibitem{bebendorf2003adaptive}
M.~Bebendorf, S.~Rjasanow, Adaptive low-rank approximation of collocation
  matrices, Computing 70~(1) (2003) 1--24.

\bibitem{yoshikawa20152}
H.~Yoshikawa, S.~Yamamoto, A fast method of time domain biem for scalar wave
  propagation in 2d using aca, Transactions of the Japan Society for
  Computational Methods in Engineering 15 (2015) 79--84.

\bibitem{borm2003hierarchical}
S.~B{\"o}rm, L.~Grasedyck, W.~Hackbusch, Hierarchical matrices, Lecture notes
  21 (2003) 2003.

\bibitem{chaillat2017theory}
S.~Chaillat, L.~Desiderio, P.~Ciarlet, Theory and implementation of h-matrix
  based iterative and direct solvers for helmholtz and elastodynamic
  oscillatory kernels, Journal of Computational physics 351 (2017) 165--186.

\bibitem{aki2002quantitative}
K.~Aki, P.~G. Richards, Quantitative seismology, University Science Books,
  2002.

\bibitem{gonzalez2002digital}
R.~C. Gonzalez, R.~E. Woods, Digital image processing prentice hall, Upper
  Saddle River, NJ.

\bibitem{tada2000non}
T.~Tada, E.~Fukuyama, R.~Madariaga, Non-hypersingular boundary integral
  equations for 3-d non-planar crack dynamics, Computational Mechanics 25~(6)
  (2000) 613--626.

\bibitem{eringen1975elastodynamics}
A.~C. Eringen, E.~Suhubi, Elastodynamics: linear theory, vol. 2, New York:
  Academic.

\bibitem{tada2001dynamic}
T.~Tada, R.~Madariaga, Dynamic modelling of the flat 2-d crack by a
  semi-analytic biem scheme, International Journal for Numerical Methods in
  Engineering 50~(1) (2001) 227--251.

\bibitem{tada2006stress}
T.~Tada, Stress green's functions for a constant slip rate on a triangular
  fault, Geophysical Journal International 164~(3) (2006) 653--669.

\bibitem{cochard1994dynamic}
A.~Cochard, R.~Madariaga, Dynamic faulting under rate-dependent friction, pure
  and applied geophysics 142~(3) (1994) 419--445.

\bibitem{ida2014parallel}
A.~Ida, T.~Iwashita, T.~Mifune, Y.~Takahashi, Parallel hierarchical matrices
  with adaptive cross approximation on symmetric multiprocessing clusters,
  Journal of information processing 22~(4) (2014) 642--650.

\bibitem{segall2010earthquake}
P.~Segall, Earthquake and volcano deformation, Princeton University Press,
  2010.
  
\bibitem{colton2013integral}
D.~Colton, R.~Kress, Integral equation methods in scattering theory, SIAM,
  2013.

\bibitem{ergin1998fast}
A.~A. Ergin, B.~Shanker, E.~Michielssen, Fast evaluation of three-dimensional
  transient wave fields using diagonal translation operators, Journal of
  Computational Physics 146~(1) (1998) 157--180.

\bibitem{pelties2012three}
C.~Pelties, J.~Puente, J.-P. Ampuero, G.~B. Brietzke, M.~K{\"a}ser,
  Three-dimensional dynamic rupture simulation with a high-order discontinuous
  galerkin method on unstructured tetrahedral meshes, Journal of Geophysical
  Research: Solid Earth 117~(B2).

\bibitem{andrews1976rupture}
D.~Andrews, Rupture velocity of plane strain shear cracks, Journal of
  Geophysical Research 81~(32) (1976) 5679--5687.

\bibitem{grasedyck2005adaptive}
L.~Grasedyck, Adaptive recompression of-matrices for bem, Computing 74~(3)
  (2005) 205--223.

\bibitem{ida1972cohesive}
Y.~Ida, Cohesive force across the tip of a longitudinal-shear crack and
  griffith's specific surface energy, Journal of Geophysical Research 77~(20)
  (1972) 3796--3805.

\bibitem{ohtani2011fast}
M.~Ohtani, K.~Hirahara, Y.~Takahashi, T.~Hori, M.~Hyodo, H.~Nakashima,
  T.~Iwashita, Fast computation of quasi-dynamic earthquake cycle simulation
  with hierarchical matrices, Procedia Computer Science 4 (2011) 1456--1465.

\bibitem{noda2020comparison}
H.~Noda, D.~S. Sato, Y.~Kurihara, Comparison of two time-marching schemes for
  dynamic rupture simulation with a space-domain biem, Earth, Planets and Space
  72~(1) (2020) 1--12.

\bibitem{desiderio2017h}
L.~Desiderio, H-matrix based solver for 3d elastodynamics boundary integral
  equations, Ph.D. thesis, Paris Saclay (2017).

\bibitem{kame2012proposal}
N.~Kame, T.~Kusakabe, Proposal of extended boundary integral equation method
  for rupture dynamics interacting with medium interfaces, Journal of Applied
  Mechanics 79~(3) (2012) 031017.

\bibitem{bonnet1998symmetric}
M.~Bonnet, G.~Maier, C.~Polizzotto, Symmetric galerkin boundary element
  methods.

\bibitem{fischer2004multipole}
M.~Fischer, U.~Gauger, L.~Gaul, A multipole galerkin boundary element method
  for acoustics, Engineering analysis with boundary elements 28~(2) (2004)
  155--162.

\bibitem{bebendorf2009}
M.~Bebendorf, S.~Kunis,
  \href{http://dx.doi.org/10.1216/JIE-2009-21-3-331}{Recompression techniques
  for adaptive cross approximation}, J. Integral Equations Applications 21~(3)
  (2009) 331--357.
\newblock \href {http://dx.doi.org/10.1216/JIE-2009-21-3-331}
  {\path{doi:10.1216/JIE-2009-21-3-331}}.
\newline\urlprefix\url{http://dx.doi.org/10.1216/JIE-2009-21-3-331}

\bibitem{otani2007fast}
Y.~Otani, T.~Takahashi, N.~Nishimura, A fast boundary integral equation method
  for elastodynamics in time domain and its parallelisation, in: Boundary
  Element Analysis, Springer, 2007, pp. 161--185.

\bibitem{hackbusch2002data}
W.~Hackbusch, S.~B{\"o}rm, Data-sparse approximation by adaptive $\mathcal
  h^2$-matrices, Computing 69~(1) (2002) 1--35.

\bibitem{oseledets2008tucker}
I.~V. Oseledets, D.~Savostianov, E.~E. Tyrtyshnikov, Tucker dimensionality
  reduction of three-dimensional arrays in linear time, SIAM Journal on Matrix
  Analysis and Applications 30~(3) (2008) 939--956.

\end{thebibliography}

\appendix
\section{Quantization Method}
\label{FDPHA}
\setcounter{figure}{0}
The quantization method (Quantization) is detailed below. 
Its implementation is in \S\ref{FDPHA1}. 
Its cost and accuracy are in \S\ref{FDPHA2}, particularly for the case where Quantization is singly applied to the ST-BIEM.  
The $\epsilon_Q$ dependence of FDP=H-matrices is in \S\ref{FDPHA3}.

\subsection{Method Detail}
\label{FDPHA1}
Quantization is applied to a temporal convolution (the value of which is denoted by $T_n$ here) which is evaluated in each time step $n$, where a variable (the slip- and opening-rate $D_{n-m}$ in the body text) and kernel $K_m$ are convolved over $m$ as 
\begin{equation}
T_n=\sum_{m=M_{init}}^{M_{fin}-1} K_mD_{n-m},
\end{equation}
where $M_{init}$ and $M_{fin} (\leq M)$ denote the start and end, respectively, of the original temporal convolution to be quantized. When employing Quantization alone, 
we have set $M_{init}$ at the minimum of the time steps $m$ that give non-zero kernel $K_m$ values, and $M_{fin}$ at the start from which the static approximation is applied over $M_{fin}\leq m<M$. 
The following discussion applies to the temporal convolution in the spatiotemporal BIE for respective source-receiver pairs. 

\subsubsection{Implementation of Quantization}
\label{FDPHA11}
For the staircase approximation, a time range $b_q \leq m <b_{q + 1}$ of the quantization number $q$ $(=0,1,2,...)$ is recursively determined as the maximum time domain that entirely satisfies the error condition $|K_m - \hat K_q | \leq \epsilon_Q | \hat K_q|$ [or $|K_m - \hat K_q | \leq\min( \epsilon_Q | \hat K_q|,\epsilon_{st})$], where $\epsilon_Q$ and $\epsilon_{st}$ are the parameters of Quantization, $\hat K_q$ is the representative value of the kernel in $b_q \leq m <b_{q + 1}$, and $b_q$ is the time step inserting the partition of $q$. The initial partition position $b_0$ is set at $M_{init}$. The recursion ends at the last time step of the convolution to be quantized with returning the last time step number $M_{fin}$ as the time step of the last partition of Quantization $b_Q$, where $Q$ denotes the maximum number of $q+1$.

We can set the $\hat K$ value arbitrarily. 
Kernel value $K_{b_q}$ at the start of the ($q$-th) sampling cluster can be an option of the $\hat K_q$ value ($\hat K_q=K_{b_q}$). For this case, we can detect the set of the quantization partitions with the $\mathcal O(M_{fin}-M_{init})$ time complexity, by defining $b_{q+1}$ as the minimum time step $m$ that breaks the error condition (or equivalently, that fulfills $|K_m - \hat K_q |> \epsilon_Q | \hat K_q |$) for each given $b_q$. Likewise, when $\hat K$ is chosen as the kernel at the end of the sampling cluster ($\hat K_q=K_{b_{q+1}-1}$), desired clusters can be obtained with the $\mathcal O(M_{fin}-M_{init})$ complexity by the sequential partition detection starting from the maximum time step. 
In the anti-plane problem simulated in this paper, we defined a $\hat K$ value as an approximate kernel average, $\hat K=(K_{b_q}+\hat K_{b_{q+1} -1})/2$, and partition $b_{q+1}$ was set at the minimum of $m$ that satisfies $|K_m - K_{b_q}|/2 $ $>\epsilon_Q|\hat K_q|$. This choice of $\hat K_q$ compromises the above two partition selection conditions and satisfies their error conditions of two times larger $\epsilon_Q$. %Our choice worked in the cases of both time-increasing and time-decreasing kernels as far as we examined.

Quantization computes the temporal convolution as 
%\begin{equation}
%\sum_{m=M_{init}}^{M_{fin}-1} K_m D_{n-m}\simeq \sum_{q=0}^{Q-1}\hat K_q\hat D_{n,q}
%\end{equation} 
%that is
\begin{equation}
T_n\simeq \sum_{q=0}^{Q-1}\hat K_q\hat D_{n,q}
\label{FDPHeq:quantA}
\end{equation} 
with 
\begin{equation}
\hat D_{n,q} := \sum_{m=b_q}^{b_{q+1}-1} D_{n-m}.
\label{FDPHeq:Dhat}
\end{equation}
$\hat D_{n,q}$ is computed at each time step $n$ with the incremental time evolution rule of $\hat D$:
\begin{equation}
\hat D_{n,q} =\hat D_{n-1,q} +(D_{n-b_q}-D_{n-b_{q+1}}). 
\label{FDPHeq:quantB}
\end{equation}
The required memory cost and time complexity 
for computing $T_n$ and $\hat D_{n,q}$ by Eqs.~(\ref{FDPHeq:quantA}) and (\ref{FDPHeq:quantB}) are $\mathcal O(Q)$.

We note that the cumulative rounding errors in the update process of the quantized $\hat D_q$ may require some error handling particularly when the sampling interval is near one ($b_{q+1}-b_q\sim 1$).  
When Quantization was applied singly, we avoided such an error by using the definition of the $q$-th slip Eq. (\ref{FDPHeq:Dhat}) in computing $\hat D$ for small sampling intervals.

\subsubsection{Cost Estimates of Quantization}
\label{FDPHA12}

The associated memory and complexity to compute the convolution are of order the number of partitions in Quantization. 
The number of partitions is strictly $\mathcal O[(a/\epsilon_Q) \log(M_{fin}-M_{init})]$ under relative error regulation $|K - \hat K| < \epsilon_Q|\hat K|$ when the kernel is the power function $K_m \sim m^a$ of exponent $a$ with regard to time step $m$. The logarithmic order was kept basically even when the kernel was a sum of the power functions in our investigation of the 2D elastodynamic problems, shown in \S\ref{FDPHA21}. 
The absolute error condition is asymptotically negligible at a distance, and hence the costs become of $\mathcal O(1)$ under the absolute error condition. 
When multiple error criteria are imposed, the asymptotic costs are determined by the asymptotically dominant criterion.

\subsection{Performance Evaluation of Quantization}
\label{FDPHA2}
The cost and the accuracy of Quantization are investigated below. 
Regarding the cost evaluation, we focus on whether Quantization successfully drops the $M$-factor in the original cost; for example, the $\mathcal O(N^2M)$ costs (the memory consumption and time complexity per time step) of the ST-BIEM are expected to reduce to almost $O(N^2)$. 
For simplicity, we solve a 2D planar crack problem as an example with structured elements. 
The kernel for the planar boundary is written as $K_{i,i,m} = K_{i-j,m}$ because of the translational symmetry of the kernel, where we use the same symbol between $K_{i,j,m}$ and $K_{i-j,m}$. 
For simplifying the problem, only in this subsection, 
we utilize this translational symmetry on the planar fault and reduce the costs of the ST-BIEM to $\mathcal O(NM)$ and investigate whether Quantization can achieve the expected almost $\mathcal O(N)$ costs on planar boundaries; note this almost $\mathcal O(N)$ achieved by Quantization alone is limited to the planar boundary case and is different from the $\mathcal O(N\log N)$ scaling achieved by FDP=H-matrices applicable to the nonplanar boundary at the same cost order, discussed in the text.

The normalization units of the following anti-plane problem are the same as in \S\ref{FDPH6} in the text. 
The following in-plane problem adopts $\alpha =1$, instead of $\beta =1$ in the anti-plane problem, with setting $\beta$ at $\beta =\alpha/\sqrt{3}$ and adopting $\alpha\Delta t/\Delta x=1/2$ for the CFL parameter.

%Figure A.1_separated_a. âš
\begin{figure}[tbp]
	 \begin{center}
	\includegraphics[width=78mm]{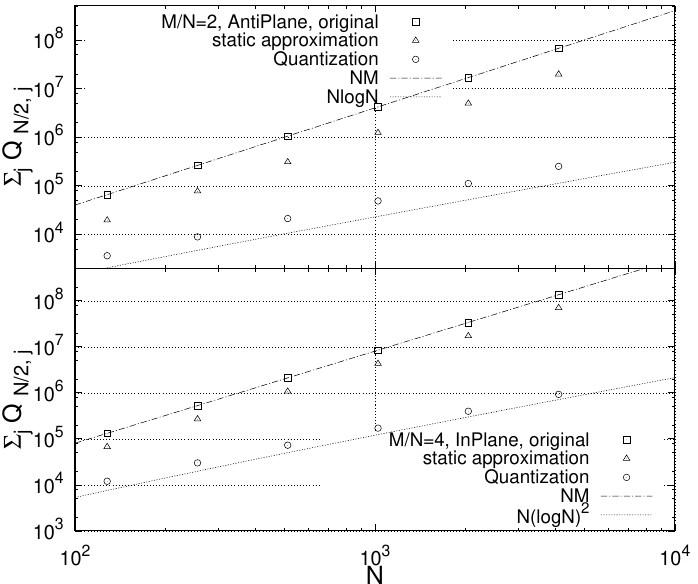}
%\\
%	\includegraphics[width=75mm]{fig_kernelrelerror_multi.eps}
	 \end{center}
\caption{
$N$ versus the number of partitions (corresponding to the cost of Quantization) per receiver, when Quantization is singly used with $\epsilon_Q=0.1$. Lines show some asymptotic scalings. 
}
\label{FDPHfig:A1_a}
\end{figure}

\subsubsection{Cost Reduction}
\label{FDPHA21}

By regarding the original ST-BEIM as a special case of $\epsilon_Q\to0$, we can measure the costs of both Quantization and the original BIEM by the number $\sum_{i,j}Q_{i,j}$ of partitions. 
In the planar fault, the estimated order of the number of partitions is further reduced to $\mathcal O(N\sum_jQ_{N/2,j})$ due to the translational symmetry above-mentioned.

Fig.~\ref{FDPHfig:A1_a} shows $\sum_{j}Q_{N/2,j}$ that expresses the typical number of partitions per receiver. The case of $\epsilon_Q=0.1$ is considered in the figure. The cost of  Quantization is shown to achieve almost $\mathcal O(N)$. This result is consistent with the estimated cost [$\mathcal O(\log L)$ that is $\mathcal O(\log N)$ per source-receiver pair] of Quantization mentioned in \S\ref{FDPHA12}. The log factor in the in-plane cases seems slightly larger $\mathcal O(\log^2 L)$ although the almost $\mathcal O(1)$ scaling holds; it would be due to that the 2D kernel in Domain I is not purely proportional to the power of time as it is actually the sum of the powers of time, namely the time-decaying wavefront and the asymptotic statics. 

\subsubsection{Kernel Accuracy} 
\label{FDPHA22}

%Figure A.1_separated_b. âš
\begin{figure}[tbp]
	 \begin{center}
	%\includegraphics[width=75mm]{unitnumber_multi.eps}
%\\
	\includegraphics[width=75mm]{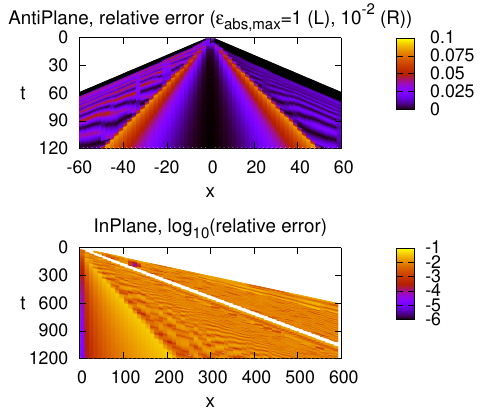}
	 \end{center}
\caption{
Errors in the quantized kernel. 
(Top) Error distribution of the quantized anti-plane kernel over space $x$ and time $t$, when the absolute error is 
regulated within $\epsilon_{\rm abs,max}$ after normalized by $K^{\rm max}_{i,j} := \max_m |K_{i,j,m}|$ for each $i,j$ pair, as in Ref.~\cite{ando2007efficient}. 
Values of $\epsilon_{\rm abs,max}$ are indicated in the figure ($x < 0$ and 0.01 at $x > 0$), and $\epsilon_Q=0.1$ applies concomitantly. 
The color bar indicates the relative error.
(Bottom) Error distribution of the quantized in-plane kernel over space $x$ and time $t$, when $\epsilon_Q=0.05$.
The color bar indicates the common logarithm of the relative error.
}
\label{FDPHfig:A1_b}
\end{figure}

%Figure A.2. âš
\begin{figure}[tbp]
	 \begin{center}
  %\includegraphics[width=40mm]{solutionapprox_ep0000001mabiki.eps}
  %\includegraphics[width=40mm]{solutionapprox_ep001mabiki.eps}
  %\\
  %\includegraphics[width=40mm]{solutionapprox_ep1000mabiki.eps}
  %\includegraphics[width=40mm]{solutionapprox_ep005abep0001mabiki_2.eps}
%
  %\includegraphics[width=78mm]{Quant_DR_summary.png}
  \includegraphics[width=100mm]{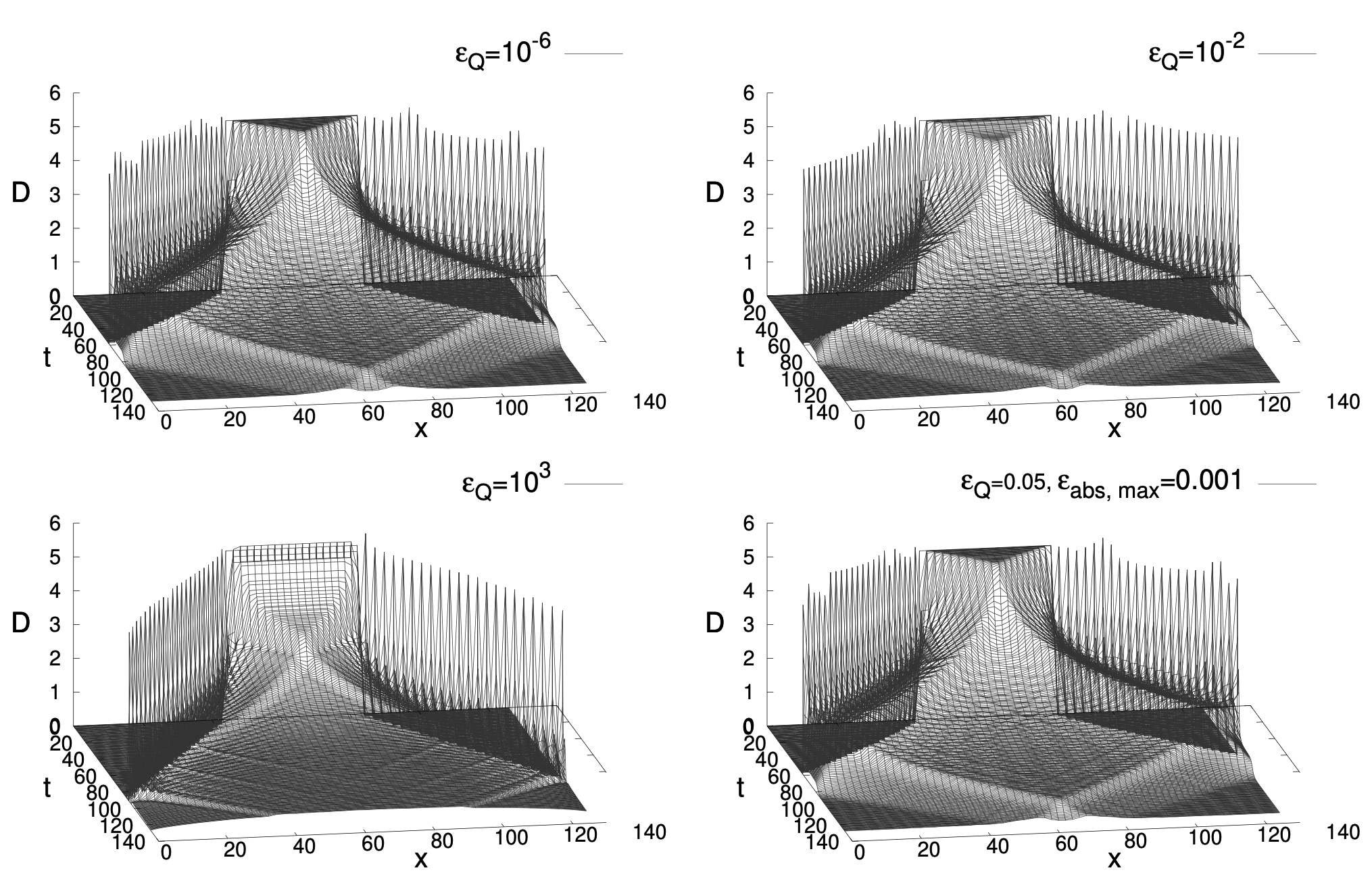}
	 \end{center}
\caption{
Slip rate $D$ evolving over space $x$ and time $t$ solved with Quantization, thinned out for visualization. 
Parameter values used for Quantization are shown in the figure. 
}
\label{FDPHfig:A2}
\end{figure}

Fig.~\ref{FDPHfig:A1_b} shows the error distributions in the kernels approximated by Quantization ($\sum_{b_q\leq m<b_{q+1}}|K_{ijm}-\hat K_{ijq}|/\sum_{b_q\leq m<b_{q+1}}|K_{ijm}|$) in respective $q$-th intervals. 
The stripes corresponds to the partitions given by Quantization schematically illustrated in Fig.~\ref{FDPHfig:5}. 
The widths of stripes are broadened as the source-receiver distance increases or the elapsed time increases in Domains I and S as expected. That in Domain S is purely a 2D feature as mentioned in \S\ref{FDPH22}. 
We see the assigned error criterion met. 
We also observe the relative error is zero around wavefronts, and Quantization automatically avoids approximating the kernel around such rapidly varying wavefronts. 

\subsubsection{Dynamic Rupture Problems}
\label{FDPHA23}
We here investigate the accuracy of the solutions simulated with the quantized kernel. We solved the dynamic rupture problems of the simple static-dynamic frictional boundary condition; the shear traction there suddenly drops to dynamic frictional strength $T_{dy}$ after the shear traction reaches yielding strength $T_{th}$. The initial stress distribution was set as in the single asperity model of Ref.~\cite{cochard1994dynamic}, where initial stress $T_{0}$ is given as the sum of background stress $T_{bg}$ and piecewise perturbation such that $T_0 (x) = T_{bg} + (T_{th} - T_{bg} + 0)H(x - x_-)H(x_+ - x)$, where $x_++x_-$ and $x_+-x_-$ are parameters determining the location and size, respectively, of the initial rupture. 

Fig.~\ref{FDPHfig:A2} shows the results obtained when $x_+ - x_-=40\Delta x$, $x_++x_-=N\Delta x$, $T_{th} =5$, and $T_{bg} =T_{dy}=0$. The increase of $\epsilon_Q$ accelerated the decrease in the slip rate in the initially fractured area. The rupture speed became smaller as $\epsilon_Q$ increased. These suggest $\epsilon_Q$ may damp the solution as artificial damping does. 
It is reasonable because the quantized solution with large $\epsilon_Q$ approaches to that of the quasi-dynamic approximation that replaces the kernel with the sum of the radiation damping term and the static kernel~\cite{ohtani2011fast}; the quasi-dynamic approximation neglects the radiated kinetic energy so that the decrease of the rupture speed and slip- and opening-rate naturally follows. 
Besides, the solution accuracy increased significantly when we set the absolute error bound despite its irrelevance at a distance. 

\subsection{$\epsilon_Q$ dependence of FDP=H-matrices}
\label{FDPHA3}
%Figure A.3. âš
\begin{figure}[tb]
  \includegraphics[width=70mm]{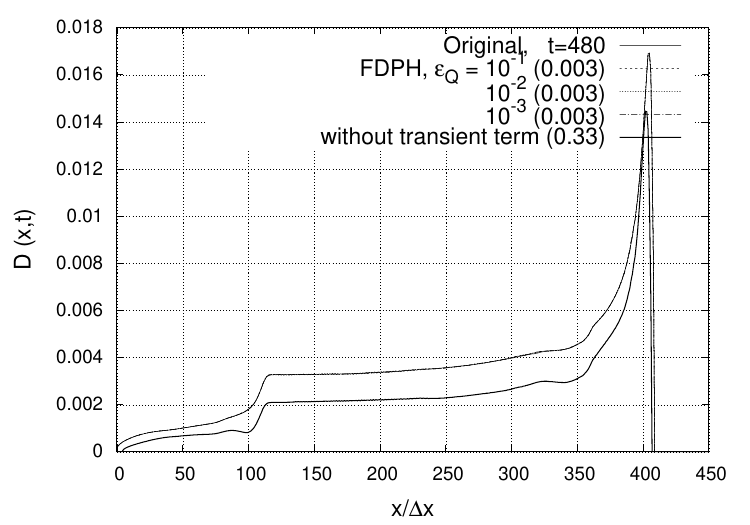}
\\
  \includegraphics[width=70mm]{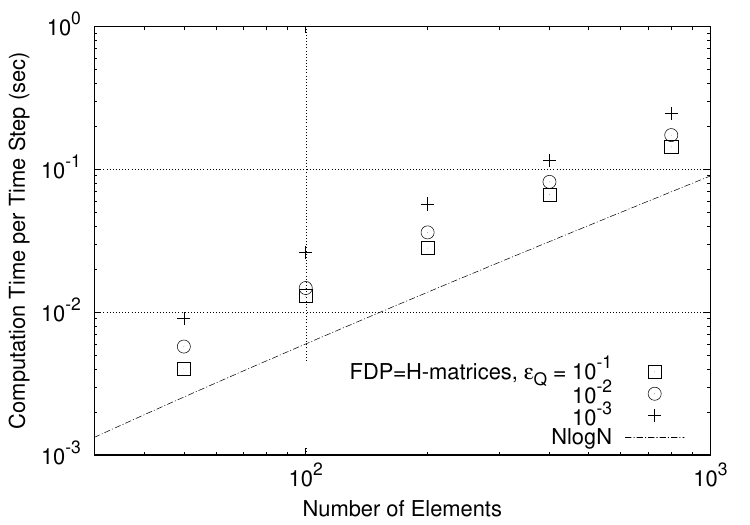}
\caption{
Dependencies of accuracy and costs on $\epsilon_Q$ in FDP=H-matrices. 
Unspecified parameter values in the left and right panels are respectively the same as those in the left and right panels in Fig.~\ref{FDPHfig:19}. 
(Top) Snapshots of slip rate $D$ over space $x$ at $t=480$ in the case of $\epsilon_Q$ ranging from $10^{-1}$ to $10^{-3}$ or without transient term, where FDP=H-matrices are abbreviated as FDPH. 
(Bottom) Computation costs of FDP=H-matrices. The $\mathcal O(1/\epsilon_Q)$ asymptote is shown by a dotted line. 
}
\label{FDPHfig:A3}
\end{figure}

We here investigate the $\epsilon_Q$ dependence of FDP=H-matrices 
by quantizing the transient term in Domain S, in a nonplanar problem studied in \S\ref{FDPH62}. The same property of Quantization is expected to the quantized Domain I kernel in the 3D problems.

Fig.~\ref{FDPHfig:A3} (top) shows the snapshots of the slip rates with several $\epsilon_Q$ values, compared with the case of erasing the transient term. Even with 100-fold increase of $\epsilon_Q$ within the range of 0.1 to 0.001, the accuracy degradation was negligible at the first digit of the relative errors. The accuracy deterioration seen in the case applying Quantization alone (\S\ref{FDPHA2}) did not occur in FDP=H-matrices even at relatively large value $\epsilon_Q=0.1$. 
%The irrrelevance of $\epsilon_Q$ in the accuracy is probably due to the combination of the absolute error condition adopted in FDP=H-matrices. 
Meanwhile, when the transient term was dropped, the solution accuracy was deteriorated by 33\%. It indicates the significance of the transient term. 
Since the transient term is significant for the accuracy while the value of the relative error bound $\epsilon_Q$, which affects the time step from which we drop the transient term, is irrelevant, 
the observed approximate $\epsilon_Q$-independence of the accuracy is probably caused by the absolute error condition added to the quantization condition (detailed in \S\ref{FDPH721}). As $\epsilon_{st}$ required much smaller values $\epsilon_{st}=10^{-6}$ to handle the 2D specific errors (detailed in \ref{FDPHH}), secondarily $\epsilon_Q$ would become irrelevant. 

Fig.~\ref{FDPHfig:A3} (bottom) shows the cost, typified by the computation cost measured by the computation time per time step, for the case of several $\epsilon_Q$ values, which are roughly proportional to the cost on $1/\epsilon_Q$. It is consistent with the theoretical estimates in \S\ref{FDPHA1}. Having said that, the cost change of FDP=H-matrices was within three-fold while $\epsilon_Q$ varies 100-fold. This relatively small dependence of the cost on $\epsilon_Q$ would suggest that the internal modules other than Quantization dominated the numerical costs. %(Probably, the absolute error condition of Quantization was not the candidate of the cause of the weak $\epsilon_Q$ dependence of the costs, since $\epsilon_{st}$ is comparable to the kernel value times $\epsilon_Q$ in Fig.~\ref{FDPHfig:A3} (bottom).) 

For both cost and accuracy, $\epsilon_Q$ was found to be a relatively irrelevant factor in FDP=H-matrices. 

\section{Arithmetics of FDP=H-Matrices in Domains I and S}
\label{FDPHB}

Below, we explain the arithmetics in Domains S and I of the $\mathcal O(N\log N)$ costs. Their main operations are respectively described in \S\ref{FDPHB1} and \S\ref{FDPHB2}, which include the associated temporal discretizations. 
The aritmetics for the 2D specific transient terms (introduced in \ref{FDPHH}) in Domains S and I are developed in similar ways in \S\ref{FDPHB3} and \S\ref{FDPHB4}, respectively. 
Related computational simplification will appear in \S\ref{FDPHdiscDomI} and \ref{FDPHC32}, and the supplemental information on the cost order is shown in \ref{FDPHB5}. 

\subsection{Domain S}
\label{FDPHB1}
The stress associated with Domain S, $T^S$, is written as
\begin{equation}
T_i^S(t) =\sum_j
\hat K^S_{i,j} \int_{t_{ij}^\beta+\Delta t_j^{\beta+}}^{\infty} d\tau D(t-\tau).
\end{equation}
After the ART and H-matrices are applied to it and Domain F (or precisely, the set of $\delta t_i$ and $\bar t_j^\pm$) is discretized in the way shown in \S\ref{FDPH43}, $T_i^S$ is discretized as 
\begin{align}
&T^S_i ((n+1)\Delta t+\delta t^\beta_i)
\\=&f_i^S
\sum_{j} g^S_{j}\sum_{m=-1}^{\infty}D_{j,n-m-\bar m_j^{\beta+}} 
\int_{\max(\bar t_j^{\beta+} , (m+\bar m_j^{\beta +})\Delta t)}^{(m+1+\bar m_j^{\beta+})\Delta t}d\tau
\\=&
f_i^S \sum_j g_j^S \left[\Delta t
\sum_{m=0}^\infty D_{j,n-m-\bar m_j^{\beta +}}
\right.
\nonumber
\\
&\left.
+ (\bar m_j^{\beta+}\Delta t-\bar t_j^{\beta+})
D_{j,n-(\bar m_j^{\beta +}-1)}\right].
\label{FDPHeq:unsimplifiedDomSstress}
\end{align}
%where the index $\beta$ of $\delta m_i$, $\bar m^+_j$ represents $\delta m_i$, $\bar m^+_j$ are those of S-waves. 
We below suppose the case of interpolating the left-hand side as 
$T^S_i ((n+1)\Delta t+\delta t^\beta_i)= T^S_{i,n+\delta m_i^\beta}$ (without loss of generality, as mentioned in \S\ref{FDPH432}).

The first term (denoted by $T_{i,n}^{S,asyc}$) is computed in the following manner. 
We introduce the increment $\Delta T_{i,n}^{S,asyc}$ of $T_{i,n}^{S,asyc}$ as 
\begin{equation}
\Delta T^{S,asyc}_{i,n}:=T_{i,n}^{S,asyc}-T_{i,n-1}^{S,asyc},
\label{FDPHeq:relationSasymp}
\end{equation}
which satisfies 
\begin{equation}
\Delta T^{S,asyc}_{i,n}
=f_i^S \sum_j g_j^S \Delta t D_{j,n-\bar m_j^{\beta +}-\delta m_i^\beta}.
\label{FDPHeq:relationdelSasymp}
\end{equation}
Eq.~(\ref{FDPHeq:relationdelSasymp}) is the same as Eq.~(\ref{FDPHeq:33}) evaluating $T_{i,n}^F$ in Domain F (appearing in \S\ref{FDPH5}) when $\hat D$ in Eq.~(\ref{FDPHeq:33}) is regarded as $D$. Hence, $\Delta T^{S,asyc}_{i,n}$ can be computed with the arithmetic of Domain F described in \S\ref{FDPH5}. 
$\Delta T^{S,asyc}_{i,n}$ evaluated in that arithmetic increments $T^{S,asyc}_{i,n}$ via Eq.~(\ref{FDPHeq:relationSasymp}) at each time step $n$ for all the receivers $i$. 

The second term in Eq.~(\ref{FDPHeq:unsimplifiedDomSstress}) becomes exactly zero (i.e. $T_{i,n}^{S}=T_{i,n}^{S,asyc}$) when we impose
\begin{equation}
\bar m_j^{\beta+}\Delta t-\bar t_j^{\beta+}=0,
\end{equation}
by utilizing the arbitrariness of $\Delta t_j^\pm$ (mentioned in \S\ref{FDPH431}). An implementation for satisfying this condition is shown in \ref{FDPHC32}. 
We skipped the evaluation of the second term in that way in the numerical experiments. 
Otherwise, we compute it with the same arithmetic as that of Domain F.

\subsection{Domain I}	
\label{FDPHB2}
The kernel of Domain I in continuous time is a sum of functions all of which separate into the corresponding spatial parts and temporal parts~\cite{bonnet1999boundary,tada2000non}. 
For the stress nuclei of the double-layer potential, 
that kernel is decomposed into two spatiotemporally separable functions, and one of the temporal part is time-invariant as in the kernel in Domain S while the other is proportional to the power of the elapsed time~\cite{ando2016fast}. For notational simplicity, hereafter we abbreviate the summation over these two time dependencies. 
The other nuclei of the double-layer potential and single-layer parts also follow the similar decomposition, and then the following arithmetic holds for them, excluding the specific expression of the semi-analytic BIE. 

After the ART and H-matrices are applied as in \S\ref{FDPH4}, the stress associated with Domain I, $T^I$, is written as
\begin{equation}
T^I_i(t)
=f_i^I \sum_j g_j^I \int^{\delta t_i^\beta+\bar t_j^{\beta+}}_{\delta t_i^\alpha+\bar t_j^{\alpha-}} d\tau h^I(\tau)D_j(t-\tau-\bar t_j).
\label{FDPHeq:origundiscdomI}
\end{equation}
After Domain F (or precisely, the set of $\delta t_i$ and $\bar t_j^\pm$) is temporally discretized as in \S\ref{FDPH43},
we obtain a partially discretized form of Eq.~(\ref{FDPHeq:origundiscdomI}):
\begin{equation}
T_{i,n}^I=f_i^I \sum_j g_j^I \sum_{m=\delta m_i^\alpha+\bar m_j^{\alpha +}}^{\delta m_i^\beta +\bar m_j^{\beta -}-1} h^I_m D_{j,n-m} 
+\mbox{decimal part},
\label{FDPHeq:discdomI}
\end{equation}
where $h_m^I$ is the temporal part in Domain I for discretized kernel $K_{i,j,m}$ (introduced in \S\ref{FDPH21}) discretized with the constant time step $\Delta t$; the first term is defined as the time steps the time ranges belong to which is fully within the original time range of Domain I, and the second term (called a decimal part hereafter in \S\ref{FDPHB2}) is that partly within Domain I while partly in Domain F. %This decimal parts is caused to keep the consistency between all domains after the approximation of the ART.
Duration of the decimal part is $(\delta t_i^\beta+\bar t_j^{\beta +})-(\delta t_i^\alpha+\bar t_j^{\alpha -})$ minus the integer multiple of $\Delta t$, and thus the temporal dependence of the kernel is modified from $h_m^I$ in it, as explicitly shown in \S\ref{FDPHB25}. 

Below, we first develop the computational procedures of the first term in Eq.~(\ref{FDPHeq:discdomI}) through \S\ref{FDPHB21}, \S\ref{FDPHB22}, \S\ref{FDPHB23}, and \S\ref{FDPHB24}. Second, we deal with the decimal part in \S\ref{FDPHB25}. 
In this paper, we assume Domain I [or more strongly, the first term in Eq.~(\ref{FDPHeq:discdomI})] exists in all the admissible leaves for simple implementation. One way of handling this assumption is detailed in \S\ref{FDPHdiscDomI}. 

\subsubsection{Decomposition of the Convolution}
\label{FDPHB21}
To begin with, the first term in Eq.~(\ref{FDPHeq:discdomI}) is represented by 
\begin{flalign}
T_{i,n}^I=&f_i^I \sum_j g_j^I \left[
\sum_{m=m_0^I}^{\delta m_i^\beta +\bar m_j^{\beta -}-1} 
-
\sum_{m=m_0^I}^{\delta m_i^\alpha+\bar m_j^{\alpha +}-1}
\right]
h^I_m D_{j,n-m} 
\nonumber
\\&+\mbox{decimal part},
\label{FDPHeq:decompdomI}
\end{flalign}
where $m_0^I$ is an appropriate constant such that $m_0^I\leq\min[\delta m_i^\alpha+\bar m_j^{\alpha+}]$. 
The first and second terms within brackets in Eq.~(\ref{FDPHeq:decompdomI}) are respectively the time integral from the onset $m_0^I$ to the time step of the P- and S-wave passage completion, 
and both are computed in the same way. Their common computational procedure is explained below by using the following irreducible expression of them:
\begin{equation}
T_{i,n}^{Ii}=f_i\sum_j g_j
\sum_{m=m_0}^{\delta m_i +\bar m_j-1} 
h_m D_{j,n-m},
\label{FDPHeq:irreddomI}
\end{equation}
where we omitted indices for notational simplicity. 

Eq.~(\ref{FDPHeq:irreddomI}) is further separated into two parts;
\begin{flalign}
T_{i,n}^{Ii}&=f_i\sum_j g_j
\sum_{m=0}^{\tilde m^{Ii1}_j + \tilde m^{Ii2}_i-1} 
h_{m+m_0} D_{j,n-m-m_0}
\\&=
T_{i,n}^{Ii1}+T_{i,n}^{Ii2}
\label{FDPHeq:irreddecompdomI}
\end{flalign}
with 
\begin{flalign}
T_{i,n}^{Ii1}&=
f_i \sum_j g_j
\sum_{m=0}^{ \tilde m^{Ii1}_j-1} 
h_{m+m_0} D_{j,n-m-m_0}
\\
T_{i,n}^{Ii2}
&=
f_i \sum_j g_j
\sum_{m=0}^{\tilde m^{Ii2}_i-1} 
h_{m+m_0+\tilde m^{Ii1}_j} D_{j,n-m-m_0-\tilde m^{Ii1}_j}
\end{flalign}
where $\tilde m^{Ii1}_j$ and $\tilde m^{Ii2}_i$ are some (arbitrary) positive constants that satisfy $\tilde m^{Ii1}_j+\tilde m^{Ii2}_i+m_0=\delta m_i+\bar m_j$. Hereafter, $\tilde m^{Ii1}_j$ and $\tilde m^{Ii2}_i$ are respectively abbreviated to $\tilde m_j$ and $\tilde m_i$.

Two scalars $\tilde m_i$ and $\tilde m_j$ are introduced for each $i$ and $j$ in each admissible leaf to make the integral lengths of the first and second terms in Eq.~(\ref{FDPHeq:irreddecompdomI}) non-negative; these are for handling $\delta m_i$ becoming negative frequently. %$T^{Ii}$ separated in the above-mentioned way is computed with an almost $\mathcal O(N)$ cost in the procedure detailed in \S\ref{FDPHB22}. 
A simple choice to obtain $\tilde m_i$ and $\tilde m_j$ will be using $\delta t_i^\prime=(r_{ij_*}-r_{i_*j_*}/2)/c$, $\bar t_j^\prime=(r_{i_*j}-r_{i_*j_*}/2)/c$ instead of using $\delta t_i$ and $\bar t_j$ giving $\delta m_i$ and $\bar m_j$ in the paper. Hereafter, $\tilde m_i$ and $\tilde m_j$ are both supposed to be of $\mathcal O(dist)$ for simplicity.
%The following will efficiently work Hereafter we assume that $\tilde m_i$ and $\tilde m_j$ are adjusted so that their average values are the same value (except for the remainder of the bisection) in each admissible leaf. 

Below, $T^{Ii1}$ and $T^{Ii2}$ in Eq.~(\ref{FDPHeq:irreddecompdomI}) are computed separately. 
The computation procedure of $T^{Ii1}$ is explained in \S\ref{FDPHB22} and \S\ref{FDPHB23}. That of $T^{Ii2}$ is in \S\ref{FDPHB24}. 

\subsubsection{$T^{Ii1}$ Computation in Eq.~(\ref{FDPHeq:irreddecompdomI}) Without Quantization}
\label{FDPHB22}
First, like the $T^F$ computation in Domain F, the $T^{Ii1}$ computation separates into a conversion from representative stress $\bar T$ to stress $T$ and that from slip- and opening-rate $D$ to representative stress $\bar T$: 
\begin{flalign}
T^{Ii1}_{i,n}&=f_i\bar T^{Ii1}_n
\label{FDPHeq:barTIi1toTIi1}
\\
\bar T^{Ii1}_n&:= \sum_j g_j\sum_{m=0} ^{\tilde m_j-1}  h_{m+m_0}  D_{j,n-m-m_0}.
\label{FDPHeq:DtobarTIi1}
\end{flalign}
Eq.~(\ref{FDPHeq:barTIi1toTIi1})
is computable with the almost $\mathcal O(N)$ costs as in H-matrices. On the other hand, Eq.~(\ref{FDPHeq:DtobarTIi1}) contains the time integral whose length is $\mathcal O(dist)$ for each $j$. It means that 
computing Eq.~(\ref{FDPHeq:DtobarTIi1}) can require the $\mathcal O(NL)$ costs both in terms of memory and computation time at every time step.
We then focus on reducing the numerical costs for evaluating Eq.~(\ref{FDPHeq:DtobarTIi1}). 

A subtask for the efficient computation of Eq.~(\ref{FDPHeq:DtobarTIi1}) is to separate $j$ and $m$ dependencies of $D_{j,n-m-m_0}$.
We then take the subsets of sources $j$, in each admissible leaf, that share the same value of $\tilde m_j=p$. As the number of sources is of $\mathcal O(diam^{D_b})$ in a leaf while that of the possible values is of $\tilde m_j-1$ of $\mathcal O(diam)$, 
such a subset of $j$ gives a computationally efficient decomposition of the summation over $j$ in Eq.~(\ref{FDPHeq:DtobarTIi1}); 
\begin{flalign}
\bar T^{Ii1}_n&= 
\sum_j \left(\sum_{p=\min_j \tilde m_j}^{\max_j \tilde m_j}\delta_{p,\tilde m_j}\right)g_j\sum_{m=0} ^{\tilde m_j-1}  h_{m+m_0}  D_{j,n-m-m_0}
\\&= 
\sum_{p=\min_j \tilde m_j}^{\max_j \tilde m_j}\sum_{m=0} ^{p-1}  h_{m+m_0} 
\sum_{j|\tilde m_j=p} g_jD_{j,n-m-m_0},
\end{flalign}
where $\sum_{j|\tilde m_j=p}=\sum_j\delta_{\tilde m_j,p}$ is introduced.
This comprises two computations: 
\begin{equation}
\bar T^{Ii1}_n= 
\sum_{p=\min_j \tilde m_j}^{\max_j \tilde m_j}\sum_{m=0} ^{p-1}  h_{m+m_0} 
\Delta \bar T_{proj,n-m-m_o,p}^{Ii1}
\label{FDPHeq:TpartialtoT}
\end{equation}
and 
\begin{equation}
\Delta \bar T_{proj,m^\prime,p}^{Ii1}:= \sum_{j|\tilde m_j=p} g_jD_{j,m^\prime},
\label{FDPHeq:defofTpartial}
\end{equation}
where $\min_j\tilde m_j$ and $\max_j\tilde m_j$ represent the minimum and maximum values of $\tilde m_j$ in an admissible leaf. 
$\Delta \bar T_{proj}$ expresses the partial sum of the inner product between $g$ and $D$, gathering the contribution from $j$ of the same $\tilde m_j=p$ in Eq.~(\ref{FDPHeq:DtobarTIi1}).
Physically, $\Delta \bar T_{proj,m^\prime,p}$ corresponds to the stress due to a wavefront that assembles the source contributions of the same travel time $p$ and the same launch time $m^\prime$.

Next we decompose the summations over $p$ and $m$ in Eq.~(\ref{FDPHeq:TpartialtoT}).
Since the range of summation $\sum_{p=\min_j\tilde m_j}^{\max_j\tilde m_j}\sum_{m=0}^{p-1}$ in Eq.~(\ref{FDPHeq:TpartialtoT}) is equivalent with intersection $(\min \tilde m_j\leq p\leq \max\tilde m_j)$ $\cap$ $(0\leq m< \max\tilde m_j)$ $\cap$ $(m<p)$, we can rewrite Eq.~(\ref{FDPHeq:TpartialtoT}) as
\begin{flalign}
\bar T^{Ii1}_n&= 
\sum_{p=\min \tilde m_j}^{\max \tilde m_j}\sum_{m=0} ^{\max \tilde m_j-1} H(p-m-0) h_{m+m_0} \Delta \bar T^{Ii1}_{proj, n-m-m_0,p}
\\&= 
\sum_{m=0} ^{\max \tilde m_j-1} h_{m+m_0} 
\sum_{p=\max(m+1,\min \tilde m_j)}^{\max \tilde m_j}
\Delta \bar T^{Ii1}_{proj, n-m-m_0,p}.
\label{FDPHeq:noquantIi1beforedecomp}
\end{flalign}
That is,
\begin{equation}
\bar T^{Ii1}_n= \sum_{m=0} ^{\max \tilde m_j-1} h_{m+m_0} 
\Delta \bar T^{Ii1}_{sum, n-(m+m_0),m}
\label{FDPHeq:noquantIi1}
\end{equation}
and
\begin{equation}
\Delta \bar T^{Ii1}_{sum, m^\prime,m}:=\sum_{p=\max (m+1, \min \tilde m_j)} ^{\max \tilde m_j} \Delta \bar T^{Ii1}_{proj, m^\prime,p}.
\label{FDPHeq:noquantIi1additional}
\end{equation}

The definitional identity $\Delta \bar T^{Ii1}_{sum, m^\prime,m}$ separates its $m$-dependence [$\bar T^{Ii1}_{sum, n-(m+m_0),m}$] into two parts in Eq.~(\ref{FDPHeq:noquantIi1}) in a deliberate fashion. 
The first subscript $m^\prime=n-(m+m_0)$ of $\bar T^{Ii1}_{sum, n-(m+m_0),m}$ corresponds to the time shift of $\Delta \bar T^{Ii1}_{proj,n-m-m_0,p}$ in Eq.~(\ref{FDPHeq:noquantIi1beforedecomp}); 
the second subscript $m$ expresses the start of summation [$p=\max(m+1,\min \tilde m_j)$] over $p$. 
This redundancy broadening the functional space from $m$ to $m,m^\prime$ gives the following useful recurrence relation: 
\begin{equation}
\Delta \bar T^{Ii1}_{sum,m^\prime,m}=\Delta \bar T^{Ii1}_{sum,m^\prime,m+1}+ \Delta \bar T^{Ii1}_{proj, m^\prime,m+1}.
\label{FDPHeq:nonquantrecurrence}
\end{equation}

Eqs.~(\ref{FDPHeq:barTIi1toTIi1}), (\ref{FDPHeq:defofTpartial}), (\ref{FDPHeq:noquantIi1}), and (\ref{FDPHeq:nonquantrecurrence}) constitute the computation of $\bar T^{Ii1}$, for the case without Quantization. 
Eq.~(\ref{FDPHeq:defofTpartial}) shows the first operation, 
which converts ${\bf D}_n$ ($D_{j,n}$ of any $j$ belonging to an admissible leaf) to 
$\Delta \bar T_{proj,n,p}$ of $p\in[\min\tilde m_j,\max\tilde m_j]$ at each time step $n$.
It can be rewritten as
\begin{equation}
{\bf \Delta \bar T}^{Ii1}_{proj, n}={\bf G}^{Ii1}{\bf D}_n
\label{FDPHeq:DtobarTIi1matrix}
\end{equation}
with
\begin{equation}
G^{Ii1}_{p,j} := \delta_{p,\tilde m_j} g_j,
\end{equation}
where ${\bf \Delta \bar T}^{Ii1}_{proj, n}:=(\Delta \bar T_{proj,n,\min \tilde m_j},...,\Delta \bar T_{proj,n,\max\tilde m_j})^{\mbox{T}}$ contains $\Delta \bar T_{proj,n,p}$ at the $p$-th component. Eq.~(\ref{FDPHeq:DtobarTIi1matrix}) parallels the conversion from $\hat D$ to $\bar T$ in Domain F shown in \S\ref{FDPH52}. 
Eq.~(\ref{FDPHeq:nonquantrecurrence}) of $m^\prime=n$ represents the second operation, which converts ${\bf \Delta \bar T}^{Ii1}_{proj, n}$ to $\bar T_{sum,n,m}^{Ii1}$ of $m\in[\min\tilde m_j-1,\max\tilde m_j)$, recursively from $\Delta T^{Ii1}_{sum,n,\max \tilde m_j}=0$ [obtained from Eq.~(\ref{FDPHeq:noquantIi1additional})]. 
Note $\Delta T^{Ii1}_{sum,n,\max \tilde m_j}=
\Delta T^{Ii1}_{sum,n,\max \tilde m_j+m}$ and 
$\Delta T^{Ii1}_{sum,n,\min\tilde m_j-m}= 
\Delta T^{Ii1}_{sum,n,\min\tilde m_j-1}$ for $m\in\mathbb N$.
% for $m+1\geq \min\tilde m_j$. 
$\Delta \bar T^{Ii1}_{sum,n-m^\prime,m}$ is updated by the following relation, noticed from Eq.~(\ref{FDPHeq:noquantIi1additional}):
\begin{equation}
{\bf \Delta \bar T}^{Ii1\prime}_{sum,n,m} =\mathcal M {\bf \Delta \bar T}^{Ii1\prime}_{sum,n-1,m},
\label{FDPHeq:noquantIi1marching}
\end{equation}  
where ${\bf \Delta \bar T}^{Ii1\prime}_{sum,n,m}$ is a vector that stores $\Delta \bar T^{Ii1}_{sum,n-m^\prime,m}$ at the component $m^\prime\in[0,m_0+\max_j\tilde m_j)$. 
The lower and upper bounds of the $m^\prime$ range of the stored ${\bf \Delta \bar T}^{Ii1\prime}_{sum,n,m}$ components are determined by the second operation [Eq.~(\ref{FDPHeq:nonquantrecurrence}) of $m^\prime=n$] 
and the third one [Eq.~(\ref{FDPHeq:noquantIi1})], and its other $m^\prime$ components are not stored. 
Unlike $\bar T^\prime_{n,m}$ in Domain F, $\Delta \bar T^{Ii1\prime}_{sum,n,m,m^\prime}$ $=\Delta \bar T^{Ii1}_{sum,n-m^\prime,m}$ holds everywhere in ${\bf \Delta \bar T}^{Ii1\prime}_{sum,n,m}$. 
Eq.~(\ref{FDPHeq:noquantIi1}) shows the third operation that converts
${\bf \Delta \bar T}^{Ii1\prime}_{sum,n,m}$ ($\Delta \bar T^{Ii1}_{sum,n-m^\prime,m}$) to $\bar T^{Ii1}_n$.
Eq.~(\ref{FDPHeq:barTIi1toTIi1}) does the fourth one that converts
$\bar T^{Ii1}_n$ to $T^{Ii1}_{i,n}$ of all the receivers $i$ belonging to the associated admissible leaf at each time step $n$.

\subsubsection{$T^{Ii1}$ Computation in Eq.~(\ref{FDPHeq:irreddecompdomI}) with Quantization}
\label{FDPHB23}
Given that the two subscripts of $\Delta \bar T^{Ii1}_{sum,n-m^\prime,m}$ range over
$m^\prime\in[0,m_0+\max_j\tilde m_j)$
and 
$m\in[\min\tilde m_j-1,\max\tilde m_j)$, 
the computation of $T^{Ii1}$ without Quantization, shown in \S\ref{FDPHB22}, requires the $\mathcal O [diam\cdot dist/(c\Delta t)^2]$ memory ($c=\alpha,\beta$) 
for $\Delta \bar T^{Ii1}_{sum,n-m^\prime,m}$ (or $\mathcal O [diam^2/(c\Delta t)^2]$, detailed in \ref{FDPHB5}). 
Even in the constant $\eta$ scheme, 
such a memory cost is totally of almost $\mathcal O(N^{2/D_b})$, which becomes almost $\mathcal O(N^2)$, not an almost linear order, at $D_b=1$ while it is almost $\mathcal O(N)$ i.e. $\mathcal O(N\log N)$ for $D_b=2$ being our main concern. 
Below, we quantize the temporal integral in Eq.~(\ref{FDPHeq:TpartialtoT})
to make such $\mathcal O(diam^2)$ history of $\Delta \bar T^{Ii1}_{sum}$ unnecessary. 

First we quantize $h$.
Quantization of the function $h_{m+m_0}$ determines the positions $b_0, ..., b_Q$ in the maximum temporal integration range of $T^{Ii1}$, $m \in [0, \max_j \tilde m_j)$ in a $j$-independent manner.
Quantized variable $\Delta \hat T^{Ii1}_{n,q}$ of quantization number $q$ is next defined for current time step $n$, 
so as to reduce the $\bar T^{Ii1}_{n}$ convolution in Eq.~(\ref{FDPHeq:TpartialtoT}) to
\begin{equation}
\bar T^{Ii1}_n\simeq \sum_q \hat h_q\Delta \hat T^{Ii1}_{n,q},
\label{FDPHeq:ThattobarT}
\end{equation}
where $\hat h_q$ is a quantized $h_{m+m_0}$ value at the $q$-th interval.
By considering the $p$-dependent summation range 
of $\Delta \bar T^{Ii1}_{proj,n-m-m_0,p}$ over $m$ in Eq.~(\ref{FDPHeq:TpartialtoT}), 
we obtain
the explicit form of $\Delta \hat T^{Ii1}_{n,q}$ as
\begin{equation}
\Delta \hat T^{Ii1}_{n,q} =
\sum_{p=\min_j \tilde m_j}^{\max_j \tilde m_j} 
\sum_{m|(b_q\leq m< b_{q+1}) \cap (0\leq m<p)} 
\Delta \bar T^{Ii1}_{proj, n-m-m_0,p}.
\end{equation}

The quantized variable $\Delta \hat T_{n,q}$ is stored only for current time step $n$, 
and we evolve it with computing its time increment, defined as 
\begin{equation}
\delta \hat T^{Ii1}_{n,q} :=
\Delta \hat T_{n, q} -\Delta \hat T_{n-1, q}. 
\label{defofhatTIi1}
\end{equation}
The explicit form of $\delta \hat T_n$ is calculated by using the following another form of $\Delta \hat T^{Ii1}_{n,q}$:
\begin{flalign}
\Delta \hat T^{Ii1}_{n,q} &=
\sum_{p=\min_j \tilde m_j}^{\max_j \tilde m_j} 
H(p-b_q-0)
\nonumber\\&
\times
\sum_{m=b_q}^{\min (b_{q+1},p)-1} 
\Delta \bar T^{Ii1}_{proj, n-m-m_0,p}. 
\label{FDPHeq:hatTlikehatD} 
\end{flalign}
We note $b_q\geq0$, and that $H(p-b_q-0)$ takes the nonzero value when $p>b_q$. 
Comparing 
Eq.(\ref{FDPHeq:hatTlikehatD})
with Eq.~(\ref{FDPHeq:Dhat}) (in the original Quantization) 
regarding the range of the summation over $m$, 
the increment of $\Delta \hat T^{Ii1}_{n,q}$ [i.e. $\delta\hat T_n$ in Eq.~(\ref{defofhatTIi1})] 
is noticed to be made of the contributions from the end points of its summation range
as in Eq.~(\ref{FDPHeq:quantB});
\begin{flalign}
&\delta \hat T^{Ii1}_{n,q} 
\nonumber\\
&=\sum_{p=\min_j \tilde m_j}^{\max_j \tilde m_j} 
H(p-b_q-0)
\nonumber\\&\times
(\delta_{m,b_q}-\delta_{m,\min(b_{q+1},p)})
\Delta \bar T_{proj,n-m-m_0,p}
\\&=\sum_{p=\min_j \tilde m_j}^{\max_j \tilde m_j} 
[H(p-b_q-0)\delta_{m,b_q}-H(p-b_{q+1}-0)\delta_{m,b_{q+1}}
\nonumber
\\&+H(p-b_q-0)H(b_{q+1}-p+0)\delta_{m,p}]
\Delta \bar T_{proj,n-m-m_0,p}
\end{flalign}
where $\min(b_{q+1},p)$ is conditioned into two cases, $p>b_{q+1}(>b_q)$ and $p\leq b_{q+1}$, in the transform to obtain the last line.
By using $\bar T^{Ii1}_{sum,m^\prime,m}$ [defined in Eq.~(\ref{FDPHeq:noquantIi1additional})], this becomes
\begin{flalign}
\delta \hat T^{Ii1}_{n,q} 
&=\Delta \bar T^{Ii1}_{sum,n-(b_q+m_0),b_q}-\Delta \bar T^{Ii1}_{sum,n-(b_{q+1}+m_0),b_{q+1}}
\nonumber\\&
+H(b_{q+1}-\min_j\tilde m_j+0)
\nonumber\\&\times
\sum_{p=\max(b_q+1,\min_j \tilde m_j)}^{b_{q+1}}\Delta \bar T_{proj,n-p-m_0,p}.
\end{flalign}
Note $\forall (a,b)$, $\sum_{p=a}^b=\sum_p H(p-a+0)H(b-p+0)$ and  $\forall q$, $b_q<\max_j\tilde m_j$.

$\delta \hat T^{Ii1}$ is computed by using the sparse matrices as $\bar T$ in Domain F. The explicit form of the sparse matrix computation is derived by comparing 
%Eq.~(\ref{FDPHeq:tensorialdeltaT}) 
the following tensorial expression of $\delta \hat T^{Ii1}_{n,q}$, 
\begin{flalign}
&\delta \hat T^{Ii1}_{n,q} 
\nonumber
\\
&=
\sum_{q^\prime,m} (\delta_{q,q^\prime}-\delta_{q+1,q^\prime})\delta_{-m,b_{q^\prime}+m_0}
\Delta \bar T^{Ii1}_{sum,n+m,b_{q^\prime}}
\nonumber
\\
&+H(b_{q+1}-\min_j \tilde m_j+0)\sum_{p,m} \delta_{-m,p+m_0}
\nonumber
\\
&\times
H(p-\max(b_q+1,\min_j\tilde m_j)+0)H(b_{q+1}-p+0) 
\nonumber
\\
&\times
\Delta \bar T^{Ii1}_{proj,n+m,p}
\label{FDPHeq:tensorialdeltaT}
\end{flalign}
with $\bar T_n=\sum_{j,m}g_j\delta_{m,-\bar m_j}\hat D_{j,n+m}$ [Eq.~(\ref{FDPHeq:barTdcomptointersection})] giving the sparse matrix computation of Eq.~(\ref{FDPHeq:42_tensorial}). 
From that comparison we notice correspondence between $q^\prime$ in the first term of Eq.~(\ref{FDPHeq:tensorialdeltaT}) and $j$ in
Eq.~(\ref{FDPHeq:barTdcomptointersection}), and similarly between $p$ in the second term of Eq.~(\ref{FDPHeq:tensorialdeltaT}) and $j$ in
Eq.~(\ref{FDPHeq:barTdcomptointersection}). 
Then, after we define 
\begin{equation}
{\bf \Delta \bar T}^{Ii1}_{sumQ,n} :=(\Delta \bar T^{Ii1}_{sum,n,b_0},...,\Delta \bar T^{Ii1}_{sum,n,b_{Q-1}})^{\mbox{T}}
\end{equation} 
that contains $\Delta \bar T^{Ii1}_{sum,n,b_q}$ at the $q$-th component, 
and conditionally-predicted representative stress vector ${\bf \delta \hat T}^{Ii1\prime}_{n,q}=
(...,
\delta \hat T^{Ii1\prime}_{n,0,q}
,
\delta \hat T^{Ii1\prime}_{n,1,q}
,
...
)^{\mbox T}
$ 
in the same manner as that of ${\bf \bar T}^\prime_n$ in \S\ref{FDPHFcost} (the $m$-th component $\delta \hat T^{Ii1\prime}_{n,m,q}$ of which is associated with $\delta \hat T^{Ii1}_{n-m,q}$), 
the computation of ${\bf \delta \hat T}^{Ii1\prime}_{n,q}$ 
at time step $n$ for the $q$-th quantization number 
is expressed as 
\begin{flalign}
&{\bf \delta \hat T}^{Ii1\prime}_{n+1,q}=\mathcal M[
{\bf \delta \hat T}^{Ii1\prime}_{n,q} +\mathcal T_q {\bf \Delta \bar T}^{Ii1}_{sumQ,n}
+ \mathcal P_q {\bf \Delta \bar T}^{Ii1}_{proj,n}
 ],
%{\bf \Delta \hat T}_{n+1,q}=\mathcal M[{\bf \Delta \hat T}_{n,q} 
%+(G_q^{Ii1} \mathcal S_q -G_{q+1}^{Ii1} \mathcal S_{q+1} +G_p^{Ii1}(\mathcal S_q -\mathcal S_{q+1}) {\bf D}_n].
\label{FDPHeq:deltaTIi1sparse}
\end{flalign}
with sparse matrices: 
\begin{flalign}
&(\mathcal T_{q})_{m,q^\prime} 
:=
\delta_{-m^\prime,b_{q^\prime}+m_0}
 (\delta_{q,q^\prime}-\delta_{q+1,q^\prime})
\\
&(\mathcal P_q)_{m,p} 
:=
H(b_{q+1}-\min_j \tilde m_j+0)
\delta_{-m,p+m_0} 
\nonumber\\
&
\times
H(p-\max(b_q+1,\min_j\tilde m_j)+0)H(b_{q+1}-p+0). 
\end{flalign}

The arithmetic for $T^{Ii1}$ computations with Quantization is as follows. 
$\Delta \bar T_{proj,n,p}$ and $\bar T_{sum,n,m}^{Ii1}$ are computed 
for all $p$ and $m$ in each time step $n$, as in the computations without Quantization (explained in \S\ref{FDPHB23}).
Next, instead of storing $\bar T^{Ii1}_{sum,n-m^\prime,m}$, 
%$\bar T$
%First, Eq.~(\ref{FDPHeq:DtobarTIi1matrix}) converts ${\bf D}_n$ to ${\bf \Delta \bar T}^{Ii1}_{partial,n}$ at the current time step $n$ for any $q$. 
%Second, $\Delta \bar T^{Ii1}_{cum,n,q}$ at the current time step $n$ for any $q$ is sequentially computed by the following equation:
%\begin{equation}
%\Delta \bar T^{Ii1}_{cum,n,q}= \Delta \bar T^{Ii1}_{cum,n,q+1} +\sum_{p=b_q}^{b_{q+1}-1} \Delta \bar T^{Ii1}_{partial,n,p}.
%\end{equation}
${\bf \delta \hat T}^{Ii1\prime}_{n,q}\in\mathbb R^{\max_j\tilde m_j+m_0+1}$ is updated to ${\bf \delta \hat T}^{Ii1\prime}_{n+1,q}$ by using Eq.~(\ref{FDPHeq:deltaTIi1sparse}); the numerically required $m$ range of $\delta \hat T^{Ii1^\prime}_{n,m,q}$ is within $m\in [0,\max_j \tilde m_j+m_0]$ given Eqs.~(\ref{FDPHeq:ThattobarT}) and (\ref{FDPHeq:deltaTIi1sparse}).
$\Delta \hat T^{Ii1}_{n,q}$ of all $q$ then evolves to $\Delta \hat T^{Ii1}_{n+1,q}$ by using $\Delta \hat T^{Ii1}_{n+1,q}=\Delta \hat T^{Ii1}_{n,q}+\delta \hat T^{Ii1}_{n+1,q}$ [Eq.~(\ref{defofhatTIi1})]. Eq.~(\ref{FDPHeq:ThattobarT}) converts $\hat T^{Ii1}_{n+1,q}$ to $\bar T^{Ii1}_{n+1}$ at time step $n+1$.
Finally, Eq.~(\ref{FDPHeq:barTIi1toTIi1}) converts $\bar T^{Ii1}_{n+1}$ to $T^{Ii1}_{i,n+1}$ for any $i$ at time step $n+1$. By using $T^{Ii1}_{n+1,i}$ for any $i$, we evaluate slip- and opening-rate ${\bf D}_{n+1}$ at time step $n+1$. Then the same procedure computing $T_{i,n+1}$ follows at time step $n+1$.

\subsubsection{$T^{Ii2}$ Computation in Eq.~(\ref{FDPHeq:irreddecompdomI})}
\label{FDPHB24}
%Next we explain the $T^{Ii2}$ computations. 
The $i$-th component of $T^{Ii2}$ at time step $n$ is written as
\begin{equation}
T_{i,n}^{Ii2}=f_i\sum_{m=0}^{\tilde m_i-1} h_{m+m_0+\tilde m_j} \sum_j g_j D_{j,n-m-m_0-\tilde m_j}.
\end{equation}
As mentioned earlier, the variable separation of the kernel in Domain I gives the time-invariant part $h(t)=1$ and the power function of the time [$h(t)=t^2$ for the case of the stress nucleus of the double-layer potential mainly considered here]~\cite{ando2007efficient,ando2016fast}; the summation over two $fgh$ of different $h$ is omitted throughout the paper for brevity. 

Using such $t$ dependence of $h(t)$, we separate the $m,j$ dependencies of $h_{m+m_0+\tilde m_j}g_j$ in the following manner. 
In the time-invariant part, $h_{m+m_0+\tilde m_j}=\Delta t$ is independent of $m$, and $h_{m+m_0+\tilde m_j}g_j$ only depends on $j$. 
The time-dependent part of $h(t)$ is discretized as $h_m=\int ^{(m+\epsilon_t)\Delta t}_{(m+\epsilon_t-1)\Delta t}dt h(t)$ under the temporal discretization adopted in \S\ref{FDPH21} (which is associated with the definition of $K_{i,j,m}$), and $h_{m+m_0+\tilde m_j}g_j$ can be expressed by the separable form: for example for $h(t)=t^2$, we have
\begin{flalign*}
&
h_{m+m_0+\tilde m_j}g_j=
g_j[(m+\epsilon_t)^2-(m+\epsilon_t)+1/3](\Delta t)^3 
\\
&
+(m_0+\tilde m_j)g_j(2m+2\epsilon_t-1)(\Delta t)^3+(m_0+\tilde m_j)^2g_j(\Delta t)^3. 
\end{flalign*}
This can be written as $\sum_{d=1}^3 g_{d,j} h_{d,m}$ with newly defined coefficients $g_{d,j},h_{d,m}$ ($d=1,2,3$).
By using such a separation of variables, 
we can rewrite the computation of $T^{Ii2}$ as
\begin{equation}
T_{i,n}^{Ii2}=f_i\sum_{m=0}^{\tilde m_i-1} \sum_{d=1}^{d_{max}} h_{d,m} \sum_j g_{d,j} D_{j,n-m-m_0-\tilde m_j},
\label{FDPHeq:separableTIi2}
\end{equation}
with coefficients $h_{d,m}, g_{d,j}$ for $d=1,...,d_{max}$, where $d_{max}$ is $1$ for the time-invariant part (where $g_{1,j}=g_j, h_{1,m}=1$), and $3$ for $h(t)\propto t^2$. 

Eq.~(\ref{FDPHeq:separableTIi2}) is decomposed into three equations:
\begin{flalign}
\Delta \bar T_{d,n-m} &:= \sum_j g_{d,j}D_{j,n-m-m_0-\tilde m_j}
\label{FDPHeq:DtoDeltaTbarIi2}
\\
\bar T^{Ii2}_{n,\tilde m} &:= \sum_{m^\prime=0}^{\tilde m-1}\sum_{d=1}^{d_{max}} h_{d,m^\prime}\Delta \bar T^{Ii2}_{d,n-m^\prime}
\label{FDPHeq:DeltaTbarIi2toTbarIi2}
\\
T_{i,n}^{Ii2}&=f_i \bar T^{Ii2}_{n,\tilde m_i}.
\label{FDPHeq:TbarIi2toT}
\end{flalign}
We hereafter introduce the conditionally-predicted representative stress $\Delta \bar T_{d,n,m}^{Ii2\prime}$ associated with $\Delta \bar T_{d,n-m}^{Ii2}$, in a similar manner to that of $T^{\prime}_{n,m}$ defined in \S\ref{FDPH52}. 
Its vector expression
${\bf \Delta \bar T}^{Ii2\prime}_{d,n}=$ $(...,$$\Delta \bar T_{d,n,0}^{Ii2\prime},$ $\Delta \bar T^{Ii2\prime}_{d,n,1}$$,...)^{\mbox T}$ is also introduced for each $d$ as a vector storing $\Delta \bar T_{d,n,m}^{Ii2\prime}$ at its $m$-th component, 
in a parallel manner to that for ${\bf T}^{\prime}_n$ defined in \S\ref{FDPHFcost}. 

Eqs.~(\ref{FDPHeq:DtoDeltaTbarIi2}), (\ref{FDPHeq:DeltaTbarIi2toTbarIi2}), and
(\ref{FDPHeq:TbarIi2toT}) are computed in the following procedure at each time step $n$. 
First, we compute $\bar T^{Ii2}_{n,m}$ for $\tilde m\in(0,\max_i\tilde m_i]$ by recursively using 
the alternative form of Eq.~(\ref{FDPHeq:DeltaTbarIi2toTbarIi2}):
\begin{equation}
\bar T^{Ii2}_{n,\tilde m+1}=\bar T^{Ii2}_{n,\tilde m}+\sum_{d=1}^{d_{max}} h_{d,\tilde m+m_0}\Delta \bar T^{Ii2\prime}_{d,n,\tilde m},
\label{FDPHeq:recursionTbarIi2}
\end{equation}
where $\max_i\tilde m_i$ represents the maximum of $\tilde m_i$ in the leaf. 
Note $\bar T^{Ii2}_{n,0}=0$, which is obtained from Eq.~(\ref{FDPHeq:DeltaTbarIi2toTbarIi2}).
$\bar T^{Ii2}_{n,\tilde m}$ is stored over $\tilde m$ for current time step $n$ with discarding its history ($\bar T^{Ii2}_{n-m,\tilde m}$ of $m\in\mathbb N$).  
Second, Eq.~(\ref{FDPHeq:TbarIi2toT}) computes $T^{Ii2}_{i,n}$ for all the receivers $i$ at time step $n$, and ${\bf D}_n$ is determined.
Third, 
the following relation given by Eq.~(\ref{FDPHeq:DtoDeltaTbarIi2}) updates 
${\bf \Delta \bar T}^{Ii2\prime}_{d,n}$ to ${\bf \Delta \bar T}^{Ii2\prime}_{d,n+1}$ for all $d$ by using ${\bf D}_n$ in each step $n$:
\begin{equation}
{\bf \Delta \bar T}^{Ii2\prime}_{d,n+1} =
\mathcal M[{\bf \Delta \bar T}^{Ii2\prime}_{d,n} +
{\bf G}^{Ii2}_d{\bf D}_n]
\label{FDPHeq:timemarchingexpressionofDtoDeltaTbarIi2}
\end{equation}
with
\begin{equation}
G^{Ii2}_{d,m,j} := \delta_{-m,m_0+\tilde m_j} g_{d,j}.
\end{equation}
The above relation is obtained from Eq.~(\ref{FDPHeq:DtoDeltaTbarIi2}) in 
a similar manner to that of Eq.~(\ref{FDPHeq:42_tensorial}) from Eq.~(\ref{FDPHeq:34}).
%and $\Delta \bar T^{Ii2}_{n,d,0}$ is stored in ${\bf \Delta \bar T}^{Ii2}_{d,n}$. evolves ${\bf \Delta \bar T}^{Ii2}_{d,n}=\mathcal M{\bf \Delta \bar T}^{Ii2}_{d,n-1}$ in each step $n$. 
The required $m$ range of $\Delta \bar T^{Ii2\prime}_{d,n,m}$ is $m\in[-(m_0+\max_j\tilde m_j),\max_i\tilde m_i)$; its lower bound is given by the operation of Eq.~(\ref{FDPHeq:timemarchingexpressionofDtoDeltaTbarIi2}), and the upper bound is by that of Eq.~(\ref{FDPHeq:recursionTbarIi2}).

\subsubsection{Decimal Part Computation in Eq.~(\ref{FDPHeq:discdomI}) }
\label{FDPHB25}
The decimal part of the stress associated with Domain I, $T^I$, in Eq.~(\ref{FDPHeq:discdomI}) is expressed as 
\begin{flalign}
&\mbox{decimal part}=
\nonumber\\&
f_i^I\sum_jg_j^I
\left[
\int
^{\bar t_j^{\beta-}+\delta t_i^\beta}
_{(\bar m_j^{\beta-}+\delta m_i^\beta)\Delta t}
-
\int
^{\bar t_j^{\alpha+}+\delta t_i^\alpha}
_{(\bar m_j^{\alpha+}+\delta m_i^\alpha)\Delta t}
 \right]ds
h^I(s)D_j(t-s).
\end{flalign}
It corresponds to the difference between the continuous [$(\bar t_j^{\alpha+}+\delta t_i^\alpha,\bar t_j^{\beta-}+\delta t_i^\beta)$] and discretized [$((\bar m_j^{\alpha+}+\delta m_i^\alpha)\Delta t,(\bar m_j^{\beta-}+\delta m_i^\beta)\Delta t)$] time ranges of Domain I.

The decimal part of Domain I vanishes when 
the $\delta t_i^c$ and $\bar t_j^c$ values satisfy the following conditions: 
\begin{flalign}
\delta t_i^c&= \delta m_i^c\Delta t
\\
\bar t_j^\alpha&= \bar m_j^{\alpha+}\Delta t-\Delta t_j^{\alpha+}
\\
\bar t_j^\beta&= \bar m_j^{\beta-}\Delta t+\Delta t_j^{\beta-}.
\end{flalign}
These are satisfied in the implementation 
in \S\ref{FDPH43} [specifically, Eqs.~(\ref{eq:discretedelti}), (\ref{eq:discretizationofbartj}), and (\ref{eq:discretizationofDeltj}), which will be satisfactory for the constant $\eta$ scheme, as also mentioned in \S\ref{FDPH43}]; the adjustment of $\delta t_i$ involves the discretization error while that of $\bar t_j^\pm$ can be error-free (\S\ref{FDPH43} and \ref{FDPHC32}). 
Therefore, the decimal part computation would be required mainly for $\delta t_i\neq \delta m_i\Delta t$ especially in considering the constant $\eta^2dist$ scheme, and otherwise we can skip it. 
 
For evaluating the decimal part, if needed,
%the constant $\eta^2 dist$ scheme, we can utilize the property of $h^I (t)$ as shown in \S\ref{FDPHB24}.
 we separate $i,j$ dependence of temporally integrated $h$ as done in \S\ref{FDPHB24};
\begin{flalign}
&\mbox{decimal part}=
f^I_{i}\sum_jg_{j}^I
\sum_d\times
\nonumber\\
&
[
h^{I,\alpha,r}_{d,i}h^{I,\alpha,s}_{d,j}
D_{j,n-\delta m_i^\beta-\bar m_j^{\beta-}}
-
h^{I,\beta,r}_{d,i}h^{I,\beta,s}_{d,j}
D_{j,n-\delta m_i^\alpha-\bar m_j^{\alpha+}}
]
\label{FDPHeq:exactsepdeciminI}
,
\end{flalign}
where $h^{I,c,r}_{d,i},h^{I,c,s}_{d,j}$ ($c=\alpha,\beta$) denote respectively $d$-th coefficients depending on receiver $i$ and source $j$. These can be obtained in the similar ways as $h_{d,m},g_{d,j}$ in \S\ref{FDPHB24}. 
All the terms in Eq.~(\ref{FDPHeq:exactsepdeciminI}) are computed for respective $d$ values by the arithmetic in Domain F, described in \S\ref{FDPH52}. 

%Other contributions caused by the difference in the receiver-averaged travel time $\bar t_j^{\alpha+}-\delta m_i^{\alpha+}\Delta t$ and $\bar t_j^{\beta-}-\delta m_i^{\beta-}\Delta t$ can also treated in the travel time $\bar t_j^c$; this change only affects the handling of the stress of Domain S (detailed in \ref{FDPHC32}). 

\subsection{Transient Terms in Domain S}
\label{FDPHB3}
The stress caused by the transient terms (the remaining from the asymptotic form), existing in the 2D problems only, in Domain S is written in the following form:
\begin{equation}
T^{S,tr}_{i,n}= f_i^{S,tr}\sum_j g_j^{S,tr}\sum_{m=0}^{\Delta m_{S,tr}-1}  h_{m}^{S,tr}  D_{j,n-m-\bar m_j^{\beta+}-\delta m_i^\beta}.
\end{equation}
Cutoff $\Delta m^{S,tr}$ is determined by the given error conditions explained in \ref{FDPHH}. When the $\Delta m^{S,tr}$ value given by the error conditions is larger than the number of the whole time step ($M$), $\Delta m^{S,tr}$ can be set at $M$. In this paper, such truncation is done in \S\ref{FDPH63} (and also in \S\ref{FDPHA3} and \ref{FDPHH}) to carefully check the parameter dependence of the cost. 

$T^{S,tr}$ is decomposed by the similar way to that of Domain I (\S\ref{FDPHB2}) as
\begin{flalign}
T^{S,tr}_{i,n}&= f_i^{S,tr}\bar T^{S,tr}_{n,\delta m_i^\beta }
\label{FDPHeq:barTTinS}
\\
\Delta \bar T^{S,tr}_{n,m}&:= \sum_j g_j^{S,tr}  D_{j,n-m-\bar m_j^{\beta+}}
\label{FDPHeq:DdelbarTinS}
\\
\bar T^{S,tr}_{n,m}&:= \sum_{m^\prime=0}^{\Delta m_{S,tr}-1}  h_{m^\prime}^{S,tr} \Delta \bar T^{S,tr}_{n-m, m^\prime}. 
\label{FDPHeq:delbarTbarTinS}
\end{flalign}
%We can check this decomposition just by the substitution of Eq. (B.28) into Eq. (B.27) and of it to Eq. (B.26). 
The computations of Eq.~(\ref{FDPHeq:barTTinS}) ($\bar T\to T$) 
and of Eq.~(\ref{FDPHeq:DdelbarTinS}) ($D \to \Delta \bar T$) 
are respectively the same as those of $\bar T\to T$ and $\hat D\to\bar T$ in Domain F, detailed in \S\ref{FDPH52}. 
Here we omitted trivial superscript: $S,tr$.
We thus focus on the new computation Eq.~(\ref{FDPHeq:delbarTbarTinS}).

$\Delta \bar T^{S,tr}_{n,m}$ is evaluated by its direct computation of the definitional identity Eq.~(\ref{FDPHeq:delbarTbarTinS}). We first compute the temporal convolution of $\Delta \bar T\to \bar T$ in Eq.~(\ref{FDPHeq:delbarTbarTinS}) at every time step only at particular $m$ that is the latest (or properly later) time completing the summation of $\Delta \bar T$; the latest one is $m = \min_j \bar m_j^-$, as far as 
Eq.~(\ref{FDPHeq:delbarTbarTinS}) is computed after the evaluation of Eq.~(\ref{FDPHeq:DdelbarTinS}). 
Before such a time step, 
the summation of the conditionally predicted representative stress of $\bar T$ (executed in the same way as in Domain F) is incomplete, and the computation of Eq.~(\ref{FDPHeq:delbarTbarTinS}) cannot be executed. 
The components of $\bar T$ at $m> \min_j \bar m_j$ are computed by the time marching rule: $\bar T_{n+1,m}=\bar T_{n,m-1}$ (${\bf \bar T}_{n+1}=\mathcal M {\bf \bar T}_n$).

Quantization can apply to $h^{S,tr}$ in Eq.~(\ref{FDPHeq:delbarTbarTinS}). Although it does not change the cost order, the memory access becomes more efficient by Quantization. In \S\ref{FDPHA3}, Quantization is applied to the transient term in Domain S to check the error property of Quantization applied to FDP=H-matrices. 

\subsection{Transient Terms in Domain I}
\label{FDPHB4}
Since the kernel is non-singular in Domain I (in-between the P- and S-waves), the remaining terms (existing in the 2D case only) from the asymptotic ones in the kernel of Domain I, called the transient terms in Domain I, is well approximated by the LRA for the third-rank tensor, such as the Tucker cross approximation (the TCA)~\cite{oseledets2008tucker}. When the TCA is applied to the transient terms (or the original kernel) in Domain I, the resultant reduced kernel takes the same algebraic form $f_ig_jh_m$ as the asymptotic factorized kernel in Domain I; $g_{d,j}h_{d,m}$, analytically obtained for the asymptotic part in Domain I in \S\ref{FDPHB2}, is also obtainable for the transient one by using the TCA once again.
Further, such a transient time dependence is well approximated by Quantization like the asymptotic part, as collectively shown in \S\ref{FDPHA2}. 
Their difference in the data-sparse approximation is as above only the above-mentioned modification of the LRA method (from the semi-analytic BIE of the FDPM to the numerical TCA).
The corresponding arithmetic then becomes the same as that for the asymptotic Domain I kernel in \S\ref{FDPHB2}.

\section{Summary of the Time Complexity and Memory Consumption} 
\label{FDPHB5}
We here summarize the cost estimates of respective domains. That of the total costs is also supplemented. 

\subsection{Computational Procedures, Required Variables, and Costs in Domain F}
\label{FDPHFcost}
The costs and required variables in Domain F are summarized below.
It is useful for this purpose 
to simply present the computations of Eqs.~(\ref{FDPHeq:34}) and (\ref{FDPHeq:35}).
We introduce a vector expression of $T^\prime_{n,m}$, ${\bf \bar T^\prime}_n:= (\bar T^\prime_{n,-\max_j\bar m_j^-+1},\bar T^\prime_{n,-\max_j\bar m_j^-+2},...,$ $
\bar T^\prime_{n,\max_i\delta m_i})^{{\mbox T}}$, which stores nonzero $T^\prime_{n,m}$ [required in Eqs.~(\ref{FDPHeq:34}) and (\ref{FDPHeq:35})] at the $m$-th component. 
We also gather $\hat D^F_{j,n}$ at current time step $n$ into a vector, ${\bf \hat D}^F_n:=(\hat D^{F}_{j_{init},n},\hat D^{F}_{j_{init}+1,n},...,$ $\hat D^{F}_{j_{fin},n},)^T$, by supposing that the sources in an admissible leaf are sorted as $j=j_{init},j_{init+1},...,j_{fin}$ as in an ordinary implementation of H-matrices, e.g., Refs.~\cite{yoshikawa20152,ida2014parallel}. 

Using ${\bf \bar T^\prime}_n$,
$\bar T^\prime\to T$ computations are written as
\begin{equation}
{\bf T}_n={\bf F  \bar T^\prime}_n.
\label{FDPHeq:39}
\end{equation}
Eq.~(\ref{FDPHeq:39}) is a vector-to-vector projection by a sparse matrix while the corresponding procedure is a scalar-to-vector computation in H-matrices. 
Using ${\bf \bar T^\prime}_n$ and ${\bf \hat D}_n$,
$\hat D\to\bar T^\prime$ computations are written as
\begin{equation}
{\bf \bar T^\prime}_{n+1} = \mathcal M \left[{\bf \bar T^\prime}_n+{\bf G\hat D}_n\right]
\label{FDPHeq:42}
\end{equation}
Eq.~(\ref{FDPHeq:42}), or equivalently ${\bf \bar T^\prime}_{n+1}-\mathcal M {\bf \bar T^\prime}_n=\mathcal M{\bf G}{\bf \bar T^\prime}_n$, is comparable with Eq.~(\ref{FDPHeq:39}).

As above, computation of Eq.~(\ref{FDPHeq:33}) is reduced to those of Eqs.~(\ref{FDPHeq:39}) and (\ref{FDPHeq:42}). Combination of Eqs.~(\ref{FDPHeq:39}) and (\ref{FDPHeq:42}) with Eq.~(\ref{FDPHeq:32}) gives the arithmetic of FDP=H-matrices in Domain F [evaluating Eq.~(\ref{FDPHeq:31})].
First, ${\bf D}_{n-m}\in \mathbb R^{N}$ of $m\in[0,\max_j\Delta m_j)$ is converted to ${\bf \hat D}_n\in \mathbb R^{N_{s,a}}$ by Eq.~(\ref{FDPHeq:32}) at each time step $n$ in all the admissible leaves, $a$, where $N_{s,a}$ denotes the number of the sources in leaf $a$. 
Second, 
 ${\bf \hat D}_n$ is converted to ${\bf \bar T^\prime}_n\in \mathbb R^{\max_{i,j}(\delta m_{a,i}+\bar m_{a,j}^-)}$ by Eq.~(\ref{FDPHeq:39}); the leaf $a$ dependencies of the receiver-dependent travel-time difference $\delta m_i$ and receiver-averaged travel time step $\bar m_j^-$ are shown only here as $\delta m_{a,i}$ and $\bar m_{a,j}^-$. 
Third, ${\bf \bar T^\prime}_n$ is converted to ${\bf T}_n\in \mathbb R^{N}$ by Eq.~(\ref{FDPHeq:42}) summed over all the admissible leaves. 

Note that sparse matrices ${\bf F}^a$ and ${\bf G}^a$ in Eqs.~(\ref{FDPHeq:39}) and (\ref{FDPHeq:42}) are expressed by vectors ${\bf f}^a\in \mathbb R^{N_{r,a}}$, ${\bf g}^a\in\mathbb R^{N_{r,a}}$ and $\delta m_i^a$, $\bar m_j^{-a}$ in each admissible leaf $a$; the leaf number dependence of ${\bf F}, {\bf G}, {\bf f}, {\bf g}, \delta m_i, \bar m_j^{-}$ is explicitly shown only here for counting the costs. 
Computations utilizing $\mathcal M$, $F$ ($S^{receiver}$), $G$ ($S^{source}$) in Eqs.~(\ref{FDPHeq:39}) and (\ref{FDPHeq:42}) 
can be coded as functions (giving updated ${\bf \bar T^\prime}$ by using $\delta m_i,\bar m_j^-,{\bf f},{\bf g}$), as well as being stored as sparse matrices.

By counting the number of components appearing in the above computational procedure, the memory and time complexity per time step to evaluate Eqs.~(\ref{FDPHeq:39}) and (\ref{FDPHeq:42}) are found to be proportional to (of order) $dist_a$, the number ($N_{s,a}$) of sources, or that ($N_{r,a}$) of receivers, in each admissible leaf $a$. Therefore, the costs become $\mathcal O(N\log N)$ in total, given the explanation related to Fig.~\ref{FDPHfig:9_b}. 
In the computation of ${\bf \hat D}_n$ ($D\to\hat D$) [ Eq.~(\ref{FDPHeq:32})], the time length of the required history of the slip and opening becomes $\mathcal O(\Delta m_j)=\mathcal O(\Delta t_j/\Delta t)=\mathcal O(1)$, so that the costs to evaluate Eq.~(\ref{FDPHeq:32}) is also $\mathcal O(N\log N)$.
By considering these, all the required memory and time complexity per time step are $\mathcal O(N\log N)$ in the arithmetic of Domain F.

\subsection{Numerical Costs in Domain I}
The numerical costs in Domain I is summarized below. We omit these of decimal parts, because they are exactly the same as those of Domain S by following the same logic for Domain S. 

Cost estimates for the $T^{Ii1}$ computation are as follows when Quantization does not apply.
The time complexity to evaluate Eqs.~(\ref{FDPHeq:barTIi1toTIi1}), (\ref{FDPHeq:defofTpartial}), (\ref{FDPHeq:noquantIi1marching}), and (\ref{FDPHeq:nonquantrecurrence}) of $m=m^\prime$ 
is of $\mathcal O(dist_a,N_{r,a},N_{f,a})$ at each time step $n$ in each admissible leaf as in the arithmetic for Domain F; the leaf $a$ dependence of the quantities is shown here for clarity of the estimate. This becomes $\mathcal O(N\log N)$ in the constant $\eta$ scheme as mentioned in the text. 
The required variables in admissible leaf $a$ are 
${\bf T}^{Ii1,a}_n \in \mathbb R^{N_{r,a}}$, ${\bf f}^a \in \mathbb R^{N_{r,a}}$, ${\bf g}^a \in \mathbb R^{N_{s,a}}$, $m_0^a \in \mathbb R$, $\tilde m_j^a \in \mathbb R$ for each $j$ belonging to leaf $a$, ${\bf \Delta\bar T}^a_{proj,n}\in \mathbb R^{\max_j\tilde m_j^a-\min_j\tilde m_j^a+1}$, and ${\bf \Delta \bar T}^{Ii1\prime}_{sum,n,m}\in\mathbb R^{m_0^a+\max_j \tilde m_j^a}$ 
of $m\in[\min_j\tilde m_j^a-1,\max_j \tilde m_j^a)$.
Among them, dominant memory consumption is to store $\bar T^{Ii1}_{sum,n,m,m^\prime}$ in 
$m\in[\min_j\tilde m_j^a-1,\max_j \tilde m_j^a)$
and 
$m^\prime\in[0,m_0^a+\max_j\tilde m_j^a)$, which is $\mathcal O[diam_adist_a/(c\Delta t)^2]$ ($c=\alpha,\beta$). 
Such a memory is estimated to be almost $\mathcal O(N^{2/D_b})$ in the constant $\eta$ scheme 
and 
$\mathcal O(N^{1+3/(2D_b)})$ in the constant $\eta dist^2$ scheme, in light of the same scale analysis as in \S\ref{FDPHC2}. 
We note that the memory for $\bar T^{Ii1}_{sum,n,m^\prime,m}$ can be 
$\mathcal O[diam_a^2/(c\Delta t)^2]$ [$\mathcal O(N^{1+1/D_b})$ in total for the constant $\eta^2dist$ scheme]
when we use the arbitrariness of the decomposition of $\tilde m_i$ and $\tilde m_j$, mentioned in \S\ref{FDPHB21}, and set $\tilde m_j=\mathcal O(diam_a)$. 
The other memory costs are $\mathcal O(dist_a,N_{r,a},N_{f,a})$ as the computational complexity per time step is.

Cost estimates for the $T^{Ii1}$ computation are then modified as below when Quantization applies.
In each leaf $a$,
the $\mathcal O [diam_adist_a/(c\Delta t)^2]$ ($c=\alpha,\beta$) memory required in the case without Quantization, to store $\Delta \bar T^{Ii1,a}_{sum,n-m^\prime,m}$, 
is reduced to the memory for storing $\Delta \hat T^{Ii1\prime,a}_{n,q}\in \mathbb R$ at $q=0,...,Q_a-1$, $\Delta \bar T^{Ii1\prime,a}_{sumQ,n}\in \mathbb R^{Q_a}$ 
$\delta \hat T^{Ii1\prime,a}_{n,q}\in \mathbb R^{\max \tilde m_j^a}$ for arbitrary $n$; the leaf $a$ dependence of the variables is shown here for clarity of the estimate. 
The memory consumption to store them is estimated at 
 $\mathcal O [Q_a dist_a/(c\Delta t)]$, given the number of components in $\Delta \hat T^{Ii1,a}_{n,q}$ of all $q=0,...,Q_a-1$, $\Delta \bar T^{Ii1,a}_{sumQ,n}$, and 
$\delta \hat T^{Ii1,a}_{n,q}$. 
$\mathcal O [Q_a dist_a/(c\Delta t)]$
means $\mathcal O(N\log^2 N)$ at $D_b=1$ in the constant $\eta$ scheme and $\mathcal O(N\log N)$ at $D_b=2,3$, which are primarily intended applications of FDP=H-matrices, given $Q_a=\log [dist_a/(c\Delta t)]$. 
$\mathcal O [Q_a dist_a/(c\Delta t)]$ is found to be almost $\mathcal O(N^{1+1/D_b})$ in the constant $\eta^2dist$ scheme in light of the same scale analysis in \S\ref{FDPHC2}. 
Additionally, the time complexity per time step also includes 
an $\mathcal O [Q_a dist_a/(c\Delta t)]$ factor, due to the evaluation of Eq.~(\ref{FDPHeq:deltaTIi1sparse}), as well as the $\mathcal O(dist_a,N_{r,a},N_{f,a})$ factors that are contained in the arithmetic of $T^{Ii1}$ without Quantization in \S\ref{FDPHB22}; this $\mathcal O [Q_a dist_a/(c\Delta t)]$ factor in the complexity is purely from $\mathcal M{\bf \delta \hat T}^{Ii1\prime}_{n,q}$ in Eq.~(\ref{FDPHeq:deltaTIi1sparse}) and can be erased out (mentioned in the later subsection), so that the $\mathcal O [Q_a dist_a/(c\Delta t)]$ increase in the complexity substantially does not exist. 

The cost for the $T^{Ii2}$ computation is estimated as follows.
In each admissible leaf $a$,
the memory is required to store $f_i$, $g_{d,j}$, $h_{d,m}$, $\bar T^{Ii2,a}_{n,\tilde m}\in \mathbb R$ of $\tilde m\in(0,\max_i\tilde m_i^a]$, $T^{Ii2,a}_{i,n}\in \mathbb R$, and ${\bf \Delta \bar T}^{Ii2\prime}_{d,n}\in\mathbb R^{\max_{i,j}(\tilde m_i^a+\tilde m_j^a)+m_0^a}$ of $1 \leq d \leq 3$, and amounts to $\mathcal O(N_{r,a},N_{s,a},dist_a)$ as in Domain F; the leaf $a$ dependence of the variables is shown here for clarity of the estimate. 
The time complexity per time step is also $\mathcal O(N_{r,a},N_{s,a},dist_a)$.

With respect to $T^{Ii2}$ computation, the $\mathcal O [dist_a/(c\Delta t)]$ factor in the complexity comes from $\mathcal M{\bf \Delta \bar T}^{Ii2\prime}_{d,n}$ and Eq.~(\ref{FDPHeq:recursionTbarIi2}) and is erasable in the following ways. The former can be erased out in a way mentioned in \ref{FDPHB5}. The latter can be erased out by using Quantization. 

\subsection{Numerical Costs in Domain S}
The numerical costs in Domain S are estimated in the same manner as those of Domain F given the coincidence of their arithmetics. 

\subsection{Numerical Costs in Total}
The cost estimates in all the domains is summarized below. 
We here introduce normalized lengths $L^\prime$$:=L$$/(\beta\Delta t)$, $dist_a^\prime$$:=dist_a$$/(\beta\Delta t)$, and $diam_a^\prime$$:=diam_a$$/(\beta\Delta t)$ to supplement them. 

The time complexity per time step in FDP=H-matrices is totally estimated to be $\mathcal O[l_a(N_a+Q_a+dist_a^\prime)]$ in an admissible leaf $a$, where $l_a$ is the rank of $\hat K^W$ summed over W = Fp, I, Fs, S, and
$N_a$ is the number of sources and receivers in an admissible leaf $a$; $Q_a$ is the number of the sampling in Quantization. 
The $Q_a dist^\prime_a$ dependent cost is caused only from Domain I. 
The memory in total is $\mathcal O[l_a(N_a+dist_a^\prime)]$ in an admissible leaf ($a$).
If Quantization is not used for $T^{Ii1}$ in Domain I, 
the time complexity is $\mathcal O[l_a(N_a+dist_a^\prime)]$ per time step, and  
the memory is $\mathcal O[l_a(N_a+dist_a^\prime+dist_a^\prime diam_a^\prime)]$, 
in admissible leaf $a$. 

We note that the $\mathcal O(Q_adist_a^\prime)$ factor included in the computation costs can become unnecessary, and hence we excluded it from the cost estimate in the last paragraph. 
This $Q_adist_a^\prime$ factor is caused by the multiplication of the matrix $\mathcal M$ (defined in \S\ref{FDPH5}) or the time integration in Domains I, and then we erase them separately as below. 
The multiplication of $\mathcal M$ to $\bar T^\prime$ can be coded as an increment of the base address of the $\bar T$ vector (the location of the first element of the $\bar T$ vector) in an implementation, and the related factor of $dist_a^\prime$ is obviated [reduced to $\mathcal O(1)$]. A similar coding manner is seen in Ref.~\cite{lapusta2000elastodynamic}, where the above-mentioned increment of $n$ is implemented with an explicitly introduced scalar incremented in each time step (as $n$ itself). 
The costs for directly evaluating the temporal integration in Domain I (included in the $T^{Ii1}$ computation without Quantization, and also in the $T^{Ii2}$ computation shown in \S\ref{FDPHB24}) is erasable by Quantization (as for $T^{Ii1}$ in \S\ref{FDPHB23}). 
We can also erase $\mathcal O(\sum_a dist_a^\prime)$ from the time complexity per time step in those ways; erasing $\mathcal O(\sum_a dist_a^\prime)$ is not relevant for $D_b\geq 1$ [where $\mathcal O(\sum_a dist_a^\prime)\lesssim \mathcal O(\sum_a N_a)$] while cancels the leading order of the complexity when $D_b< 1$ [where $\mathcal O(\sum_a dist_a^\prime)> \mathcal O(\sum_a N_a)$]. 
 
$Q_a$ is $\mathcal O(\log dist_a^\prime)$ (See \S\ref{FDPHA1}), and $l_a$ is $\mathcal O(1)$ (See \S\ref{FDPH41}). 
Although $Q_a$ is of order $1/\epsilon_Q$, as shown in \S\ref{FDPHA3}, $\epsilon_Q$ can be set at a relatively large value such as $\epsilon_Q=0.1$, by using the absolute error condition as done in this paper (supplemented in \S\ref{FDPH71}). 
$\sum_a dist_a^\prime$ is $\mathcal O(N\log N)$ given $L^\prime=\mathcal O(N^{1/D_b})$ and $diam^\prime=\mathcal O(N_a^{1/D_b})$
 at $D_b\geq 1$ in the constant $\eta$ scheme.
%Note that the effect of $\epsilon_Q$ is reflected in $Q_a$, which becomes of $\mathcal O(dist_a^\prime)$ when $\epsilon_Q\to0$. 
%Even when $\epsilon_Q$ is so small that the computation cost is unchanged, since the above estimate contains $\mathcal O(dist_a)$ for the first time (as long as $L=\mathcal O(N^{1/D_b}$). When $\epsilon_Q$ is excessively small, the memory cost is of $\mathcal O(N\log N+(L^\prime)^2\log N)$, which can be of $\mathcal O(N^2\log N)$ when $D_b=1$, yet of $\mathcal O(N\log N)$ when $D_b=2,3$. 

By considering these estimates of $Q_a,l_a,diam_a^\prime$, the above-mentioned costs become $\mathcal O$$(N$$\log N)$ in the constant $\eta$ scheme, and $\mathcal O(N^{3/2}+NL^\prime)$ in the constant $\eta^2dist$ scheme, for the case of $D_b>1$ [which is typical in the 3D problems firstly intended in this study]; these can be achieved even without Quantization as noticed from the above estimate. 
In the case of $D_b=1$ (typical for the 2-D problems), where the $\eta^2dist$ scheme is not quite necessary (See \S\ref{FDPHschemedepcosts}) and Quantization becomes useful certainly (mentioned in \S\ref{FDPHB23}), 
the time complexity per time step is $\mathcal O(N\log N)$ and the total memory becomes $\mathcal O$$(N$$\log N+$$L^\prime$$\log N$$\log L^\prime)$ for the constant $\eta$ scheme; among the 2D cases, the anti-plane problems have no Domain I [that induces $L^\prime\log N\log L^\prime$ factors in the 3D and 2D in-plane problems when $D_b=1$], and thus in the anti-plane problems, the cost estimates for $D_b=1$ are the same $\mathcal O(N\log N)$ for the constant $\eta$ scheme, as for $D_b>1$. 
In the case of $D_b <1$, e.g., excessively distant two objects, the total memory requirement becomes almost $\mathcal O(L^\prime)$ rather than $\mathcal O(N\log N)$ or $\mathcal O(N^{3/2})$, while the time complexity per time step is the same as that of $D_b=1$. 
We last note that the computational complexity for executing the LRA is on the same order as that of the stress computation per time step [$\mathcal O(N\log N)$ or $\mathcal O(N^{3/2})$], when we consider the partially-pivoting ACA, ACA+, and the TCA. It is negligible in the total computational complexity, given that the LRA is executed just once in the simulation while the stress computation is iterated $M$ times.

\section{Parameter Range Bounds For Simple Domain Setting} 
We here introduce some useful conditions on $\eta,l_{min}$ for 
simplifying the implementation of FDP=H-matrices. 

\subsection{To Satisfy Discretized Causality}
\label{FDPHdisccausality}
Going through the following procedure, we can reduce
the condition  $\delta m_i+\bar m_j^->0$ of Eq.~(\ref{eq:discretizedcausality}) for all the $i,j$ pairs in the admissible leaves to the requirement for the parameters $(\eta,l_{min})$ of H-matrices.

The definitions of $i_*,j_*,dist, diam$ in our definition shown in \S\ref{FDPH421}, 
yields an inequality concerning the approximation of the travel time,
\begin{equation}
r_{ij_*},r_{i_*j}\geq r_{i_*j_*}-diam/2.
\end{equation}
Using this inequality, 
we find 
the approximated travel time given in the continuous forms of
Eqs.~(\ref{FDPHeq:19}) and (\ref{FDPHeq:18})
satisfies
\begin{equation}
c(\delta t_i+\bar t_j)
=r_{ij_*}-r_{i_*j_*}+r_{i_*j}
\geq dist,
\label{eq:causalityafterapprox}
\end{equation}
where we used $r_{i_*j_*}=\bar r=dist-diam$ met in our definitions of $i_*,j_*,dist, diam$. 

Besides, when $\delta t_i+\bar t_j^-$ (where $\bar t_j^-=\bar t_j-\Delta t_j^-$) is discretized as $\delta m_i+\bar m_j^-$, as in \S\ref{FDPH43}, 
$(\delta m_i+\bar m_j^-)\Delta t$ can be smaller than $\delta t_i+\bar t_j^-$ by $2\Delta t$ at most, given the {\it twice} roundings involved with the definitions of two values $\delta m_i$ and $\bar m_j^-$;
\begin{equation}
(\delta m_i+\bar m_j^-)\Delta t\geq  \delta t_i+\bar t_j^--2 \Delta t.
\label{eq:condistracom}
\end{equation}
Here, we supposed $\delta C^{c-}$ [a positive safe coefficient for $\Delta t_j^-$ in Eq.~(\ref{FDPHeq:7})] to be smaller than $1$ in considering the rounding process of $\Delta t_j^-$, as we adopted in this paper as Eq.~(\ref{eq:safecoeffordiscDeltj}). 

Eqs.~(\ref{eq:causalityafterapprox}) and (\ref{eq:condistracom}) give
\begin{equation}
(\delta m_i+\bar m_j^-)\Delta t\geq dist/c-\Delta t_j^- -2 \Delta t.
\label{eq:discretizedbound}
\end{equation}
Therefore, the discretized causality, 
$\delta m_i+\bar m_j^->0$, 
that is $\delta m_i+\bar m_j^-\geq 1$,
is satisfied all the pairs of the sources and receivers in the admissible leaves when
\begin{equation}
l_{min}/\eta \geq c(\max_j \Delta t_j^- +3 \Delta t).
\end{equation}
We here replaced $dist$ in the right-hand side of 
Eq.~(\ref{eq:discretizedbound}) with its minimum $l_{min}/\eta$.

\subsection{To Define Domain I for All the Source-Receiver Pairs in the Admissible Leaves}
\label{FDPHdiscDomI}
For the simple implementation, we assumed that Domain I exists for all the pairs of the sources and receivers in the admissible leaves before and after the approximation of the ART and the discretization (in \S\ref{FDPHB2}). This corresponds to separating Domains Fp and Fs for all of them. 
We can express such a postulate as additional requirements for all the receivers ($i$) and sources ($j$) in the admissible leaves: 
\begin{equation}
t_{ij}^{\alpha+}< t_{ij}^{\beta-}    
\end{equation}
before the ART and the discretization and  
\begin{equation}
\bar t_{j}^{\alpha+}+C_s\Delta t< \bar t_{j}^{\beta-} 
\end{equation}
after the ART with the discretization, where the factor $C_s$ is a safe coefficient of $\mathcal O(1)$ to deal with the temporal discretization; $C_s\geq 2$ (corresponding to the twice roundings in \S\ref{FDPHdisccausality}) gives the separation between the discretized Domain Fp and discretized Domain Fs. 

We can reduce the above separation conditions between Fp and Fs (both before and after applying the ART and discretization) to a constraint on $l_{min}$ and $\eta$ by considering its most demanding configuration where a source and a receiver come the closest. In the way of clustering we adopted (defined in \S\ref{FDPH42}), the possible shortest distance between the collocation points of the source and receiver elements is given by $dist$ for receiver $i$ and source $j$ in each admissible leaf, and $dist$ is bounded by $l_{min}/\eta$ for all the admissible leaves. Then we have 
\begin{equation}
l_{min}/\eta>\frac{\max_j(\Delta t_j^{\alpha+}+\Delta t_j^{\beta-})}{\beta^{-1}-\alpha^{-1}}
\end{equation}
as the most demanding form of $t_{ij}^{\alpha+}< t_{ij}^{\beta-}    
$. Similarly, as that setting gives $r_{i_*j},r_{ij_*}>dist+diam/2$ with $r_{i_*j_*}=diam+dist$, we have
\begin{equation}
l_{min}/\eta>\frac{\max_j(\Delta t_j^{\alpha+}+\Delta t_j^{\beta-})+C_s\Delta t}{\beta^{-1}-\alpha^{-1}}
\end{equation}
for $\bar t_{j}^{\alpha+}+C_s\Delta t< \bar t_{j}^{\beta-}$. The latter gives the stricter bound than the former and describes the constraint on $\eta$ and $l_{min}$ independent of the element configuration.
The $\eta$ value in the above evaluation is modified as $\eta\to\eta_0$ for the constant $\eta^2dist$ scheme explained in \S\ref{FDPH422}.

\section{Arithmetic of FDP=H-Matrices in Inadmissible Leaves}
\label{FDPHE}

In the inadmissible leaves, we partitions the time range of the convolution just into Domain S and the others (regarded as Domain F hereafter). 
This is to deal with that all the Domains F, I, and S in continuous time are inevitably contaminated in one time step in some inadmissible leaves. 
After the kernel for the inadmissible leaves separates into Domains S and F, the kernel is replaced with the time-independent static asymptotic form in Domain S by the FDPM. 
The discretized kernel for the inadmissible leaves are not spatially approximated with the LRA in FDP=H-matrices, as in H-matrices of the spatial BIEM.
Besides, the ART is not applied. As Domain I is not considered in the inadmissible leaves, Quantization is not applied either.

With regard to Domain F,
the way of computing the stress in an inadmissible leaf is the same as that in the original ST-BIEM.
The computation of Domain S in an inadmissible leaf is unchanged from that of the FDPM~\cite{ando2016fast}.

Since the above substituted kernel is independent from the number ($M$) of the time steps, 
we find
the computational complexity per time step and the memory consumption are strictly $\mathcal O(N)$ in the inadmissible leaves, considering a similar logic to that of H-matrices in the spatial BIEM, mentioned in \S\ref{FDPH23}. 

\section{Slight Error Reduction When Using Eq.~(\ref{eq:discretizationofbartj})}
\label{FDPHC32}
We introduced Eq.~(\ref{eq:discretizationofbartj}) as a slight modification of the definition of $\bar t_j$ from Eq.~(\ref{FDPHeq:19}), and then a small (negligible in the constant $\eta$ scheme) discretization error of the travel time arises. 
On the other hand, we have one remaining degree of freedom in 
($\delta C^{c+}, \delta C^{c-}$) after they satisfy Eq.~(\ref{eq:safecoeffordiscDeltj}); it implies that by adjusting ($\delta C^{c+}, \delta C^{c-}$) while defining $\bar t_j$ by Eq.~(\ref{FDPHeq:19}), we can meet Eqs.~(\ref{eq:discretizationofbartj}) and (\ref{eq:discretizationofDeltj}) without inducing any discretization errors of $\bar t_j$ and $\Delta t_j$. We show such another discretization process of Domain F below. 

As seen in \S\ref{FDPH431}, we meet 
the time range $t\in (\bar m_j^-\Delta t, (\Delta m_j +\bar m_j^-)\Delta t)$ involved in the discretized Domain F with the original continuous time range $t\in (\bar t_j^-,\bar t_j^-+\Delta t_j)$ of Domain F. 
That requirement gives a special suite of ($\delta C^{c+}, \delta C^{c-}$) or equivalently $\Delta t_j^\pm$ such that
\begin{flalign}
\bar t_j^-
&=\bar m_j^-\Delta t,
\label{FDPHeq:specialdeltminus}
\\
\bar t_j^-+\Delta t_j^+
&=(\bar m_j^-+\Delta m_j)\Delta t
\label{FDPHeq:specialdeltplus}
\end{flalign}
or in another suite of expressions,
\begin{flalign}
\bar t_j-\Delta t_j^-
&=\lceil(\bar t_j-\Delta t_j^-)/\Delta t\rceil\Delta t.
\\
\bar t_j+\Delta t_j^+
&=\lfloor(\bar t_j+\Delta t_j^+)/\Delta t\rfloor\Delta t
\end{flalign}
where $\bar t_j=t_{i_*j}$, $\Delta t_j^\pm$ are given by Eqs.~(\ref{FDPHeq:6}) and (\ref{FDPHeq:7}), and $\bar m_j^-$ and $\Delta m_j$ are given by Eqs.~(\ref{eq:origdefofbarmj}) and (\ref{eq:origdefofDelmj}), respectively; the latter expressions are comparable with the $\bar m_j^-$ and $\Delta m_j$ values seen in \S\ref{FDPHeq:29_prev}. 
That is, 
we require the discretization conditions on $\bar t_j^-$ [Eq.~(\ref{FDPHeq:specialdeltminus})] and $\bar t_j^-+\Delta t_j$ (that can be denoted by $\bar t_j^+$) [Eq.~(\ref{FDPHeq:specialdeltplus})]
while we introduced those on $\bar t_j^-$ [Eq.~(\ref{eq:discretizationofbartj})] and $\Delta t_j$ [Eq.~(\ref{eq:safecoeffordiscDeltj})] in \S\ref{FDPH431}; both give the discrete Domain F compatible with the approximation of the ART. 
Then substituting the expressions of $\bar m_j^-$ and $\Delta m_j$ [Eqs.~(\ref{eq:origdefofbarmj}) and (\ref{eq:origdefofDelmj}), respectively], 
we find the minimum non-negative integers $\delta C_j^{c\pm}\geq 0$ that suffice the above conditions:
\begin{flalign}
\delta C^{c-}_j
&=
\frac{r_{i_*j} -\Delta x_j/2}{c\Delta t}
-
\left\lfloor
\frac{r_{i_*j} -\Delta x_j/2}{c\Delta t}
\right\rfloor
\\
\delta C^{c+}_j
&=
\left\lceil 
\frac{r_{i_*j} +\Delta x_j/2}{c\Delta t}
\right\rceil
-
\frac{r_{i_*j} +\Delta x_j/2}{c\Delta t}.
\end{flalign}
For such $\delta C^{c\pm}_j$ values, 
we have
\begin{flalign}
\bar t_j^{-}&= 
\left\lfloor
\frac{r_{i_*j} -\Delta x_j/2}{c\Delta t}
\right\rfloor\Delta t
\label{FDPHeq:specialdeltminus_simpler}
\\
\bar t_j^{-}+\Delta t_j&= 
\left\lceil 
\frac{r_{i_*j} +\Delta x_j/2}{c\Delta t}
\right\rceil\Delta t,
\label{FDPHeq:specialdeltplus_simpler}
\end{flalign}
or equivalently,
\begin{flalign}
\bar m_j^{-}&= 
\left\lfloor
\frac{r_{i_*j} -\Delta x_j/2}{c\Delta t}
\right\rfloor
\\
\Delta m_j&= 
\left\lceil 
\frac{r_{i_*j} +\Delta x_j/2}{c\Delta t}
\right\rceil -\bar m_j^{-}
. 
\end{flalign}
These expressions are similar to the original Eqs.~(\ref{eq:origdefofbarmj}) and (\ref{eq:origdefofDelmj}) for $\bar m_j^-$ and $\Delta m_j$, with dropping $\delta C^{c\pm}_j$ and flipping the floor and ceil functions in the right hand sides of Eqs.~(\ref{eq:origdefofbarmj}) and (\ref{eq:origdefofDelmj}).
The above are suitable for the 3D cases, and $\delta C^{c+}_j$ is further incremented for the error control in the 2D cases (while $\delta C^{c-}_j$ is dimension-independent), as detailed in \S\ref{FDPHH}. 
We used such a choice of $\delta C^{c\pm}_j$ in the numerical experiments of the anti-plane problem in the text. 
%%%%%%%%%%%%%%%This is $\bar m_j^-$ and $\bar m_j^-+\Delta m_j$ [given by Eqs.~(\ref{eq:origdefofbarmj}) and (\ref{eq:origdefofDelmj})] and 
%When $\Delta t_j^\pm$ (and $\bar t_j$) satisfy these conditions, the integral range becomes
%\begin{eqnarray}
%\min[t_{i_*j}^+, (m+\bar m_j^{-}+1)\Delta t+(t_{i_*j}-\bar t_j)]
%%{\min(\Delta t_j^{+}, (m+\bar m_j^{-}+1)\Delta t-\bar t_j)} 
%&\to&
% (m+\bar m_j^{-}+1)\Delta t
%\\
%\max[t_{i_*j}^- , (m+\bar m_j^{-})\Delta t +(t_{i_*j}-\bar t_j)]
%% {\max(- \Delta t_j^{-}, (m+\bar m_j^{-})\Delta t-\bar t_j)}
%&\to&
%(m+\bar m_j^{-})\Delta t. 
%\end{eqnarray}
%Note that the use of special $\Delta t_j^\pm$ [Eq.~(\ref{FDPHeq:specialdeltminus}) and (\ref{FDPHeq:specialdeltplus})] simultaneously simplifies the integral equations in the Domains I and S (respectively mentioned in \S\ref{FDPHB25} and \S\ref{FDPHB1}).
%$\delta C^{c+}$ and $\delta C^{c-}$ are the minimum non-negative values that satisfy the above relation in the 3D cases; $\delta C^{c+}$ is further increased by positive integers in the 2D cases, as detailed in \S\ref{FDPHH}, while $\delta C^{c-}$ is dimension-independent. The numerical experiments of the anti-plane problems in texts showed the results in adopting the above special set of $\delta t_j^\pm$. 

The above conditions Eqs.~(\ref{FDPHeq:specialdeltminus_simpler}) and (\ref{FDPHeq:specialdeltplus_simpler}) indicate that $\Delta t_j^\pm$ (and $\delta C^{c\pm}_j$) for source $j$ become leaf-dependent given the leaf dependence of $r_{i*j}$; it is naturally expected from the original FDPM where the $\Delta t_j^\pm$ values also depend on receiver $i$ (thus precisely given as $\Delta t_{ij}^\pm$). Meanwhile, the $\delta C^{c\pm}_j$ values can be leaf independent, as originally shown in \S\ref{FDPH431}; $\hat K^F$ and $h^F$ are determined depending on such a choice of $\delta C^{c\pm}_j$, and the error order is mostly independent of $\mathcal O(1)$ variations in $\delta C^{c\pm}_j$ for the constant $\eta$ scheme, as also mentioned in \S\ref{FDPH431}. 
We saw in \ref{FDPHB} (especially in \S\ref{FDPHB1} and \S\ref{FDPHB25}) that the arithmetics for Domains I and S require additional considerations on the correction terms unless the above conditions Eqs.~(\ref{FDPHeq:specialdeltminus_simpler}) and (\ref{FDPHeq:specialdeltplus_simpler}) are met, and then the above conditions will be rather for the simplification of the arithmetics for Domains I and S. 

Note that even after erasing the discretization error due to Eq.~(\ref{eq:discretizationofbartj}) of $\bar t_j^-$, we have another discretization error on the same order in using Eq.~(\ref{eq:discretedelti}) of $\delta t_i$. To reduce its error order, we can consider more accurate interpolation for $\delta t_i$ than mere rounding. 

%We note that the rounding of the decimals may be numerically erroneous due to the rounding errors of $\bar t_j^{\pm}$ especially when $\bar t_j^{\pm}$ are near integer numbers. To deal with that, it is useful to change the minimum of safe coefficients $\delta C^{c\pm}$ from $0$ to $1$; we adopted it in the numerical experiments in the paper. 

%We also reamrk that it is also possible to discard the small decimal differences in $\bar t_j^\pm/\Delta t-\bar m_j^\pm$ as the small fraction of $\bar t_j$ in the travel time separation $t_{ij}\approx \delta t_i+\bar t_j$; this change does not affect the computation of the degenerating normalized waveform, since the travel time separation is applied only to $t_{ij}$ in the slip rate [See Eq.(\ref{FDPHeq:beforeapproxDomF})]. However, such handlings using $\bar t_j$ only treat $\bar t_j/\Delta t-\bar m_j^+$ or $\bar t_j/\Delta t-\bar m_j^-$. Therefore, without choosing $\Delta t_j^\pm$ satisfying Eqs.~(\ref{FDPHeq:specialdeltminus}), (\ref{FDPHeq:specialdeltplus}), we need to compute one decimal part at least either in Domain I or in Domain S.

\section{A Case Where the Partially Pivoting ACA Erroneously Works}
\label{FDPH_ACAcheckapp}
\setcounter{figure}{0}
%Figure 12_2. âš
\begin{figure}[tbp]
   \begin{center}
  \includegraphics[width=100mm]{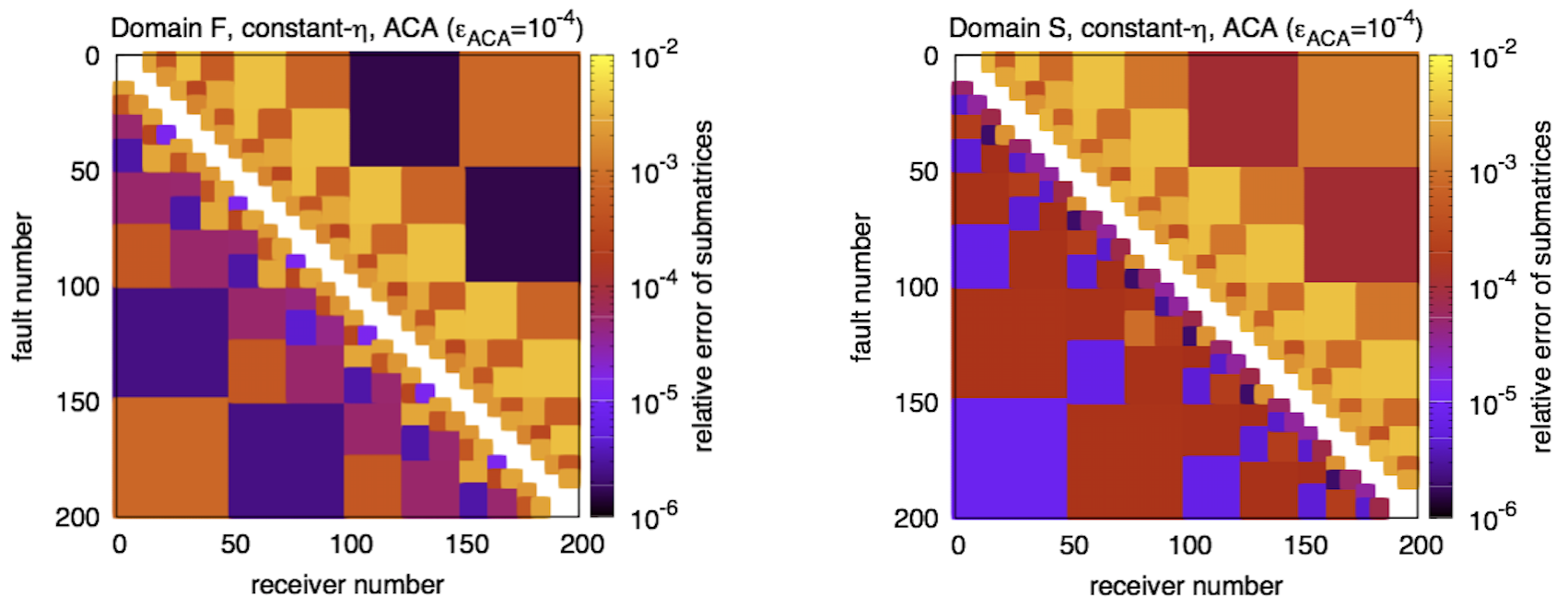}
   \end{center}
\caption{
Error and rank distributions of submatrices approximated by the partially-pivoting ACA. The other settings are the same as in Fig.~\ref{FDPHfig:12}. 
(Left) Error distribution in $\hat K^F$ for constant $\eta$.
%Some submatrices have errors $1\sim2$ digits larger than the required approximate upper bound ($\epsilon_{ACA}$).
(Right) Error distribution in $\hat K^S$ for constant $\eta$.
}
\label{FDPHfig:12_2}
\end{figure}

We saw in \S\ref{FDPH611} that ACA+ achieved the $\mathcal O(1)$ ranks of the kernel submatrices. 
That means the LRA itself functions even for the kernel in Domain F. 
Meanwhile, we sometimes also observed that the most standard technique, the partially pivoting ACA, did not satisfy the required accuracy [Fig.~\ref{FDPHfig:12_2} (top left)]; the setting in the following is the same as ACA+ cases in \S\ref{FDPH611}. 
Even when $\epsilon_{ACA}$ was set at $10 ^ {-4}$, the approximated matrix contained the errors of order $10 ^{-3}\sim10 ^ {-2}$; it means that $\epsilon_{H}$ was $10 ^{-3}\sim 10 ^ {-2}$. 
This accuracy degradation was also observed in the asymptotic kernel in Domain S (the static kernel) [Fig.~\ref{FDPHfig:12_2} (top right)]. 
As ACA+ worked in both domains, these accuracy degradation are ascribed to the problems of the partially pivoting ACA as the LRA method, rather than to the principal limitation of the LRA. 
This accuracy problem seems consistent with the indication of several previous studies of H-matrices in the spatial BIEM~\cite{borm2003hierarchical,grasedyck2005adaptive}. 

The reason of these problems seems related to the Taylor series, what usually guarantees the degenerate form of the discretized kernel for H-matrices and is substantially executed in the partially pivoting ACA. 
The point will be that 
the Taylor series in the source-receiver distance cannot get a fast convergent series if the source and receiver are too close (closer than some sort of a threshold, approximately the value of $diam$). 
Along this line, 
the problem will be ascribable to the source-receiver distance selected as the initial basis function of the LRA (substantially imposed with ${\bf f}_{a0}, {\bf g}_{a0}$), which corresponds to that at the initial pivoting point~\cite{bebendorf2003adaptive}.

Fig.~\ref{FDPHfig:12_2} supports the above consideration by indicating that the partially pivoting ACA erroneously stopped improving the LRA at the upper triangular side of the matrix, where the distance between the source and receiver were relatively smaller at the initial pivoting point (than that of the lower triangular side where the partially pivoting ACA works successfully), given the location of the ordinary (and our) initial pivoting point set at the top-left apex of the submatrix. 
This problem then seems apter to occur as $\eta$ gets larger, because its root will be the non-convergence of the Taylor series applied to the close source receiver pairs.

\section{Handling of the 2D-Specific Errors in Spatiotemporally Separating the Kernel}
\label{FDPHH}
\setcounter{figure}{0}
 
Below, we detail the way of handling the errors specifically arising in the 2D problems. The 3D problems do not have such errors, and the following error handling becomes unnecessary. 

We first introduce the design of the 2D error handling in \S\ref{FDPH721}. It contains two tuning parameters for the error suppression: the duration of Domain F (more precisely $\Delta t_j^+$) and the upper bound of the absolute error, $\epsilon_{st}$. Their tunings are detailed in \S\ref{FDPH722} and \S\ref{FDPH723}, respectively.

\subsection{Two Techniques for Handling 2D Specific Errors}
\label{FDPH721}
%Figure 18. âš
\begin{figure*}[tbp]
\includegraphics[width=125mm]{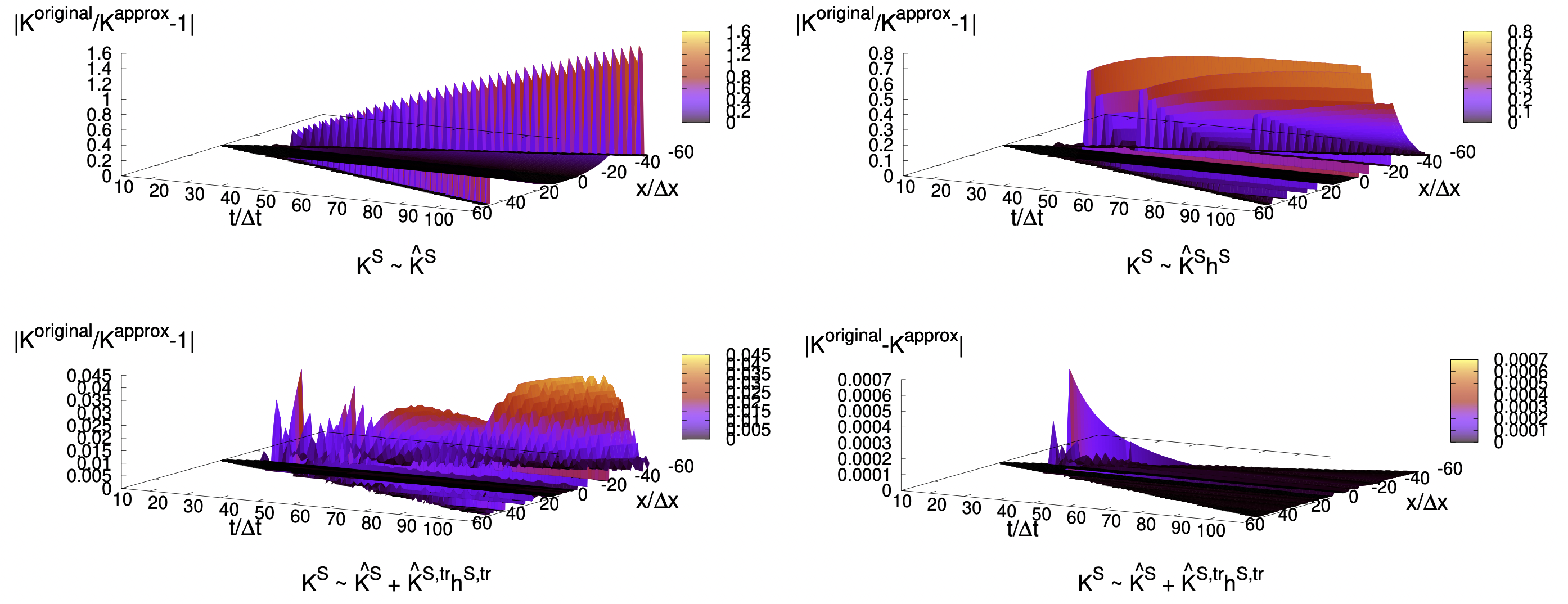}
%   \begin{center}
%  \includegraphics[width=60mm]{fig_relativekernelerror_lesstr.eps}
%  \includegraphics[width=60mm]{fig_relativekernelerror_trtost.eps}
%\\
%  \includegraphics[width=60mm]{fig_relativekernelerror.eps}
%  \includegraphics[width=60mm]{fig_abskernelerror.eps}
%   \end{center}
\caption{
Error distributions in approximate kernels $K^{\rm approx}$ of different temporal ranks, in a 2D planar boundary case. 
Errors are quantified by the difference between $K^{\rm approx}$ and original kernel $K^{\rm original}$. 
Used values of the approximation parameters are $\epsilon_{Q} =\epsilon_{ACA} =\epsilon_{st}= 10^{-3}$, $l_{min} / \Delta x = 5$, and $\eta = 5.67$ and the temporal distance between the travel time and the end of Domain F is enlarged by $3\Delta x / \beta$. 
(Top left) Relative error of the asymptotic kernel, being one example case of the temporally first rank. 
(Top right) Relative error, for the case of the temporally first rank, where the temporal pivot point is set at the start of Domain S. 
(Bottom left) Relative error, for the case of the temporally second rank. 
(Bottom right) Absolute error, for the case of the temporally second rank, normalized by the radiation damping term. 
}
\label{FDPHfig:18}
\end{figure*}

In the original FDPM, Ref.~\cite{ando2007efficient} dealt with the error caused by the spatiotemporal separation of the 2D kernel by enlarging the temporal distance [$\Delta t^+_{(j)}$, represented by Eq.~(\ref{FDPHeq:6})] between the travel time and the end of Domain F.
The increment of $\Delta t^+_j$ is called an additional width of Domain F in this paper. 
The additional width of Domain F allows the FDPM to regulate the error with keeping the computational speed mostly~\cite{ando2007efficient}.
 
However, introducing additional width of Domain F can enhance another error in using degenerating normalized waveform [Eq.~(\ref{FDPHeq:21})] in FDP=H-matrices. 
This is because the approximation of normalized waveforms by the ART, Eq.~(\ref{FDPHeq:21}), depends on the duration, $\Delta t_j$, of Domain F. As the ART does not apply to the inadmissible leaves, 
its error is only related to the admissible leaves giving relatively smaller kernel values and then may not be much crucial, but handling this error trade-off is preferable in terms of the error control. 

We then utilize a property of the elastodynamic kernel that its time dependence reduces to a sum of power functions of time. 
This property is kept even in the analytic form of the 2D kernel, e.g., in Ref.~\cite{tada2001dynamic}, although the 2D specific transient time dependence is associated with the reduced time (elapsed time from the passage of the wavefronts) unlike the original asymptotic one in Domain I depending on the original time from the origin.

Considering that property, we also adopt a temporal LRA that contrasts with the spatial LRA in H-matrices. 
This temporal LRA is applied to the kernel in Domains I and S in the admissible leaves, 
and the suite of the temporal LRA and spatial H-matrices is implemented by the Tucker cross approximation (the TCA)~\cite{oseledets2008tucker}, known as a fast approximate LRA technique for the third-order tensors. The TCA approximates the discretized kernel of the receivers, sources, and time steps to a sum of the products of the vectors depending on any of them. 
The spatiotemporal variable separation of the FDPM can be regarded as a part of an (analytic) example of the TCA, where the number of vectors in the temporal direction (hereafter called the rank in the temporal direction) is one in Domain S and two in Domain I, for the case of the double-layer potential we considered in the text. 
By increasing the temporal rank, the TCA allows us to avoid using the excessively widened Domain F.

Fig.~\ref{FDPHfig:18} shows the error in the kernel tensor associated with the spatiotemporal separation of the kernel, reduced by the TCA. The case of a planar fault is considered in the figure, and the adopted parameter values are listed in its caption. 
We computed the case of $\Delta t^+/(\beta\Delta x)=\mathcal O(1)$ that we want to adopt in FDP=H-matrices. 
The static approximation (denoted by $K^S \sim\hat K^S$ in the figure) the original FDPM adopted includes almost $100\%$ relative errors in that case. 
Another case of the temporally first rank (denoted by $K^S\sim\hat K^{S\prime}h^{S\prime}$), 
where the temporal pivot point is set at the start of Domain S, also does almost $100\%$ relative errors. 
The case of the temporally second rank (denoted by $K^S \sim\hat K^S +\hat K^{S,tr}h^{S,tr} $), considering the temporal pivot point at the start of Domain S for approximating the transient part, then 
reduces such numerical errors greatly. The relative error becomes order $1\%$, and the absolute error becomes order $10^{-5}$.
This remarkable accuracy improvement of $K \sim \hat K^S +\hat K^{S,tr} h^{S,tr}$ in Fig.~\ref{FDPHfig:18} (bottom) may be consistent with that the 2D kernel in Domain S comprises the static term ($\hat K^S$) and the long temporal tails decaying roughly in proportion to the inverse root of the elapsed time, as seen in its analytic form, e.g., of Ref.~\cite{cochard1994dynamic}.

Given the above result, we adopted the TCA of the temporally second rank ($K^S\sim\hat K^S +\hat K^{S,tr} h^{S,tr}$) in Domain S for the admissible leaves, as well as the tuning of the additional width of Domain F.
We did not apply the TCA in the inadmissible leaves as the approximation of the ART is not applied there. 
We determined the additional width of Domain F (defined containing Domain I in the inadmissible leaves, as mentioned in \ref{FDPHE}) in the inadmissible leaves by the error regulation rule similar to that of Quantization, which sets the initial time step of Domain S (the end of Domain F plus 1) at a time step after which the absolute errors are smaller than $\epsilon_Q$ and $\epsilon_{st}$, respectively, between the original kernel and the asymptotic one. 
Besides, in order to introduce the transient time-dependent kernel in Domain S with a finite cost in the admissible leaves, we determined the time step after which the transient time-dependent part of the kernel in the admissible leaves is discarded. Such a time step is set at a time step under the same condition as that for determining the start of Domain S in inadmissible leaves; in summary, all the staircase approximations in the paper is regulated by $\epsilon_Q$ and $\epsilon_{st}$, except for the TCA in the admissible leaves. 
We did not introduce further higher ranks of the TCA, because the error was mostly caused by the spatially close block clusters corresponding to the inadmissible leaves [Fig.~\ref{FDPHfig:18} (bottom right)] to which the TCA does not apply. 
We also note that the enlargement of Domain F does not affect the asymptotic cost scaling of $\mathcal O (N\log N)$ because the duration of Domain F is independent of $N$.

For estimating the error caused by Quantization applied in Domain I (explained in \S\ref{FDPH323}) in the 3D cases, 
we additionally applied Quantization to the transient term of Domain S in our implementation of FDP=H-matrices. 
It also gave a measurable acceleration of the computation related to the memory access in our numerical experiments while the asymptotic size scaling of the cost is unchanged for that case. 
The error due to Quantization in the 2D Domain S will be the upper bounds for that in the 3D Domain I, because the absolute value of the 2D-specific transient term is comparable to that of the kernel in Domain F while the 3D kernel takes much smaller value in Domain I than in Domain F.

\subsection{$\Delta t^+_j$ Dependence}
\label{FDPH722}
%Figure 19. âš
\begin{figure*}[tb]
\centering
\includegraphics[width=125mm]{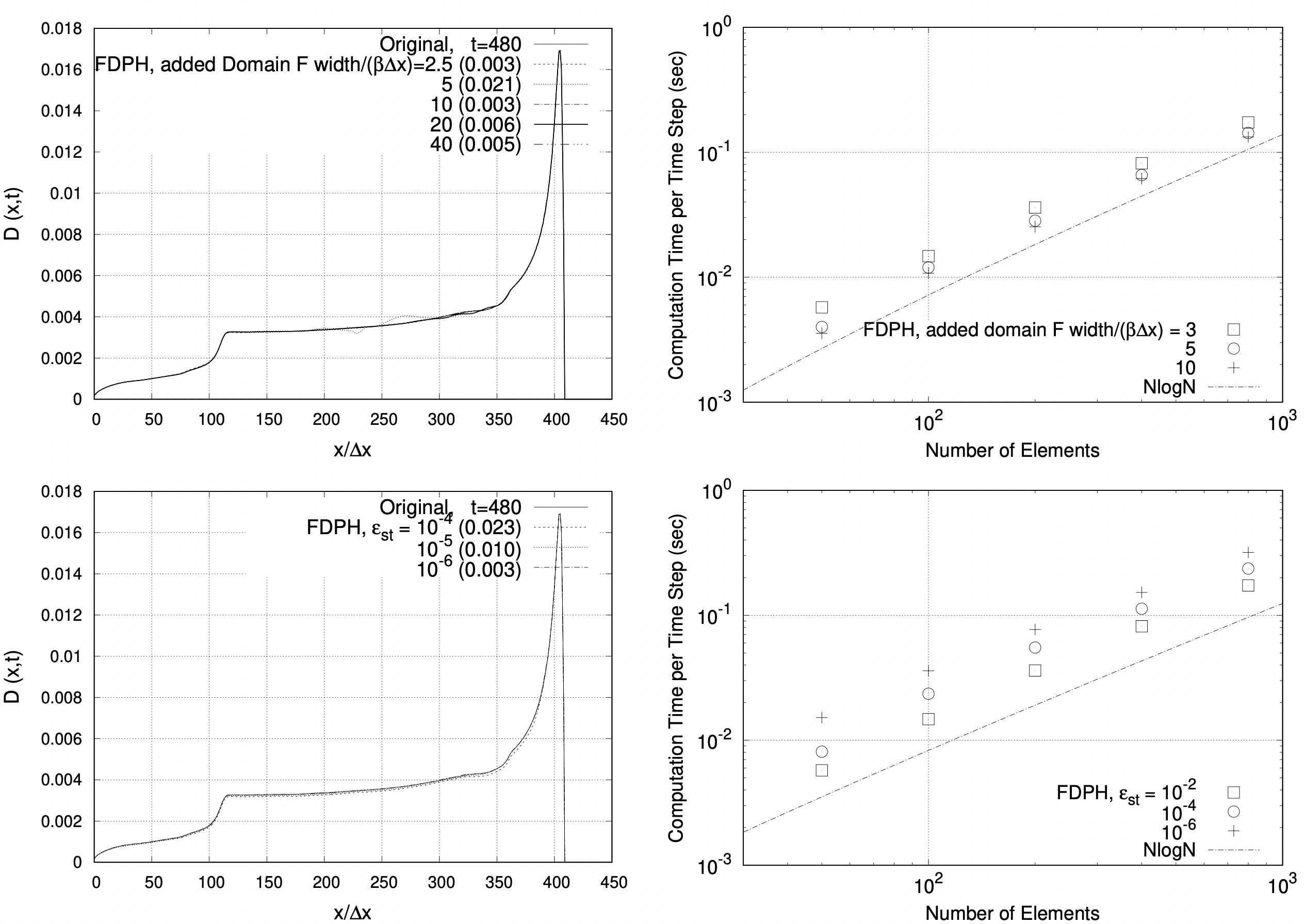}
%\begin{minipage}{1.0\hsize}
%\centering
%  \includegraphics[width=60mm]{fig_Dx_snapshot_nonplanar_deltatplusF.pdf}
%  \includegraphics[width=60mm]{fig_timeconsumption_mac_addwidthdepoffastness.pdf}
%\end{minipage}
%\\
%\centering
%\begin{minipage}{1.0\hsize}
%\centering
%  \includegraphics[width=60mm]{fig_Dx_snapshot_nonplanar_epST.pdf}
%  \includegraphics[width=60mm]{fig_timeconsumption_mac_epsilonstdep.pdf}
%\end{minipage}
\caption{
Dependence of the error and cost on the width of Domain F and $\epsilon_{st}$. FDP=H-matrices are abbreviated to FDPH in the figure. 
Bracketed numbers in the legends of the left panels indicate the relative errors of FDP=H-matrices in the same manner as in Fig.~\ref{FDPHfig:15}.
Problem and parameter settings in the left panels and the right panels are respectively the same as in Fig.~\ref{FDPHfig:16} and \ref{FDPHfig:14}. 
(Top left) Snapshots of slip rate $D$ over space $x$ at $t=480$ with several values of the width of Domain F. 
(Top right) The size scaling of the computation time per time step, defined in \S\ref{FDPH621}, with several values of the width of Domain F. 
The $N \log N$ cost scaling is indicated by a dotted line. 
(Bottom left) Snapshots of slip rates $D$ over space $x$ at $t=480$ under several $\epsilon_{st}$ values. 
(Bottom right) Size scaling of the computation time per time step for several $\epsilon_{st}$ values. The cost scaling of $N \log N$ is indicated by a dotted line. 
}
\label{FDPHfig:19}
\end{figure*}

Here, we investigate the dependence of the accuracy and cost 
on tuning parameter $\Delta t^+_j$ (abbreviated to $\Delta t^+$ below) for handling the errors due to the spatiotemporal separation specifically arising in the 2D problems. 

Fig.~\ref{FDPHfig:19} (top left) shows the accuracy of the solution with several additional widths of Domain F. 
The error is shown to be suppressed below $1\%$ except for the case of adding $5\beta\Delta x$ to $\Delta t^+$.

The error causes related to $\Delta t^+$ 
comprise the variable separation in Domain S and the approximation of the normalized waveform.
Among them, 
the approximation error related to the normalized waveform seems not relevant, 
since the observed error is not proportional to the duration of Domain F unlike its analytic evaluation given in Eq.~(\ref{FDPHeq:21}). 
Most errors would then be ascribed to the variable separation of the Domain S kernel. Consistently, we also observe the accuracy improvement following the width increase in the parameter range where the added width is larger than $5\beta\Delta x$. Such an error reduction is also an expected property for the factorized kernel of the FDPM. 

%This also implies that the error related to $\Delta t^+$ is easily reduced with additional width wide enough for the width dependence of variable separation errors to become monotonous (as long as $\Delta t^+$ is too large), because the absolute error of the variable separation decreases as the absolute value of the transient term decreases. 

%The $\mathcal O (N\log N/N_*)$ scaling of the numerical costs is maintained against the $\Delta t^+$ increase [Fig.~\ref{FDPHfig:19} (top right), measured by the computation time per time step]. The change in $\Delta t^+$ seemed to slightly affect the proportionality coefficient between the computation cost and the number of elements. 
%The cost in the case adding $3\Delta x/\beta$ to $\Delta t^+$ is reduced to about half when adding $10\Delta x/\beta$ to $\Delta t^+$.

Fig.~\ref{FDPHfig:19} (top right) shows the computation time per time step with various $\Delta t^+$ values. It indicates that the associated cost variation is within a factor, and the $\mathcal O (N\log N/N_*)$ cost scaling is maintained. 
The computation seems to become slightly faster when we impose moderately large $\Delta t^+$, probably due to the difference by factors between the computational complexities for the transient term in Domain S and for Domain F. 
We note that the taken computation time increases as $\Delta t^+$ grew when $\Delta t^+$ was of 100 $\Delta t$ or larger (excessively large values yet possibly required in the case of the temporally first rank, not plotted). 

As above, as far as we set the additional width at not excessively large values, the error in the normalized waveform can be irrelevant. Good convergence of the variable separation for such a case of a narrow Domain F would owe to the above-mentioned TCA. 

\subsection{$\epsilon_{st}$ Dependence}
\label{FDPH723}
Below, we investigate the dependence of the solution accuracy and cost on $\epsilon_{st}$, the absolute error bound for the separation of variables in Domain S (corresponding to the static approximation in the original ST-BIEM e.g., Ref.~\cite{tada2001dynamic}) and Quantization.
To appropriately evaluate the $\epsilon_{st}$ dependence of the computational time, we here impose a related acceleration technique for computing the transient term in Domain S (explained in \S\ref{FDPHB3}). 

The solution accuracy is shown in Fig.~\ref{FDPHfig:19} (bottom left).
The relative error increases roughly in proportion to the logarithm of $\epsilon_{st}$ within the range $\epsilon_{st}=10 ^{-4}\sim 10^{-6}$. 
This error gives the systematic decrease in the slip- and opening-rates. 
It is consistent with the nature of the static approximation and is also observed in the accuracy evaluation (\S\ref{FDPHA2}) of Quantization alone which employs a kind of the static approximation. 

The computation time per time step is shown in Fig.~\ref{FDPHfig:19} (bottom right). 
The cost is roughly inversely proportional to $\epsilon_{st}$. Even if $\epsilon_{st}$ changes $10^4$-fold, the computation speed changes only about 3 times, and the effect of $\epsilon_{st}$ to the cost was quite small. It is consistent with that the absolute error condition is negligible for a large source-receiver distance as in the admissible leaves. 

The bound $\epsilon_{st}$ of the absolute error dominantly controls the accuracy while it does not affect the cost largely. This tendency will be inherited to FDP=H-matrices in the 3D problems applying Quantization to Domain I.

\section{Supplemental Calculations}
\subsection{The Amplitude Term and Its Degenerate Form}
\label{FDPHF2}
The $abe$ component of $\hat K_{ij}^{F_P}$ in Domain Fp of the admissible leaves, obtained from the P-wave part and near-field part of the elastodynamic Green's function, is calculated as
\begin{flalign*}
&
(\hat K_{ij}^{F_P})_{abe}=
\\
&-
\int_{\Gamma_j} d\Sigma(\boldsymbol\xi) C_{abcd}\nu_f(\boldsymbol\xi)C_{efgh}
\frac 1{4\pi\rho\alpha^2} 
\frac{\partial^2 }{\partial\xi_h\partial x_c} 
\\&
\left[
\frac{\gamma_d\gamma_g}{|{\bf x}-\boldsymbol\xi|}
\int ^{t_{ij}+\Delta t_j^+}_{t_{ij}-\Delta t_j^-}d\tau^\prime
H\left(\frac{|{\bf x}_i-\boldsymbol\xi|}{\alpha}-\tau^\prime\right)
\right.
\\
&+
\left.
\frac{3\gamma_d\gamma_g-\delta_{d,g}}{|{\bf x}_i-\boldsymbol\xi|^3}
\int ^{t_{ij}+\Delta t_j^+}_{t_{ij}-\Delta t_j^-}d\tau^\prime
\int^{|{\bf x}_i-\boldsymbol\xi|/\beta}_{\tau^\prime} dt^\prime
t^\prime
H\left(t^\prime-\frac{|{\bf x}_i-\boldsymbol\xi|}{\alpha}\right)
\right]
\end{flalign*}
for respective stress fields due to the motion of source $j$ that covers $\Gamma_j$, when collocated at ${\bf x}_i$ for receiver $i$. 
The first term is purely impulsive, as seen in Ref.~\cite{ando2016fast}. 
The second term is the near-field term contaminated in Domain F due to the discretization. 
In the brackets, the time-($\tau^\prime$- or $t^\prime$-) dependence of the integrands is replaced by the dependence on $t_{ij}\pm \Delta t^{\pm}_{j}$ after the integrands are integrated over Domain F, and hence $\hat K_{ij}^{F_P}$ is surely time-independent. 

Since travel time $t_{ij}=r_{ij}/\alpha$ is proportional to distance $r_{ij}$ like the static kernel, the above can be expanded (after the analytic execution of the differentiation) in $dist/(diam+dist)$ except the small factors of $\mathcal O(\Delta t_j^\pm,\Delta x_j)$; $\mathcal O(\Delta t_j^\pm/diam)$ factor is treated as additional source $j$ dependence like $\mathcal O(\Delta x_j/diam)$ factors that exist even in the static problem. 

The same holds in the S-wave cases where the kernel comprises the impulsive S-wave part and the contaminated near-field and static terms.

\subsection{Error Evaluation of Degenerating Normalized Waveforms Including Stress-Traction Projection}
\label{FDPHC1}
The following evaluates the error of the expansion that reduces the normalized waveform depending on both the receivers and sources to the degenerating normalized waveform depending on the sources. 

The kernel is the sum of the function expressing the tensorial radiation pattern [the orientation dependence, such as $\gamma$ in Eq.~(\ref{eq:3DGreen})] and the geometrical spreading (the distance dependence) with depending on time. 
We can roughly separate the error cause to that of the orientation dependence and that of the geometrical spreading and time-dependence.
The error associated with the orientation dependence of respective terms is estimated at the amount of the variation in the orientation. It is $\mathcal O(\delta r/\bar r)$, equals to $\mathcal O[1/(1+\eta^{-1})]$ given $\delta r/\bar r<1/(1+\eta^{-1})$ for an admissible leaf; it further becomes 0 on a line boundary as in the travel-time approximation Eq.~(\ref{FDPHeq:17}). 
The estimate for the other error cause is twofold. 
When the (orientation-independent) geometrically-spreading time-dependent part takes a staircase form or is delta-functional temporally, like the impulsive and static effects of the P- and S-waves, we have no errors incurred by them, as the plane-wave approximation predicts. 
On the other hand, we can also consider the case the associated space-time dependence is given by a scaling function $f(ct/r)$, like the near-field term and the 2-D P- and S-waves; for that case,  substituting $t=r/c+t_R$ with the reduced time $t_R$ and expanding $f(ct/r)$ in $r$ near $\bar r$, we find $f(ct/r)=f[1+ct_R/\bar r+\mathcal O((1+\eta^{-1})^{-1})]$, where we used 
$\mathcal O(diam/\bar r)=\mathcal O[1/(1+\eta^{-1})]$. 
It is always the error cause even on a line boundary unlike the orientation dependence. 
Given these, the error caused by the use of the degenerating normalized waveform is $\mathcal O[1/(1+\eta^{-1})]$ on an arbitrary boundary geometry at most; excluding this part is rather related to the far-field approximation than the plane-wave, and it rapidly decreases in the 3D cases while it remains to certain extent even at a distance in the 2D cases. We note that the error order becomes further smaller given the normalization condition of the normalized waveform, as mentioned in the text, related to Eq.~(\ref{FDPHeq:21}). 

Additionally, we would emphasize that the above error estimate implicitly relies on that the kernel is independent of the orientation of the receiver element. 
The stress nucleus, $K_{abe}$ in Eq.~(\ref{FDPHeq:1}) to give the $ab$ component of the stress after convolved with the $e$-component of the slip and opening, is such a case; the evaluation of the displacement is also included in it. 
On the other hand, since the traction is significantly depends on the receiver even at infinite distance ($diam/\bar r \to 0$), 
if $K_{abe}$ in the definition of the degenerating normalized waveform Eq.~(\ref{FDPHeq:29}) is replaced with the traction nucleus $K_{T,ae}$ such that $T_a=\int d\Gamma \int d\tau K_{T,ae}\Delta u_e$, 
the error order is not $\mathcal O[1/(1+\eta^{-1})]$ and is $\mathcal O(1)$ even for infinitesimal $\eta\to0$. 
To evade this error cause, we first compute the stress tensors (the traction vectors for virtual elements oriented in $x_1,x_2,x_3$ directions) at the receiver locations in evaluating the traction vectors of the receivers with FDP=H-matrices. The traction vector ${\bf T}$ for the original receiver boundary is 
then computed from the stress tensor $\boldsymbol\sigma$ as ${\bf T}=\boldsymbol\sigma\boldsymbol\nu$ from the definitional identity. 
The above holds also for the single-layer potential case. 

\subsection{Scale Analysis for the Cost Scaling of FDP=H-Matrices}
\label{FDPHC2}
A scale analysis is here conducted to obtain the typical $N$ dependence of the numerical costs in FDP=H-matrices shown in Fig.~\ref{FDPHfig:9_b}. 
We focus on the cost scaling of the constant $\eta^2dist$ scheme, as that of the constant $\eta$ scheme of $\mathcal O(N\log N)$ is obvious by considering that of H-matrices~\cite{borm2003hierarchical} in the spatial BIEM, as mentioned in the text related to Fig.~\ref{FDPHfig:9_b}. 
We here normalize the length scale by $\Delta x_j$ and assume $\Delta x_j$ of any elements $j$ is on the order of constant $\Delta x$.

As shown in Fig.~\ref{FDPHfig:9_a}, most of the kernel tensor components are covered by the largest-scale block clusters in the constant $\eta^2dist$ scheme. It also means that the numerical costs are dominated by theirs. This observation is a starting point for the following cost order estimates of the constant $\eta^2dist$ scheme. 

Let us first estimate the number of leaves at the smallest level. Those leaves have the longest sides, which is $\mathcal O(\eta L)$, independent of the dimension of the fault. 
Moreover, $\mathcal O(\eta L)=\mathcal O(\sqrt L)$ holds in the constant $\eta^2dist$ scheme. Therefore, by supposing that the largest-class block clusters occupy most of the spatial regions as mentioned above, we obtain the estimate of the number of the largest-class block clusters: $\mathcal O(L^{2D_b}/\sqrt{L}^{2D_b})=\mathcal O(L^{D_b})=\mathcal O(N)$. 

The costs are then estimated as the product of the number of clusters and the costs per clusters.
Since the values of $N_{s,a}+N_{r,a}$ (the number of elements in block cluster $a$) are $\mathcal O(diam^{D_b})$, that is $=\mathcal O[(\eta L)^{D_b}]=\mathcal O(L^{D_b/2})=\mathcal O(N^{1/2})$ in the largest-class clusters,
the costs regarding the spatial integral $\sum_a (N_{s,a}+N_{r,a})$ are $\mathcal O(N)\times \mathcal O(N^{1/2})=\mathcal O(N^{3/2})$. 
On the other hand, 
since the values of $dist$ are $\mathcal O(L)$ in the largest block clusters, 
the temporal ones $\sum_a dist$ [$=\mathcal O(\sum_a\bar r)$] (the sum of the temporal integration lengths) are $\mathcal O(N)\times \mathcal O(L)=\mathcal O(NL)$. 
These estimates of the costs successfully capture the leading orders, that is except the log factors, of the typical costs in the constant $\eta^2 dist$ scheme, shown in Fig.~\ref{FDPHfig:9_b}.

\subsection{Discretization of Domain F After the ART}
\label{FDPHC3}
We here detail the discretization of the right-hand side of Eq.~(\ref{FDPHeq:22}) appearing in \S\ref{FDPH431}. 
The BIE for Domain F has originally been 
\begin{equation}
T_i^F(t) =
%\sum_j
%\int^{t_{ij}+\Delta t_j^{+}}_{t_{ij}-\Delta t_j^{-}}
%d\tau K_{i,j}(\tau)D(t-\tau)
%\\&=&
\sum_j\hat K^F_{i,j}\int^{\Delta t_j}_0
d\tau h_{i,j}(\tau)D(t-t^-_{ij}-\tau).
\label{FDPHeq:beforeapproxDomF}
\end{equation}
$\hat K^F_{i,j}h_{i,j}(\tau)$ constitutes $K_{i,j}(t)$. 
After the approximation of the ART, this becomes
\begin{equation}
T_i^F(t) =
\sum_j\hat K_{i,j}\int^{\Delta t_j}_0
d\tau^\prime h^F_{j}(\tau^\prime)D(t-\delta t_i -\bar t_{j}^--\tau^\prime),
\label{FDPHeq:beforediscDomF}
\end{equation}
as shown in \S\ref{FDPH42}. 
Below, we disretize Eq.~(\ref{FDPHeq:beforediscDomF}). 
The approximation of $\hat K$ is not discussed here. 
Hereafter, we alter $t$ into $t+\delta t_i$ for erasing $\delta t_i$ from the right-hand side.

By interpolating the slip- and opening-rate as Eq.~(\ref{FDPHeq:Dinterpolation}) in a piecewise-constant manner, 
and substituting the collocation time $t=(n+1)\Delta t$ of time step $n$,
we can calculate Eq.~(\ref{FDPHeq:beforediscDomF}) 
as  
%\begin{eqnarray}
%&&T_i^F(t+\delta t_i) 
%%\\ &=&
%%\sum_{j,m}\hat K^F_{i,j}D_{j,n-m} \int^{\Delta t_j^{+}}_{-\Delta t_j^{-}}
%%d\tau^\prime h^F_{j}(\tau^\prime)
%%[
%%H((n-m+1)\Delta t-((n+1)\Delta t-\bar t_{j}-\tau^\prime) )
%%\nonumber
%%\\&&
%%-H((n-m)\Delta t-((n+1)\Delta t-\bar t_{j}-\tau^\prime) )
%%]
%\\&=&
%\sum_{j,m}\hat K^F_{i,j}D_{j,n-m} \int^{\Delta t_j^{+}}_{-\Delta t_j^{-}}
%d\tau^\prime h^F_{j}(\tau^\prime)
%\nonumber
%\\&&
%[
%H(\tau^\prime+\bar t_{j}-m\Delta t)-H(\tau^\prime+\bar t_{j}-(m+1)\Delta t )
%]
%\end{eqnarray}
\begin{flalign}
&T_i^F(t+\delta t_i) 
%\\
%&=&
%\sum_{j,m}\hat K^F_{i,j}D_{j,n-m} 
%[
%H(\bar t_j-\Delta t_j^{-} -m\Delta t+0)
%-H(\bar t_j+\Delta t_j^{-}-(m+1)\Delta t+0)
%]
%\nonumber
%\\&&
%\int^{\Delta t_j^{+}}_{-\Delta t_j^{-}}
%d\tau^\prime h^F_{j}(\tau^\prime)
%[
%H(\tau^\prime+\bar t_{j}-m\Delta t)-H(\tau^\prime+\bar t_{j}-(m+1)\Delta t )
%]
\\=&
\sum_{j,m}\hat K^F_{i,j}D_{j,n-m} 
[
H(\bar t_j^{-} -m\Delta t)
-H(\bar t_j^{+}-(m+1)\Delta t)
]
\\&
\times
\int^{\min[\Delta t_j, (m+1)\Delta t-\bar t_j^-]}_{\max[0, m\Delta t-\bar t_j^-]}
d\tau^\prime h^F_j(\tau^\prime).
\end{flalign}
The function $[
H(\bar t_j^{-} -m\Delta t(+0))
-H(\bar t_j^{+}-(m+1)\Delta t(+0))
$]
takes nonzero values only within 
$(\bar t_j^{-}\leq m\Delta t )\cap (\bar t_j^{+}> m\Delta t )$, i.e., 
$ \lceil \bar t_j^{-} /\Delta t\rceil \leq m < \lfloor \bar t_j^{+} /\Delta t\rfloor $. 
%Such a range of $m$ is %$m\in[\lceil \bar t_j^{-} /\Delta t\rceil,\lfloor \bar t_j^{+} /\Delta t\rfloor)$.

%where $\bar m_j^{-}:= \lceil \bar t_j^{-} /\Delta t\rceil$ and $\Delta m_j:= \lfloor \bar t_j^{+} /\Delta t\rfloor -\bar m_j^{-}$ are exactly what defined in \S\ref{FDPH431}. Although these are respectively rounding-up of $\bar t_j^-/\Delta t$ and rounding-down of $\bar t_j^+/\Delta t-\bar m_j^-$ as long as $\bar t_j$ is given by Eq.~(\ref{FDPHeq:19}) [and neglecting the safe coefficients $\delta C_j^\pm$ in Eqs.~(\ref{FDPHeq:6}) and (\ref{FDPHeq:7})], we avoided writing do so in the paper, by considering the replacement of $\bar t_j$ value from such a special value.

By using $\lceil \bar t_j^{-} /\Delta t\rceil$
 and 
$\lfloor \bar t_j^{+} /\Delta t\rfloor$,
we express the discretized BIE for Domain F as follows:
\begin{flalign}
%&&T_i^F(t+\delta t_i) 
%\\&=&
%\sum_{j}\hat K^F_{i,j}\sum_{m=\bar m_j^{-}}^{\bar m_j^{-}+\Delta m_j^--1}D_{j,n-m} 
%\int^{\min(\Delta t_j^{+}, (m+1)\Delta t-\bar t_j)}_{\max(- \Delta t_j^{-}, m\Delta t-\bar t_j)}
%d\tau^\prime h^F_{j}(\tau^\prime)
%&=&
%\sum_{j}\hat K^F_{i,j}\sum_{m=0}^{\Delta m_j^{-}-1}D_{j,n-m-\bar m_j^{-}} 
%\nonumber\\&&
%\int^{\min(\Delta t_j^{+}, (m+\bar m_j^{-}+1)\Delta t-\bar t_j)}_{\max(- \Delta t_j^{-}, (m+\bar m_j^{-})\Delta t-\bar t_j)}
%d\tau^\prime h^F_{j}(\tau^\prime)
%\\&=&
&T_i^F(t+\delta t_i)= 
\sum_{j}\hat K^F_{i,j}\sum_{m=\lceil \bar t_j^{-} /\Delta t\rceil
}^{
\lfloor \bar t_j^{+} /\Delta t\rfloor-1}h^F_{j,m-\lceil \bar t_j^{-} /\Delta t\rceil} D_{j,n-m} 
\label{FDPHeq:31_prev}
\end{flalign}
with 
\begin{flalign}
&h^F_{j,m}:=
\int^{\min[\Delta t_j, (m+1+\lceil \bar t_j^{-} /\Delta t\rceil)\Delta t-\bar t_j^-]}_{\max[0, (m+\lceil \bar t_j^{-} /\Delta t\rceil)\Delta t-\bar t_j^-]}
d\tau h^F_{j}(\tau).
\end{flalign}
By using 
$h_j(\tau)=h_{i_*j}(\tau)=K^F_{i_*j}(\tau+t_{i_*j}^-)/\hat K^F_{i_*j}$, 
we obtain
\begin{flalign}
h^F_{j,m}
=
\frac 1 {\hat K^F_{i_*,j}}
\int^{\min[\Delta t_j, (m+1+\lceil \bar t_j^{-} /\Delta t\rceil)\Delta t-\bar t_j^-]}_{\max[0, (m+\lceil \bar t_j^{-} /\Delta t\rceil)\Delta t-\bar t_j^-]}
d\tau K_{i_*j}(\tau^\prime+t_{i_*j}^-).
\label{FDPHeq:29_prev}
%\\&=&
%\frac 1 {\hat K^F_{i_*,j}}
%\int^{\min(t_{i_*j}^+, (m+\bar m_j^{-}+1)\Delta t+(t_{i_*j}-\bar t_j))}_{\max[t_{i_*j}^- , (m+\bar m_j^{-})\Delta t +(t_{i_*j}-\bar t_j)]}
%d\tau K_{i_*j}(\tau)
\end{flalign}
We used $K^F_{i_*j}(\tau+t_{i_*j}^-)=K_{i_*j}(\tau+t_{i_*j}^-)$ in $t\in(t_{i_*j}^-,t_{i_*j}^-+\Delta t_j)$. 
%where $\bar t_j=t_{i_*j}$ [Eq.(\ref{FDPHeq:19})] is used for deriving the last line from the second last.
Eqs.~(\ref{FDPHeq:31_prev}) and (\ref{FDPHeq:29_prev}) generally hold. 
We see the definitions of $\bar m_j^-$ and $\Delta m_j$ [Eqs.~(\ref{eq:origdefofbarmj}) and (\ref{eq:origdefofDelmj}), respectively] in Eq.~(\ref{FDPHeq:31_prev}), and hence Eqs.~(\ref{FDPHeq:31}) in \S\ref{FDPH431} is met; note $\bar t^\pm_j=\bar t_j\pm\Delta t_j^\pm$ and $\bar t_{j}=t_{i_*j}$. 
As far as we meet $\bar t_j^-=\bar m^-_j\Delta t$ [Eq.~(\ref{eq:discretizationofbartj})] and $\Delta t_j=\Delta m_j\Delta t$ [Eq.~(\ref{eq:discretizationofDeltj})], assumed in \S\ref{FDPH431} (the parameter choice for satisfying which is also in  \S\ref{FDPH431}), 
we have
$\lceil \bar t_j^{-} /\Delta t\rceil\Delta t=\bar t_j^-$ and 
$ \lfloor \bar t_j^{+} /\Delta t\rfloor-\lceil \bar t_j^{-} /\Delta t\rceil=\Delta m_j$,
and thus Eq.~(\ref{FDPHeq:29_prev}) for $h_{j,m}^F$
reduces to Eq.~(\ref{FDPHeq:29}), shown in \S\ref{FDPH431}.

\section{List of Key Formulas}
\label{sec:keyformulas}
\setcounter{table}{0}

%%%%%%key formulas
\begin{table*}[tb]
   \caption{Key formulas of FDP=H-matrices for the data-sparse approximations in Domain F. The notation in each equation follows the text. The leaf $a$ dependencies of the parameters are here indicated explicitly.}
\begin{tabular}{c}
\hline
Key Formulas in Data-Sparse Approximations
\\
\hline
Travel time between the collocation points of receiver $i$ and source $j$:
\\
\vbox{
\begin{equation}
t_{ij}^c = \frac{r_{ij}} {c}. \tag{15}
\end{equation} 
}
\\
\hline
Temporal distances from the travel time to the leading(-)- and trailing(+)-edges of the wave: 
\\
\vbox{
\begin{equation}
\Delta t_{aj}^{c\pm} = \frac{\Delta x_j}{2c} +\delta C_{aj}^{c\pm}\Delta t. \tag{(16) and (17)}
\end{equation}
\centering
Their optional leaf $a$ dependencies were added in \S\ref{FDPH431} and \ref{FDPHC32}.
}
\\
\hline
Amplitude term in Domain F: 
\\
\vbox{
\begin{equation}
\hat K^F_{a,i,j}=\int^{t^c_{ij}+\Delta t_{aj}^{c+}}_{t^{c}_{ij}-\Delta t^{c-}_{aj}} d\tau K_{i,j} (\tau).
\tag{51}
\end{equation}
\centering
Parameters $t^{c\pm}_{aij}=t^c_{ij}\pm \Delta t_{aj}^{c\pm}$ in Eq.~(\ref{FDPHeq:28}) are defined around Eq.~(\ref{FDPHeq:durationofDomF}).
}
\\
\hline
Discretized degenerating normalized waveform:
\\
\vbox{
\begin{equation}
h^F_{a,j,m}=\frac 1 {\hat K^F_{a,i_*^a,j}} 
\int_{m\Delta t+t_{i_*^aj}^c-\Delta t^{c-}_{aj}}
^{(m+1)\Delta t+t_{i_*^aj}^c-\Delta t^{c-}_{aj}}
d\tau K_{i_*^a,j} (\tau).
\tag{50}
\end{equation}
\centering
Representative receiver $i_*^a$ is set for each admissible leaf $a$.
}
\\
\hline
Receiver-dependent travel-time-step difference:
\\
\vbox{
\begin{equation}
\delta m_{a,i}^c=
\left\lfloor 
\frac{r_{ij_*^a}-r_{i_*^aj_*^a}}{c\Delta t}
\right\rfloor.
\tag{42}
\end{equation}
\centering
Representative source $j_*^a$ is set for each admissible leaf $a$.
}
\\
\hline
Receiver-averaged travel time step and discretized duration of Domain F for $j$ in $a$: 
\\
\vbox{
\begin{flalign}
\bar m_{a,j}^{c-}&= 
\left\lceil 
\frac{r_{i_*^aj} -\Delta x_j/2}{c\Delta t}
\right\rceil
\tag{44}
\\
\Delta m_{a,j}^c&= 
\left\lfloor 
\frac{r_{i_*^aj} +\Delta x_j/2}{c\Delta t}
\right\rfloor -\bar m_{a,j}^{c-};
\tag{45}
\end{flalign}
\centering 
$\Delta m^c_{a,j}$ and also $\delta C^{c+}_{aj}$ increase by a integer number for the 2D problem (\S\ref{FDPH431}).
}
\\
\hline
    \end{tabular} 
   \label{KeyFormulas1}
\end{table*}

\begin{table*}[tb]
   \caption{Key formulas of FDP=H-matrices for the arithmetic in Domain F. The notation in each equation follows the text. The leaf $a$ and rank $l$ dependencies of the variables are here indicated explicitly. $\bar T^{F\prime}_{n,m}$ is expressed as $\bar T^{F\prime}_{a,l,m,n}$ for uniformity of notation.}
   \label{KeyFormulas2}
\begin{tabular}{c}
\hline
Key Formulas in Arithmetic 
\\
\hline
Conversion from $D$ to $\hat D^F$: 
\\
\vbox{
\begin{equation}
\hat D^F_{a,j,n} = \sum_{m=0}^{\Delta m^{c}_{a,j}-1} h^F_{a,j,m} D_{j,n-m}.
\tag{52}
\end{equation}
}
\\
\hline
Conversion from $\hat D^F$ to $\bar T^{F\prime}$: 
\\
\vbox{
\begin{equation}
\bar T^{F\prime}_{a,l,m,n+1}=
\sum_{m^\prime}
\delta_{m,m^\prime+1}
\left[\bar T^{F\prime}_{a,l,m^\prime,n}
+\sum_j g^F_{a,l,j}\delta_{m,-\bar m_{a,j}^{c-}}\hat D^F_{a,j,n}\right]. 
\tag{(56), (60), (67), and (68)}
\end{equation}
}
\\
\hline
Conversion from $\bar T^{F\prime}$ to $T^F$: 
\\
\vbox{
\begin{equation}
T^F_{i,n}=\sum_a\sum_l\sum_m f_{a,l,i}^F \delta_{\delta m^c_{a,i},m} \bar T^{F\prime}_{a,l,m,n}. 
\tag{(57), (59), and (71)}
\end{equation}
}
\\
\hline
    \end{tabular} 
\end{table*}

\end{document}